\documentclass{aa}  
\usepackage{twoopt}
\usepackage{multicol}
\bibpunct{(}{)}{;}{a}{}{,}             %% natbib format for A&A and ApJ
  \newcommandtwoopt{\citeads}[3][][]{\href{http://adsabs.harvard.edu/abs/#3}%
    {\def\hyper@linkstart##1##2{}%
     \let\hyper@linkend\@empty\citealp[#1][#2]{#3}}}
  \newcommandtwoopt{\citepads}[3][][]{\href{http://adsabs.harvard.edu/abs/#3}%
    {\def\hyper@linkstart##1##2{}%
     \let\hyper@linkend\@empty\citep[#1][#2]{#3}}}
  \newcommandtwoopt{\citetads}[3][][]{\href{http://adsabs.harvard.edu/abs/#3}%
    {\def\hyper@linkstart##1##2{}%
     \let\hyper@linkend\@empty\citet[#1][#2]{#3}}}
  \newcommandtwoopt{\citeyearads}[3][][]%
    {\href{http://adsabs.harvard.edu/abs/#3}
    {\def\hyper@linkstart##1##2{}%
     \let\hyper@linkend\@empty\citeyear[#1][#2]{#3}}}
\makeatother

\usepackage{graphicx}
\usepackage{footnote}
\usepackage{amsmath}
\usepackage{amssymb}
\usepackage{xcolor}
\usepackage{pdflscape}
\usepackage{url}
\usepackage[varg]{txfonts}
\usepackage{bm}		% Bold maths symbols, including upright Greek
\usepackage{booktabs}
\usepackage{txfonts}
\usepackage[pagewise]{lineno}
\newcommand{\gtsim}{\mbox{{\raisebox{-0.4ex}{$\stackrel{>}{{\scriptstyle\sim}}$}
}}}

\DeclareGraphicsExtensions{.pdf,.png,.jpg}
\begin{document} 

   \title{ALMACAL: XIII. Evolution of the CO luminosity function and the molecular gas mass density out to $z \sim 6$}

   % \subtitle{}

   \author{Victoria~Bollo\inst{1\thanks{Email: victoria.bollo@eso.org}},
          C\'eline~P\'eroux\inst{1, 2},
          Martin~Zwaan\inst{1},
          Aleksandra~Hamanowicz\inst{3},
          Jianhang~Chen\inst{4},
          Simon Weng\inst{5, 6},\\
          Claudia~del~P.~Lagos\inst{6, 7, 8},
          % Ian Smail, \inst{10}
          Mat\'ias~Bravo\inst{9}, 
          R.\,J.~Ivison\inst{1,6,10,11}
          \and
          Andrew~Biggs\inst{12}
          }

   \institute{European Southern Observatory, Karl-Schwarzschild-Str. 2, 85748 Garching near Munich, Germany \label{1}
              \and
             Aix Marseille Univ., CNRS, LAM, (Laboratoire d’Astrophysique de Marseille), UMR 7326, F-13388 Marseille, France \label{2}
            \and
            Space Telescope Science Institute, 3700 San Martin Drive, Baltimore, MD 21218, USA \label{3}
            \and
            Max-Planck-Institut für Extraterrestrische Physik (MPE), Giessenbachstrasse 1, D–85748 Garching, Germany \label{4}
            \and
            Sydney Institute for Astronomy, School of Physics A28, University of Sydney, NSW 2006, Australia \label{5}
            \and
            ARC Centre of Excellence for All Sky Astrophysics in 3 Dimensions (ASTRO 3D) \label{6}
            \and
            International Centre for Radio Astronomy Research (ICRAR), M468, University of Western Australia, 35 Stirling Hwy, Crawley, WA 6009, Australia \label{7}
            \and 
            Cosmic Dawn Center (DAWN), Denmark \label{8}
            \and
            Department of Physics \& Astronomy, McMaster University, 1280 Main Street W, Hamilton, ON L8S 4M1, Canada \label{9}
%            Centre for Extragalactic Astronomy, Department of Physics, Durham University, South Road, Durham DH1 3LE, UK \label{10}
            \and 
            Institute for Astronomy, University of Edinburgh, Royal Observatory, Blackford Hill, Edinburgh EH9 3HJ, UK \label{10}
            \and
            School of Cosmic Physics, Dublin Institute for Advanced Studies, 31 Fitzwilliam Place, Dublin D02 XF86, Ireland \label{11}
            \and
            UK Astronomy Technology Centre, Royal Observatory, Blackford Hill, Edinburgh EH9 3HJ, UK \label{12}}

   \date{Received XXX; accepted XXX}

\titlerunning{ALMACAL XIII --- CO LF and molecular gas mass density}
\authorrunning{V.~Bollo et al.}
% \abstract{}{}{}{}{} 
% 5 {} token are mandatory
 
\abstract{Cold molecular gas, largely traced by CO emission, is the primary fuel for star formation, making it essential for understanding galaxy evolution. 
ALMA has made significant progress in the study of the cosmic evolution of cold molecular gas. 
% So far, the results are still limited by the small survey sizes and cosmic variance effects. 
Here, we exploit the ALMACAL survey to address issues relating to small sample sizes and cosmic variance, utilising calibration data from ALMA to compile a statistically significant and essentially unbiased sample of CO-selected galaxies.
By employing a novel statistical approach to emission-line classification using semi-analytical models, we place strong constraints on the CO luminosity function and the cosmic evolution of molecular gas mass density ($\rho_{\text{H}_2}$) back to $z \sim 6$.
The cosmic molecular gas mass density increases with redshift, peaking around $z\sim 1.5$, then slowly declines towards higher redshifts by $\sim 1$\,dex. 
Our findings confirm the key role of molecular gas in fueling star formation.
{The new $\rho_{\text{H}_2}$ estimates allow us to revisit the cosmic baryon cycle, showing that the ratio of molecular gas-to-stellar mass density is consistent with the so-called `bathtub model' of baryons, which implies a continuous replenishment of gas.}
The cosmic gas depletion timescale, estimated on a global scale, is shown to be fairly constant at all redshifts. 
We emphasise the importance of surveys using multiple small fields rather than a single contiguous area to mitigate the effects of cosmic variance.
}
  % % aims heading (mandatory)
  %  {}
  % % methods heading (mandatory)
  %  {}
  % % results heading (mandatory)
  %  {}
  % % conclusions heading (optional), leave it empty if necessary 
  %  {}

   \keywords{molecular gas -- galaxy evolution -- star formation -- ISM -- high-redshift}

   \maketitle
%
%-------------------------------------------------------------------
\section{Introduction} \label{sec:introduction}

% General introduction, context
The baryon cycle is a key driver of galaxy evolution, involving the inflow and outflow of the gas that regulates star formation over cosmic time \citep{perouxCosmicBaryonMetal2020, walterEvolutionBaryonsAssociated2020}.
Galaxies form and grow within the framework of the baryon cycle, where gas is accreted from the inter- and circum-galactic media, primarily in the form of neutral atomic hydrogen (H\,{\sc i}), which eventually cools and condenses into molecular hydrogen (H$_2$), the raw material for star formation \citep{tumlinsonCircumgalacticMedium2017, tacconiEvolutionStarFormingInterstellar2020}.

Understanding the baryon cycle is crucial for interpreting two fundamental cosmic quantities: the cosmic star-formation rate density (SFRD) and the stellar mass function.
The cosmic SFRD, $\psi_{\star}(z)$, tracks the rate at which stars are formed across the Universe as a function of redshift. It peaks around two billion years after the Big Bang ($z\sim 2$), then declines towards the present day \citep{madauCosmicStarFormationHistory2014}. 
The cosmic stellar mass function, $\rho_{\star}(z)$, represents the cumulative mass of stars within galaxies, providing insights into the efficiency of star formation and how galaxies build up their stellar content over time. Together, these two observables are key to understanding the history of star formation in the Universe.

Cold gas -- atomic and molecular -- plays a critical role in the evolution of these cosmic quantities \citep{carilliCoolGasHighRedshift2013, saintongeColdInterstellarMedium2022}. 
Atomic hydrogen (H\,{\sc i}) is the largest, more diffuse gas reservoir in galaxies, which must cool and condense to form molecular clouds from which stars ultimately form. 
At low redshift, H\,{\sc i} is typically observed through its $21-$cm emission (e.g.~\citealt{zwaanHIPASSCatalogueOHI2005, jonesALFALFAMassFunction2018}) while at high redshift it is detected through absorption lines (e.g.~\citealt{perouxEvolutionOHIEpoch2003}). 
These methods provide a direct way to map the distribution of atomic gas and its role in galaxy evolution.
However, molecular gas is a more direct precursor to star formation, which takes place in the dense core of cold gas clouds \citep{kennicuttStarFormationMilky2012, tacconiEvolutionStarFormingInterstellar2020}. 
Because H$_2$ lacks a permanent dipole moment, we generally rely on carbon monoxide (CO) as its tracer \citep{carilliCoolGasHighRedshift2013, bolattoCOtoH2ConversionFactor2013}.{ Observatories such as the Atacama Large Millimeter Array (ALMA), the Northern Extended Millimeter Array (NOEMA) operated by the Institute for Radio Astronomy in the Millimetre Range (IRAM), the Very Large Array, the Atacama Pathfinder EXperiment (APEX), and the Owens Valley Radio Observatory (OVRO), among others,} have enabled the detection of CO across a wide range of redshifts, allowing us to estimate the molecular gas content in galaxies \citep{hodgeHighredshiftStarFormation2020}.
% Overall, studies have shown that the cosmic molecular gas density, $\rho_{\text{H}_2} (z)$, is less or equal to the atomic hydrogen density, $\rho_{\text{HI}}(z)$, throughout cosmic history \citep{perouxCosmicBaryonMetal2020}. At any time in the history of the Universe, there is more mass locked up in atomic gas than in molecular gas.

% Summary of relevant literature
Another key observable is the CO luminosity function (CO LF), which quantifies the distribution of CO luminosities in galaxies. It provides a valuable diagnostic for studying the molecular gas properties and its role in galaxy evolution. 
By constructing the CO LF across different cosmic epochs, previous works have traced the evolution of the molecular gas mass density across different redshifts, $\rho_{\text{H}_2}(z)$. 
For instance, the ALMA Spectroscopic Survey in the Hubble Ultra Deep Field (ASPECS) survey {\citep{decarliALMASPECTROSCOPICSURVEY2016, walterALMASPECTROSCOPICSURVEY2016}} detected CO in galaxies across $z=1$--3 using ALMA bands 3 and 6, covering areas of 4.6 and 2.9~arcmin$^2$, respectively {\citep{decarliALMASpectroscopicSurvey2019, decarliALMASpectroscopicSurvey2020}}.
\cite{boogaardNOEMAMolecularLine2023} combined the results of ASPECS with NOEMA observations to constrain the bright-end of the CO LF up to $z\sim 6$, based on observations of the Hubble Deep Field North (HDFN), spanning $8.5$~arcmin$^2$.
\citet{lenkicPlateauBureHighz2020} explored CO emission lines in 110 main-sequence galaxies using fields from the original PHIBBSS survey \citep{guilloteauIRAMInterferometerPlateau1992}, covering a total area of $\sim130$~arcmin$^2$. 
The COLDz survey detected CO in galaxies at $z= 2$--3 and at $z = 5$--7 using more than 320\,hr of time with the Very Large Array \citep{pavesiCOLuminosityDensity2018, riechersCOLDzShapeCO2019, riechersCOLDzHighSpace2020}, spanning $\sim 60$~arcmin$^2$. 
\cite{audibertCOALMARadioSource2022} exploited the ALMA calibrator catalogue to study the CO LF up to redshift $z \sim 2.5$ and to examine the role of {radio emission} in galaxy evolution.
In addition to CO-based measurements, molecular gas masses have also been inferred from dust emission using scaling relations based on stellar mass or star-formation rates \citep{bertaMolecularGasMass2013, scovilleEvolutionInterstellarMedium2017, magnelliALMASpectroscopicSurvey2020} and from the [C\,{\sc ii}] emission line at higher redshifts {(e.g.~\citealt{aravenaALMAReionizationEra2024, dessauges-zavadskyALPINEALMAIISurvey2020, zanellaIIEmissionMolecular2018}).}
Despite these advancements, {the molecular gas mass density at} several redshift bins in these surveys exhibit significant measurement uncertainties, sometimes exceeding 1~dex, highlighting ongoing challenges in accurately measuring molecular gas content.

% Another challenge stems from the indirect nature of CO observations. CO does not always accurately trace the entire molecular gas content, particularly in low-metallicity environments where ultraviolet (UV) radiation can dissociate CO molecules. In such cases, molecular hydrogen (H$_2$) survives in regions where CO is absent, forming "CO-dark" molecular gas. This discrepancy introduces uncertainties into estimates of the total molecular gas content, affecting the accuracy of CO LF measurements.
% The CO emission line serves as one of the most common molecular gas tracers at low redshift, but still pose some challenges. The CO molecule can be easily dissociated in low metallicity and dust abundance environments \citep{maddenTracingTotalMolecular2020}. 
% At reshifts higher than 5, the effect of the increased cosmic microwave background (CMB) temperature that represents a stronger background against which CO lines are observed -> out of the scope of this paper

% Gap in knowledge 
Accurate measurements of the cosmic molecular gas mass density face two primary obstacles: cosmic variance and Poisson uncertainties. 
Cosmic variance reflects variations in galaxy properties across different regions of the Universe, while Poisson errors arise from the low number of detected galaxies in each luminosity and redshift bin.  
For instance, bright sources -- which are sparsely distributed -- can significantly increase shot noise.
On the other hand, faint sources contribute to clustering noise, where their uneven distribution over large cosmic structures impacts measurements of the matter distribution of the Universe. 

The choice of the observed region can introduce significant uncertainties when measuring cosmic properties, such as the number density of galaxies and LF.
For example, \cite{driverQuantifyingCosmicVariance2010} used the Sloan Digital Sky Survey (SDSS) with galaxies at redshift $z = 0.03$--0.1 to study how the cosmic variance changes with survey volume and field shape, providing a formula to quantify these variations. 
Similarly, \cite{mosterCOSMICVARIANCECOOKBOOK2011} used $N$-body simulations to calculate the excess variance caused by cosmic effects, showing that it significantly exceeds Poisson variance, which only accounts for random sampling.
Both studies concluded that surveys that only probe limited volumes inevitably provide biased estimates of cosmic statistics.
This makes it challenging to derive accurate and representative measurements of molecular gas over cosmic time.

Simulations play a crucial role in overcoming observational limitations by providing theoretical predictions of CO emission and molecular gas distribution in galaxies. Hydrodynamical simulations and semi-analytical models (SAMs) have successfully reproduced a wide range of galaxy properties such as UV-to-FIR LFs of galaxies, the number counts, and the redshift distribution of sub-millimetre (sub-mm) galaxies \citep[SMGs;][]{narayananGeneralModelCO2012,somervillePhysicalModelsGalaxy2015}.

Current cosmological simulations do not directly include a cold gas component due to the complexity of the underlying physics, We therefore rely on post-processing to derive molecular gas properties.
There has been a growing interest in using simulations to model the CO LF and the molecular gas density at several redshifts \citep{lagosSimulationsModellingISM2012, lagosMolecularHydrogenAbundances2015, poppingSubmmEmissionLine2016, valliniCOLineEmission2018, lagosSharkIntroducingOpen2018, poppingArtModellingCO2019, poppingALMASpectroscopicSurvey2019, maioAtomicMolecularGas2022,betherminCONCERTOHighfidelitySimulation2022, bisigelloSPRITZSparklingSimulated2022, garcia$textttslick$ModelingUniverse2023, guoNeutralUniverseMachineEmpiricalModel2023,lagosQuenchingMassiveGalaxies2024}. However, reproducing the bright end of the CO LF remains difficult, and molecular gas density predictions have not yet converged.

Recent studies have explored the impact of cosmic variance on observables like the CO LF using simulated data. For instance, \cite{poppingALMASpectroscopicSurvey2019} compared observational data from the ASPECS survey with predictions from cosmological simulations, including post-processed IllustrisTNG \citep{pillepichFirstResultsIllustrisTNG2018} and the Santa Cruz semi-analytical model \citep{somervilleNatureHighredshiftGalaxies2001}. They found that cosmic variance has a stronger impact at lower redshifts ($z < 1$) since the same area of sky at higher redshifts has a larger volume.
Similarly, \cite{keenanBiasesCosmicVariance2020} used simulations to find that for a survey of $\sim 50$~arcmin$^2$ with CO luminosities of around $10^{10}$ K\,km\,s$^{-1}$~pc$^{2}$, the uncertainties from Poisson noise and cosmic variance become comparable.
% Ignoring cosmic variance in such surveys can lead to an underestimation of total uncertainty by a factor of $\sqrt{2}$.
Also, \cite{gkogkouCONCERTOSimulatingCO2022} found that using only Poisson variance under-estimates the total uncertainty by up to 80\%, particularly for lower luminosity sources and larger survey areas. They reported that at redshifts $z < 3$, cosmic variance can introduce up to 40\% uncertainty in molecular gas density estimates in small surveys (e.g.~4.6\,arcmin$^2$, similar to ASPECS), but this drops to below 10\% for surveys covering more than 43.2~arcmin$^2$.

Observations have begun to acknowledge cosmic variance as a key factor in explaining discrepancies between molecular gas density estimates across different surveys \citep{lenkicPlateauBureHighz2020, boogaardNOEMAMolecularLine2023}. This marks a shift from previous studies which often neglected to include cosmic variance in uncertainty estimates, resulting in optimistic error estimates.

% This paper: 
This paper aims to alleviate the limitation introduced by cosmic variance when estimating the molecular gas mass density by exploiting ALMA calibrator data, as part of the ALMACAL$-22$ survey \citep{zwaanALMACALSurveyingUniverse2022}.
We make use of the sample selected and described in a previous paper of this series \citep{bolloALMACALXIIData2024a}, consisting of the highest-quality and deepest data compiled from 2016 to 2022.  
This study builds upon the methodology and results of the previous ALMACAL-CO survey \citep{hamanowiczALMACALVIIIPilot2022}. The proof-of-concept pilot of this earlier work successfully demonstrated the feasibility of using ALMA calibration data to characterise CO-selected galaxies and estimate their molecular gas content. We use the statistical approaches and classification techniques introduced in that study in this extended analysis. With the larger dataset now available, we can refine our results and place stronger constraints on the CO LF and the evolution of the molecular gas mass density across cosmic time, fully aligning with the expectations set by the previous work.

% Organization
This paper is organised as follows. 
In Section~\ref{sec:almacal}, we describe the ALMACAL$-22$ survey and summarise the calibration process, the selection of the sample, and the imaging. 
Section~\ref{sec:line_search} details the steps taken to search for serendipitous detections of CO emission lines, including the source finding algorithm used (\S\,\ref{sec:sofia}), the completeness correction (\S\,\ref{sec:completeness}), the estimation of the fidelity (\S\,\ref{sec:reliability}) and the redshift (\S\,\ref{sec:redshift}) and the determination of the final catalogue (\S\,\ref{sec:final-catalog}). 
Section~\ref{sec:co-lum-func} presents the volume estimates (\S\,\ref{sec:volume}), the CO LF (\S\,\ref{sec:co-lf}), the Schechter fits (\S\,\ref{sec:schechter-fits}), the molecular gas mass density (\S\,\ref{sec:h2-density}), the results of adopting the lowest CO transition possible (\S\,\ref{sec:discussion_low_j}), and the uncertainties of our results (\S\,\ref{sec:discussion_biases}).
Section~\ref{sec:discussion} compares our findings of the CO LF (\S\,\ref{sec:discussion_colf}) and the molecular gas mass density (\S\,\ref{sec:discussion_h2}) with observations and simulations, discusses the effect of cosmic variance (\S\,\ref{sec:discussion:cosmic_variance}), and provides a complete census of the baryon cycle (\S\,\ref{sec:discussion_baryons}).
In Section~\ref{sec:conclusions}, we summarise our key conclusions. 
Throughout this paper, we use $H_0 = 70$ kms$^{-1}$Mpc$^{-1}$, $\Omega_{\text{M}} =0.3$ and $\Omega_{\Lambda} = 0.7$. The present-day cosmological critical density is $\rho_{0, \text{crit}} \simeq 277.4$ $h^2$ M$_{\odot}$ kpc$^{-3}$.

%-----------------------------
% ----------------------------
% ----------------------------
\section{ALMACAL$-22$ and sample selection} \label{sec:almacal}

% Introduction to ALMACAL
ALMACAL-22 exploits the archival ALMA calibrator sources and their fields observed since 2016 to produce scientific outcomes \citep{zwaanALMACALSurveyingUniverse2022}.
ALMA calibrators are typically bright quasars, observed for a few minutes in every ALMA PI-led observing project.
Each calibration scan has a set up that matches the PI requirements, so the full ALMACAL survey has a diverse range of configurations in terms of spatial and spectral resolution, sensitivity and integration time.
Most of the calibrators in ALMACAL are bright sub-mm point sources classified as blazars (\citet{bonatoALMACALIVCatalogue2018a}, Weng et al., in preparation).

This work uses the latest data release, which comprises the ALMA calibrator data taken up until 2022 May, so-called ALMACAL$-22$ \citep{bolloALMACALXIIData2024a}.
The full ALMACAL survey comprises more than 30\,TB of calibrator data from ALMA Cycle 1 to Cycle 10, over more than 1000 fields. 
\cite{bolloALMACALXIIData2024a} defined a pruned sample to represent the highest quality data, where each entry exceeds ten minutes of accumulated integration time. 
We focus the analysis on this pruned sample, which includes only observations made with the 12-metre array and files with low root-mean-square noise values (RMS $<0.01$ mJy per channel).
We summarise the main calibration steps, sample selection and data processing below. 

% Calibration 
A dedicated ALMACAL pipeline \citep{oteoALMACALExploitingALMA2016} automates the processing of calibrator data and produces images of delivered ALMA datasets.
% using the \texttt{scriptForPI.py} to generate fully calibrated data. 
The pipeline calibrates using \texttt{scriptForPI.py}, extracts calibrator data, applies self-calibration to correct phase and amplitude variability, and removes the bright calibrator. 
A point source model is applied during self-calibration, resulting in calibrated visibilities and the creation of data cubes for the ALMACAL$-22$ dataset.

% Selection/ pruning
Due to the uneven nature of the ALMACAL data, a pruning procedure was applied to select the highest-quality data.
The pruning steps are based on key properties like integration time, frequency coverage, spatial resolution and RMS noise.
The on-source integration of observations covering the same frequencies was set to accumulate more than 10 minutes to reach a meaningful sensitivity level.
Only observations with noise levels consistent with theoretical sensitivity were selected. 
% Files with high RMS noise or artefacts were excluded. 
From an initial 34909 measurement sets (ms), only the $\sim 20$\% with the highest quality remains in the pruned sample.
{The total number of data cubes and calibrators in the pruned sample is listed in \citealt{bolloALMACALXIIData2024a}, Table 1.}
{In addition to the pruning, there is potentially a bias related to clustering effects around the calibrators, as these objects are often located in massive haloes and associated with large overdensities. The pruning fields and data in this work explicitly address this concern by removing CO lines coinciding with the redshifts of the calibrators. A detailed discussion is provided in Section \S \ref{sec:reshift-prob}.} 

% Processing
Data cubes -- consisting of two spatial and one spectral dimension -- were created by combining multiple observations. 
The \textit{uv} observations were concatenated into a single file to re-calibrate the uv weights using CASA's \texttt{statwt} \citep{casateamCASACommonAstronomy2022}.
The choices made to optimise the imaging process can be summarised as follows.
The beam size samples 3 pixels, and the image size is circular, with a ratio $1.8\times$ that of the primary beam. 
The channel width is 31.2 MHz, a value that balances spectral resolution, SNR and data volume.
Imaging is performed with CASA's \texttt{tclean}, using linear interpolation, natural weighting, and 0.5$''$ uv tapering.
{We took special care examining the shape of the synthesised beam when combining data from different arrays. By comparing the PSF shape to a 2D Gaussian fit, we found minimal deviations. As a result, the contributions to the residual map are negligible, ensuring that the flux measurements remain accurate and free from significant beam distortions.}
The pruned sample contains 1508 data cubes covering 401 different calibrator fields. Most of the cubes are in ALMA bands 3 and 6, with a maximum integration time of $\sim 7$ hours and a mean sensitivity of $\sim 0.78$ mJy per channel (see \citealt{bolloALMACALXIIData2024a} for further details).

% We carefully analysed the synthesized beam to assess any potential non-Gaussian features introduced by the combination of arrays. Cross-sections of the PSF along the vertical and horizontal axes closely matched a Gaussian profile with symmetric and minimal sidelobes. Additionally, residuals between the PSF and a 2D Gaussian fit were negligible, confirming that the beam shape remains Gaussian-like.
% These analyses indicate that our approach does not introduce significant non-Gaussian artefacts or inaccuracies in flux measurements, mitigating concerns such as the JvM effect \citep{jorsaterHighResolutionNeutral1995}.

\begin{table}[htp]
	\centering
	\caption{Properties of the CO transitions in the ALMACAL$-22$ survey}
	\label{tab:volume}
	\begin{tabular}{lccccc} % 
		\hline 
		\rule{0pt}{2.5ex}Transition & $\nu_{\text{rest}}$ & Redshift  & Volume & N$_{\text{ind}}$ & CV \\ [0.1cm]
        & [GHz] &  & $\log$[cMpc]$^3$ &  &$\%$ \\ [0.1cm]
		\hline
		\rule{0pt}{2.5ex}CO($1-0$) & 115.27 &$0 - 0.37$ & 4.85  &  173 & 4.85 \\ [0.1cm]
		CO($2-1$) & 230.54 &$0 - 1.70$ & 6.07  & 240 &  1.75 \\ [0.1cm]
		CO($3-2$) & 345.54 & $0 - 3.15$ & 6.30  & 259 & 1.51 \\ [0.1cm]
        CO($4-3$) & 461.04 & $0 - 4.40$ & 6.39  & 274 & 1.20 \\ [0.1cm]
        CO($5-4$) & 576.27& $0 - 5.80$ & 6.44  & 287 & 1.07 \\[0.1cm]
        CO($6-5$) & 661.47& $0 - 7.10$ & 6.47 & 284 & 1.40 \\ [0.1cm]
		\hline 
        \rule{0pt}{2.5ex}Total & & & 7.06 & 299 &  \\ [0.1cm]
        \hline 
	\end{tabular}
  \tablefoot{N$_{\text{ind}}$ is the number of independent sightlines in which a given transition can be detected. CV is the percentage of cosmic variance estimated using Eq.~\ref{eq:cosmic-variance} (see \S\,\ref{sec:discussion:cosmic_variance} for more details).}
\end{table}
% N - number of independent pointings (fields)

%-----------------------------
% ----------------------------
% ----------------------------
\section{Line search} \label{sec:line_search}

Detecting emission lines in more than $10^3$ cubes with significant noise variations is challenging. To address this, we use the source-finding algorithm SoFiA2 \citep{westmeierSOFIAAutomatedParallel2021}, to identify potential candidates. We assess these candidates' reliability based on the S/N and line width detected, and we estimate the completeness factor of the entire line search process.
In this section, we describe the identification of CO candidates (\S\,\ref{sec:sofia}), the estimation of the completeness (\S\,\ref{sec:completeness}), reliability (\S\,\ref{sec:reliability}), the determination of redshifts (\S\,\ref{sec:redshift}), and the final compilation of a catalogue of candidates (\S\,\ref{sec:final-catalog}).

\subsection{Source searching algorithm} \label{sec:sofia}

SoFiA2 is a fully automated pipeline designed to find 3D sources in the WALLABY survey conducted on the Australian SKA Pathfinder (ASKAP). Based on the open access source finder SoFiA 1 \citep{serraSOFIAFlexibleSource2015}, SoFiA2 has been re-written in C with multi-threading for speed, being considerably faster and more efficient than the previous version. It allows for searching on the spectral and spatial axes, offering two different algorithms: a simple threshold finder and the smooth and clip (S+C) finder. The threshold finder applies a flux threshold to the data, which is useful mainly for data adjusted for noise variations. The default S+C finder, described in \citealt{serraATLAS3DProjectXIII2012}, iteratively smooths the data cube across various spatial and spectral scales to detect significant emission signals above a set threshold. 

% Smoothing
We input the processed ALMA data cubes after cleaning, calibrating, and based on multiple observations combined in uv space. 
We perform the search on data cubes that remain uncorrected for the shape of the primary beam response function.
Before the primary beam correction, the data preserves its original sensitivity pattern to apply a uniform detection threshold to the emission line search.
We flag spectral channels where the noise deviates from the median by more than $5\sigma$.
In SoFiA2, we apply spatial and spectral smoothing based on the median absolute deviation (MAD) with a $3\sigma$ threshold. 
This helps to automatically flag corrupted data, like channels with radio-frequency interference or pixels with residual continuum emission.
SoFiA2 measures and adjusts for the local noise level using a running window of 25 pixels and 15 channels in the spatial and spectral domain.

% Finding algorithm
We chose the smooth and clip (S+C) algorithm to identify sources, while smoothing to suppress noise and outliers. 
The algorithm iteratively rejects data points that deviate beyond the $3 \sigma$ threshold, so extreme values do not affect the smoothing process. 
Then, the `linker' combines detected pixels in the binary mask into coherent detections using a friends-of-friends algorithm. It links all pixels within a merging radius of three, treating sources as three-dimensional collections of pixels. 
After the source identification, SoFiA2 delivers the properties of the sources, such as their position, and size. 
Three co-authors (VB, CP, MZ) visually inspected the spectra and moment maps of the candidates to ensure that they were not confused with noise peaks. The final catalogue contains 87 emission lines (see \S\,\ref{sec:final-catalog} for further details about the selection).

\begin{figure}[htb]
    \includegraphics[width=1.\columnwidth]{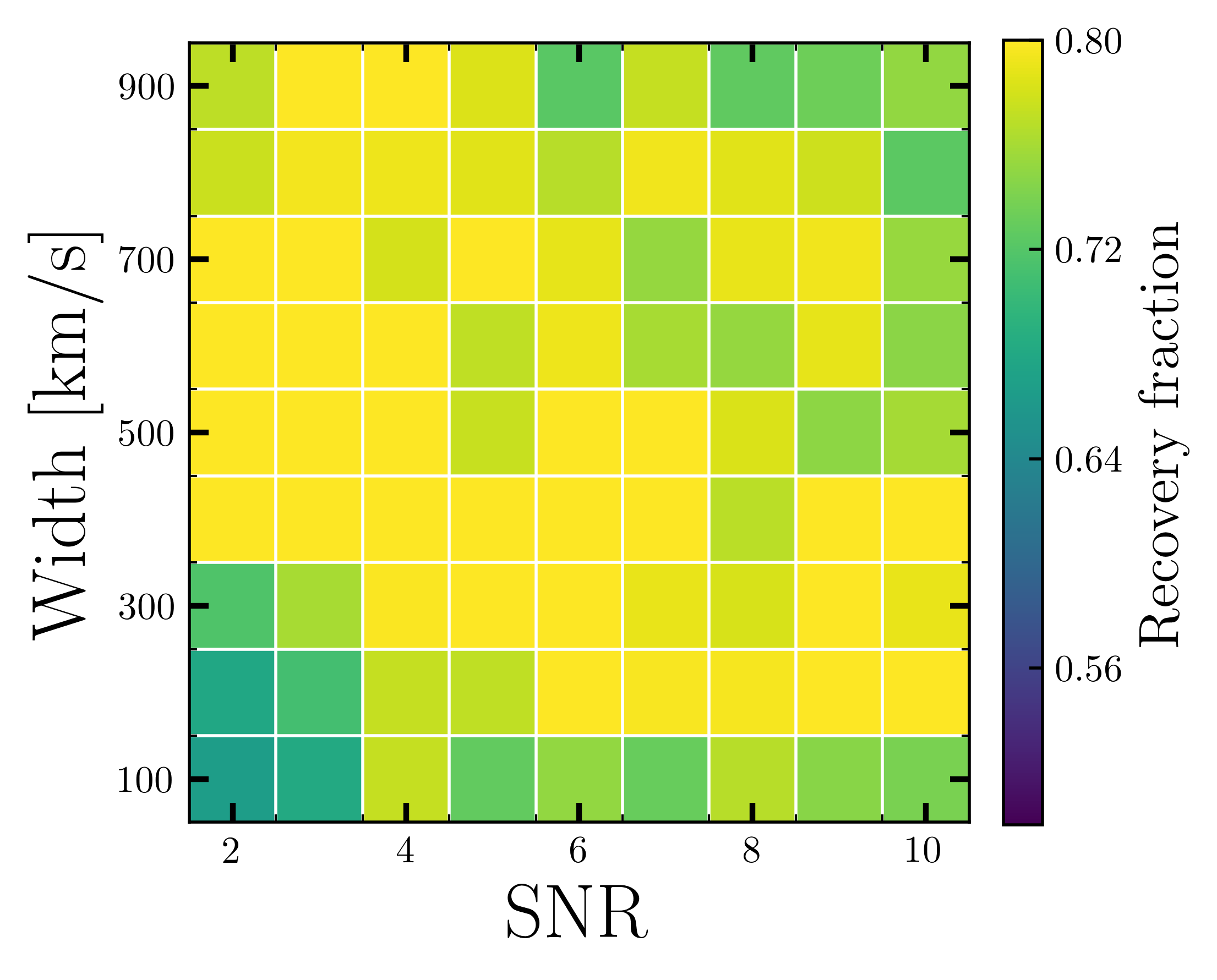}
    \caption{Completeness fraction of mock sources. The grid shows several combinations of the injected peak signal-to-noise ratio (SNR) and line width.  
    The heat map represents how successfully the algorithm, SoFiA2 \citep{westmeierSOFIAAutomatedParallel2021}, detects mock sources injected into data cubes, with the recovery fraction indicated by the colour. 
    For sources with SNR $\gtsim 4$ and line width $\gtsim 300$ km\,s$^{-1}$ we reach a completeness above $\sim 75$\%.
    }
    \label{fig:completeness}
\end{figure}

\subsection{Completeness} \label{sec:completeness}

We perform a completeness test to assess the quality of the cubes and the efficiency of the source-finding algorithm in retrieving artificial sources.
To estimate the completeness factor, we inject mock sources into each data cube.  
We add artificial sources to each data cube and run the source finder with the same parameters used to look for real emissions to quantify the recovery fraction. Iterating this process helps us to determine the quality of the cubes and algorithm's robustness under different conditions.

A set of mock sources is generated with known flux densities, positions and sizes.
The S/N values of the synthetic sources range from 2 to 18, and the line widths vary between 100 and 1600 km\,s$^{-1}$. These ranges were chosen to explore the parameter space broadly.
In each cube, 20 mock sources are randomly injected with varying properties. 
We run SoFiA2 on the data cube containing both the real and injected sources.
We estimate the percentage of recovered sources by comparing the position of detected sources with the injected mock sources.

To determine whether SoFiA successfully recovers the injected sources and records both true positives (correctly identified sources) and false negatives (missing sources).
The process is repeated ten times per cube to build up statistically significant samples, with each iteration using different mock sources to simulate varying conditions.

We analyse the detection rate, or completeness factor ($c_i$), based on the SNR and line width of the mock sources across two categories: high-completeness and low-completeness cubes. 
Some cubes consistently showed a recovery fraction of zero across all iterations. Further investigation revealed that these cubes were either extremely narrow (significantly less than 4 GHz) or exhibited structured noise. Since real detections in these cubes would likely be missed {and their missing data cannot be corrected}, we excluded them from the sample. {However, cubes with low but non-zero completeness are kept, as their incomplete information can be accounted for using the completeness factor, and they still contribute to the statistical inference in the analysis.}
Ultimately, we kept 1107 cubes, removing 401 cubes with zero completeness.

Fig.~\ref{fig:completeness} displays the completeness grid for our sample, indicating the fraction of successfully recovered injected sources as a function of signal-to-noise ratio (SNR) and line width within feasible detection limits. We find that synthetic sources with SNR $\gtrsim 4$ and line widths of $\gtrsim 300$ km\,s$^{-1}$ are successfully recovered in over 70\% of cases. Although the recovery fraction becomes roughly homogeneous after these values, there is a lack of monotonic improvement towards the top right corner.
% Low SNR lines with relatively large line widths have high completeness mainly because they fall likely in wide spectral cubes. 
To investigate why the highest bin in the grid does not reach full completeness, we examined several mock sources with high SNRs and broad velocity components. In most cases, these sources fell into narrow spectral cubes, where the continuum level is insufficient for comparison, leading the algorithm to miss them. 
% Overall, SNR remains the dominant factor in line recovery success.

\subsection{Fidelity} \label{sec:reliability}

We estimate the reliability of emission lines based on the significance of the detections compared to the noise distribution in the data cubes.
We invert the cubes and run the source finder (SoFiA2) with the same parameters, again searching for emission lines (which are, of course, absorption features in the physical cube).
As we do not expect absorption features, any emission lines in the inverted cubes must represent random noise peaks.

We examine the spatial and spectral extent of the identified sources, estimating their SNR and line width. 
This distribution of noise peaks can be defined through the fidelity coefficient, given by:
\begin{equation} \label{eq:fidelity}
    F(\text{S/N}, \sigma) = 1 - \frac{N_{\text{negative}}(\text{S/N}, \sigma)}{N_{\text{positive}}(\text{S/N}, \sigma)},
\end{equation}
\noindent
where $N_{\text{negative}}$ and $N_{\text{positive}}$ are the number of negative and positive detections for a certain combination of S/N and line width.

\begin{figure}[htb]
    \includegraphics[width=1.\columnwidth]{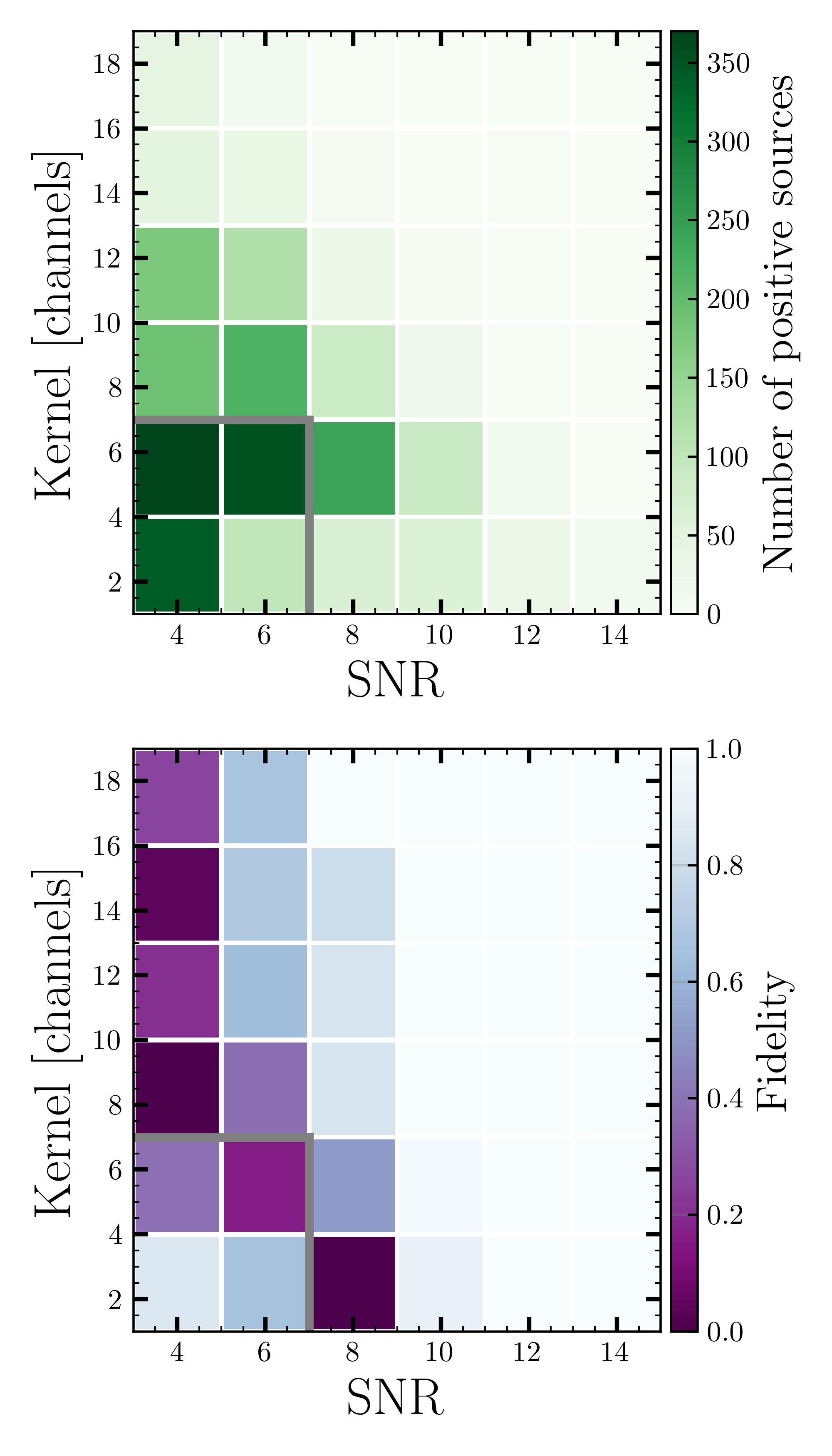}
    \caption{\textit{Top}: 2D histogram of the number of positive sources detected in all the data cubes when searching for emission lines using SoFiA2. The x-axis represents the S/N of the line and the y-axis shows the number of channels spanned by each detection. \textit{Bottom}: Fidelity grid derived from Eq.~\ref{eq:fidelity} considering the number of positive and negative sources found for a given combination of S/N and number of channels. Values close to unity represent the highest fidelity, while zero represents the least reliable parameters.  Lines with SNR above seven and spanning about seven channels usually have high fidelity values, i.e.\ beyond the grey lines of both panels.}
    \label{fig:reliabillity}
\end{figure}

Fig.~\ref{fig:reliabillity}, top panel, shows the 2D histogram of the positive candidates. The $x-$axis represents SNR, while the $y-$axis is the number of detection channels. Note that over 400 positive detections are initially found, but not all are real (see \S\,\ref{sec:final-catalog}). The bottom panel of Fig.~\ref{fig:reliabillity} shows the fidelity grid, built as indicated by Eq.~\ref{eq:fidelity}. 
{There are cases where a high fidelity value is found for low SNR and small kernel widths (bottom left corner). This may result from a limitation of the source-finding algorithm, which can favour narrow lines and lead to poorly constrained statistics in that range. However, these values do not affect our estimates, since the majority of our detections lie well above this region.}
Emission line candidates spanning at least seven channels with S/N $> 7 $ reach the highest fidelity values, as indicated beyond the grey lines in the top and bottom panels of Fig.~\ref{fig:reliabillity}. 
Most of the high-fidelity sources are outside the grey region.

\begin{figure}[htb]
    \includegraphics[width=.9\columnwidth]{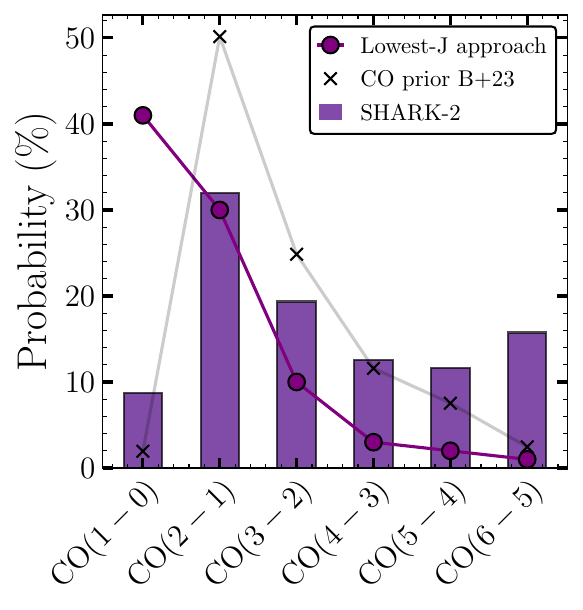}
    \caption{Probability distribution of all the CO detections in our sample. Given the measured properties of each emission line, we estimated redshifts using a probabilistic approach, based on the semi-analytical model, \texttt{SHARK-2} \citep{lagosQuenchingMassiveGalaxies2024} (see \S\,\ref{sec:reshift-prob}). Overall, the CO(2--1) transition is the most likely to be detected in our sample, while the CO(1--0) transition is the least likely. This probability is carried into the further analysis. For comparison, we also include the CO prior redshift distribution association of sources with {\it HST} counterparts reported in \cite{boogaardNOEMAMolecularLine2023}, written as B$+23$, {and the probability distribution obtained adopting the lowest-$J$ approach explained in Section \S \ref{sec:discussion_low_j}}.
    }
    \label{fig:redshift_estimation}
\end{figure}

\subsection{Redshift estimation} \label{sec:redshift}

In ALMACAL$-22$, identifying emission line candidates to determine redshifts  poses significant challenges. Some calibrator fields are observed multiple times using different ALMA bands, but the archival calibrator data provides uneven spectral coverage across different fields. 
% In a few fields, we have sufficient coverage to detect multiple CO transitions, allowing us to search for a second line to confirm the redshift. 
% More than 1 line detected 
When an emission line was detected in fields with coverage in multiple ALMA bands, the likelihood of covering the spectral range for a second CO emission line increases. By evaluating the probability of each detected line being a specific CO transition, we checked if the spectral coverage includes another transition in any of our data cubes. This analysis revealed a total of 37 fields where a second transition could have been potentially covered.
However, in most cases (29 of those 37 fields), potential detections fell outside the field of view due to the shrinking primary beam in higher ALMA bands. For the other eight sources, the sensitivity is too low to detect fainter transitions. Ultimately, we did not identify any potential transitions in these fields, so we are reliant on detections of one emission line.

We took a two-step approach to estimate the redshift. 
First, we employed a probabilistic method to estimate the redshift (\S\,\ref{sec:reshift-prob}). 
Second, we utilised known spectroscopic or photometric information (\S\,\ref{sec:redshift-info}).

\subsubsection{Probabilistic approach} \label{sec:reshift-prob}

We created a redshift probability distribution using the semi-analytical model, \texttt{SHARK-2} \citep{lagosQuenchingMassiveGalaxies2024}, in the same way as presented in \citealt{hamanowiczALMACALVIIIPilot2022} with a previous version, \texttt{SHARK-1} \citep{lagosSharkIntroducingOpen2018}. \texttt{SHARK-2} includes improvements in physical models tracking the properties of supermassive black holes and active galactic nuclei feedback, and the environmental effects of satellite galaxies.

Based on the galaxy population of the SAM, we created a 2D histogram that maps CO transition flux against redshift. 
We then calculate a probability coefficient for each combination of flux and redshift by dividing the number of objects in each bin by the total number of galaxies in the simulated sample. 
This approach allowed us to determine the likelihood of a given CO flux originating at a specific redshift.
High-level transitions ($J_{\text{up}} = 7, 8, 9, 10$) are rare since very high kinetic temperatures and densities are needed to excite the CO molecule sufficiently to populate these higher rotational levels. 
The most likely detections are transitions $J_{\text{up}} = 1, 2, 3, 4, 5$ and 6, placing our candidates between redshifts 0 and 6.
Instead of simply selecting the $J-$transition with the highest probability for each detection, we adopted a more comprehensive approach. We included all plausible $J$ transitions, each weighted according to the probabilities evaluated by \texttt{SHARK-2}. This method allows us to account for the uncertainty in identifying the exact $J$ transition.

Fig.~\ref{fig:redshift_estimation} shows the probability distribution of all the CO candidates in our sample. \texttt{SHARK-2} assigns the CO(2--1) transition most frequently in our catalogue, and CO(1--0) as the least frequent. 
This distribution closely matches the shape of the CO line prior used for redshift associations in sources without counterparts seen in \cite{boogaardNOEMAMolecularLine2023}. 
Two main factors influence the shape of this distribution.
First, the CO molecule emits radiation at different frequencies depending on its rotational energy level ($J$ level). 
Some transitions, like CO(2--1), are naturally more common, or stronger, while others are rarer or weaker. 
Second, the volume probed by our observations for each transition, especially CO(1--0), also plays a role, as lower-frequency transitions probe smaller volumes.
%The varying likelihood of detecting specific CO transitions predicted by \texttt{SHARK-2}, rather than assuming all transitions are equally probable, demonstrates its accuracy in weighing the probabilities of identifying different CO lines.

\begin{figure}[htb]
    \includegraphics[width=1\columnwidth]{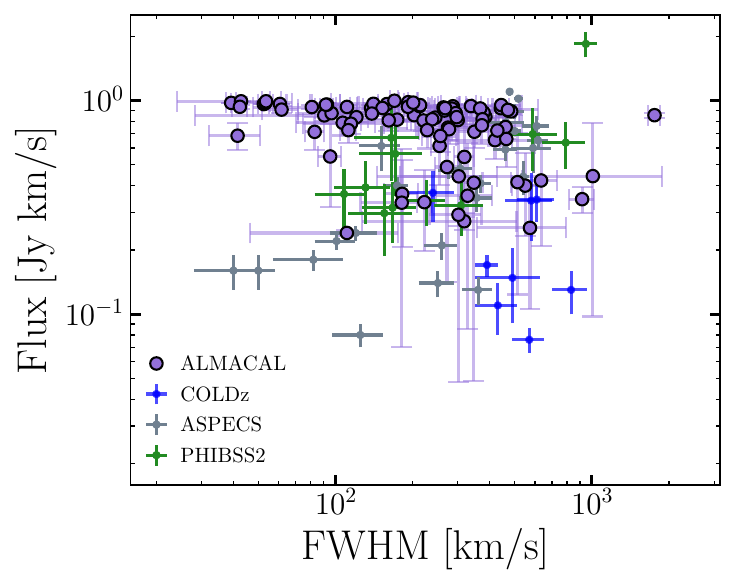}
    \caption{Integrated flux and width of the CO emitters detected in ALMACAL$-22$ (purple). For comparison, we include detections from other surveys: ASPECS \citep{decarliALMASpectroscopicSurvey2020}, COLDz \citep{riechersCOLDzHighSpace2020} and PHIBSS \citep{lenkicPlateauBureHighz2020}. Our candidates span a similar range of line widths as previous studies, and are generally brighter, as expected from the difference in depth and volume.}
    \label{fig:fluxes}
\end{figure}

\subsubsection{Existing spectroscopic and photometric data} \label{sec:redshift-info}

% Mention the optical counterpart
We inspect NED, as well as MUSE and {\it HST} archives, to find the candidates' optical/NIR counterparts. This search yields seven MUSE cubes.
From the MUSE cubes, we confirm the spectroscopic redshift of three objects, all of them at $z<1$. 
We fix the known spectroscopic redshift for three sources and set the fidelity parameter to one.
The other objects are not detected in any of the possible emission lines that MUSE could have covered. However, MUSE spectra do not cover any emission lines of sources at $1.4\lesssim z\lesssim 2.8$. Another reason for not detecting these sources could be heavy dust obscuration, making a galaxy faint in the rest-frame UV range. 

Photometric data are available for seven sources. 
We performed SED fitting with \texttt{EAZY} \citep{brammerEAZYFastPublic2008} to estimate their photometric redshifts (photo$-z$). In three cases, the photo$-z$ values match one of the highest probability redshifts given by \texttt{SHARK-2}; we adopt the photometric redshift for these sources. For the remaining objects, there are fewer than five photometric bands available, resulting in broad photo$-z$ probability distributions that cannot reliably estimate the redshift or assign a redshift probability greater than $10\%$. For these cases, we keep the redshift probability described earlier (\S\,\ref{sec:reshift-prob}). 

\subsection{Final catalogue} \label{sec:final-catalog}

The final catalogue was created after a thorough visual inspection of all detections, along with an analysis of the spectra and the integrated emission-line map across the spectral range of the detected lines.
We select candidates with S/N $> 4$ and a clear peak in the moment map compared to the background noise. Duplicate detections -- e.g.\ those offset by only a few pixels but originating from the same region -- were excluded. The final catalogue of CO emitters includes 87 sources.

We extract the spectra using the aperture from SoFiA's output catalogue. 
To ensure the accuracy of the number of channels spanned by the emission line, we re-calculated the S/N estimate by dividing the peak flux by the RMS of the moment map. 
The moment map was created by integrating across the frequency range of the emission and collapsing it into a single image.
To measure the emission line flux, we used the function \texttt{scipy.optimize.curve\_fit} to fit a Gaussian profile to the emission lines. 
We repeated this procedure 100 times for each candidate, perturbing the flux using the RMS of the cube. 
From the best fit, we use the mean values of the standard deviation and amplitude to estimate the flux and the full-width-half maximum (FWHM). Our final catalogue of CO sources contains the flux and FWHM with their uncertainties, as well as the completeness and fidelity factors estimated from the S/N and line width of each object, reported in Table~\ref{tab:detections} of the Appendix. {Generally, we see that in fields where a lower rms noise level is reached, more detections are found. However, in some cases, fields with lower sensitivity still exhibit multiple detections, which can be attributed to the stochastic nature of source distribution within the field of view.} {Also, there are a few sources with a FWHM $\gtrsim 1000$ km/s, such as J0334-4008.4, most likely explained by blended sources for which the achieved resolution is insufficient to resolve them.}

Fig.~\ref{fig:fluxes} shows the measured CO fluxes against the FWHM of each emission-line candidate. 
The FWHM values measured in ALMACAL$-22$ are comparable with previous surveys, while the fluxes are generally higher. 
This trend is expected due to the differences in the sensitivity and volume probed by different surveys. 
The ALMACAL$-22$ survey benefits from a larger observational volume, but it has an uneven sensitivity distribution among data cubes. As a result, we can identify and measure the flux of more luminous sources, while detecting fainter systems is more challenging. 

As an additional verification of the final catalogue, we estimate the probability of detecting high-redshift interlopers with bright [C\,{\sc ii}] emission, at $1900.53$ GHz, instead of a CO transition. The [C\,{\sc ii}] emission line can be detected in one of the ALMA bands for redshifts $z>4$, where it enters ALMA band {7}. 
{We estimate the volume covered by the [C\,{\sc ii}] line in Band 7 between redshifts $z = 4$ and $z=6$, to be 5858.16 cMpc$^3$, using the same methodology employed to estimate the probed volume for CO transitions (see \S\,\ref{sec:volume}). We calculated the expected number of sources within this volume based on the [C\,{\sc ii}] LF in this redshift range (\citealt{yanALPINEALMAIISurvey2020, casavecchiaCOLDSIMPredictionsII2024}). We found that a maximum of two sources can be expected in a survey covering the volume probed by the [C\,{\sc ii}] line. 
We further discuss the effect of removing two Band 7 detections in \S\,\ref{sec:discussion_biases}.}

% Redshift of the calibrator compared with other lines, move this to another section.
To confirm that our detections are independent of the calibrator, we verify that the redshifts of our sources differ from those of the calibrator. The calibrator redshift catalogue ($z_{\text{cal}}$) in ALMACAL$-22$ is based on a database from \cite{bonatoALMACALIVCatalogue2018a}, which is extended through cross-matching with optical catalogues (NED, SIMBAD and \citealt{mahonyOpticalPropertiesHighfrequency2011}) and supplemented by 70 VLT/X-Shooter spectra {(ID 111.253L.001, PI: S. Weng and ID 0101.A-0528, PI: E. Mahony)}. Our sample includes 87 CO emitters spread across 46 quasar fields, with redshift information available for 37 fields (Weng et al., in preparation).
For each emission line, we compare the calibrator’s redshift against all probable redshifts. Additionally, we verify that the estimated photometric redshift (\S\,\ref{sec:redshift-info}) is distinct from that of the calibrator. Four candidates were excluded from our sample due to probable redshifts within 2000 km\,s$^{-1}$ of the quasar redshift.
Thus, we assert that our candidates are independent of the calibrator. Despite using data centred on calibrators, this approach does not introduce bias into our survey.

\begin{table}[htb]
	\centering
	\caption{Schechter best-fit parameters of the CO LF}
	\label{tab:schechter_pars}
	\begin{tabular}{ccc} % 
		\hline \hline
		\rule{0pt}{2.5ex}Redshift & $\log_{10}\Phi^*$ & $\log_{10}L^{\prime *}$ \\ [0.1cm]
        & [Mpc$^{-3}$ dex$^{-1}$] &  [K km s${-1}$ pc$^{-2}$] \\
		\hline  
		\rule{0pt}{2.5ex}$0.0 - 0.5$ &  $-1.51^{+0.32}_{-0.37}$ &${9.31}^{+0.18}_{-0.09}$  \\ [0.2cm]
		$0.5 - 1.0$ & ${-2.48}^{+0.51}_{-0.31}$ & ${10.12}^{+0.17}_{-0.51}$  \\ [0.2cm]
		$1.0 - 2.0$ & ${-3.32}^{+0.50}_{-0.17}$ & ${10.58}^{+0.26}_{-0.18}$  \\ [0.2cm]
        $2.0 - 3.0$ & ${-3.61}^{+0.59}_{-0.08}$ & ${10.59}^{+0.30}_{-0.17}$  \\ [0.2cm]
        $3.0 - 4.0$ & ${-4.10}^{+0.91}_{-0.29}$ & ${10.62}^{+0.50}_{-0.18}$ \\ [0.2cm]
        $4.0 - 6.0$ &  ${-3.54}^{+0.40}_{-0.16}$& ${10.16}^{+0.33}_{-0.17}$ \\ [0.2cm]
        \hline
	\end{tabular}
 \tablefoot{Fixed $\alpha=-0.2$}
\end{table}

\begin{figure*}[htb]
    \centering
    \raisebox{0.0cm}{\includegraphics[width=0.66\columnwidth]{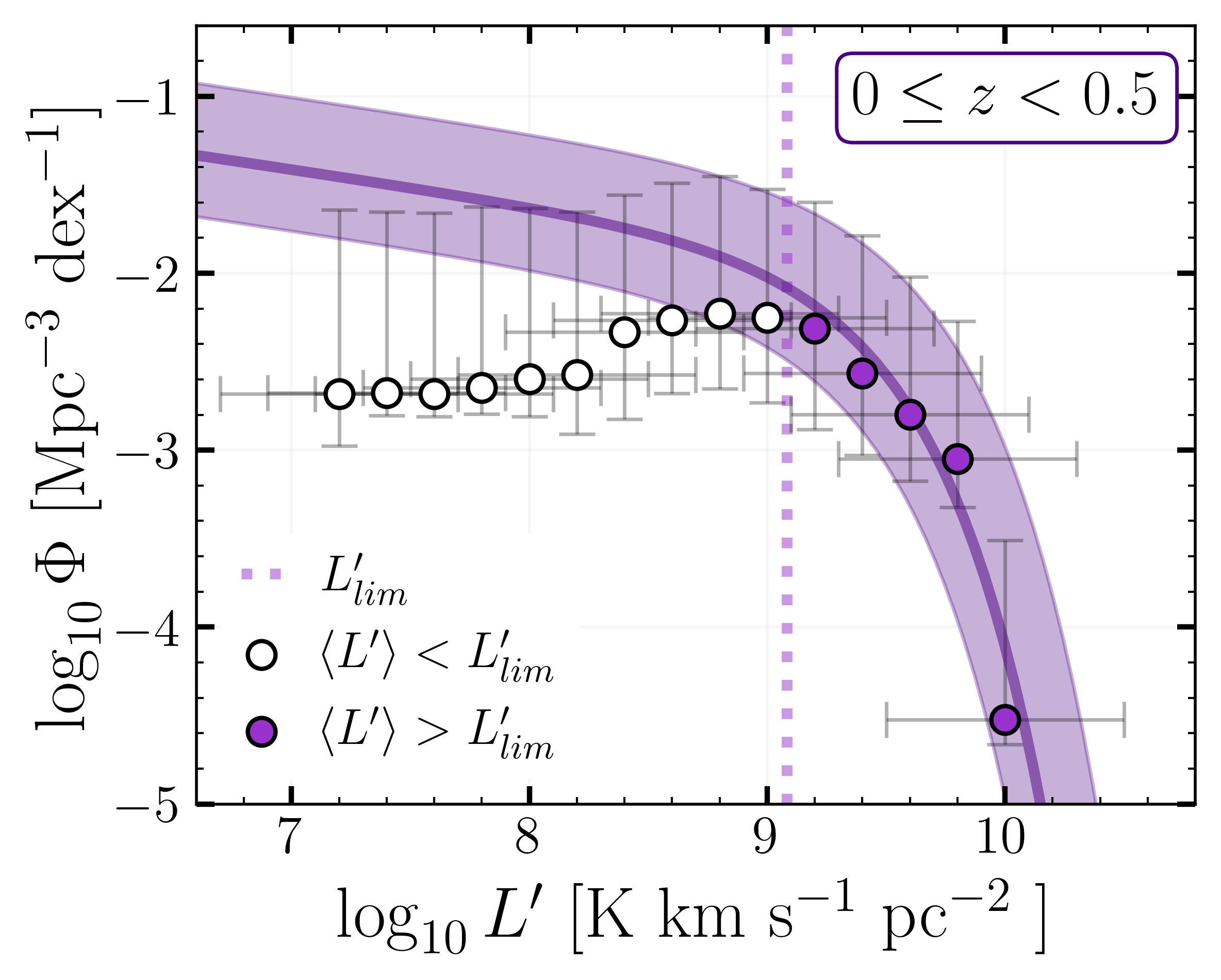}}
    \raisebox{0.0cm}{\includegraphics[width=0.66\columnwidth]{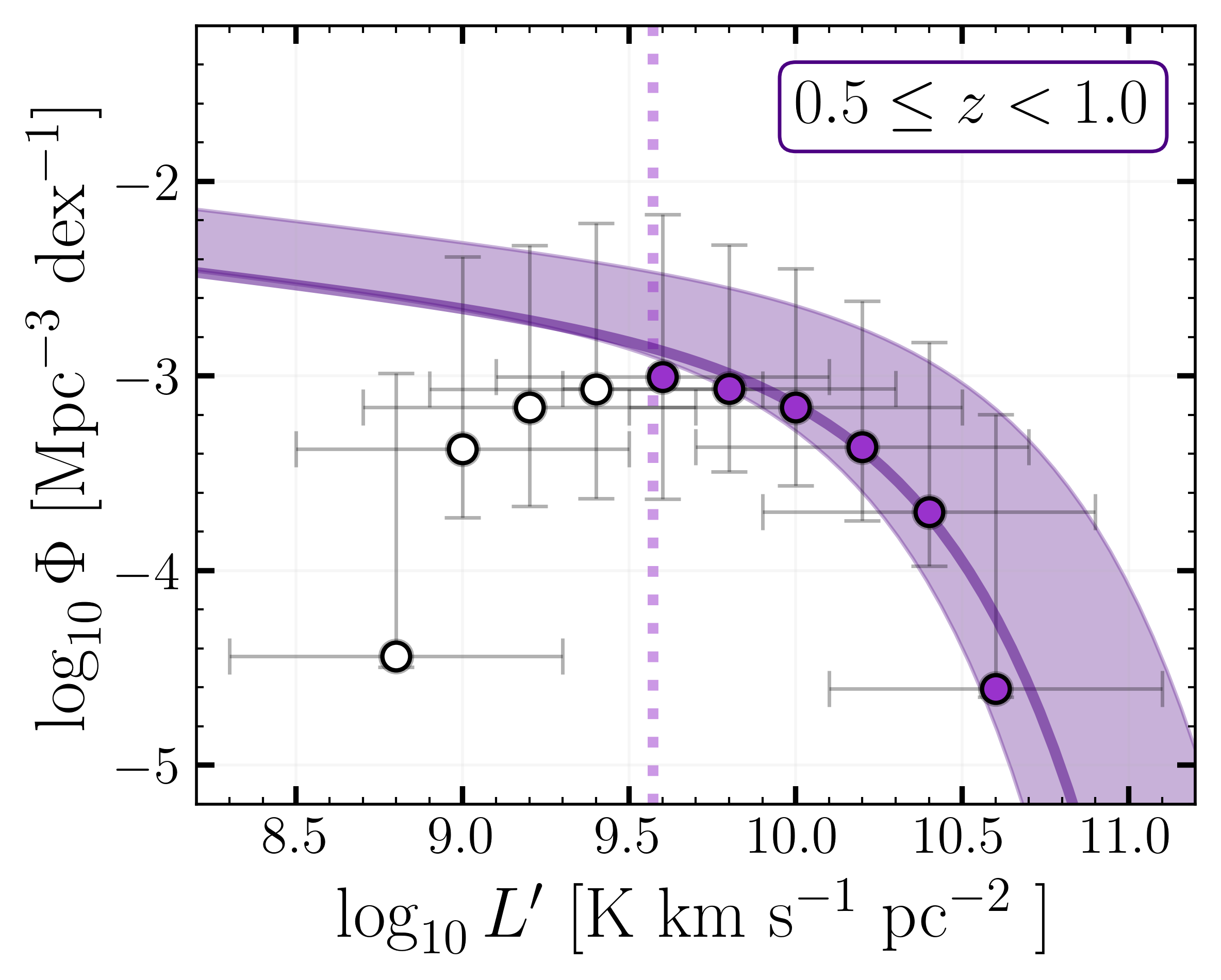}}
    \raisebox{0.0cm}{\includegraphics[width=0.66\columnwidth]{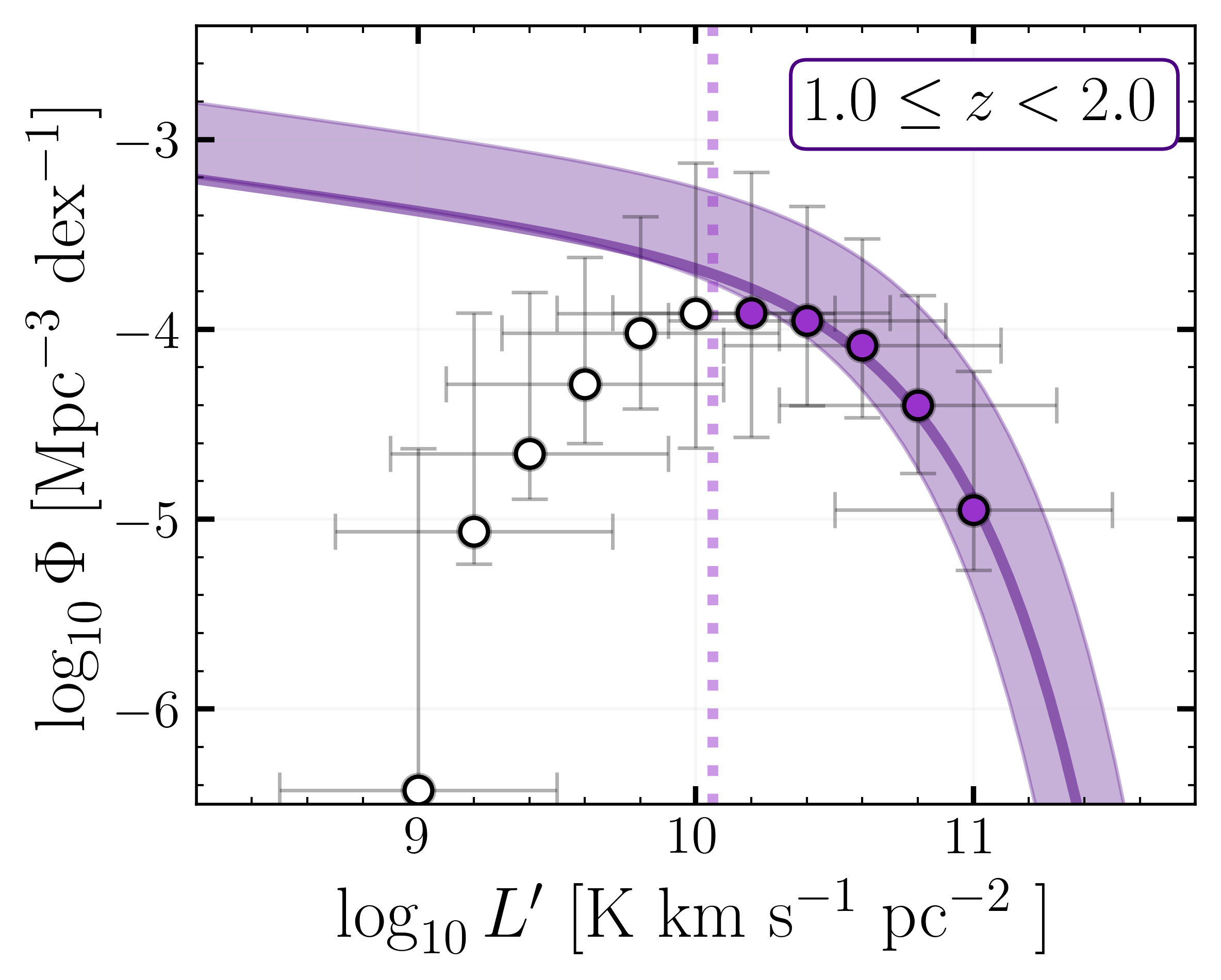}} \\
    \raisebox{0.0cm}{\includegraphics[width=0.66\columnwidth]{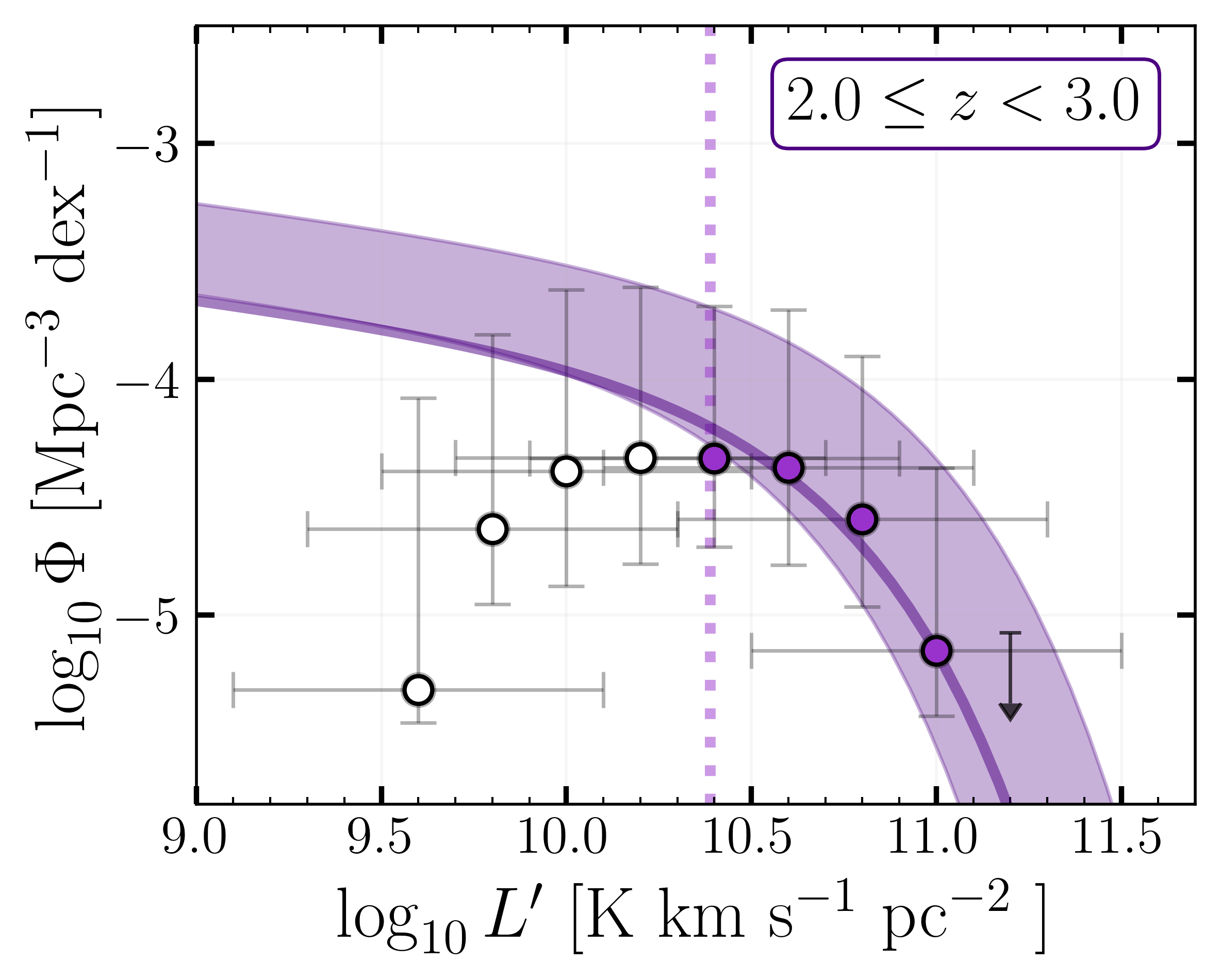}}
    \raisebox{0.0cm}{\includegraphics[width=0.66\columnwidth]{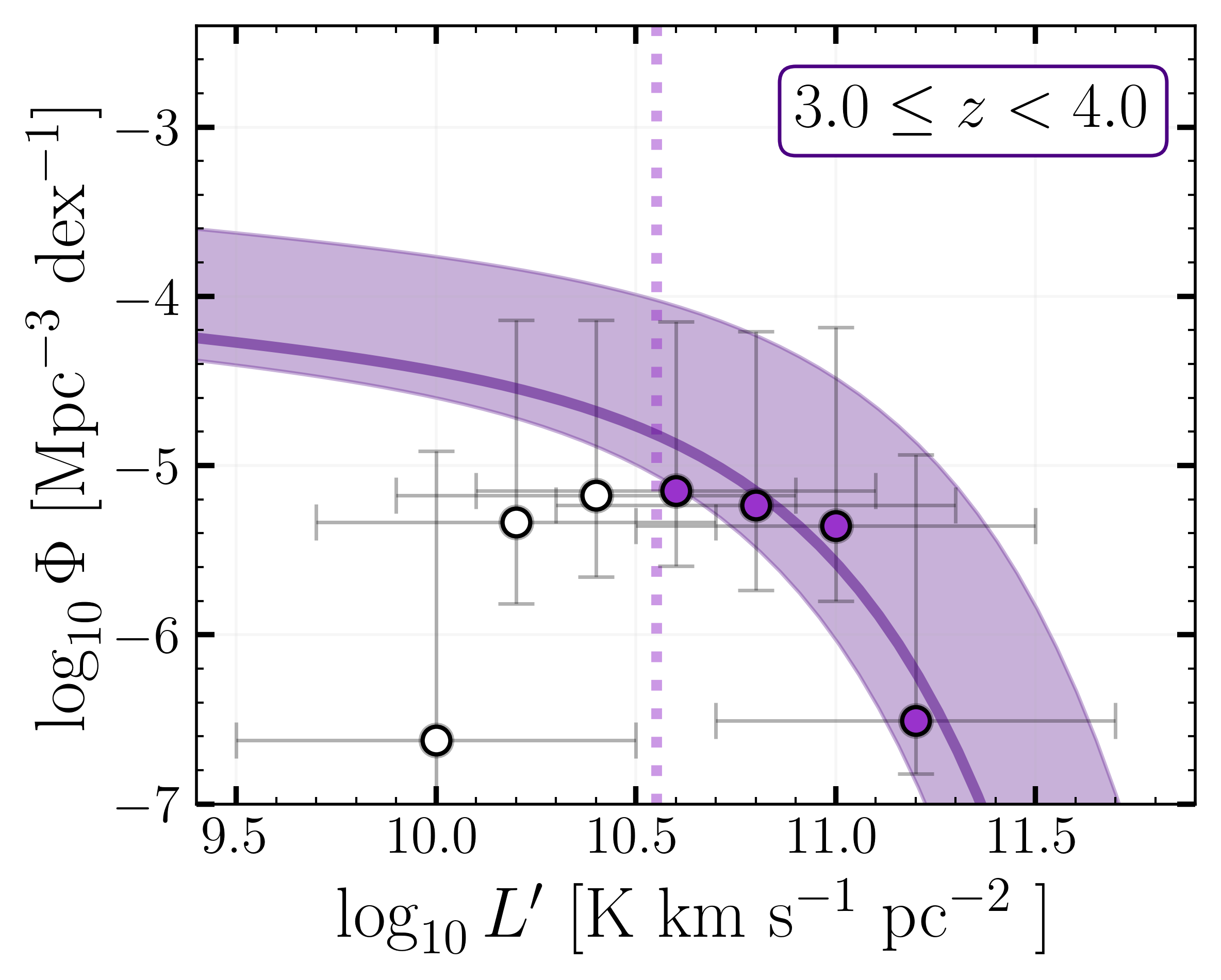}}
    \raisebox{0.0cm}{\includegraphics[width=0.66\columnwidth]{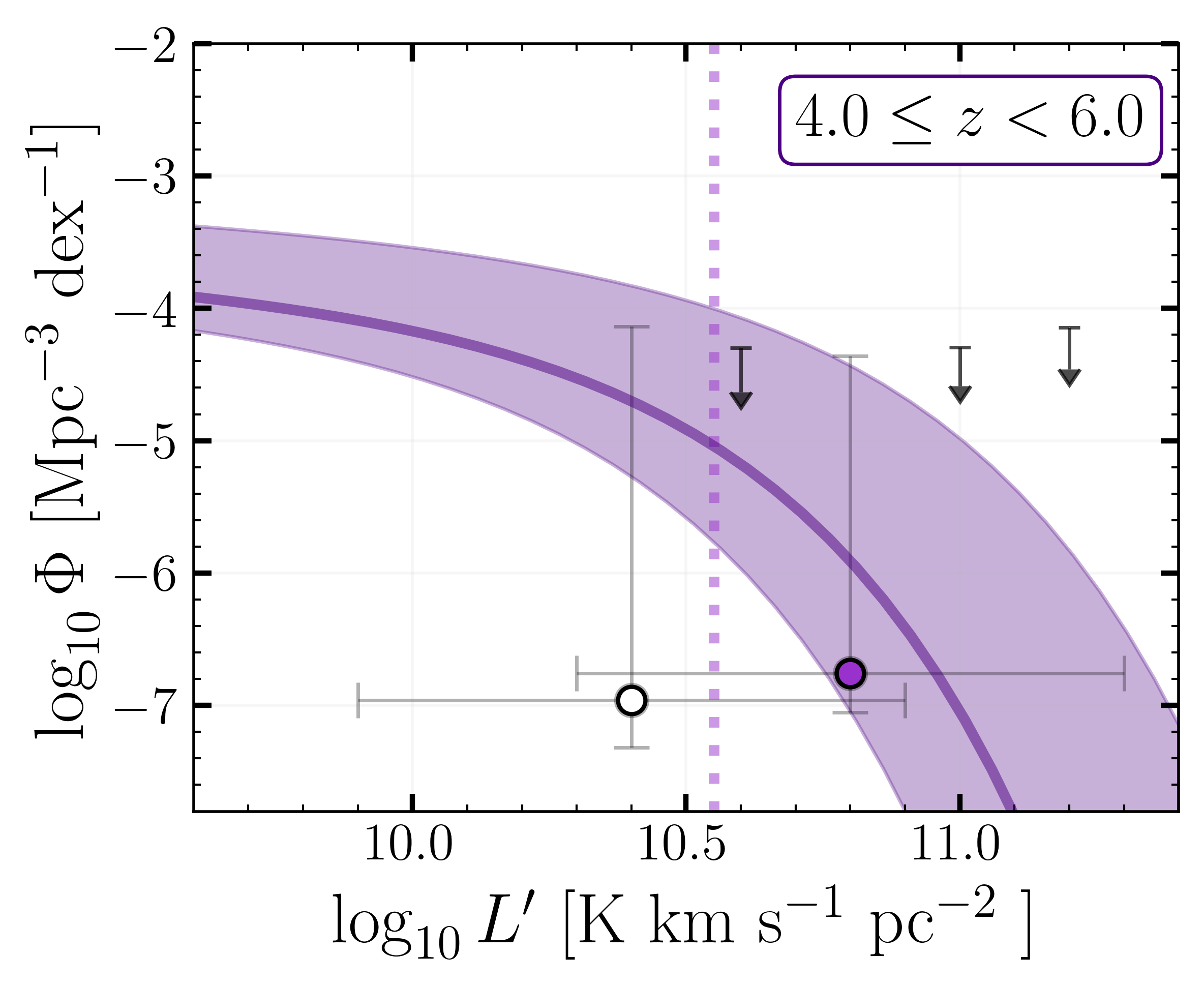}}
    \caption{CO LF across redshift bins: $z = 0$--0.5, $z = 0.5$--1, $z = 1$--2, $z=2$--3, $z=3$--4 and $z=4$--6. Detections are shown in circles and non-detection as arrows. The vertical dotted line in each panel indicates the detection limit, representing the faintest luminosity detectable at a $50\%$ reduction in primary beam sensitivity. The CO LFs are fitted using only the filled points that lie above this luminosity threshold. The best-fit Schechter function \citep{schechterAnalyticExpressionLuminosity1976} is plotted, with $\Phi^*$ and $L^{\prime *}$ allowed to vary freely, while the faint-end slope is fixed at $\alpha=-0.2$. The shaded region represents $1\sigma$ confidence intervals. Overall, the CO LF reveals a decreasing normalisation factor, while the `knee' of the function shifts to higher luminosities as redshift increases.}
    \label{fig:colf_schechter_fits}
\end{figure*}

\section{CO luminosity function and molecular gas mass density} \label{sec:co-lum-func}

A LF describes the statistical distribution of the number density of sources over a range of luminosities. In particular, the CO LF provides insight into the abundance of molecular gas in galaxies by revealing the variation in the number density of CO emitters. This section describes the methodology used to derive the CO LF and the results. First, we estimate the volume covered by the ALMACAL$-22$ survey in \S\,\ref{sec:volume}. We build the CO LF in  \S\,\ref{sec:co-lf} and fit Schechter functions in \S\,\ref{sec:schechter-fits}. We present the molecular gas mass density estimates at different redshift bins in \S\,\ref{sec:h2-density}. Finally, we explore the lowest$-J$ approach in \S\,\ref{sec:discussion_low_j} and the uncertainties associated with our measurements in \S\,\ref{sec:discussion_biases}.

\subsection{Volume estimation}\label{sec:volume}

We estimate the co-moving volume covered in our sample using the same methodology described in \cite{hamanowiczALMACALVIIIPilot2022}. 
Here, we briefly summarise the steps for this calculation.

First, we define the sky area covered by each ALMA cube, which depends on the frequency range and the field of view of the observations. 
For each CO transition ladder (from $J=6$ to $J=0$), we calculate the co-moving volume by integrating over both the redshift element and the solid angle subtended by the detectable emission lines.

Second, we integrate the sensitivity variation across ALMA data cubes. 
The sensitivity decreases from the centre of the primary beam, following a pattern that resembles a Gaussian function. We estimate the lower limits of detectable volume, modelling an emission line with CO luminosities ranging from $10^5$ to $10^{13}$ K km\,s$^{-1}$ pc$^2$. When the frequency coverage of the cube constrains the volume, we integrate the volume elements through concentric rings, each defined by the observed frequency. The resulting co-moving volumes for each CO transition are detailed in Table~\ref{tab:volume}.

% \begin{equation}
%     dV_c = D_H \frac{(1+z)^2 D_A^2}{E(z)}d\Omega dz \hspace{3px},
% \end{equation}
% where $D_H$ is the Hubble distance, $D_A$ is the angular diameter distance at the given redshift $z$, and the scale factor, $E(z)$, is defined as:
% \begin{equation}
%     E(z) = \sqrt{\Omega(1+z)^3 + \Omega_k(1+z)^2 + \Omega_{\Lambda}} \hspace{3px},
% \end{equation}

\subsection{CO LF} \label{sec:co-lf}

We calculate the CO luminosity from CO fluxes using the following equation, from \citet{solomonMolecularInterstellarMedium1997},

\begin{equation}
    L'_{\text{CO}} = 3.25 \times 10^7 \frac{S_{\text{CO}}\Delta V}{(1+z)^3} \left (\frac{D_{\text{L}}}{\nu_{\text{obs}}}\right )^2 \hspace{3px} \text{[K km s}^{-1} \text{pc}^2\text{]} \hspace{3px},
\end{equation}
\noindent
where $L'_{\text{CO}}$ is in units of K km s$^{-1}$ pc$^2$, $\nu_{\text{obs}}$ is the observed frequency of the CO line in GHz, $D_{\text{L}}$ is the luminosity distance of the galaxy in Mpc, $z$ is the redshift, and S$_{\text{CO}}\Delta V$ is the integrated flux in Jy km s$^{-1}$.
We convert the luminosity measured for mid- and high-J CO into CO(1--0), scaling by the empirical conversion factors from \cite{boogaardALMASpectroscopicSurvey2020}: $r_{j\rightarrow 1} =$\{$3.33, 5.20, 4.76, 2.70, 0.53$\} at $z<2$ and $r_{j\rightarrow 1} =$ \{$4.09, 8.24, 12.21, 14.68, 13.86$\} at $z>2$, for $J = 2, 3, 4, 5, 6$.

To construct the CO LF, we sample over the uncertainties of the integrated flux and the conversion factor used to derive the CO(1--0) flux, the completeness and fidelity factor, and the redshift probability.
The completeness factor depends on whether the emission lines were found in a cube with a high or low completeness factor (\S\,\ref{sec:completeness}).
The fidelity depends on the S/N and the line width of the emission lines. 
As noted previously, if the line has been confirmed by a counterpart with a matching spectroscopic or photometric redshift, we assume the fidelity is equal to one.

We perform 1000 realisations of the CO LF over five independent bins, shifted by 0.2 dex, where each bin spans one dex in luminosity. 
We adopted the same methodology presented by \cite{decarliALMASPECTROSCOPICSURVEY2016}, using independent shifted bins. 
This approach allows us to evaluate the dependence of the re-constructed CO LF on the bin definition.
In each iteration, the redshift of each source will depend on the redshift probability calculator described earlier (\S\,\ref{sec:redshift}). Hence, the same source will have different redshift values according to the weight of their probability in each realisation.

We bin galaxies by their CO luminosities, count the number in each bin, then normalise these counts by the co-moving volume. This gives us the number density of galaxies at different CO luminosities.
We build the normalised LF, $\Phi(L)$, also known as the number density of objects per unit luminosity, per unit co-moving volume, using the following expression: 
\begin{equation} \label{eq:co_lf}
    \Phi(\log L^{\prime}) \cdot d(\log L^{\prime}) \hspace{1px}  =  \sum_i^{} \frac{F_i}{V_i \cdot c_i} \hspace{6px} [\text{Mpc}^{-1} \text{dex}^{-1}] \hspace{3px},
\end{equation}
\noindent
where $\Phi(L^{\prime})$ is the luminosity density in units of Mpc$^{-3}$ dex$^{-1}$, $d(\log L^{\prime})$ is the luminosity bin, $V_i$ is the comoving volume accessible for each transition, $F_i$ is the fidelity and $c_i$ is the completeness factor estimated in \S\,\ref{sec:completeness} and Eq.~\ref{eq:fidelity}.

The final CO LF was divided into six redshift bins: $0\leq z < 0.5$, $0.5\leq z< 1$, $1\leq z<2$, $2\leq z <3$, $3 \leq z < 4$ , and $4 \leq z < 6$. 
Fig.~\ref{fig:colf_schechter_fits} shows the median values with their uncertainties. The error bars show the natural spread of data derived from 1000 realisations. We calculate the 16th and 84th percentiles and the $1\sigma$ Poisson confidence intervals for each bin, which accounts for the statistical fluctuations, following the method described in \cite{gehrelsConfidenceLimitsSmall1986}. When a particular luminosity bin contains fewer than one source on average, we provide a $1\sigma$ upper limit to reflect the lower reliability of the data in that bin.

\subsection{Schechter fits} \label{sec:schechter-fits}

We fit the CO LF using an analytical Schechter function \citep{schechterAnalyticExpressionLuminosity1976} in logarithmic scale as follows:
\begin{equation}
    \log \Phi(L') = \log \Phi^* + \alpha \log \left ( \frac{L'}{L^{\prime *}} \right ) - \left ( \frac{L'}{\ln(10) L^{\prime *}} \right ) + \log(\ln10) \hspace{3px},
\end{equation}
\noindent
where $\Phi(L')$ is the number of galaxies per co-moving volume with a CO luminosity between $\log L'$ and $\log L' + d(\log L')$. $\Phi^*$ is the normalisation factor, $\alpha$ is the faint-end slope, and $L^{\prime *}$ indicates the luminosity at which the LF changes from a power law to an exponential function, also known as the `knee' of the LF.

Fig.~\ref{fig:colf_schechter_fits} shows the best-fit Schechter functions, with the shaded regions representing $1\sigma$ confidence intervals. Vertical lines indicate the luminosity limits, which were determined by considering the noise level of each ALMA cube and a 50\% drop-off across the Gaussian-shaped primary beam. To ensure consistency, we adopted the limits from the shallowest cubes in the sample for each redshift bin and CO transition. 
The highest redshift panel contains only one data point above the luminosity limit with significant statistical weight. Nevertheless, we display the fit for completeness. 
To derive the Schechter functions fits, we perform 500 realisations, sampling the CO LF within its uncertainties. We only include bins where the median luminosity exceeds the calculated luminosity limit (indicated by filled points in Fig.~\ref{fig:colf_schechter_fits}).
We allow the parameters $\Phi^*$ and $L^{\prime *}$ to vary freely during the realisations. However, due to the limited number of data points at the faint end, we adopted a fixed slope at $\alpha = -0.2$, following the approach used by \cite{boogaardNOEMAMolecularLine2023}.
The median values and the uncertainties obtained from the best fits are reported in Table~\ref{tab:schechter_pars} for different redshift bins.

\begin{figure*}
    \includegraphics[width=2\columnwidth]{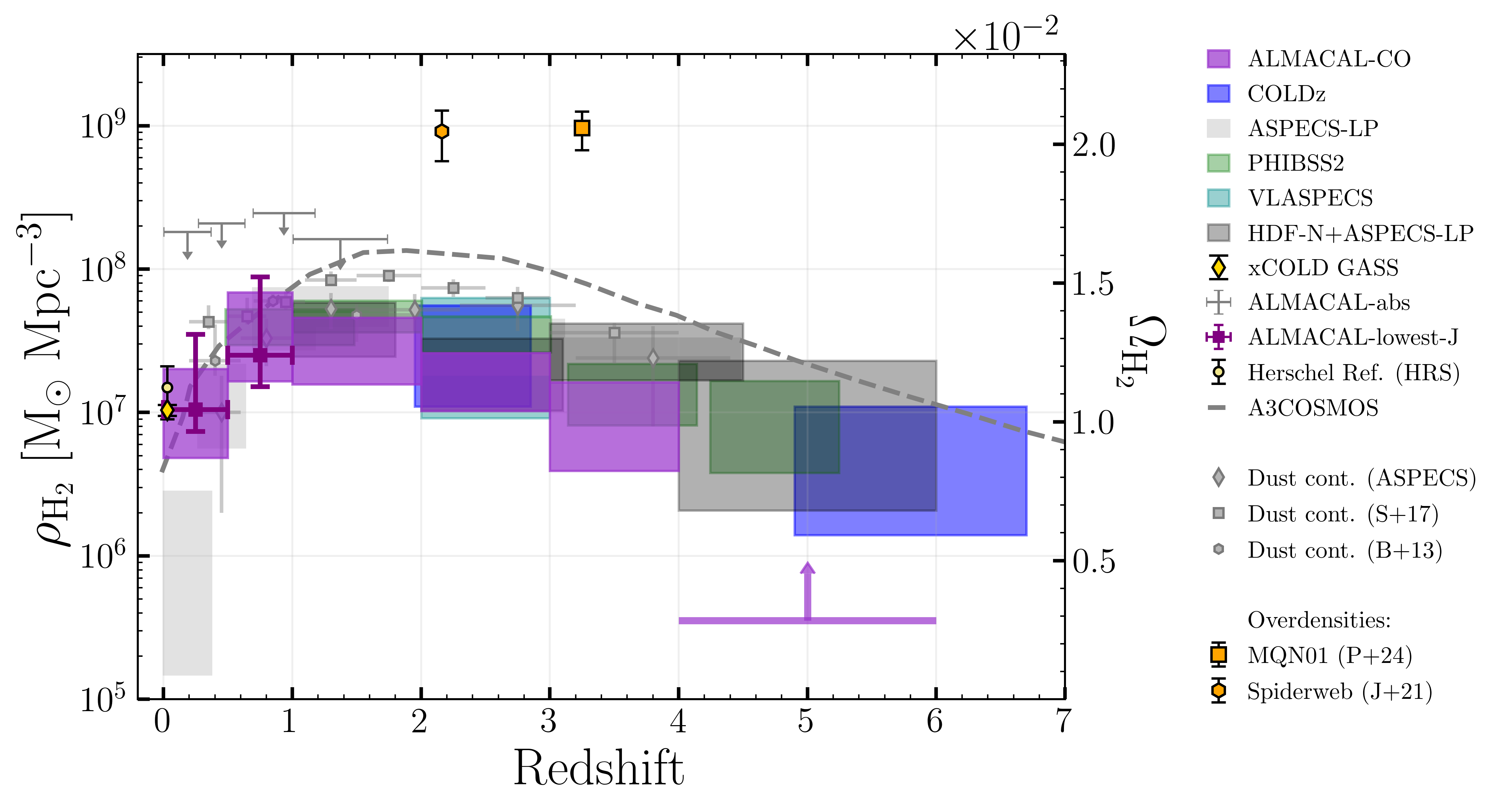}
    \caption{Cosmic molecular gas mass density evolution measured from ALMACAL$-22$ in purple. The right$-y$ axis represents the unitless density parameter for molecular gas, $\Omega_{\text{H}_2} = \rho_{\text{H}_2} / \rho_{\text{crit}, z=0}$. For comparison, we include estimates from other surveys: COLDz \citep{riechersCOLDzHighSpace2020}, ASPECS CO \citep{decarliALMASpectroscopicSurvey2020} and dust measurements \citep{magnelliALMASpectroscopicSurvey2020},  PHIBSS \citep{lenkicPlateauBureHighz2020}, VLASPECS \citep{riechersVLAALMASpectroscopic2020}, HDF-N+ASPECS-LP \citep{boogaardNOEMAMolecularLine2023}, xCOLD GASS \citep{saintongeXCOLDGASSComplete2017,fletcherCosmicAbundanceCold2021}, ALMACAL in absorption \citep{klitschALMACALAbsorptionselectedGalaxies2019}, ALMACAL$-22$ adopting the lowest$-J$ approach (this work, see \S\,\ref{sec:discussion_low_j}), from the $K-$band of the Herschel Reference Survey (HRS) \citep{andreaniMolecularMassFunction2020}, {A$^3$COSMOS \citep{liuAutomatedMiningALMA2019},} other estimates from dust continuum \citep{scovilleEvolutionInterstellarMedium2017, bertaMolecularGasMass2013}, and from recent works reporting over-densities at $z\sim 2$ in the Spiderweb galaxy protocluster \citep{jinCOALASATCACO102021} and at $z\sim 3$ in the MUSE Quasar Nebula 01 (MQN01) field \citep{pensabeneALMASurveyMassive2024}.
    % It is important to note that statistical errors mainly dominate the error bars in our measurements but overcome cosmic variance effects.
    We plot consistent estimates scaled to the same cosmology ($H_0$) used in this work and without helium contribution.
    We find overall consistency with the trend reported by other blind surveys, although slightly lower values are found at high redshift ($z>1$).
    }
    \label{fig:omega_h2}
\end{figure*}

\subsection{Molecular gas mass density} \label{sec:h2-density}

The molecular gas mass density refers to the total mass of molecular gas per unit volume in a given region of the Universe. It is often denoted as $\rho_{\text{H}_2}$ or $\Omega_{\text{H}_2}$ when normalised to the critical density of the Universe.
We derive the molecular gas mass of the CO emitters using the CO(1--0) luminosity and a conversion factor, $\alpha_{\text{CO}}$, as follows:
\begin{equation} \label{eq:omega}
    M_{\text{H}_2} = \alpha_{\text{CO}} \cdot L'_{\text{CO}(1-0)} \hspace{6px} [M_{\odot}] \hspace{3px},
\end{equation}
\noindent
where we used $\alpha_{\text{CO}} = 3.6 M_{\odot}$ (K km\,s$^{-1}$ pc$^2$) for all detections. Adopting this value makes the comparison with previous studies straightforward, since they use the same $\alpha_{\text{CO}}$ \citep{decarliALMASpectroscopicSurvey2019, riechersCOLDzShapeCO2019, lenkicPlateauBureHighz2020, riechersCOLDzHighSpace2020}. 
A different constant value affects our results linearly.

We derive the molecular gas mass density, $\rho(M_{\text{H}_2})$, by integrating $L^{\prime}_{\text{CO}} \Phi(L^{\prime}_{\text{CO}})$ up to the luminosity limit and applying the conversion factor, $\alpha_{\text{CO}}$. We exclude the same data points as before when fitting Schechter functions.
Fig.~\ref{fig:omega_h2} shows the estimates from ALMACAL$-22$ for each redshift bin, where the right $y-$axis shows the unitless density parameter for molecular gas, $\Omega_{\text{H}_2} = \rho_{\text{H}_2} / \rho_{0, \text{crit}}$. The associated uncertainties were estimated in the same way as for the CO LF, explained above.

\begin{table}[htb]
	\centering
	\caption{Cosmic molecular gas density from ALMACAL}
	\label{tab:omegah2_pars}
	\begin{tabular}{ccc} % 
		\hline \hline
		\rule{0pt}{2.5ex}Redshift & $\log \rho_{\text{H}_2}$   \\ [0.1cm]
        & [M$_{\odot}$ Mpc$^{-3}$] \vspace{3px} \\
		\hline  
		\rule{0pt}{2.5ex}$0.0 - 0.5$ &  $6.92^{+0.35}_{-0.37}$  \\ [0.2cm]
		$0.5 - 1.0$ & ${7.44}^{+0.35}_{-0.38}$ \\ [0.2cm]
		$1.0 - 2.0$ & ${7.35}^{+0.21}_{-0.51}$   \\ [0.2cm]
        $2.0 - 3.0$ & ${7.18}^{+0.13}_{-0.47}$  \\ [0.2cm]
        $3.0 - 4.0$ & ${6.84}^{+0.34}_{-0.35}$  \\ [0.2cm]
        % $4.0 - 6.0$ &  ${5.41}^{+0.69}_{-0.41}$ \\ [0.2cm]
        \hline
	\end{tabular}
 % \tablefoot{}
\end{table}

\subsection{Lowest$-J$ possible}\label{sec:discussion_low_j}

We now explore the conservative approach of assuming the lowest$-J$ CO transition detectable for all sources.
Given the frequency of each line, we classify the CO emission line at the lowest possible redshift. 
This method focuses on transitions from $J=1$ to $J=4$, which are expected to be the brightest and the most likely to be observed.
We perform 1000 realisations using the same luminosity and redshift bins. In each realisation, we perturb the luminosity and the empirical scaling factor to convert high$-J$ into CO(1--0) within linear uncertainties. We apply the same estimates of the volume and the correction factors (completeness and fidelity) described in \S\,\ref{sec:co-lf}.

We build the CO LF using Eq. \ref{eq:co_lf} and derive the molecular gas mass density using Eq.~\ref{eq:omega}. 
Our results are displayed in Fig.~\ref{fig:omega_h2} as purple boxes. These estimates match the previous results using the redshift probability calculator built from the semi-analytical model, \texttt{SHARK-2}, described in \S\,\ref{sec:redshift}. The main difference is the uncertainty, which is larger for the estimates derived using a probabilistic approach.
Using the lowest$-J$ approach, most of our detections would be classified as CO(1--0), considering that most were found in band 3. This likely skews the distribution of CO across cosmic epochs, since it does not consider the volume probed by each transition. As a result, we may be over-estimating the molecular gas content through this approach. Still, it is a conservative way to validate the use of models to estimate redshifts based on more complex CO line ratios, as previously noted by \cite{hamanowiczALMACALVIIIPilot2022}. 

\subsection{Possible biases and uncertainties} \label{sec:discussion_biases}

We identify the potential effects of the measured uncertainties on interpreting our results.
One significant factor is the Poisson error, particularly in bins with very few objects ($\lesssim 5$), as we are probing nearly one dex in luminosity at the bright end of the LF. 
Additionally, the redshift uncertainties, particularly in converting high$-J$ CO transitions into CO(1--0), could lead to the misclassification of transitions. Misinterpreting a higher$-J$ transition as CO(1--0) would imply lower gas masses. 

Another uncertainty stems from the uneven sensitivity of the data cubes. While we treat sensitivity on average, the data combined multiple observations with varying configurations, leading to changes in the noise across different spectral channels. Despite the efforts to correct the presence of false positives or missing detections through the completeness and reliability factors, we still allow for the possibility that some uncertainties are not fully understood.
In some cases, we had to exclude data cubes where no mock sources were retrieved, resulting in a zero completeness factor. Low completeness and low reliability, especially for sources with low signal-to-noise ratio and narrow line width, add further uncertainty, particularly for the brightest sources.

We also explored the possibility that we are detecting other lines than CO, particularly the bright [C\,{\sc ii}] emission line, but the probabilities are quite low. {This line is detectable in ALMA Band 7 at high redshifts ($z>4$), where the observed volume suggests a maximum of two detections. 
We observe five detections in Band 7, and even if two of them in the high-redshift bin were removed, it would not affect the overall results, as the values in that bin are lower limits. }
% Our overall results remain unchanged when excluding two sources as potential contaminants when falling at the high redshift bin, as the values provided in that bin are only lower limits . 

\section{Discussion} \label{sec:discussion}

This section presents our findings in the context of cosmic evolution. In \S\,\ref{sec:discussion_colf} and \S\,\ref{sec:discussion_h2}, we compare our CO LF and the molecular gas mass density estimates with previous works using observations and simulations. In \S\,\ref{sec:discussion:cosmic_variance}, we delve into the effects of the field-to-field variance of ALMACAL$-22$ compared with previous surveys. In \S\,\ref{sec:discussion_baryons}, we investigate the evolution of baryons associated with galaxies averaged over cosmic time and space, including the new estimates from ALMACAL$-22$. 
% \S \ref{sec:discussion_low_j} explores the lowest$-J$ approach when determining the redshift of the sources. Finally, we discuss possible biases and uncertainties or our results in \S \ref{sec:discussion_biases}. 

\begin{figure*}[htb]
    \includegraphics[width=2\columnwidth]{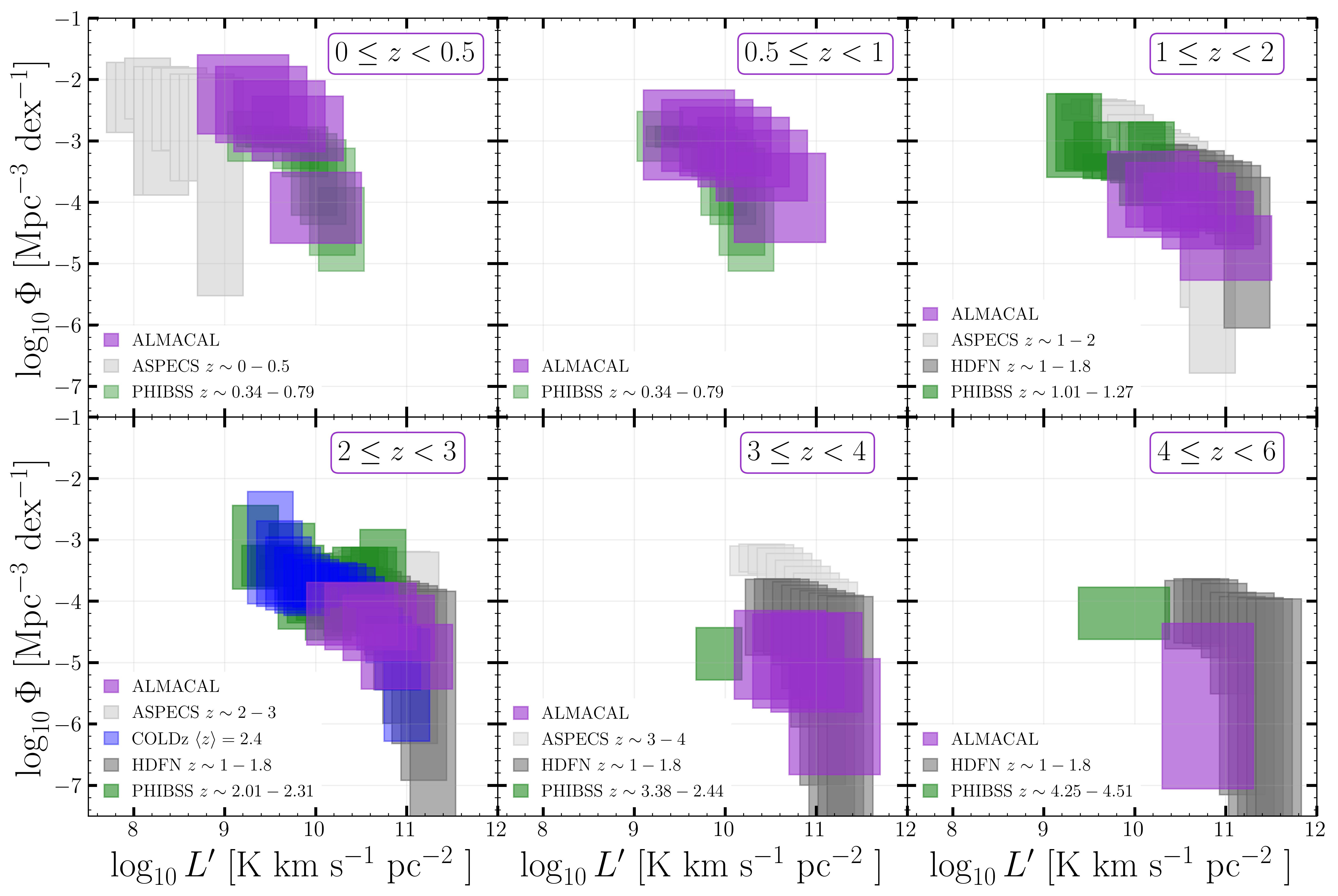}
    \caption{The CO LF in different redshift bins, from $z=0$ to $z=6$, as indicated at the top right of each panel. The measurements from ALMACAL$-22$ are shown in purple, including the uncertainty of each bin given by the extension of the boxes. We derived the CO LF based on the CO(1--0) luminosity, calculated as explained in \S\,\ref{sec:co-lf}.
    For comparison, we include observational constraints found by previous surveys at similar redshift ranges: ASPECS \citep{decarliALMASpectroscopicSurvey2019, decarliALMASpectroscopicSurvey2020}, PHIBSS2 \citep{lenkicCOExcitationHighz2023}, COLDz \citep{riechersCOLDzShapeCO2019}, HDFN \citep{boogaardNOEMAMolecularLine2023}.
    In some cases, the estimates from these surveys were derived from different CO transitions, here we converted the luminosities $L_{\text{CO(1--0)}}^{\prime}$ consistently to our methodology. The CO LF derived from the ALMACAL survey aligns well with findings from other surveys, while probing a wider redshift range.}
    \label{fig:lco_evolution}
\end{figure*}

\begin{figure}[htb]
    \includegraphics[width=1\columnwidth]{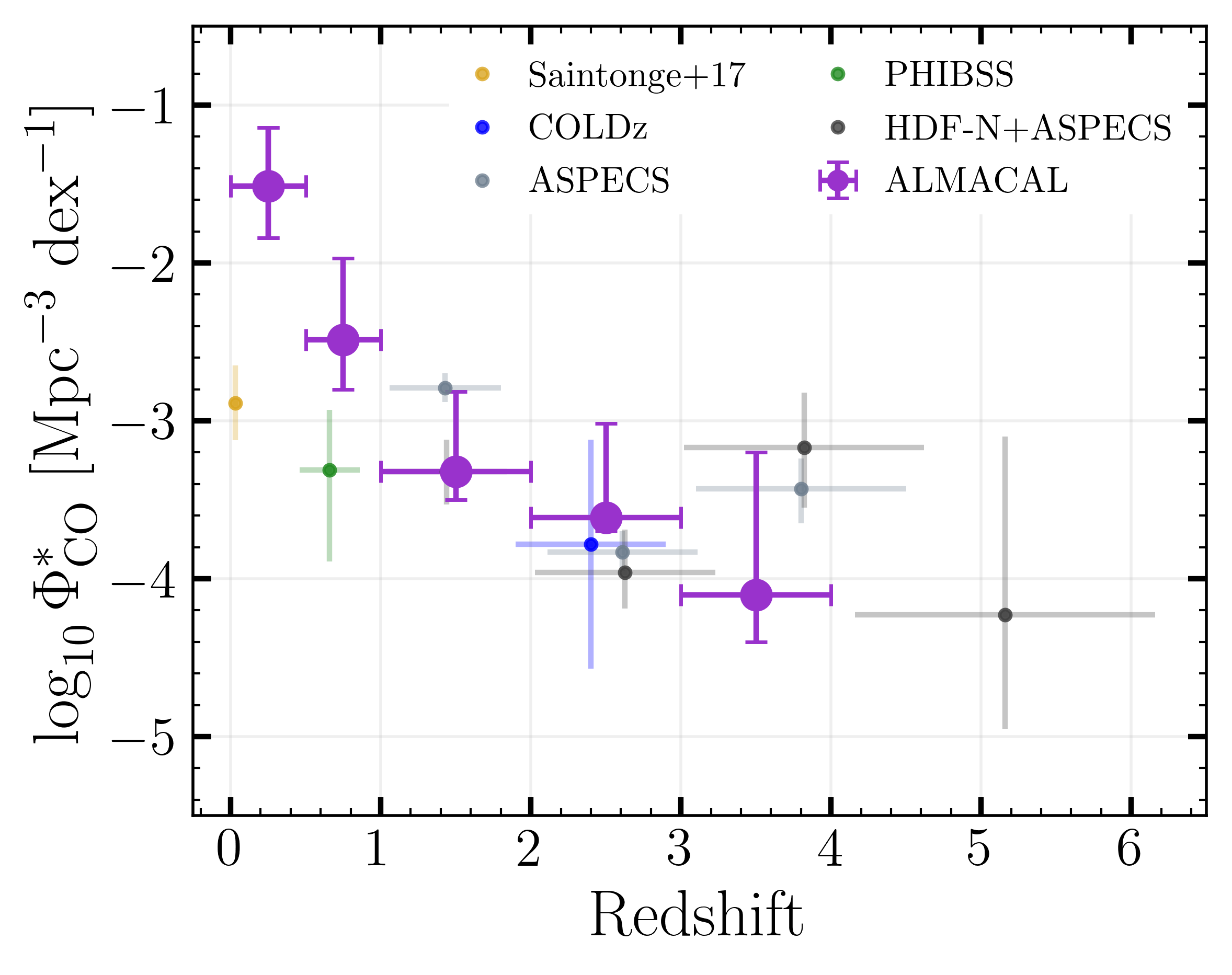}
    \includegraphics[width=1\columnwidth]{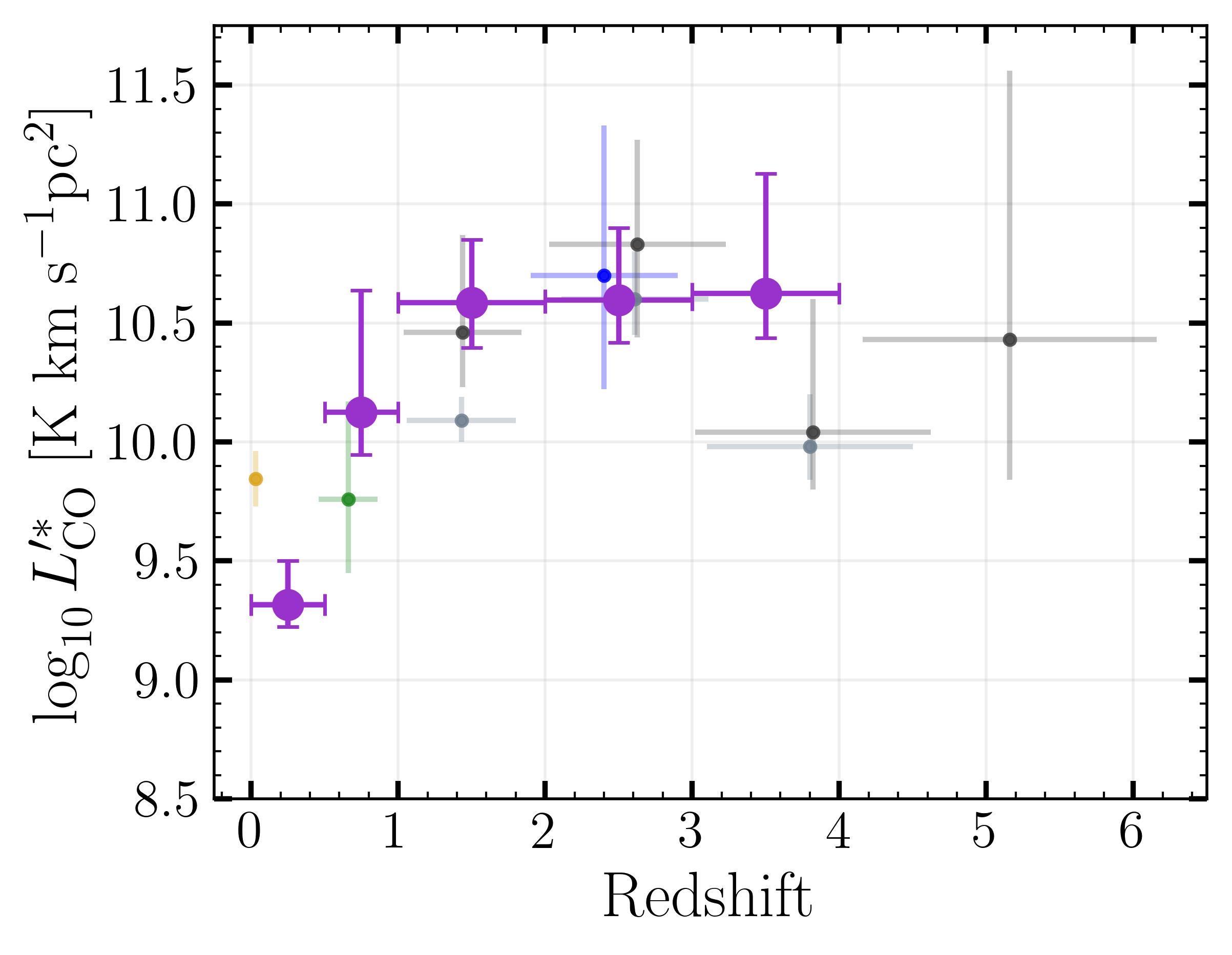}
    \caption{Evolution of the Schechter best-fit parameters for the CO LF across $z\sim0$ to $z\sim 6$. \textit{Top}: the evolution of $\Phi^*$ shows a decrease with redshift from $z=0$. \textit{Bottom}: the evolution of $L^*_{\text{CO}}$ shows a consistent increase from $z\sim 0$ to $z\sim 2$, and remains roughly constant at higher redshifts. For the fits shown in Fig.~\ref{fig:colf_schechter_fits}, the faint-end slope was fixed at $\alpha=-0.2$.}
    \label{fig:schechter_pars_evolution}
\end{figure}

\begin{figure*}[htb]
    \includegraphics[width=2\columnwidth]{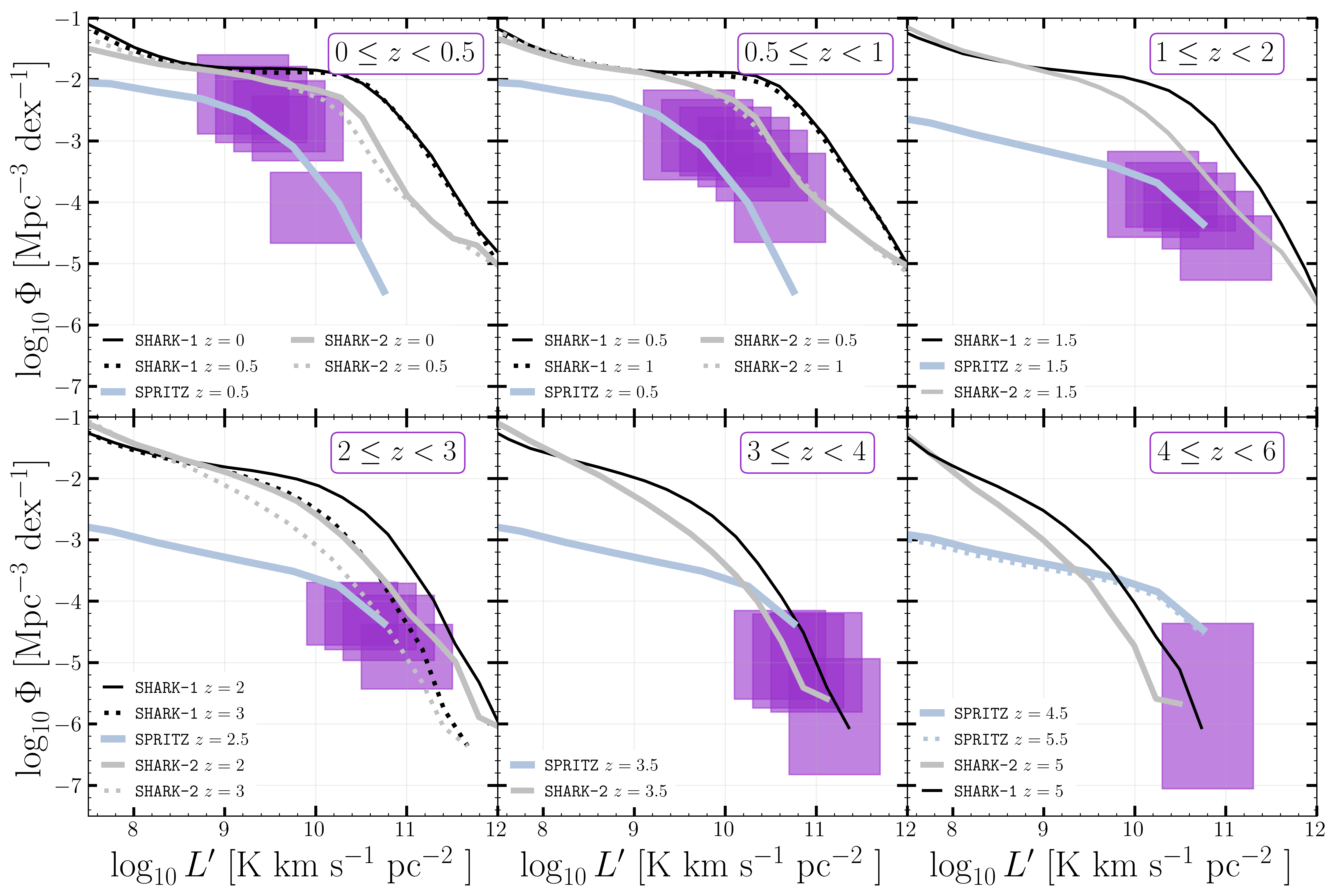}
    \caption{CO LF from redshift $z=0$ to $z=6$ derived from ALMACAL$-22$ in comparison with simulations. We include the predictions from \texttt{SHARK-1} \citep{lagosSharkIntroducingOpen2018} at $z=0$--3, from \texttt{SHARK-2} \citep{lagosQuenchingMassiveGalaxies2024} at $z=0$--5.5, and from \texttt{SPRITZ} \citep{bisigelloSPRITZSparklingSimulated2022} at $z=0.5$--5.5. We report consistency in most of the redshift ranges, and we recall the need for simulations to expand their range towards the bright end of the CO LF.}
    \label{fig:lco_evolution-sim}
\end{figure*}

\begin{figure}[h!]
    \includegraphics[width=1\columnwidth]{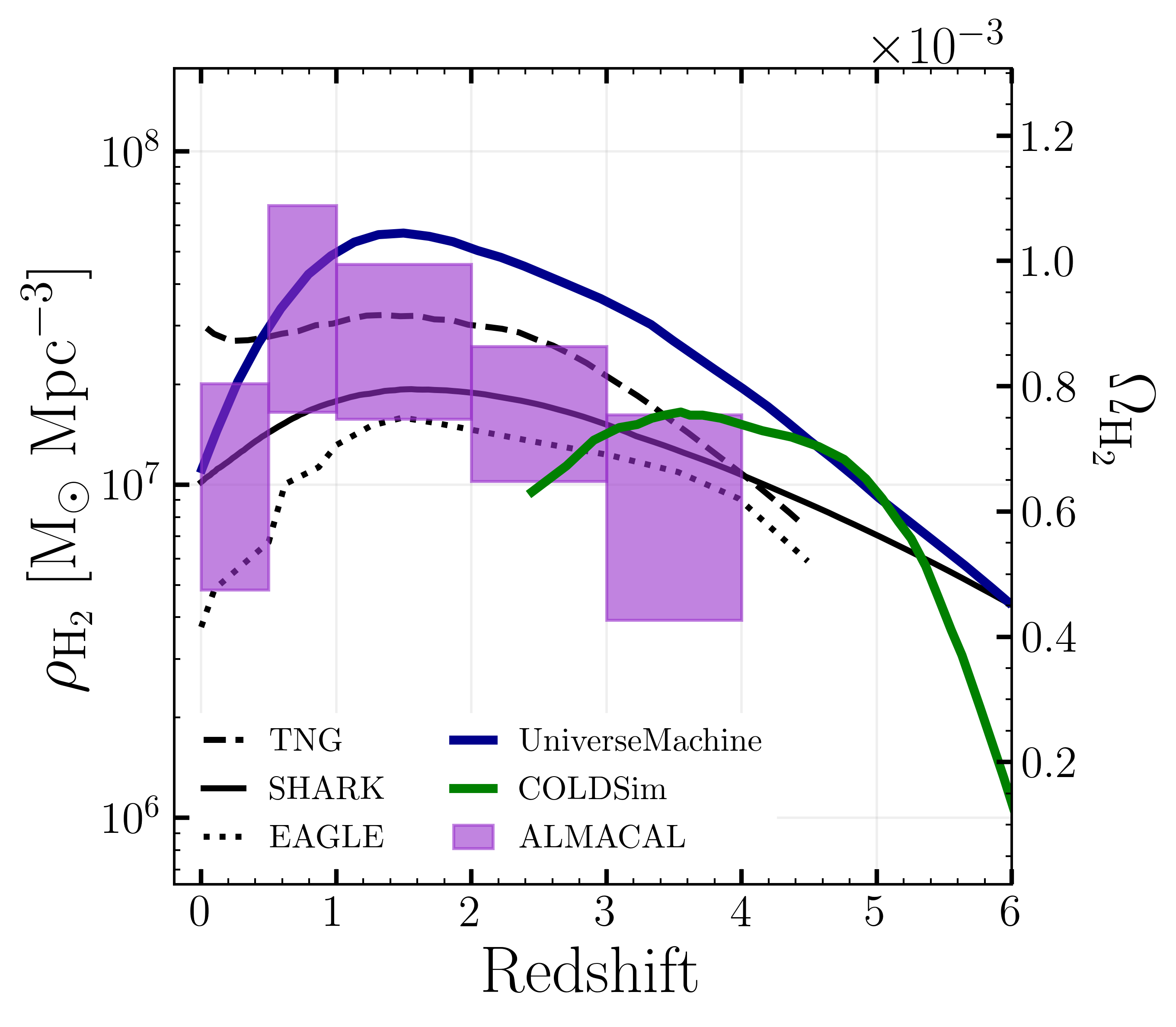}
    \includegraphics[width=1\columnwidth]{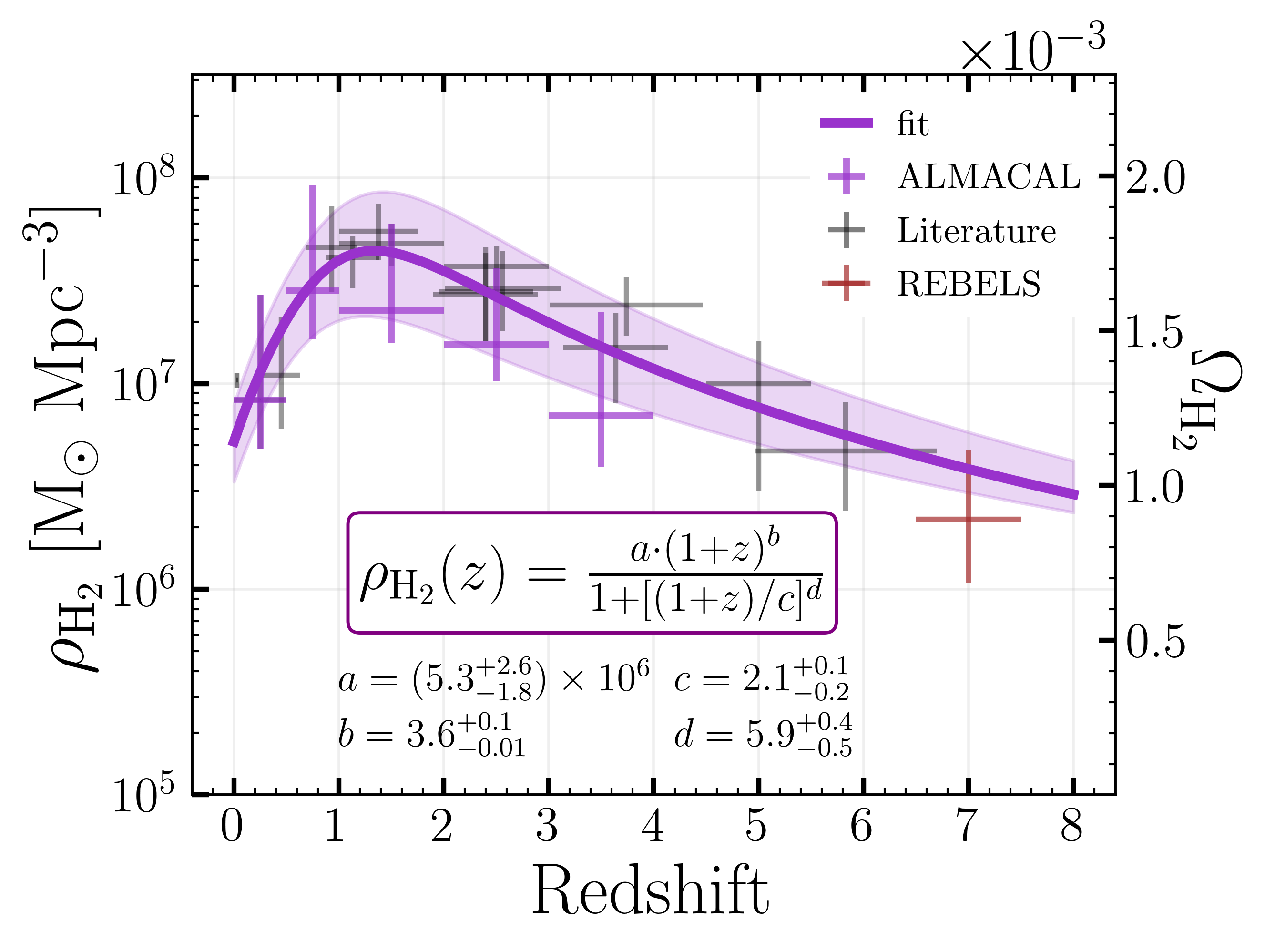}
    \caption{\textit{Top}: Evolution of the molecular gas mass density obtained from ALMACAL$-22$ (purple) in comparison with predictions from simulations. We include the results from Illustris TNG \citep{poppingALMASpectroscopicSurvey2019}, EAGLE \citep{lagosMolecularHydrogenAbundances2015}, \texttt{SHARK-2} \citep{lagosSharkIntroducingOpen2018}, UniverseMachine \citep{guoNeutralUniverseMachineEmpiricalModel2023}, and COLDSim \citep{maioAtomicMolecularGas2022}. 
    \textit{Bottom}: We compiled several measurements from previous {CO} surveys, as shown by the grey points, including ALMACAL$-22$ (see text). We present a new fit of the functional form used in \citealt{walterEvolutionBaryonsAssociated2020}, including the ALMACAL$-22$ estimate, shown by the purple line, with the $1\sigma$ uncertainty shown by the shaded region. For reference, we add the estimates of $\rho_{\text{H}_2}$ at the highest redshift measured up to date by the REBELS survey as the dark red point \citep{aravenaALMAReionizationEra2024}, which is in excellent agreement with the extrapolation of our fit.
    }
    \label{fig:omega-h2-sim}
\end{figure}

\subsection{CO LF} \label{sec:discussion_colf}

\subsubsection{Comparison with other observations}

Fig.~\ref{fig:lco_evolution} shows the redshift evolution of the CO LF derived from $z=0$ to $z=6$ from the ALMACAL$-22$ survey. Our estimates are based on the CO(1--0) luminosity. For comparison, we also include measurements from previous surveys at similar redshifts. Findings from the COLDz survey \citep{riechersCOLDzShapeCO2019} ASPECS survey are also based on the $L_{\text{CO}(1-0)}^{\prime}$. Estimates from 
PHIBSS2 \citep{lenkicCOExcitationHighz2023}, ASPECS \citep{decarliALMASpectroscopicSurvey2019, decarliALMASpectroscopicSurvey2020}, and HDFN surveys \citep{boogaardNOEMAMolecularLine2023} are derived from higher$-J$ CO transitions, scaled to the CO(1--0) consistently with our measurements as detailed above (\S\,\ref{sec:co-lf}). 

At low redshift, $z \sim 0$--1.0, our estimates indicate higher values of the number density of sources at a given luminosity, compared to findings from ASPECS and PHIBSS. 
At intermediate redshift, $z\sim 1$--2, we find values slightly lower than the results from PHIBSS but consistent with ASPECS and HDFN.
At higher redshift, $z \sim 2$--6, our results are in good agreement with COLDz and HDFN, but slightly lower than ASPECS. At these redshifts, the estimates from PHIBSS are derived from high$-J$ CO transitions, and the uncertainties converting them CO(1--0) may be larger, so they were included just for reference.

Another interesting way to explore the evolution of the CO LF is through their Schechter parameters. In Fig.~\ref{fig:schechter_pars_evolution}, we show the evolution of the Schechter parameters derived in \S\,\ref{sec:schechter-fits}. We compare our results with the surveys mentioned above, including the parameters of the well-known CO LF in the local Universe reported by xCOLD \citep{fletcherCosmicAbundanceCold2021}.
The normalisation factor (top panel), $\Phi^*$, shows a decreasing trend towards higher redshift, from $z\sim 0$ to $z\sim 6$ by a factor of $\sim 3\times$.
The characteristic luminosity (bottom panel), $L^{\prime *}$, increases from redshift $z\sim 0$ to $z\sim 2$ by a factor of $\sim 1.15\times$ and has a constant value towards higher redshift. 
This evolution suggests the CO luminosity in high-redshift galaxies is dominated by galaxies of $\log L^{\prime}_{\text{CO}} \lesssim 10.5 $ up to cosmic noon, after which brighter systems become more common.
Overall, there is a good agreement between the parameters estimated from the ALMACAL$-22$ survey and previous studies. The largest difference lies in the low-redshift bins, for which the assumption of a fixed faint slope at $\alpha =-0.2$ may not be appropriate, as low-$z$ studies have reported a different value ($\alpha \sim 1.2$, \citealt{fletcherCosmicAbundanceCold2021}). 
Further, examination of the evolution of the Schechter parameters of the CO LF is needed, particularly in the low-mass regime ($\log L_{\text{CO}}^{\prime} \lesssim 9$), where the number density values are more difficult to constrain due to observational limitations.

\subsubsection{Comparison with simulations}
% By boxes for statistics

Several approaches have been used to model the physics of the CO emission line through hydrodynamical simulations and semi-analytical models. 
Hydrodynamical simulations simultaneously model the physics of gas, dark matter and stars within galaxies, providing detailed predictions about the formation of molecular gas like CO and H$_2$ (e.g.~\citealp[]{pelupessyMOLECULARGASCO2009, obreschkowSIMULATIONCOSMICEVOLUTION2009, narayananGeneralModelCO2012, katzInterpretingALMAObservations2017, valliniCOLineEmission2018, casavecchiaCOLDSIMPredictionsII2024}). 
These simulations capture the small-scale physical processes, but their high computational cost limits the number of galaxies that can be simulated, especially over a wide range of properties. 
% As a result, hydrodynamical simulations struggle to provide large statistical samples, making it difficult to explore population-wide relations, like the CO-IR luminosity relationship. 
Also, their resolution is often insufficient to fully resolve small-scale processes like star formation \citep{lagosSimulationsModellingISM2012}.
On the other hand, semi-analytical models (SAMs) use simplified assumptions to describe the evolution of galaxies, such as symmetries in galaxies and halos. 
These models are usually run on halo catalogues which are made beforehand from an available dark-matter-only simulation \citep[e.g.~\texttt{SURFS}][]{elahiSURFSRidingWaves2018}, and then simplified models for gas physics are applied. 
The main advantage of SAMs is their computational efficiency, allowing them to explore a wide range of values in the parameter space. SAMs have been successful in modelling the interstellar medium (ISM), star formation, and CO emission across different transitions, as well as considering factors like gas metallicity and radiation fields \citep[e.g.~UV and X-ray][]{lagosSimulationsModellingISM2012, poppingEvolutionAtomicMolecular2014, poppingSubmmEmissionLine2016}.
While hydrodynamical simulations provide detailed insights into physical processes, SAMs offer broader statistical power, complementing both approaches in understanding the physics of CO emission and galaxy evolution.

We display the predictions of the CO LF from simulations in Fig.~\ref{fig:lco_evolution-sim}. We compare the results from ALMACAL$-22$ with the findings from the semi-analytical model \texttt{SHARK-1} \citep{lagosSharkIntroducingOpen2018}, \texttt{SHARK-2} \citep{lagosQuenchingMassiveGalaxies2023} and the phenomenological simulation \texttt{SPRITZ} \citep{bisigelloSPRITZSparklingSimulated2022}, which is based on empirical and theoretical relations. At low redshift, $z\lesssim 1$, \texttt{SPRITZ} produces values that are very much consistent with ALMACAL, particularly at the bright end of the LF. In this range, both \texttt{SHARK} versions produce values slightly above our estimates. 
In the rest of the panels, simulations are in good agreement with our findings, particularly with \texttt{SHARK-2}, which effectively constrains the bright end of the CO LF. {Although we use the redshift probability distribution derived from SHARK-2 as a prior in our analysis, the final results on the molecular gas mass density and luminosity function are decoupled from the simulation. This independence is demonstrated by the differences observed in the CO luminosity functions and the results from the lowest$-J$ approach, explained in Section \S \ref{sec:discussion_low_j} and further discussed in \S \ref{subsec:discussion_h2_obs}}.

\texttt{SHARK-2} implements a new AGN feedback model, compared to \texttt{SHARK-1}, which reduces gas content and enhances quenching in massive galaxies, especially at lower redshifts, driving differences seen at higher luminosities. At lower masses, the updated model accounts for additional environmental processes like ram pressure and tidal stripping, affecting gas content and causing variations in the faint-end slope of the CO LF, particularly at lower redshifts where satellite galaxies are more common.
We recall the need for hydrodynamical simulations to expand the simulated galaxy population towards the high-mass regime and probe fainter regimes of the LF. So far, the CO LF, particularly at high redshift, is still not available in a cosmological context, i.e.~big boxes \citep{bisigelloSPRITZSparklingSimulated2022, garcia$textttslick$ModelingUniverse2023}.

\subsection{Molecular gas mass density evolution} \label{sec:discussion_h2}

\subsubsection{Comparison with other observations} \label{subsec:discussion_h2_obs}

Fig.~\ref{fig:omega_h2} compares the molecular gas mass density estimates, $\rho_{\text{H}_2}$, from ALMACAL$-22$ with previous works. We include the estimates from ASPECS \citep{decarliALMASpectroscopicSurvey2019, decarliALMASpectroscopicSurvey2020}, COLDz \citep{riechersCOLDzHighSpace2020}, PHIBSS \citep{lenkicPlateauBureHighz2020}, 
VLASPECS \citep{riechersVLAALMASpectroscopic2020}, ALMACAL in absorption \citep{klitschALMACALVIMolecular2019}, dust measurement from \cite{bertaMolecularGasMass2013, scovilleEvolutionInterstellarMedium2017, magnelliALMASpectroscopicSurvey2020}, estimates in the local Universe from xCOLD GASS \citep{saintongeXCOLDGASSComplete2017, fletcherCosmicAbundanceCold2021} and \cite{andreaniMolecularMassFunction2020}, estimates at redshift $z\sim 2.2$ and $z\sim 3.2$ for over-dense systems from \citep{jinCOALASATCACO102021, pensabeneALMASurveyMassive2024}.
{We also included the estimates from the A$^3$COSMOS dataset \citep{liuAutomatedMiningALMA2019}, which uses different calibration methods to derive the molecular gas mass, including CO lines, SED-fitted dust mass and Rayleigh-Jeans(RJ)-tail dust continuum.}

In the lower redshift bins ($z\sim 0$--1), our estimates are consistent with the findings from xCOLD GASS and \cite{andreaniMolecularMassFunction2020}, and slightly above the estimate by ASPECS.
{The lowest$-J$ approach, explained in Section \S \ref{sec:discussion_low_j}, shows excellent agreement with the main methodology, which relies on redshift estimated trough \texttt{SHARK-2}, demonstrating that our results remain independent}.
At $z\sim 1$--2, we agree with the results from PHIBSS, but are below HDFN and ASPECS.
At $z\sim 2$ our estimates are consistent with VLASPECS, while PHIBSS, while HDFN and ASPECS report higher values. 
At $z\sim 3$--4 we find consistency with PHIBSS, but the other surveys show higher values.
In the highest redshift panel, the integration of the CO LF does not provide fully consistent measurement, as none of the individual measurements is completely above the luminosity limit, so we provide a lower limit.
Estimates of the molecular gas mass density from dust continuum \cite{bertaMolecularGasMass2013, scovilleEvolutionInterstellarMedium2017, magnelliALMASpectroscopicSurvey2020} are usually slightly above our estimates but still follow the same evolutionary trend.
High-redshift estimates of regions classified as over-densities are reported by \cite{jinCOALASATCACO102021} and \cite{pensabeneALMASurveyMassive2024}. 
These estimates are $\sim 1.5$ dex above the trend followed by blind surveys, including ALMACAL, as one might expect. 
In particular, \cite{pensabeneALMASurveyMassive2024} provides estimates, based on ALMA bands 3 and 6, of a sky region comparable in size to the ASPECS large programme ($\sim 4$ arcmin$^2$ vs.\ $4.2$ arcmin$^2$). 
Their detection of over-densities at such high redshifts further highlights the significant impact of cosmic variance on cosmic measurements. 
This emphasises the critical need for surveys that cover multiple sky regions, rather than the common practice of focusing on a single contiguous area. 

\subsubsection{Comparison with simulations}

Cosmological hydrodynamical simulations have also predicted the evolution of the molecular gas mass density, $\rho_{\text{H}_2}$.
These simulations infer the molecular gas mass density by post-processing the outputs of cosmological boxes. First, they simulate the large-scale distribution of galaxies and gas in cosmological boxes. Then, molecular gas content is estimated {\it a posteriori} by models of density or temperature that link the simulated gas properties to the presence of molecular gas. 
We compare our estimates with predictions from simulations in the top panel of Fig.~\ref{fig:omega-h2-sim}. We include the predictions from the semi-analytical model \texttt{SHARK-1} \citep{lagosSharkIntroducingOpen2018}, the cosmological simulation TNG \citep{poppingArtModellingCO2019}, EAGLE \citep{lagosMolecularHydrogenAbundances2015}, COLDSim \citep{maioAtomicMolecularGas2022, casavecchiaCOLDSIMPredictionsII2024}, and UniverseMachine \cite{guoNeutralUniverseMachineEmpiricalModel2023}.
Among these, COLDSim stands out by uniquely integrating time-dependent non-equilibrium chemistry with cosmological hydrodynamics, enabling direct modelling of molecular gas.
We find a relatively good agreement between our estimate and the predictions by \texttt{SHARK}, EAGLE and COLDSim, while TNG and UniverseMachine have higher values. 

% We also include previous compilations of the molecular gas density from observations presented by \cite{perouxCosmicBaryonMetal2020} and \cite{walterEvolutionBaryonsAssociated2020}. 
% Both works compiled estimates available up to that date, with the latter performing a fit to the evolutions as presented in figure \ref{fig:omega-h2-sim}.
We follow the model presented in \cite{walterEvolutionBaryonsAssociated2020} to fit the molecular gas mass density evolution with redshift, compiling the current estimates. We include all the CO observations compiled in that work together with the new estimates from ALMACAL$-22$. We perform 500 realisations of the following functional fit proposed by \cite{madauCosmicStarFormationHistory2014}:

\begin{equation} \label{eq:h1-fit}
    \rho_{\text{H}_2} = \frac{a \cdot (1+z)^{b}}{1 + [(1+z)/c]^{d}} \hspace{3px}.
\end{equation}

We find $a = (4.9 ^{+1.6}_{-3.1})\times 10^6$, $b = 3.5 ^{+0.2}_{-0.01}$ , $ c =2.2 ^{+0.2}_{-0.2} $, and $d = 6.0 ^{+0.4}_{-0.4}$.
The bottom panel of Fig.~\ref{fig:omega-h2-sim} shows our results with a purple solid line, and the $1\sigma$ uncertainty is shown in the shaded region.
The contribution of ALMACAL$-22$ to the molecular gas mass density estimation indicates lower values at high redshift. 
% The fit provided by \cite{walterEvolutionBaryonsAssociated2020} is roughly consistent with our fit in the low redshift bins and starts to differ from $z \geq 1$. 
% They also included estimates from the dust continuum, which may explain the offset compared to our fit, particularly at high redshifts.
If we extrapolate our model to high redshift, we find an excellent agreement with the estimation of $\rho_{\text{H}_2}$ at redshift $z\sim 7$ measured by the REBELS survey \citep{bouwensReionizationEraBright2022} from the [C\,{\sc ii}] emission line reported by \cite{aravenaALMAReionizationEra2024}, shown in the dark red point in the bottom panel of Fig.~\ref{fig:omega-h2-sim}.

\subsection{Cosmic variance}
\label{sec:discussion:cosmic_variance}

\begin{figure}[h!]
    \includegraphics[width=1\columnwidth]{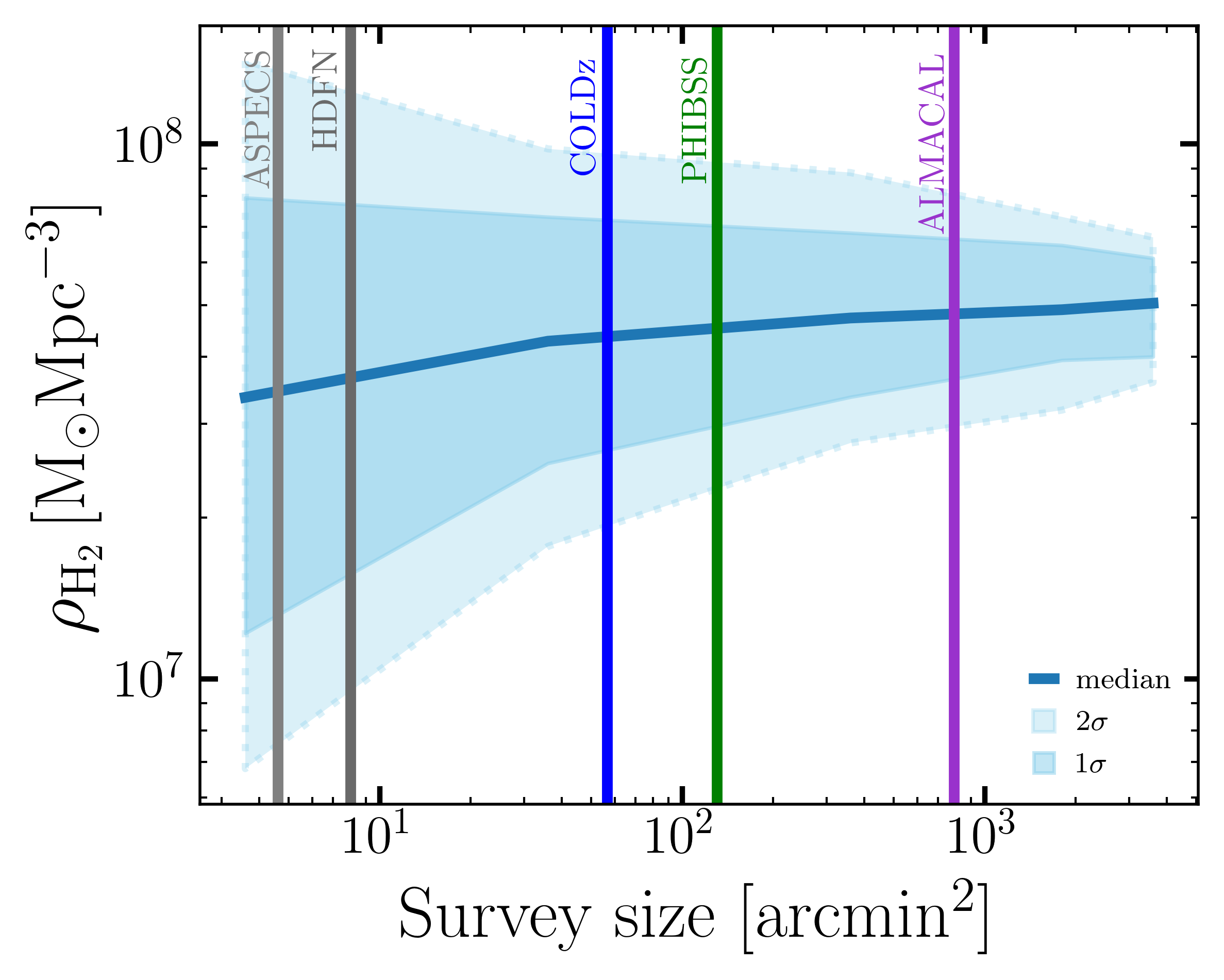}
    \includegraphics[width=1\columnwidth]{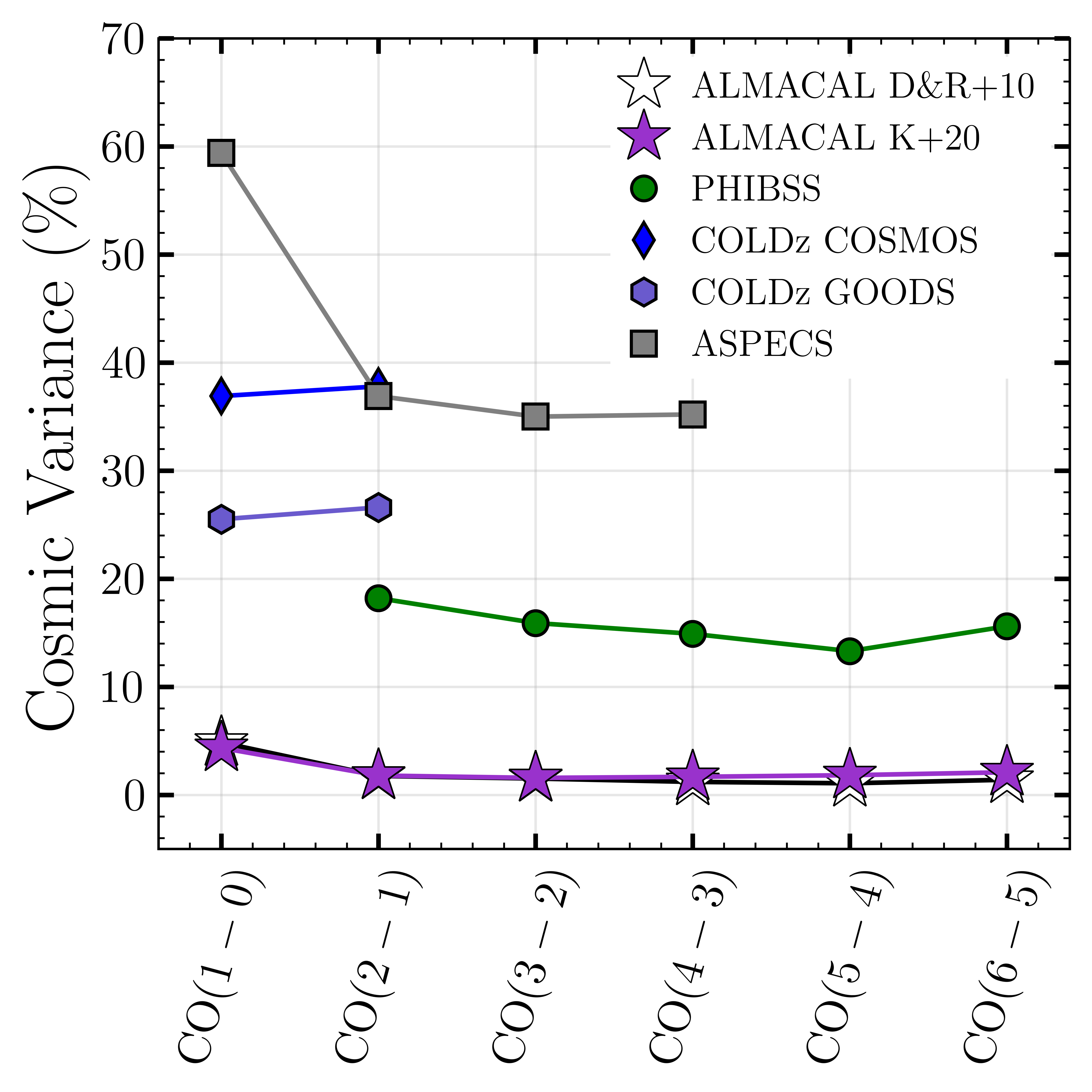}
    \caption{
    \textit{Top}: Effect of cosmic variance based purely on the survey size.
    We show the results of the molecular gas mass density variation derived from the SIDES simulation \citep{betherminCONCERTOHighfidelitySimulation2022, gkogkouCONCERTOSimulatingCO2022}, publicly available for $J=2$ at redshifts $z=0.5$--6. The solid blue line indicates the median value of $\rho_{\text{H}_2}$, and the dark and light areas represent the $1\sigma$ and $2\sigma$ variation due to the cosmic variance.
    We plot the area covered by ALMACAL$-22$ along with other CO surveys that have measured the molecular gas mass density. Overall, ALMACAL$-22$ is the survey less affected by cosmic variance to date.
    \textit{Bottom}: Cosmic variance estimation of ALMACAL$-22$ survey in purple, based on \citealt{driverQuantifyingCosmicVariance2010} (D\&R+10), for different CO transitions (Table~\ref{tab:volume}).
    We also plot the variance derived from the formula presented by \cite{keenanBiasesCosmicVariance2020} (K+20), which agrees with the previous prescription.
    The figure shows estimates from ASPECS \citep{decarliALMASpectroscopicSurvey2019, decarliALMASpectroscopicSurvey2020}, PHIBSS \citep{lenkicPlateauBureHighz2020}, COLDz GOODS and COSMOS \citep{pavesiCOLuminosityDensity2018, riechersCOLDzShapeCO2019}, previously reported in \cite{lenkicPlateauBureHighz2020}. 
    For every CO transition, ALMACAL$-22$ has the lowest effect of field-to-field variance ($\lesssim 5\%$).
    }
    \label{fig:cosmic_variance}
\end{figure}

We explore the effect of the field-to-field variance in ALMACAL$-22$ in comparison with the previous surveys.
The public simulation, SIDES \citep{betherminCONCERTOHighfidelitySimulation2022}, provides an estimate of the evolution of variance in the molecular gas mass density as a function of survey size, assuming $J=2$ and a redshift range $z=0.5$--6. 
The estimate of the cosmic variance is affected by the redshift slice, since high-redshift bins are less prone to cosmic variance effects.
We compare the effect of cosmic variance only considering the area covered by ALMACAL$-22$ and previous surveys in the top panel of Fig.~\ref{fig:cosmic_variance}.

Aside from the total area covered, the number of independent sight lines is key to estimating cosmic variance.
We adapt Equation~4 from \cite{driverQuantifyingCosmicVariance2010} for a conical survey, replacing the transverse lengths A and B by $\pi R$ as follows:
\begin{equation}\label{eq:cosmic-variance}
\begin{split}
         \zeta_{\text{CV}}(\%) = 
        & [219.7 - 52.4 \log_{10}(\pi R^2 \cdot 291.0) \\
         & + 3.21 (\log_{10}(\pi R^2 \cdot 291.0))^2] /  \sqrt{N\cdot C/291} \hspace{3px} ,
    \end{split}    
\end{equation}
\noindent
where $R$ is the transverse length at the median redshift, $C$ is the radial depth, and $N$ is the number of independent sight lines, all in units of $h_{0.7}^{-1}$.
We performed this calculation for each CO transition. We consider unique calibrator fields and data cubes where each line could be detected to account for the number of independent sight lines. Table~\ref{tab:volume} provides the estimates of the redshift range for each CO transition, the number of independent sight lines ($N_{ind}$), and the percentage of cosmic variance. 

The bottom panel of Fig.~\ref{fig:cosmic_variance} compares the cosmic variance from different surveys. We use the values of cosmic variance compiled in \cite{lenkicPlateauBureHighz2020} (Table~1 of their paper).
We see that for all CO transitions, ALMACAL$-22$ has the lowest cosmic variance seen to date by at least a factor $\sim 5\times$, reaching values lower than 5\%. The main reason that ALMACAL$-22$ has such low values of cosmic variance is that the survey comprises many independent calibrator fields instead of one contiguous area.
Previous studies have acknowledged this issue, particularly for the ASPECS survey. \cite{poppingALMASpectroscopicSurvey2019} estimated the H$_2$ cosmic density in boxes covering the same volume covered in ASPECS and
simulated the field-to-field variations.
They found that cosmic variance can lead to variations up to a factor of $3\times$. This variance is particularly significant due to the small survey area of ASPECS, 4.6 arcmin$^2$.

The formula provided by \cite{driverQuantifyingCosmicVariance2010} for estimating cosmic variance was originally intended for surveys measuring the variance in galaxy counts in fairly uniform galaxy populations. However, it has been noted by \cite{keenanBiasesCosmicVariance2020} that it does not take into account the variance in luminosity moments since they use galaxies in a restricted range of magnitude at $z\sim 0$. This issue becomes important for CO line surveys or molecular gas surveys, particularly for biased populations like bright CO emitters. A new method to estimate cosmic variance was proposed by \cite{keenanBiasesCosmicVariance2020} using simulated data. This new prescription includes the influence of galaxy number counts and how galaxies are distributed and clustered. 
We use the formula presented in Appendix~A of \cite{keenanBiasesCosmicVariance2020} for mean brightness temperature to calculate the fractional uncertainty in the cosmic variance of different area and redshift intervals. We use the mean redshift, the redshift interval and the area covered by each field for a given CO transition to estimate the fractional uncertainty. We then combine the uncertainty of all the fields as independent samples dividing by the square root of the number of independent sight lines ($\sqrt{N_{\text{ind}}}$). We obtain values consistent with the estimates derived from the formula proposed by \cite{driverQuantifyingCosmicVariance2010}, as shown in the bottom panel of Fig.~\ref{fig:cosmic_variance}.

\cite{gkogkouCONCERTOSimulatingCO2022} used the Uchuu cosmological simulation to analyse the impact of survey size on the variance of the luminosity function. They modelled the errors using both Poisson statistics and clustering contributions. The total variance is decomposed into Poisson and clustering components.
They reported the field-to-field variance of simulations that replicate the redshift range and sizes of observational data from ASPECS and COLDz. The field-to-field variance introduced in the molecular gas density ranges from 77\% to 81\% in an ASPECS-like survey and from 57\% to 73\% for a COLDz-like survey.
They also found a shift between the mean $\rho_{\text{H}_2}$ of their model and the one from \cite{keenanBiasesCosmicVariance2020}, which could be mainly explained by differences in the integration limits, the cosmological simulation used, the scaling relations, and the scatter to assign CO luminosities to galaxies.
The predictions from \cite{keenanBiasesCosmicVariance2020} may be under-estimated by an order of magnitude according to the predictions from \cite{gkogkouCONCERTOSimulatingCO2022} at redshifts $z\sim 2$--4. However, both approaches agree we need survey sizes of at least $\sim 70$ arcmin$^2$ to prove the evolution at redshifts, $z \gtrsim 3$.  

Overall, cosmic variance plays a significant role in the measurement of galaxy number density and LF, with uncertainties reaching as high as 70\% for the smallest fields and decreasing to about 25\% for the largest ones probed until now. Accounting for cosmic variance is essential to improve the accuracy of observational constraints on the cosmic star-formation rate density and to deepen our understanding of the large-scale structure of the Universe.
{Figure \ref{fig:omega_h2} shows that the ALMACAL results are consistent with previous surveys; however, the uncertainties due to cosmic variance are not displayed. In Figure \ref{fig:cosmic_variance}, we compare the impact of cosmic variance and we observe that this issue dominates other surveys. In contrast, due to its survey strategy, ALMACAL has a minimum value.}
To optimise survey strategies, it is beneficial to utilise multiple small fields instead of a single contiguous area, as this approach effectively mitigates cosmic variance. 
Future surveys are encouraged to consider the impact of cosmic variance when reporting uncertainties, moving beyond the traditional focus on Poisson uncertainties. The literature has shown that relying solely on Poisson uncertainties fails to capture the true levels of uncertainty \citep{poppingALMASpectroscopicSurvey2019, keenanBiasesCosmicVariance2020, gkogkouCONCERTOSimulatingCO2022}.
 
\begin{figure}[h!]
    \includegraphics[width=1\columnwidth]{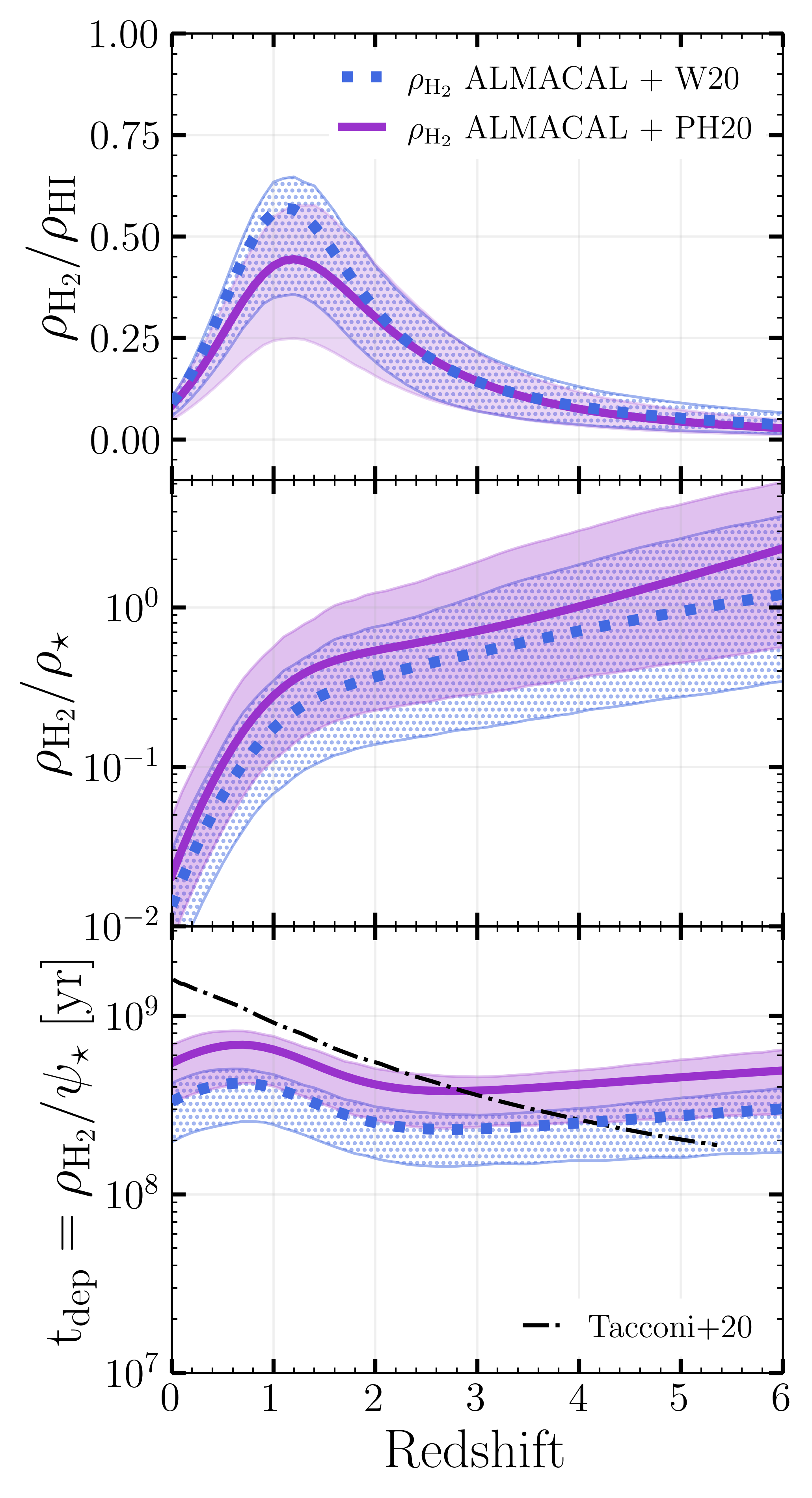}
    \caption{Redshift evolution of baryonic component in the Universe. \textit{Top}: Ratio of cosmic molecular-to-atomic gas density as a function of redshift. 
    \textit{Middle}: Ratio of molecular gas-to-stellar mass density as a function of redshift. 
    \textit{Bottom:} Cosmic gas depletion timescale, is defined as the density of molecular gas over the cosmic star-formation rate density.
    We include the molecular gas depletion times commonly derived for individual galaxies as the ratio $M_{\text{H}_2}/$SFR, as reported in  \cite{tacconiEvolutionStarFormingInterstellar2020}.
    We use the fit of $\rho_{\star}$ presented by \cite{madauCosmicStarFormationHistory2014}.
    We used the values of $\rho_{\text{HI}}$  and $\psi_{\star}$ fitted in the compilation done by \cite{perouxCosmicBaryonMetal2020} in dotted blue and \cite{walterEvolutionBaryonsAssociated2020} in purple. For the star-formation rate density and stellar mass function, the former assumes a Chabrier IMF  \citep{chabrierGalacticStellarSubstellar2003} and a return fraction $R = 0.41$, while the latter assumes a Salpeter IMF \citep{salpeterLuminosityFunctionStellar1955} and $R = 0.27$.
    The $\rho_{\text{H}_2}$ corresponds to the fit presented in \S\,\ref{sec:discussion_h2}, which includes the results from previous surveys that have measured CO along with ALMACAL. 
    % Overall, the content of the atomic hydrogen always surpasses the molecular gas.
    % The fits reported by \cite{walterEvolutionBaryonsAssociated2020} are displayed in each panel by the grey curve for reference on how the baryon cycle census would change when having lower molecular gas mass density values at high redshift ($1 \lesssim z$).
    }
    \label{fig:baryons}
\end{figure}

\subsection{Census of the baryon cycle} 
\label{sec:discussion_baryons}

% The return fraction is R = 0.27 for a Salpeter IMF and R = 0.41 for a Chabrier IMF that is more weighted towards massive stars (Madau & Dickinson 2014).

{In this section, we use the new $\rho_{\text{H}_2}$ estimates provided by ALMACAL to revisit the cosmic baryon cycle.} Specifically, we investigate how the cosmic baryon density has evolved over time, focusing on the contributions from stars, atomic hydrogen (H\,{\sc i}), and molecular gas (H$_2$).

\cite{madauCosmicStarFormationHistory2014} compiled data of two key observables: the cosmic star-formation rate density, $\psi_{\star}$, and the stellar mass density, $\rho_{\star}$. \cite{perouxCosmicBaryonMetal2020} used a Chabrier IMF \citep{chabrierGalacticStellarSubstellar2003} in their review, and a return fraction $R=0.41$, which is the fraction of the stellar mass that is immediately returned to the gas when massive stars explode, whereas \cite{walterEvolutionBaryonsAssociated2020} assumed a Salpeter IMF \citep{salpeterLuminosityFunctionStellar1955} and $R=0.27$.
For the atomic gas mass density, \cite{perouxCosmicBaryonMetal2020} modelled the evolution of neutral gas with redshift using a two-parameter power law, which uses less free parameters. On the other hand, \cite{walterEvolutionBaryonsAssociated2020} adopted a hyperbolic tangent function, originally suggested by \cite{prochaskaAstrophysicalConsequencesIntervening2018}, to fit the atomic gas. 
% This updated fi incorporates measurements gathered by \cite{neelemanCONTENTUNIVERSE102016} and new constraints at low redshift.
In summary, we adopt the best-fit models for $\psi_{\star}(z)$, $\rho_{\star}(z)$ and $\rho_{\text{HI}}(z)$ from both \cite{perouxCosmicBaryonMetal2020} and \cite{walterEvolutionBaryonsAssociated2020}.
For $\rho_{\text{H}_2}(z)$, we use the updated fit described in \S \ref{sec:discussion_h2}, which includes estimates from ALMACAL$-22$.

We analyse the gas density ratios of these three components (stars, atomic and molecular gas) in Fig.~\ref{fig:baryons}.
In each panel, the purple solid lines indicate the ratios derived using $\rho_{\text{H}_2}(z)$ from this work, and the fits for $\rho_{\text{HI}}$, $\psi_{\star}$ and $\rho_{\star}$ provided in \cite{walterEvolutionBaryonsAssociated2020}.
The dotted blue lines show the ratios using $\rho_{\text{H}_2}(z)$ including ALMACAL$-22$ and the fits for $\psi_{\star}$, $\rho_{\star}$ and $\rho_{\text{HI}}$ presented in \cite{perouxCosmicBaryonMetal2020}. 
For both models, the $1\sigma$ uncertainty is shown by the shaded regions.
% We also show the fits of $\rho_{\text{H}_2}$ provided in \cite{walterEvolutionBaryonsAssociated2020} by a grey dashed line.

The top panel shows the molecular-to-atomic gas density ratio as a function of redshift ($\rho_{\text{H}_2}/ \rho_{\text{HI}}$). 
It peaks at around $z\sim1.5$, indicating a period of high star-formation activity, characterised by the conversion of H\,{\sc i} to H$_2$ being particularly efficient. 
This efficiency is slightly lower than the one predicted by \cite{walterEvolutionBaryonsAssociated2020} but still consistent
% where they claimed that $\rho_{\text{H}_2} / \rho_{\text{HI}}$ approaches equality at $z \sim 1.5$. 
with their predicted scenario of cosmic twilight, where the star-formation activity in galaxies declines, likely due to the shutdown of the inflow and the accretion of gas. 
% In this scenario,  the total cold gas mass in galaxies will always be dominated by diffuse atomic gas rather than molecular gas at all redshifts (HI $ >$ H$_2$).

The middle panel shows the ratio of molecular gas-to-stellar mass density ($\rho_{\text{H}_2} / \rho_{\star}$) as a function of redshift.
This quantity evolves significantly from $z\sim 1.5$ down to the local universe ($z=0$), indicating that the molecular gas for star formation becomes more limited relative to the existing stellar mass. 
At high redshift ($z>1.5$), our fits indicate a lower molecular gas, which means a lower capacity for star formation compared to that observed by \cite{walterEvolutionBaryonsAssociated2020}.
The Chabrier IMF predicts about 30\% lower total stellar mass for a given population than the Salpeter IMF, since fewer low-mass stars are included. This would suggest that galaxies still have significant molecular gas reservoirs available for star formation, particularly at high redshifts, which may not be true at low redshifts.

The bottom panel shows the cosmic gas-depletion timescale, defined as the density of molecular gas over the cosmic star-formation rate density ($\rho_{\text{H}_2} / \psi_{\star}$). 
The dot-dashed line shows the functional fit to the molecular depletion timescale based on M$_{\text{H}_2}$/SFR ratio for a sample of massive galaxies \citep{tacconiEvolutionStarFormingInterstellar2020}, which roughly agrees with the global estimates from \cite{perouxCosmicBaryonMetal2020}.
At redshifts above $z\sim 2$, we find a relatively constant depletion timescale, which indicates that molecular gas is consumed by star formation steadily during this period.
The depletion timescale increases from $z\sim 2$ to $z=0$, so the consumption rate is slowing down relative to the available gas, leading to the quenching of star formation.
A Salpeter IMF would imply that galaxies deplete their gas more quickly, while a higher ratio (Chabrier IMF) would suggest galaxies have longer depletion timescales, consistent with galaxies at high redshift \citep{scovilleEvolutionInterstellarMedium2017}.
This trend is consistent with the findings from \cite{perouxCosmicBaryonMetal2020} and their claim regarding the need to continuously replenish molecular gas to sustain star formation.

These findings align with predictions from the gas regulator model, also known as the `bathtub model', as described in studies like \cite{boucheImpactColdGas2010, daveGalaxyEvolutionCosmological2011, lillyGasRegulationGalaxies2013, pengHaloesGalaxiesDynamics2014}. This model uses continuity equations to track the flow of baryons into and out of galaxies and picture them as systems in a dynamic equilibrium between gas inflow, outflows and star formation. As previously noted by \cite{perouxCosmicBaryonMetal2020}, this model provides an effective framework for understanding how the continuous cycling of baryons drives galaxy evolution. In the early stages, gas accumulates, and star formation is primarily constrained by the available cold gas reservoir. As galaxies evolve, they reach a steady state where the net gas accretion rate regulates the overall star-formation rate. 

% Although cosmic variance is less of an issue in ALMACAL compared to other studies, our low statistics contribute significantly to the overall uncertainty. 
% The fact that we are probing particularly the bright end of the luminosity function can imply that the luminosity bins are more correlated with each other \citep{gkogkouCONCERTOSimulatingCO2022}. 
% By integrating the luminosity function and neglecting the faint end to avoid any assumptions, one can still underestimate the cosmic molecular density, as seen in comparisons among simulations \citep{keenanBiasesCosmicVariance2020}.  

% Despite these uncertainties, our estimates still hold significant value, providing insight into the distribution and evolution of molecular gas across cosmic time. 
% The trends we observe are consistent with previous findings but show a peak of $\rho_{\text{H}_2}$ around $z\sim 1$ and lower values at higher redshifts. 

\section{Conclusions}\label{sec:conclusions}

In this paper, we have presented the ALMACAL-CO project, built upon the experience of the previous pilot survey presented by \cite{hamanowiczALMACALVIIIPilot2022}.
We searched for serendipitous detections of CO emission lines in the fields of ALMA calibrators. 
We used the highest available quality of data, selected as part of the ALMACAL$-22$ data release presented in \cite{bolloALMACALXIIData2024a}. 
We performed a search for CO emitters using the source-finding algorithm, SoFiA-2 \citep{westmeierSOFIAAutomatedParallel2021}.
We detected $87$ CO emitters over $299$ calibrator fields. We confirmed the redshift of three sources with optical spectroscopic counterparts and estimated the photometric redshift of three sources with optical and near-IR counterparts.
We determined redshift probabilities for the remaining 81 candidates based on the semi-analytical model, \texttt{SHARK-2} \citep{lagosQuenchingMassiveGalaxies2023}. We built the CO LF through realisations that sample the properties of the CO emitter, such as their completeness, fidelity and redshift probability. We constrained the CO LF and calculated the molecular gas mass density up to redshift, $z\sim 6$.
We summarise our main findings as follows:

\begin{enumerate}
    \item We have probed the bright end of the CO LF, from $z\sim 0$ to $z$ $\sim 6$. We find good agreement with existing observations, but with slightly higher values of the number density of sources in the low redshift bins ($z\lesssim 1$) and slightly lower values in the higher redshift bins ($z \gtsim 1$). Simulations of the CO LF \citep[\texttt{SHARK, SPRITZ}]{lagosSharkIntroducingOpen2018, bisigelloSPRITZSparklingSimulated2022} agree with our findings, but we stress the need to expand their prediction to the bright end (Figs~\ref{fig:lco_evolution} and \ref{fig:lco_evolution-sim}).

    \item We found an evolution of the Schechter parameters of the CO LF with redshift. The normalisation factor, $\Phi^{*}_{\text{CO}}$, decreases with redshift. The characteristic luminosity, defined as the knee of the luminosity function, shows an increasing trend from the local Universe to $z\sim2$ (Fig.~\ref{fig:schechter_pars_evolution}).

    \item The molecular gas mass density reported by ALMACAL$-22$ shows higher values than previous observations in the lower redshift bins ($z\lesssim1$). It presents a slight tension at higher redshift ($z\geq1$), where ALMACAL$-22$ finds lower values (Fig.~\ref{fig:omega_h2}). We found a good agreement with the simulations predicting $\rho_{\text{H}_2}$ from \texttt{SHARK}, EAGLE and COLDSim (Fig.~\ref{fig:omega-h2-sim}, top panel). 

    \item We present a new analytical molecular gas mass density fit, compiling previous estimates based on CO surveys and the new ALMACAL$-22$ results. Irrespective of what CO line $\rho_{\text{H}_2}$ is based on, the effect of cosmic variance is less than 5\% (Fig.~\ref{fig:omega-h2-sim}, bottom panel).

    \item ALMACAL$-22$ is the survey that best addresses the uncertainty introduced by cosmic variance at the brighter end (Fig.~\ref{fig:cosmic_variance}).

    \item We re-visit the redshift evolution of the baryon components, updating the molecular gas mass density with the values provided by ALMACAL. We use previous fits of the atomic gas mass density, stellar mass function and the cosmic star-formation rate density from \cite{perouxCosmicBaryonMetal2020} and \cite{walterEvolutionBaryonsAssociated2020}.  
    Overall, we report consistency with their claims regarding the evolution of the density ratios, though with smaller molecular gas content (Fig.~\ref{fig:baryons}). 
\end{enumerate}

Our findings serve as a valuable stepping stone for future work that conducts surveys less prone to cosmic variance.
We emphasise the need for additional telescope time to confirm the redshift of our sources, particularly those that may be obscured at optical wavelengths, where sub-mm observations would be crucial.

Probing the molecular gas density across different environments is a collective effort by the scientific community. The more data we gather, the clearer the picture becomes. Each survey brings unique strengths and limitations, and to fully understand the evolution of molecular gas in the Universe, we also need simulations with diverse prescriptions. Together, these approaches will help unravel the complexities of molecular gas evolution over cosmic time.

\begin{acknowledgements}
      We thank L.~Boogaard and I.~Smail for their contribution to the analysis and discussion.
      This research was supported by the International Space Science Institute (ISSI) in Bern, through ISSI International Team project \#564 (The Cosmic Baryon Cycle from Space). 
      % This paper makes use of the following ALMA data: ADS/JAO.ALMA \\

% \includeonly{project_codes_almacal}

\end{acknowledgements}

% \section*{Data Availability}
% The ALMA Calibrator Source Catalogue is available \href{https://almascience.eso.org/alma-data/calibrator-catalogue}{here}.
% ALMACAL raw data and processed data cubes are available upon reasonable request. Please contact \href{almacal@eso.org}{almacal@eso.org}.

  \bibliographystyle{aa} % style aa.bst
  \bibliography{main} % your references Yourfile.bib
\begin{appendix}
\onecolumn 
\section{Detections} \label{appendix:detections}
We display the spectra and the moment map of the 87 detections found in the ALMACAL$-22$ survey, identified as CO emitters. 
\begin{figure}[h!]
\centering
    \includegraphics[width=0.32\textwidth]{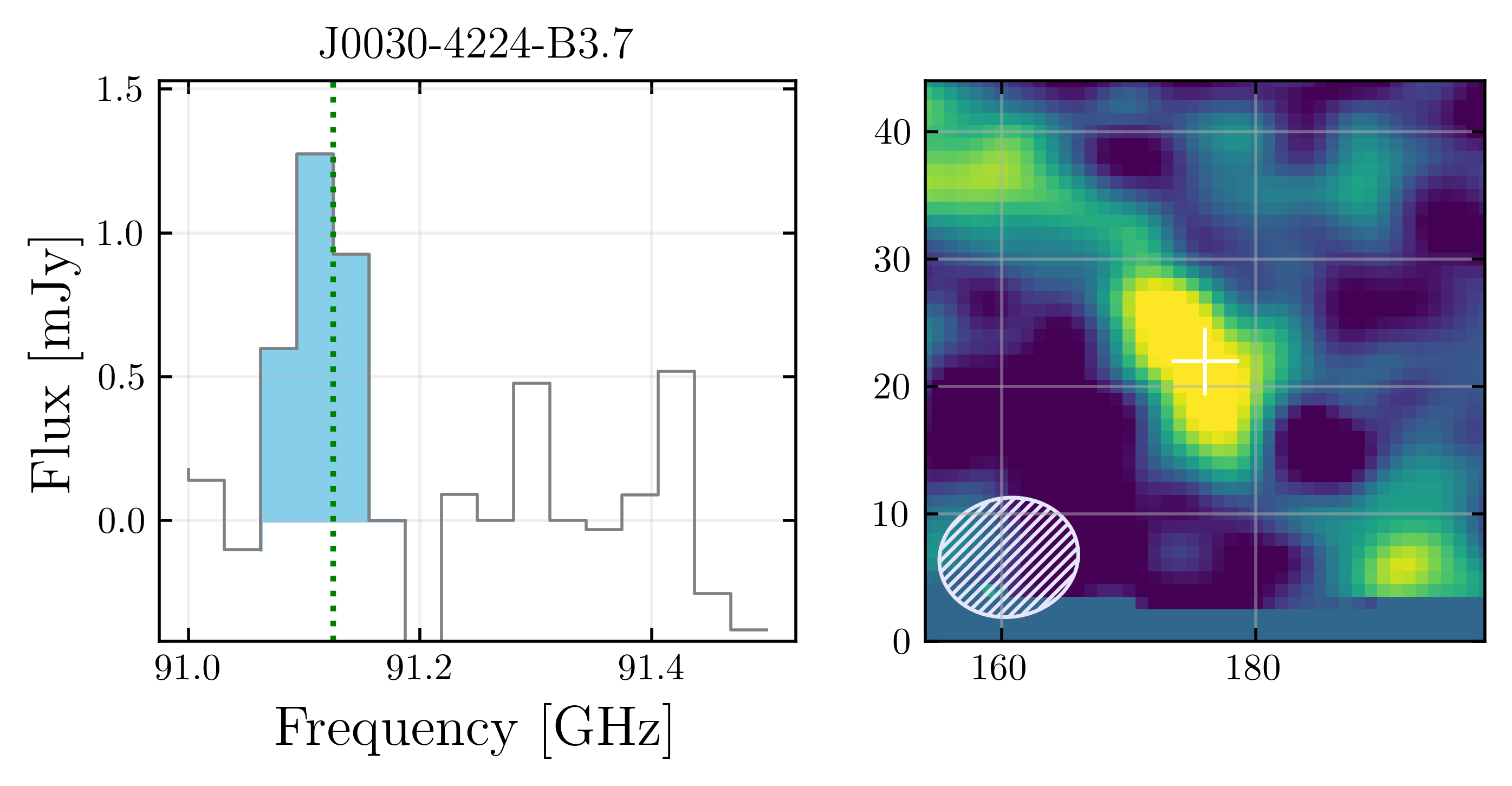}
    \includegraphics[width=0.32\textwidth]{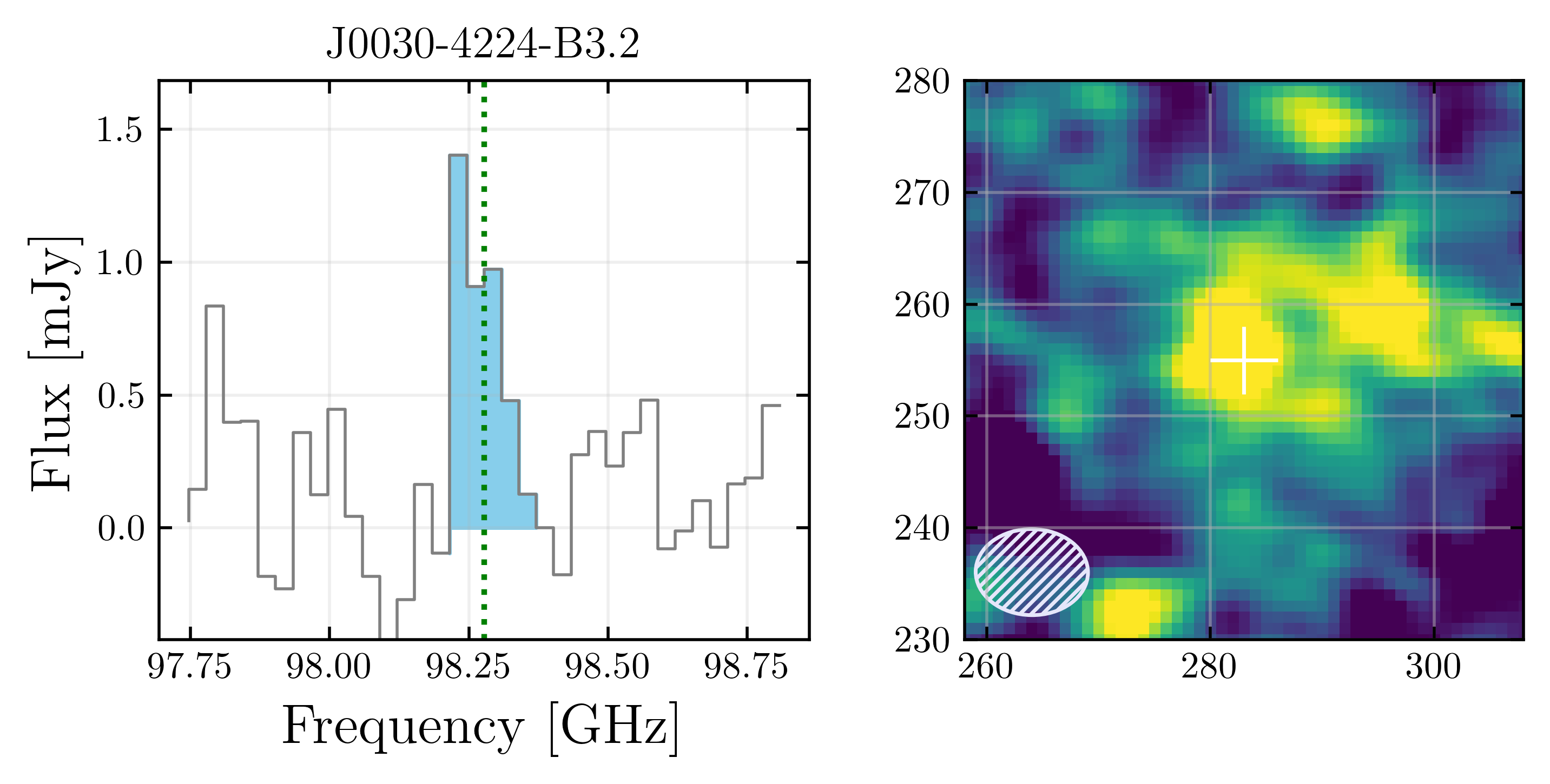}
    \includegraphics[width=0.32\textwidth]{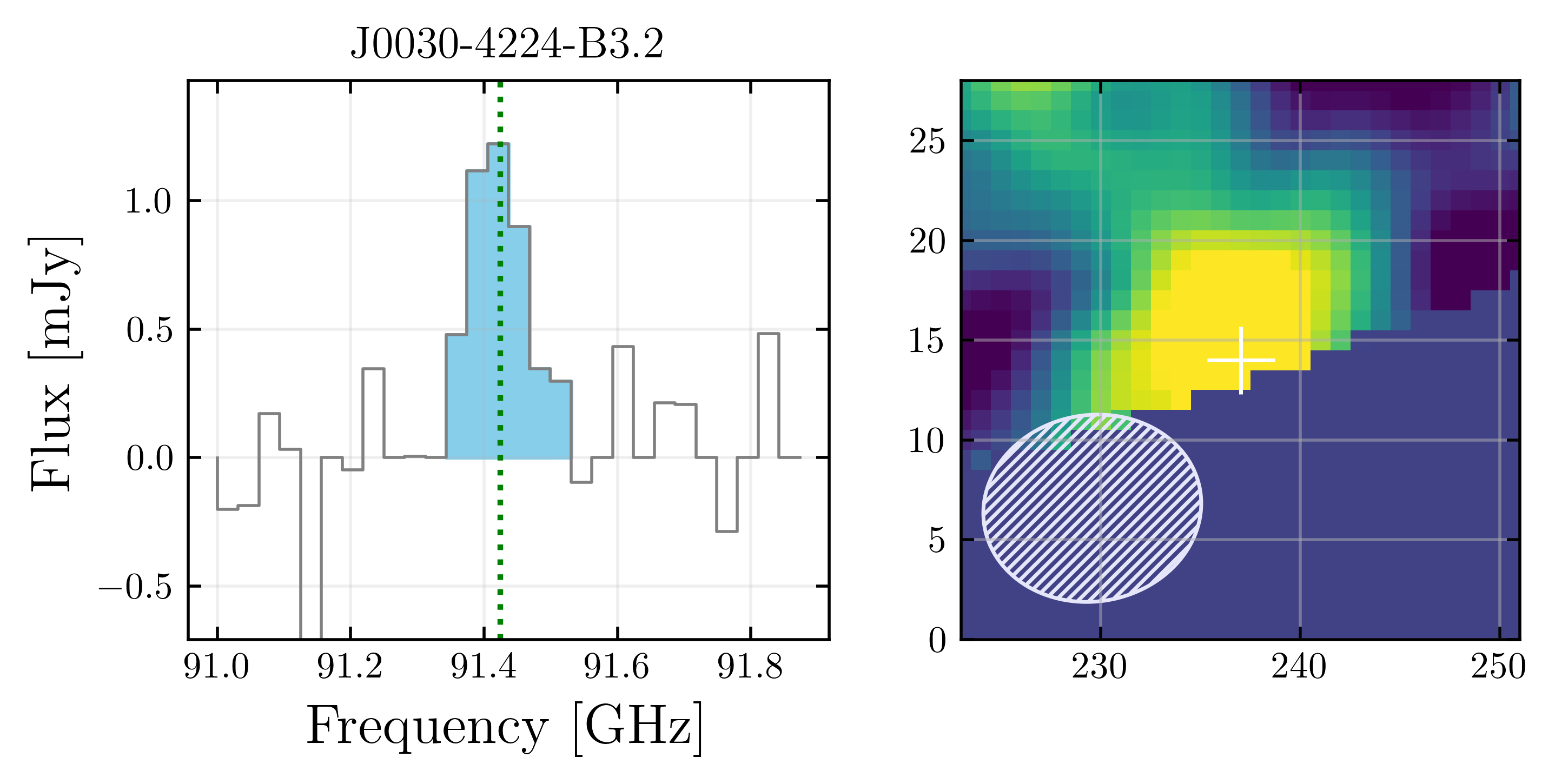}

    \includegraphics[width=0.32\textwidth]{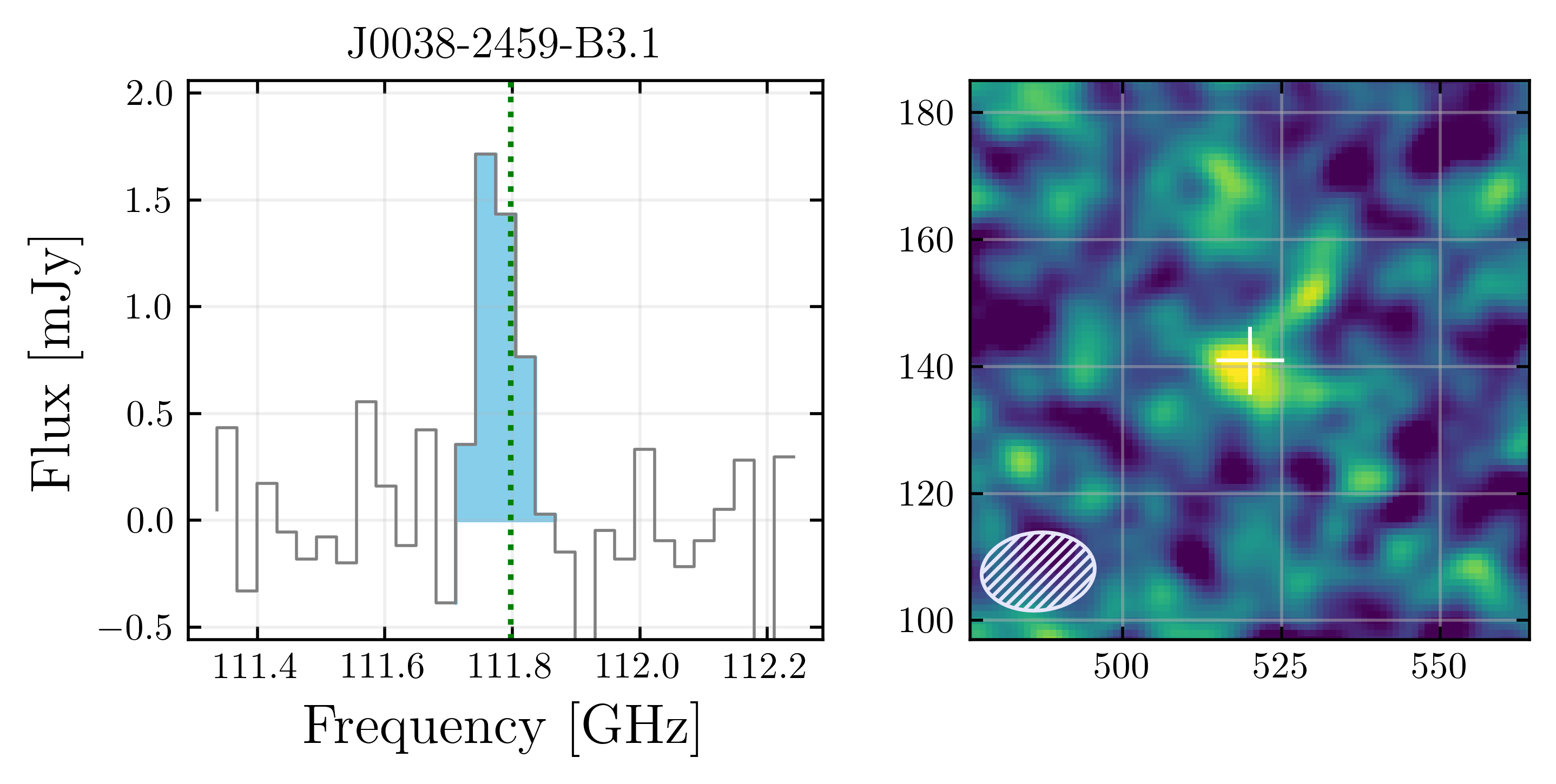}
    \includegraphics[width=0.32\textwidth]{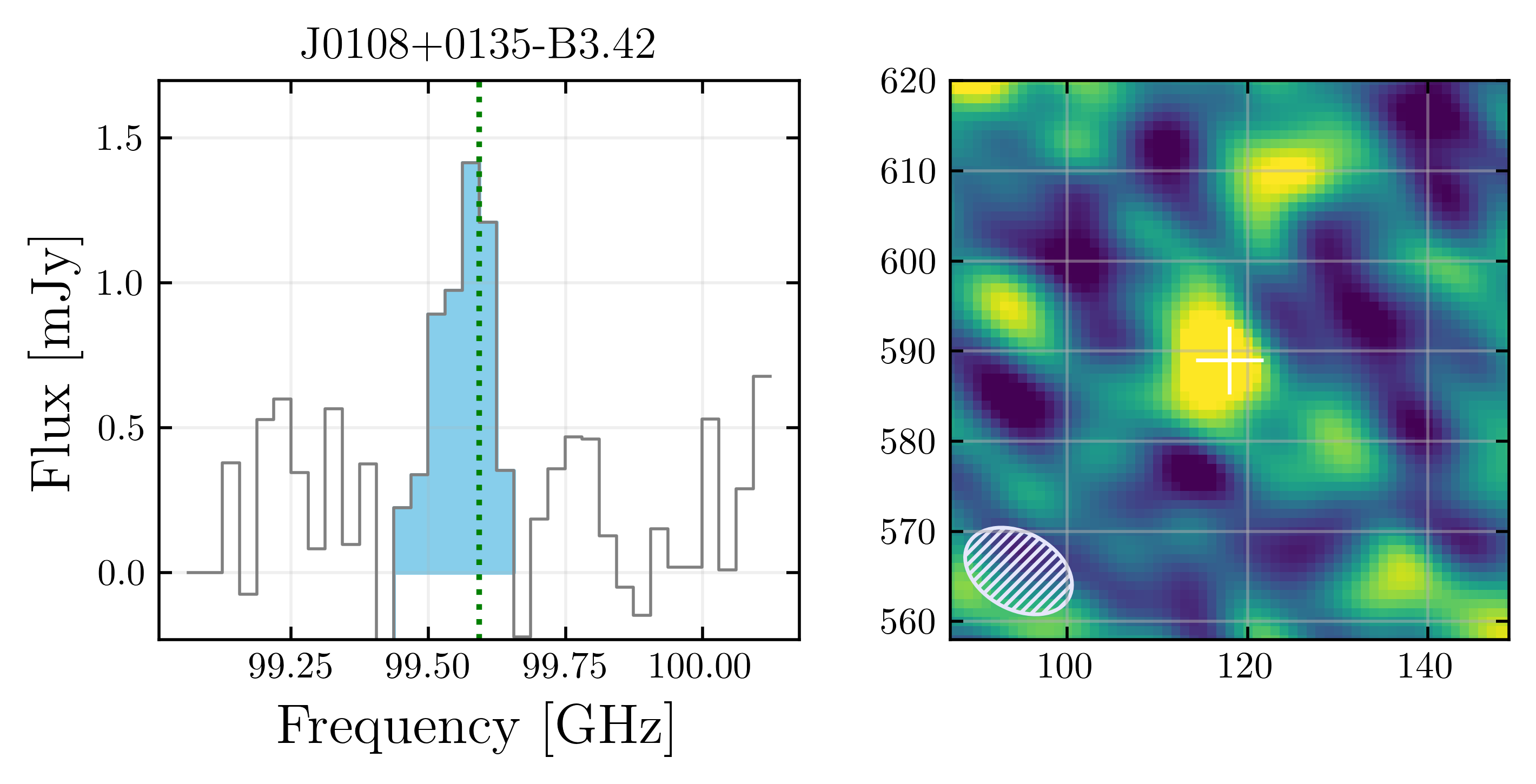}
    \includegraphics[width=0.32\textwidth]{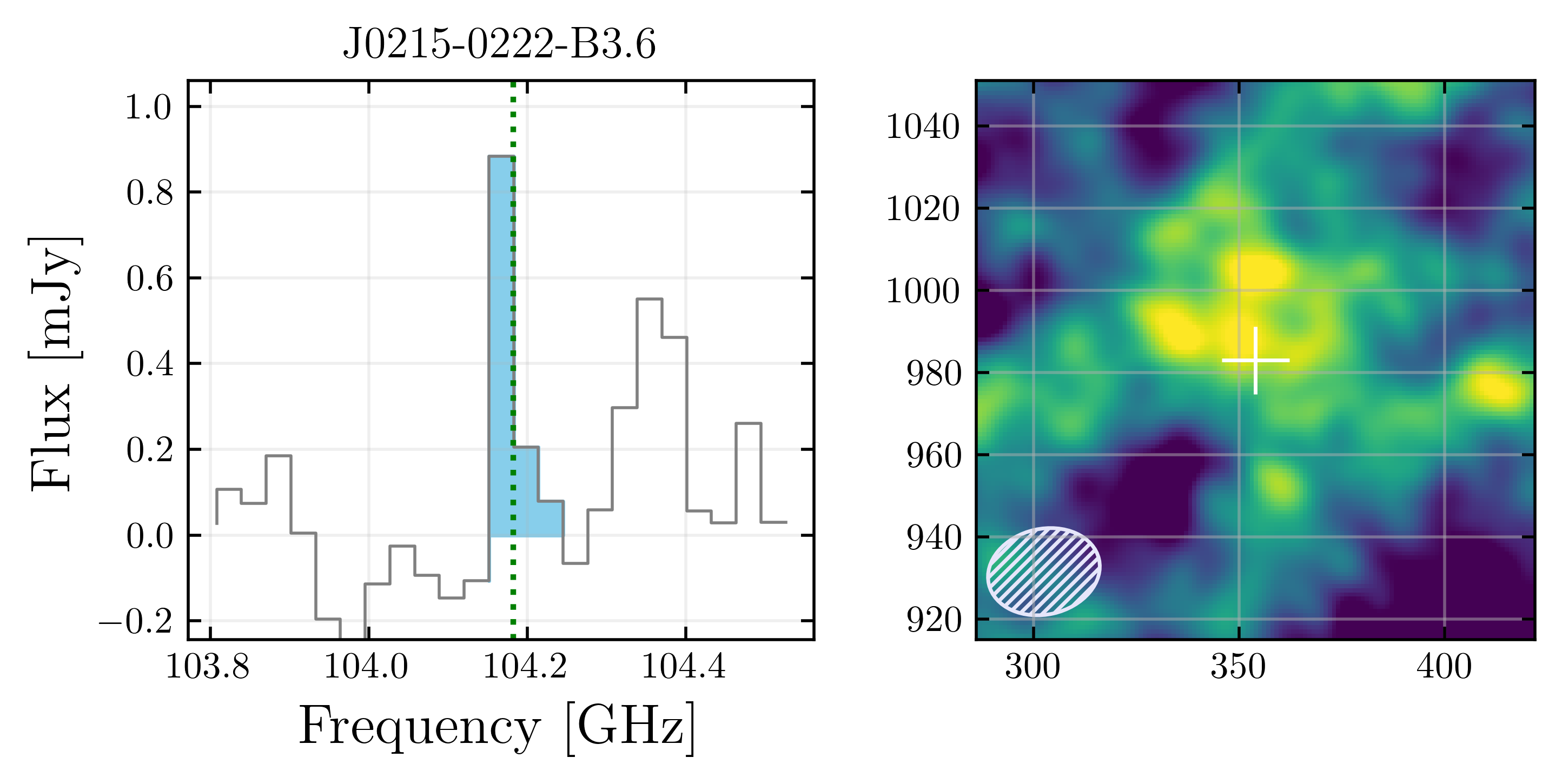}

    \includegraphics[width=0.32\textwidth]{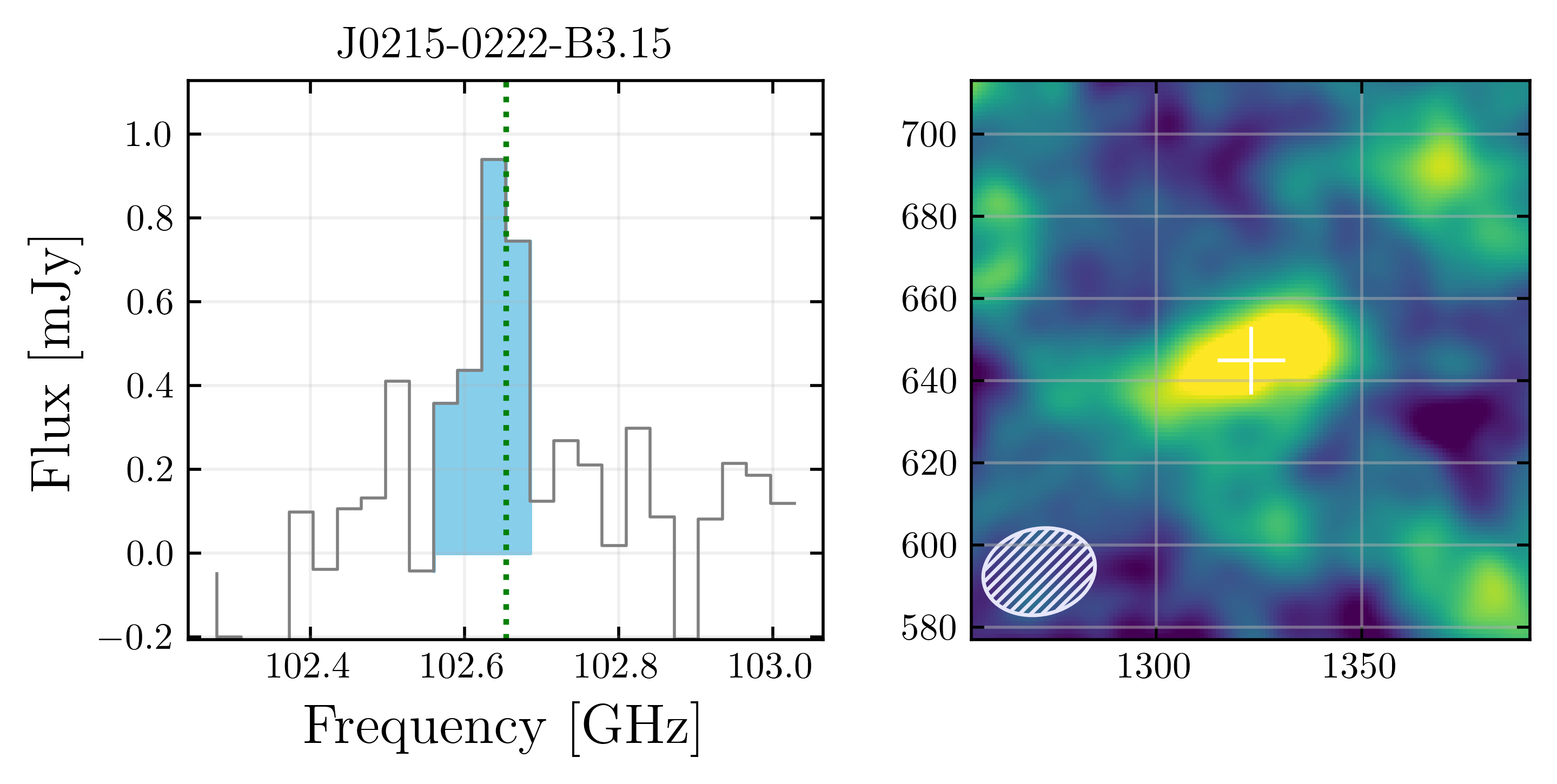}
    \includegraphics[width=0.32\textwidth]{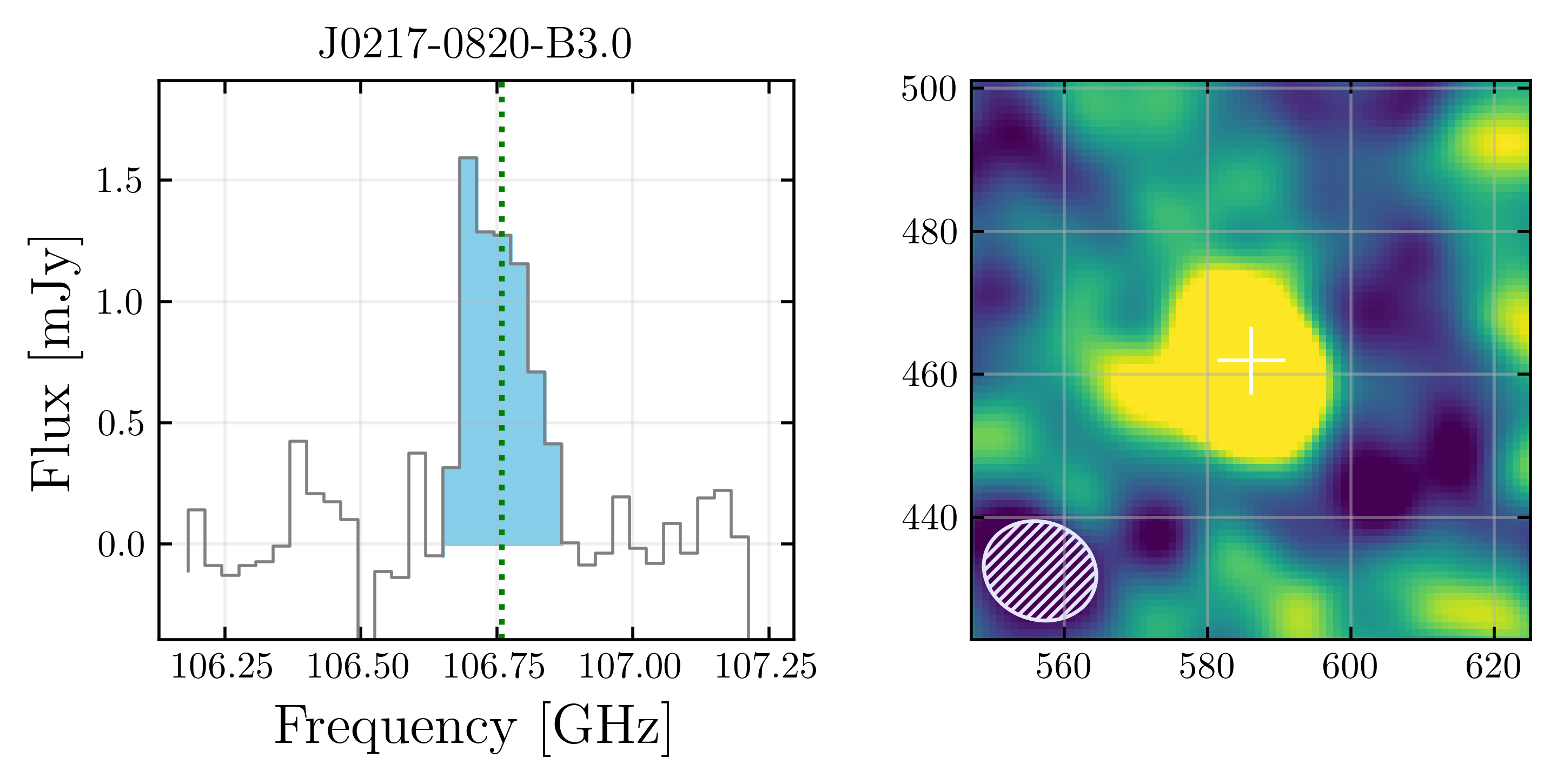}
    \includegraphics[width=0.32\textwidth]{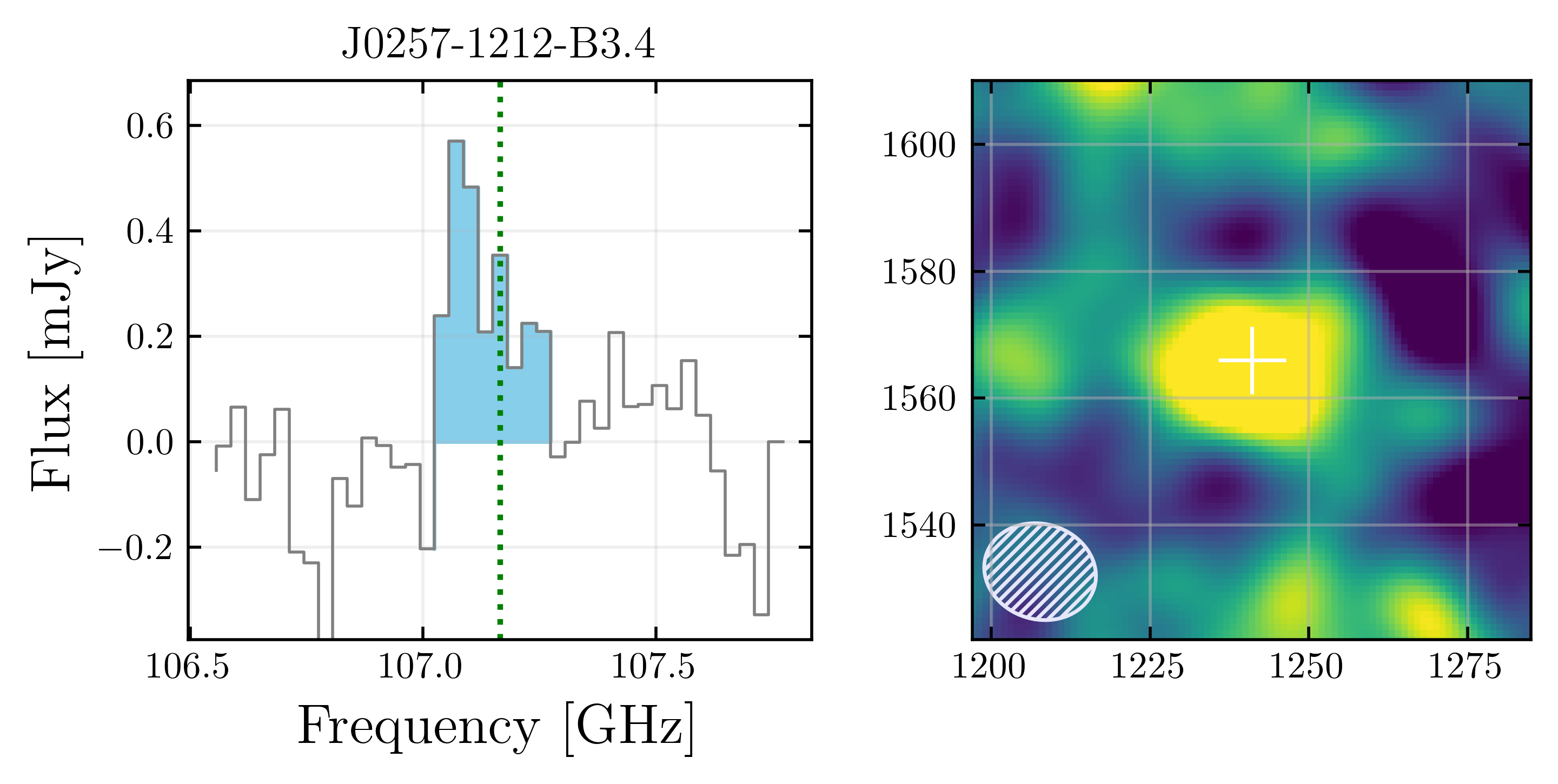}

    \includegraphics[width=0.32\textwidth]{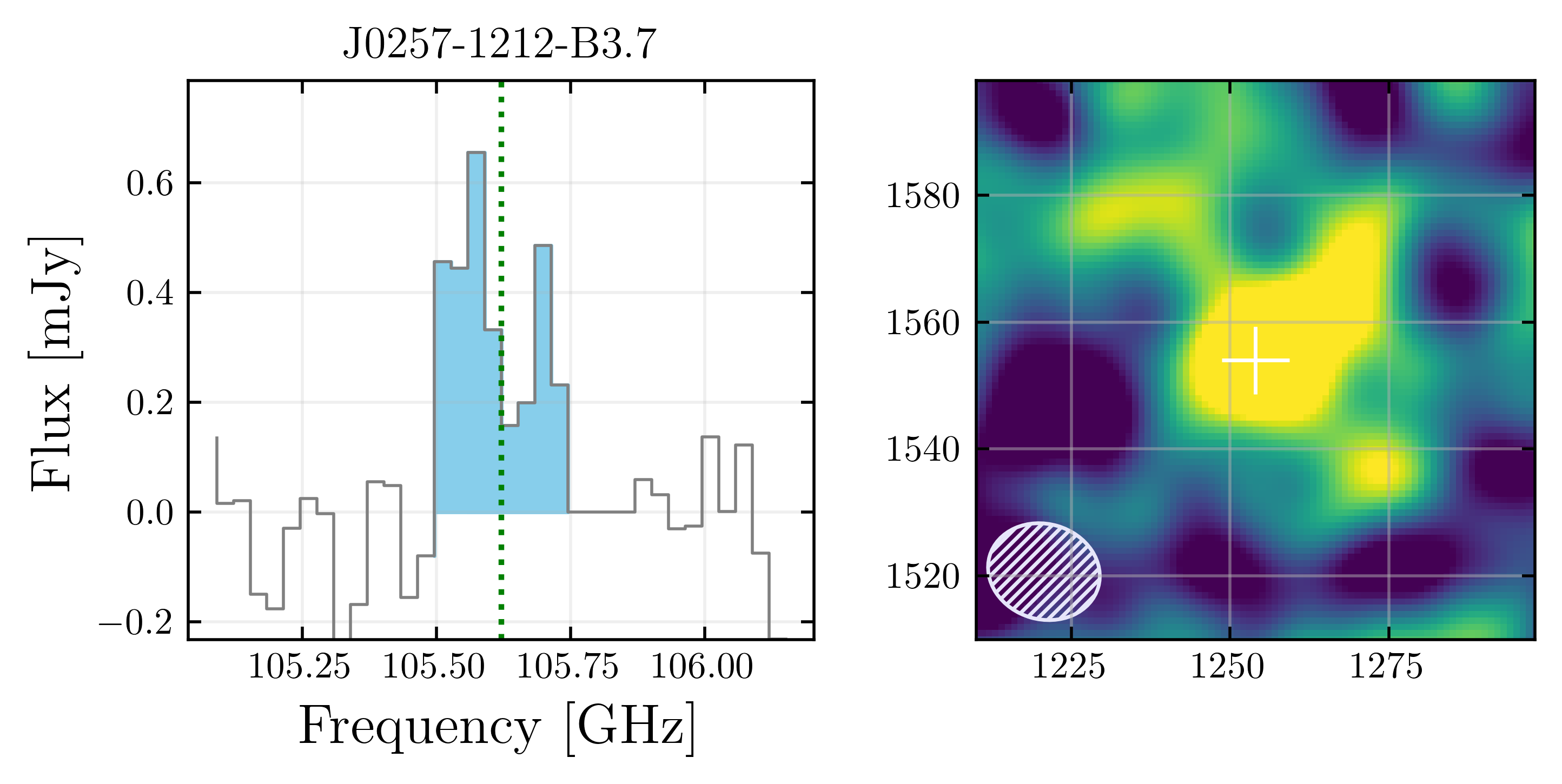}
    \includegraphics[width=0.32\textwidth]{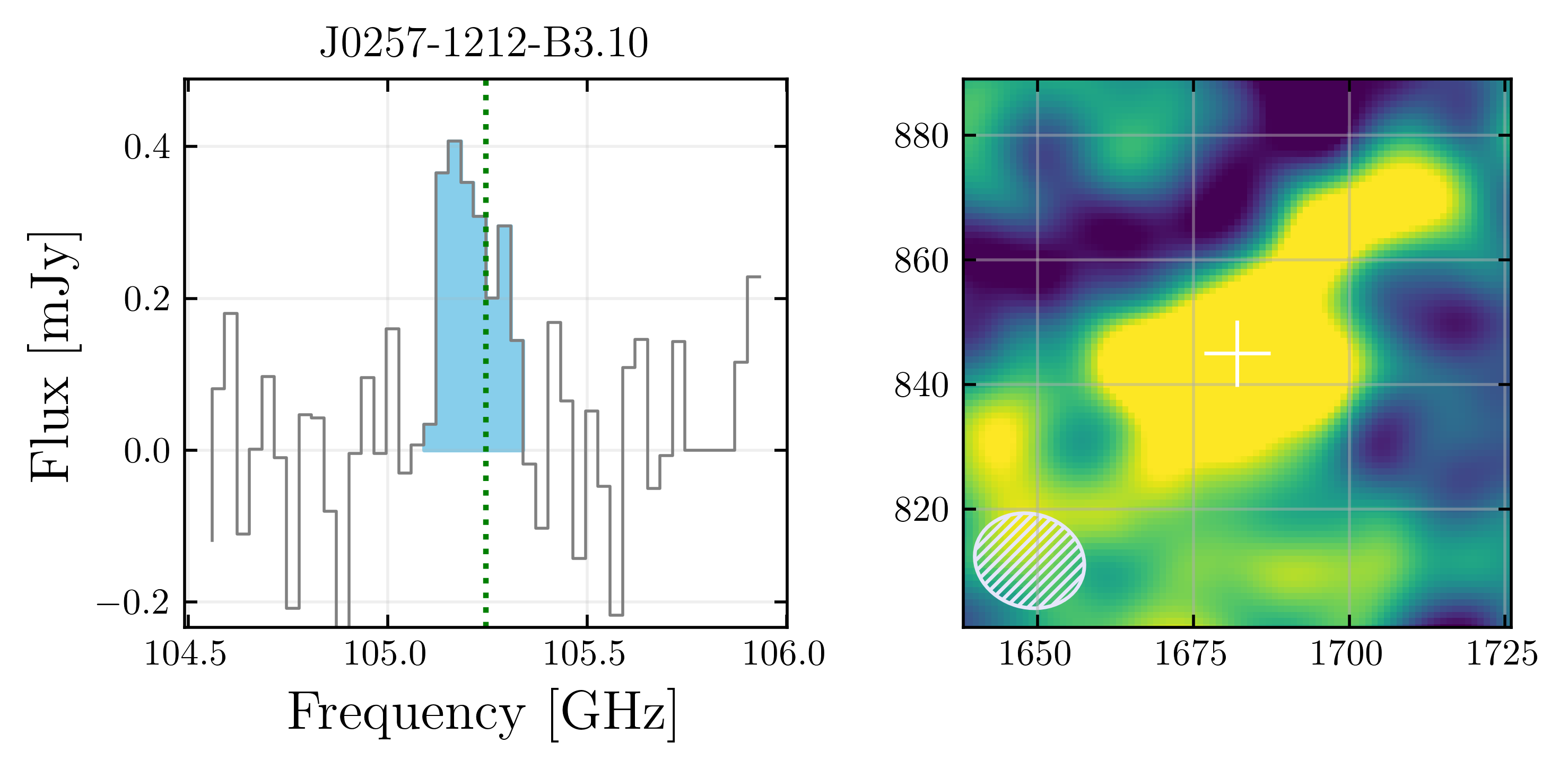}
    \includegraphics[width=0.32\textwidth]{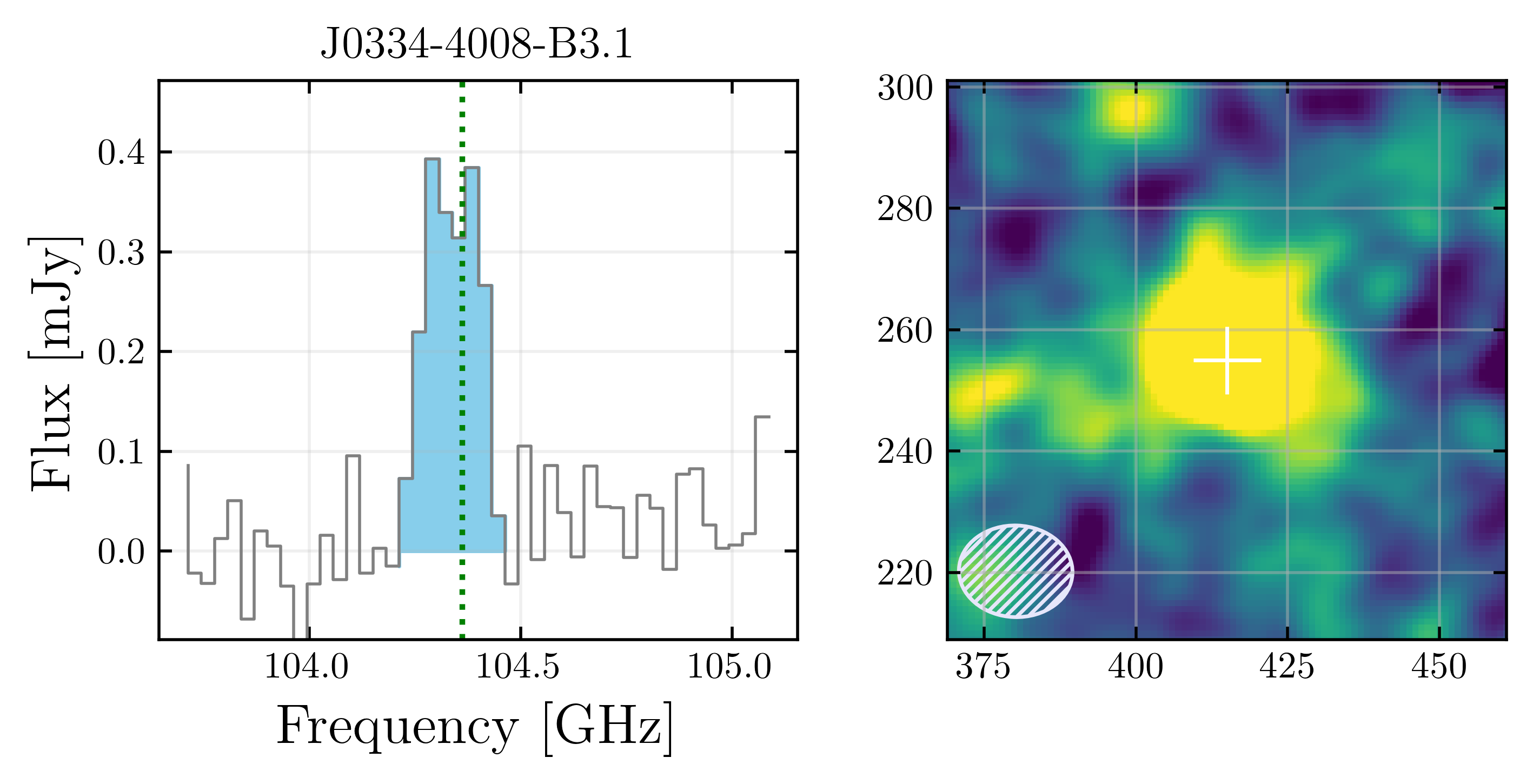}

    \includegraphics[width=0.32\textwidth]{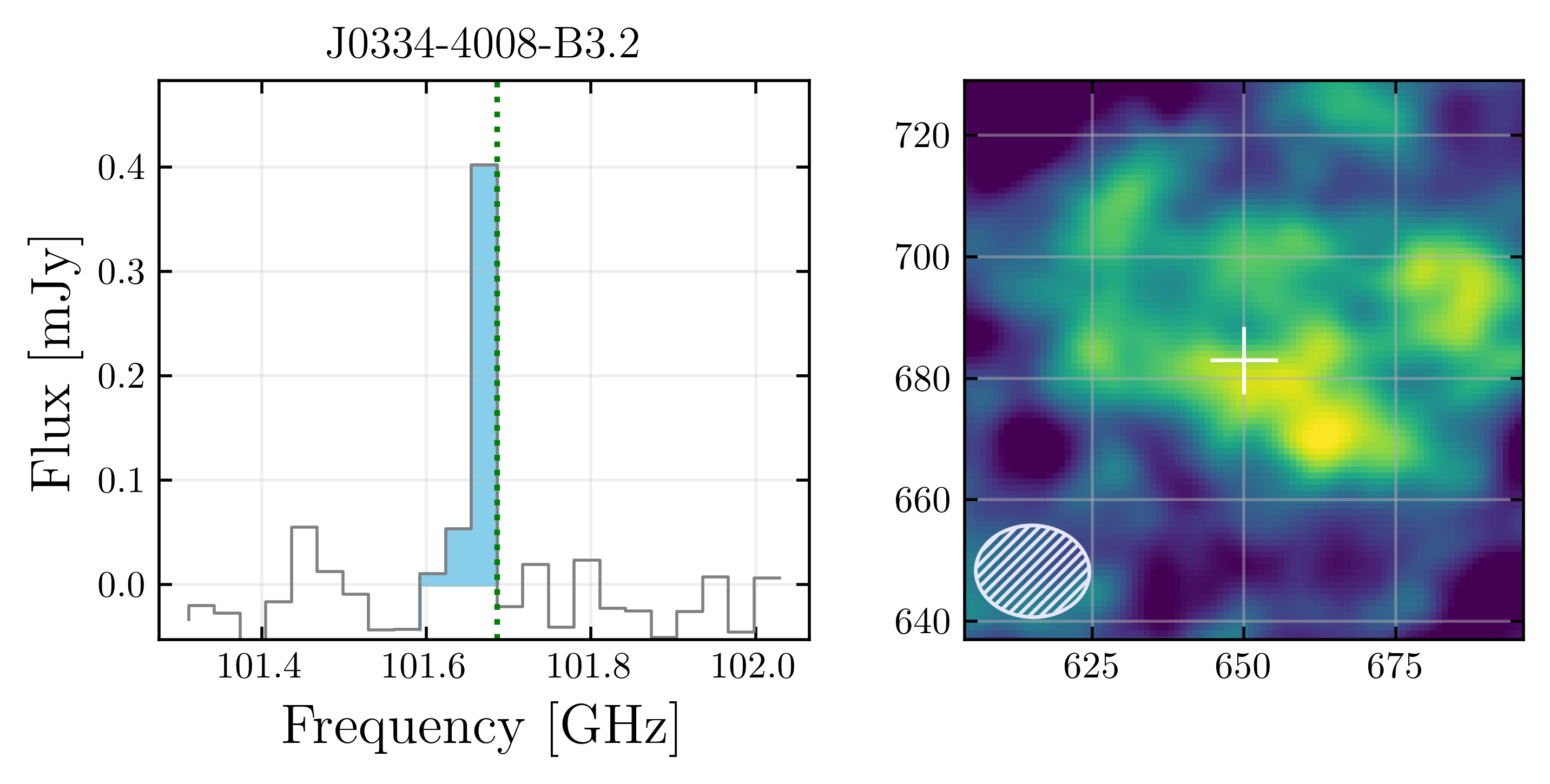}
    \includegraphics[width=0.32\textwidth]{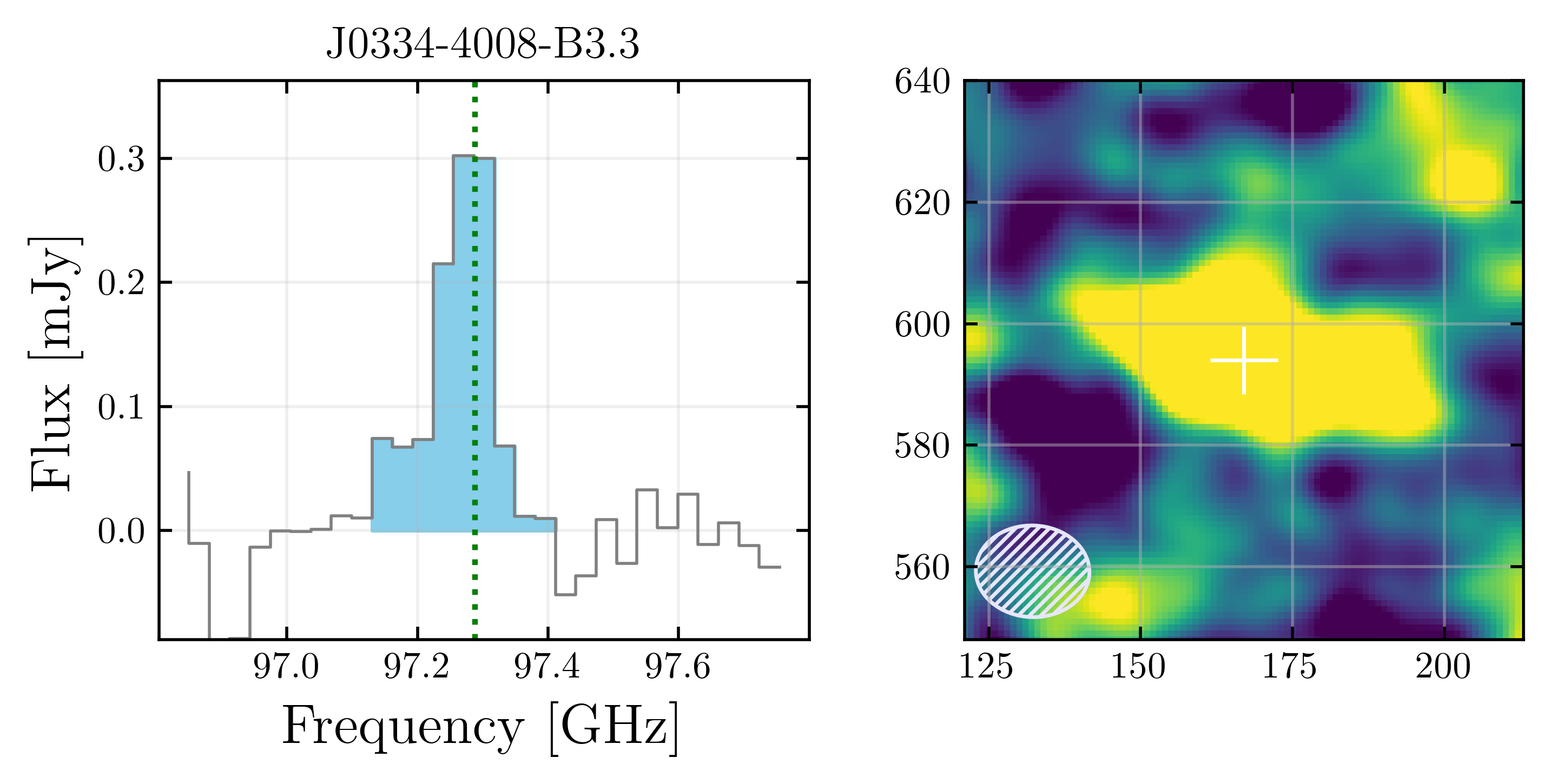}
    \includegraphics[width=0.32\textwidth]{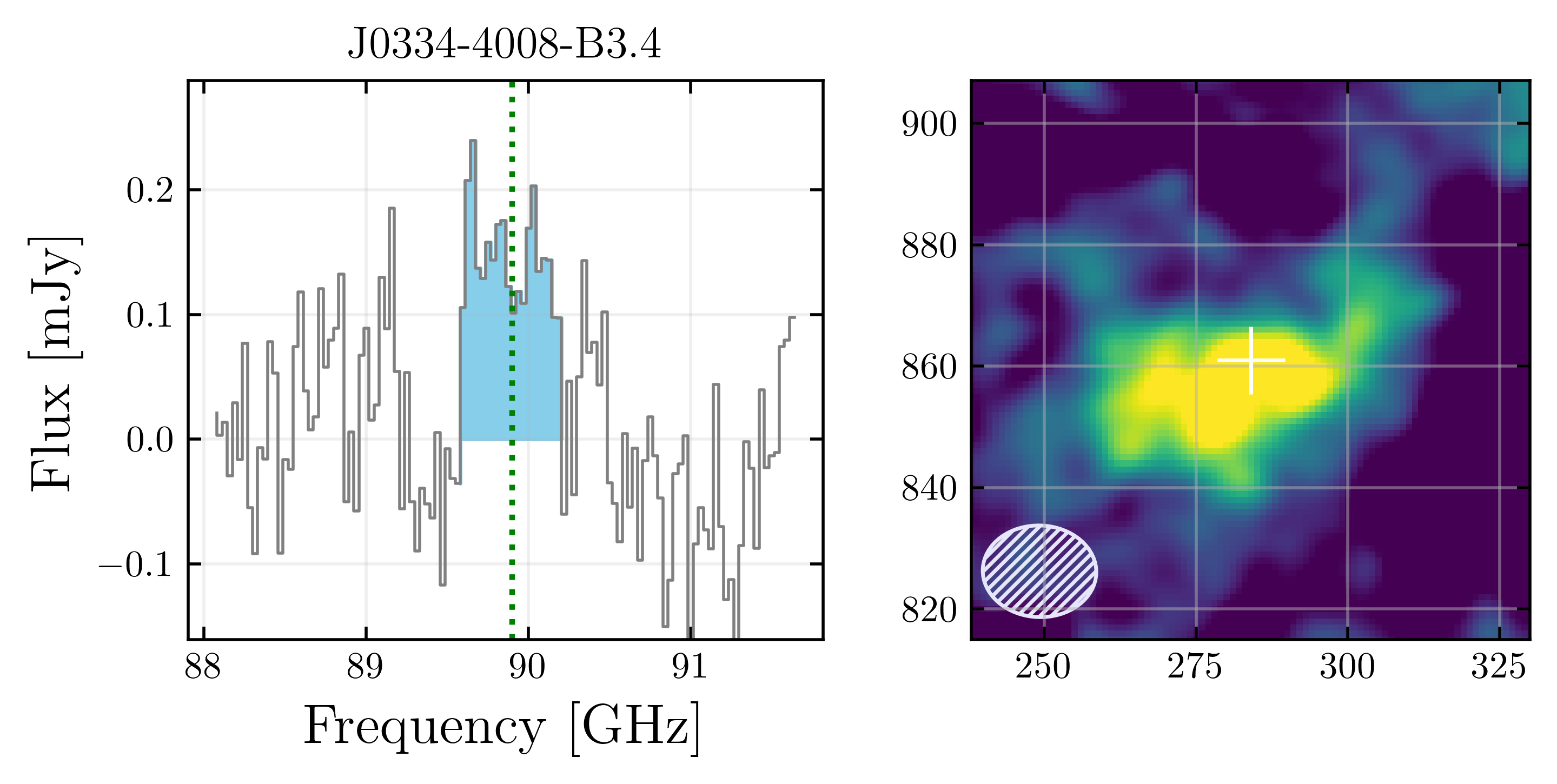}
    
    \includegraphics[width=0.32\textwidth]{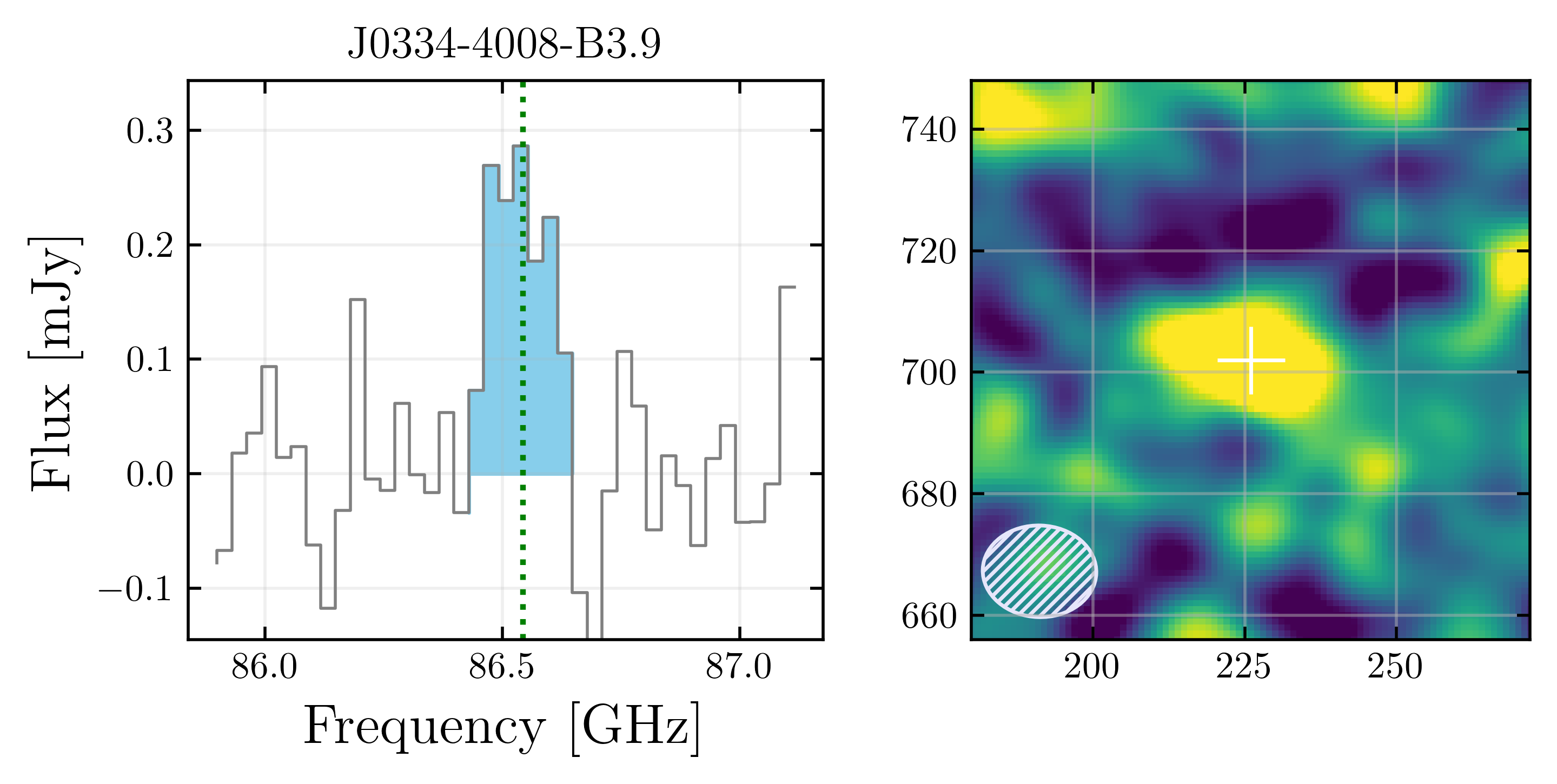}
    \includegraphics[width=0.32\textwidth]{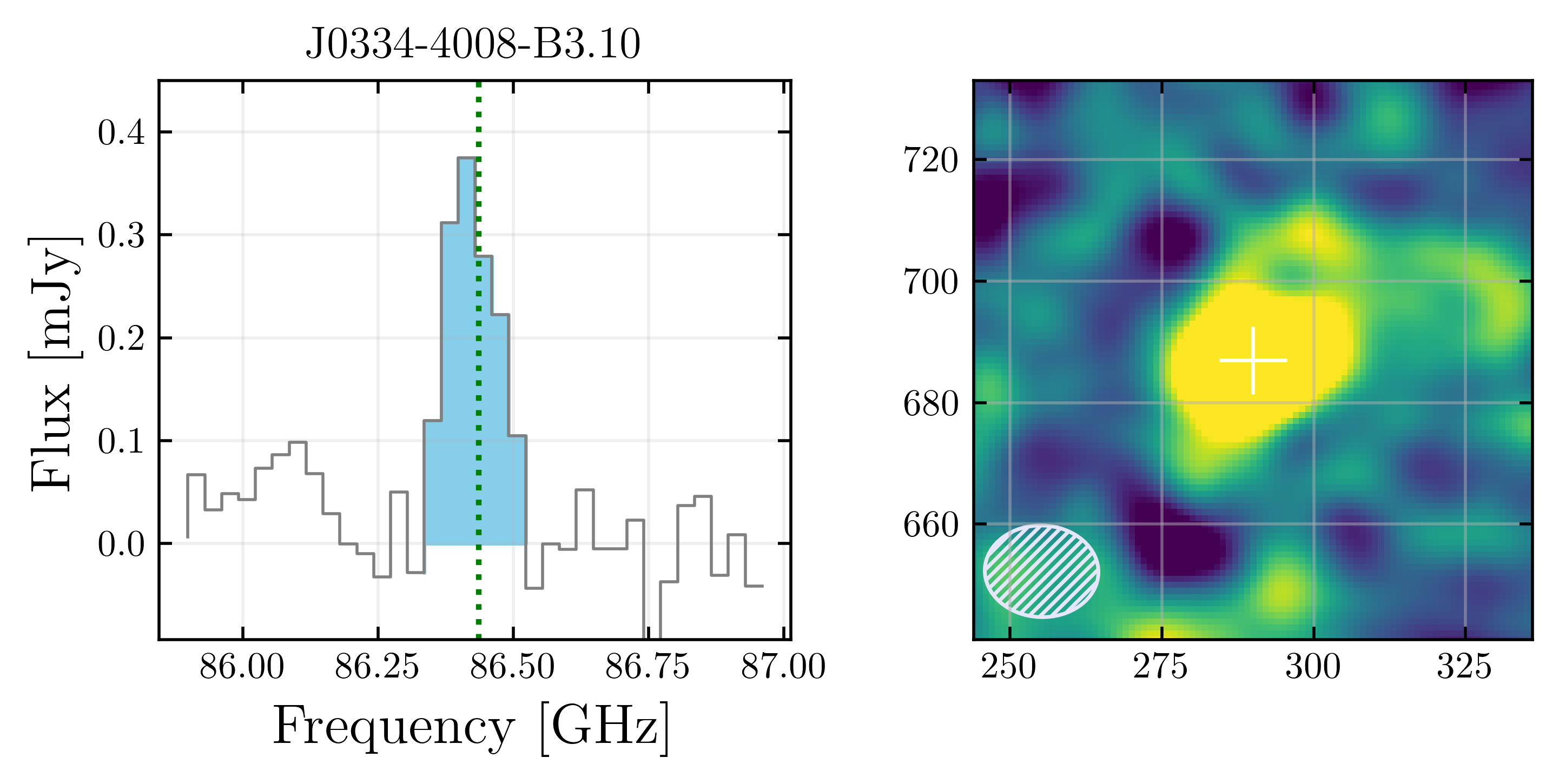}
    \includegraphics[width=0.32\textwidth]{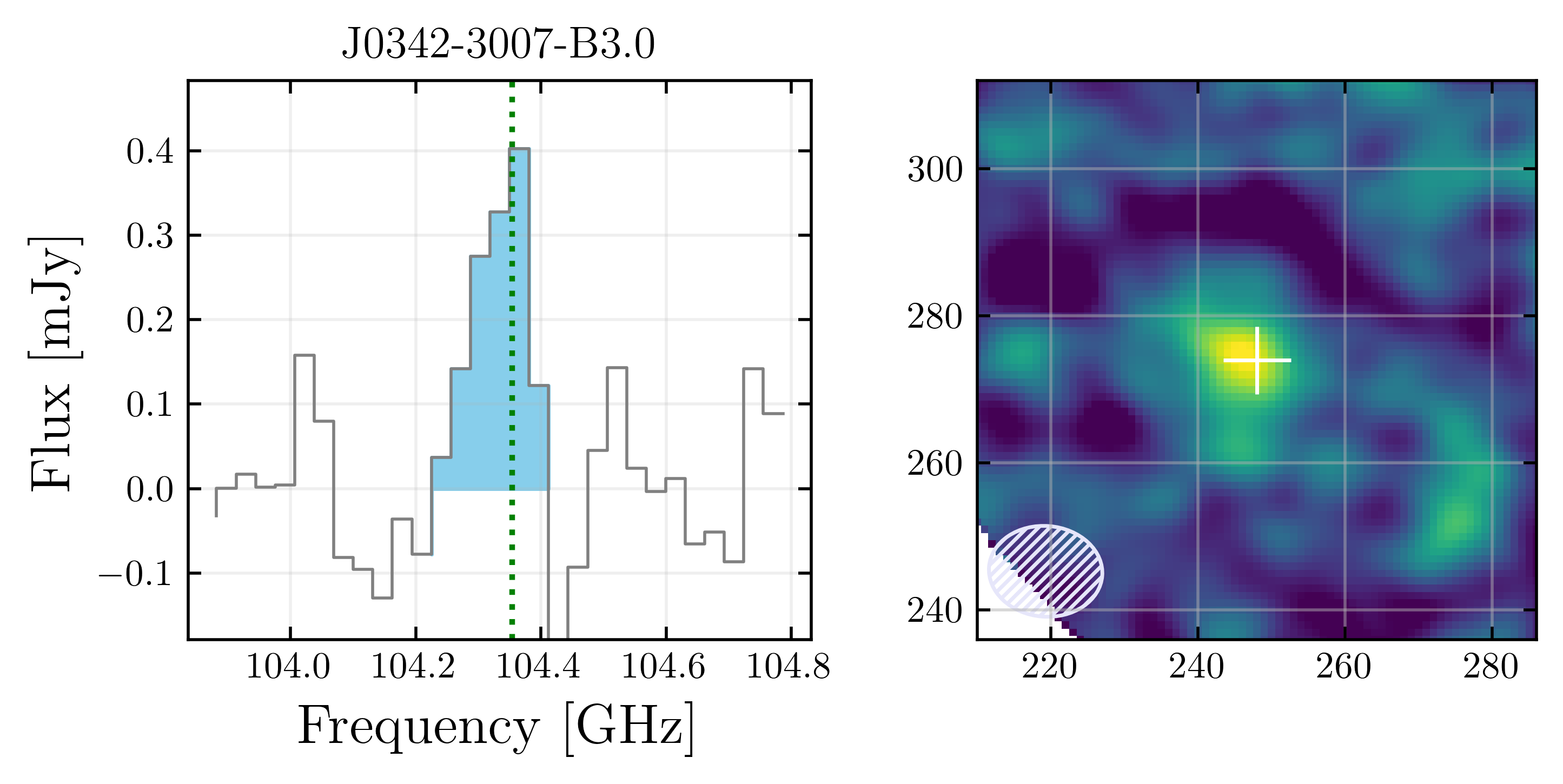}

    \includegraphics[width=0.32\textwidth]{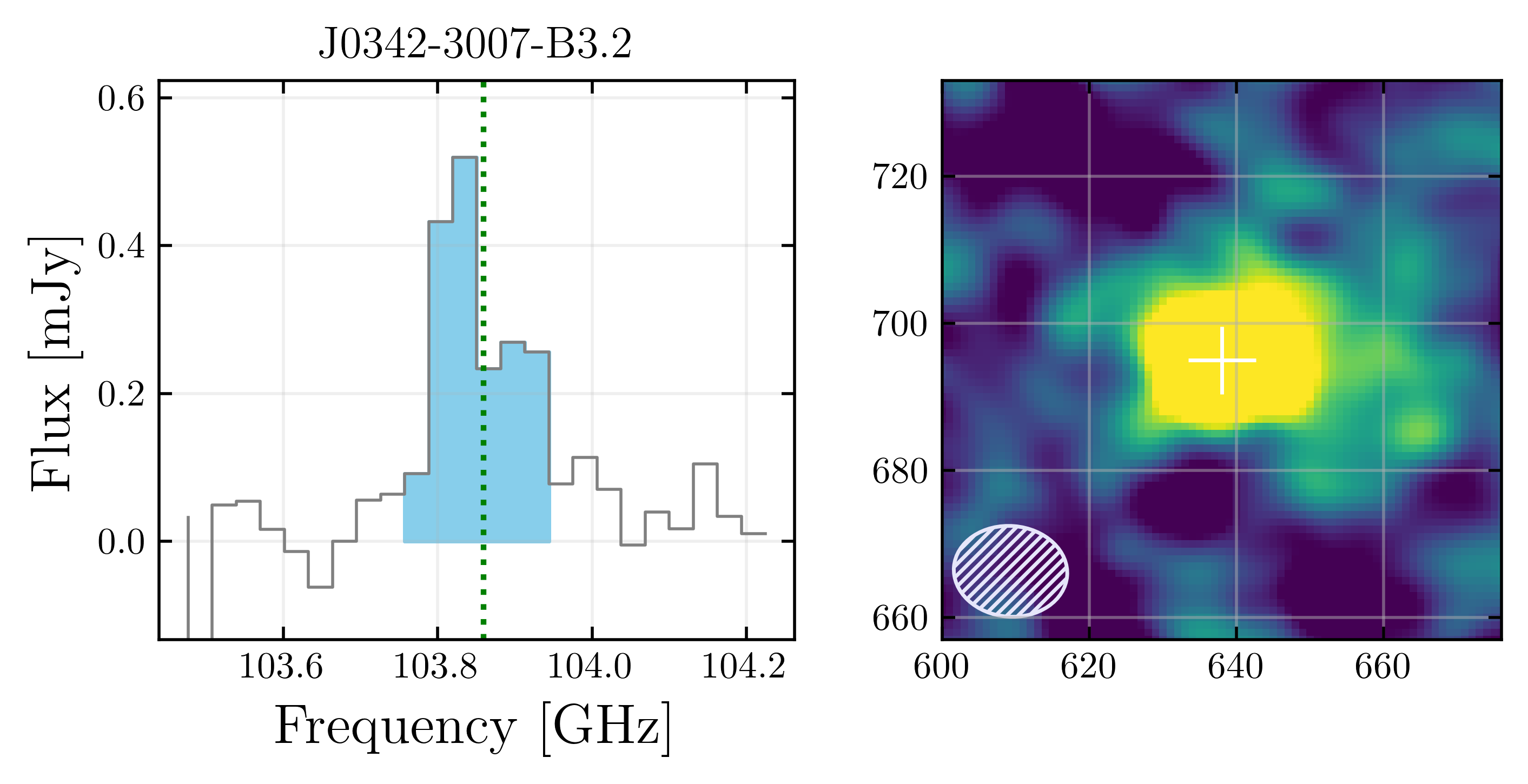}
    \includegraphics[width=0.32\textwidth]{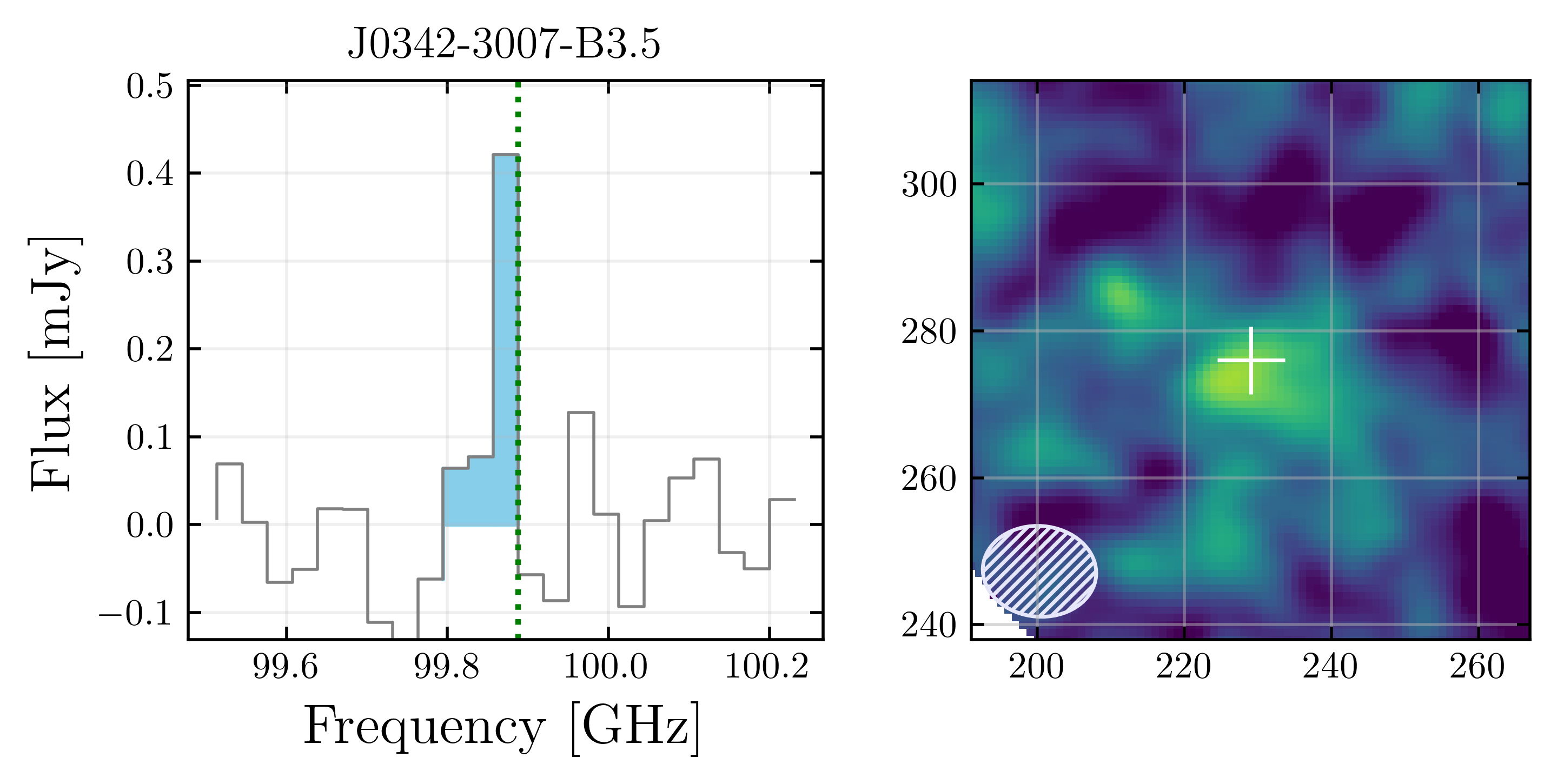}
    \includegraphics[width=0.32\textwidth]{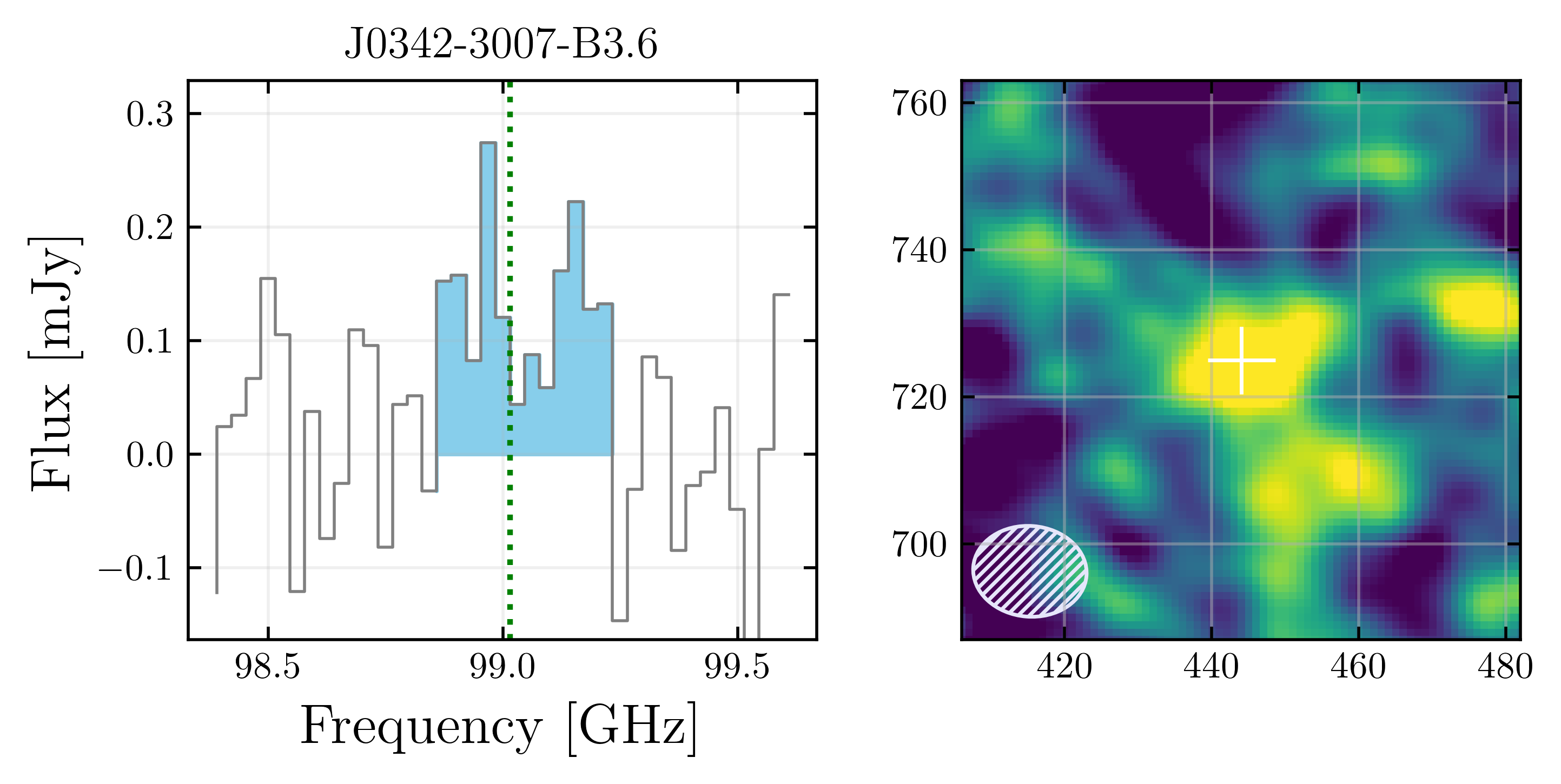}

    \caption{ALMACAL$-22$ emission lines detections. Left panel: Spectrum of the emission lines at the brightest pixel. Right panel: The emission line moment map centred on the detection range. The parameters of all candidates are summarised in Table \ref{tab:detections}}
    \label{fig:detections1}
\end{figure}

\begin{figure}
\centering
     \includegraphics[width=0.32\textwidth]{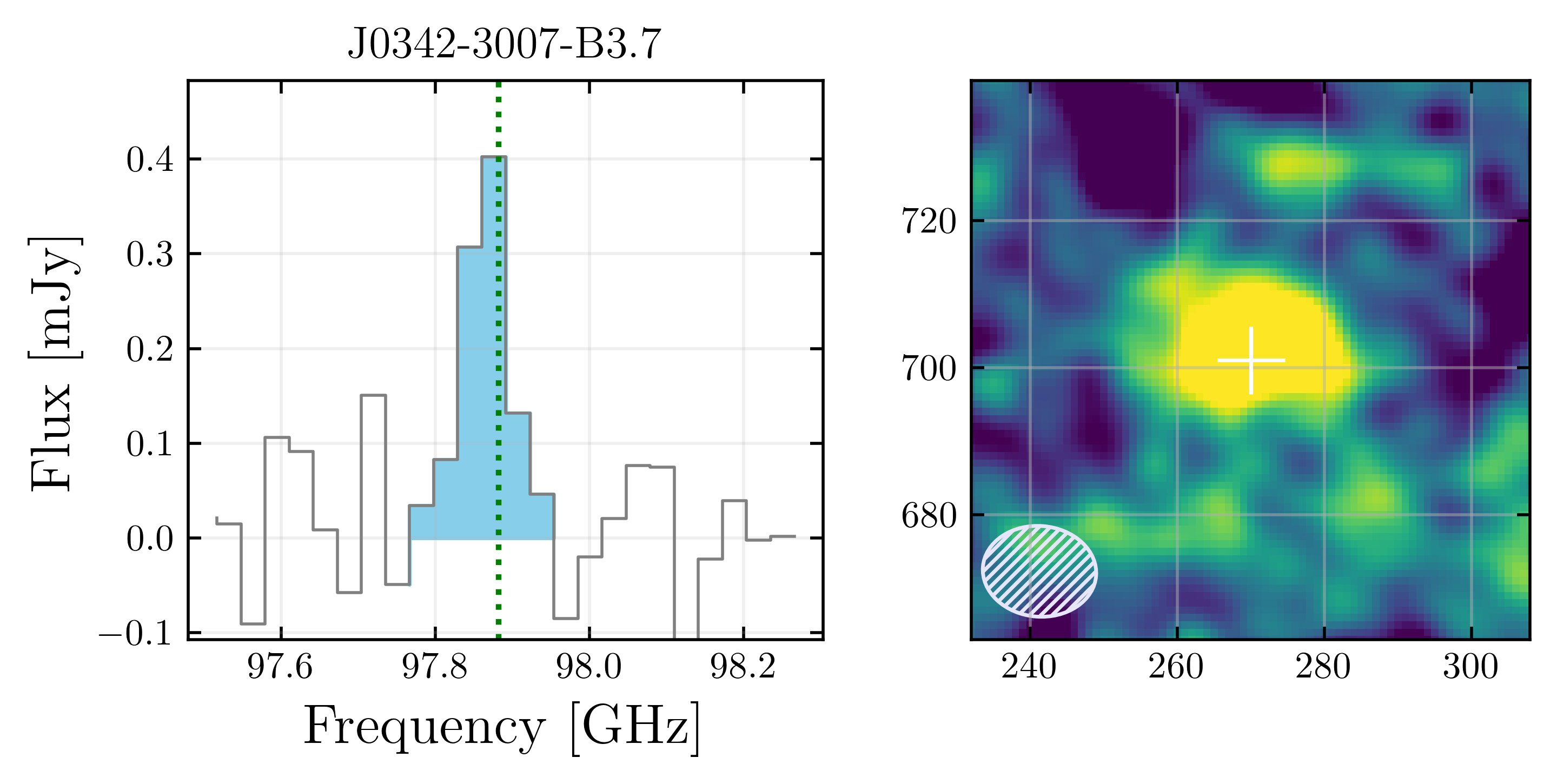}
    \includegraphics[width=0.32\textwidth]{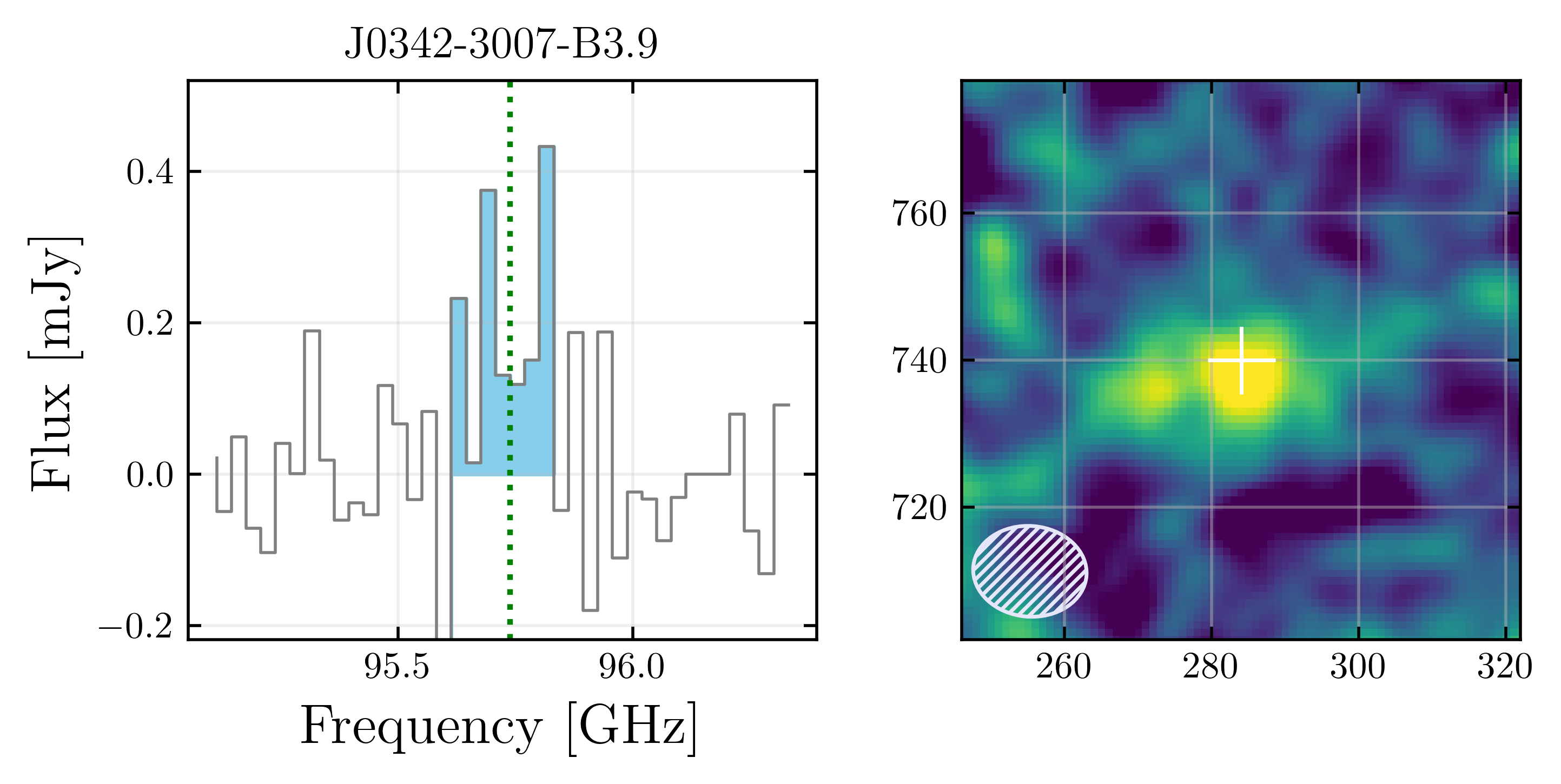}
    \includegraphics[width=0.32\textwidth]{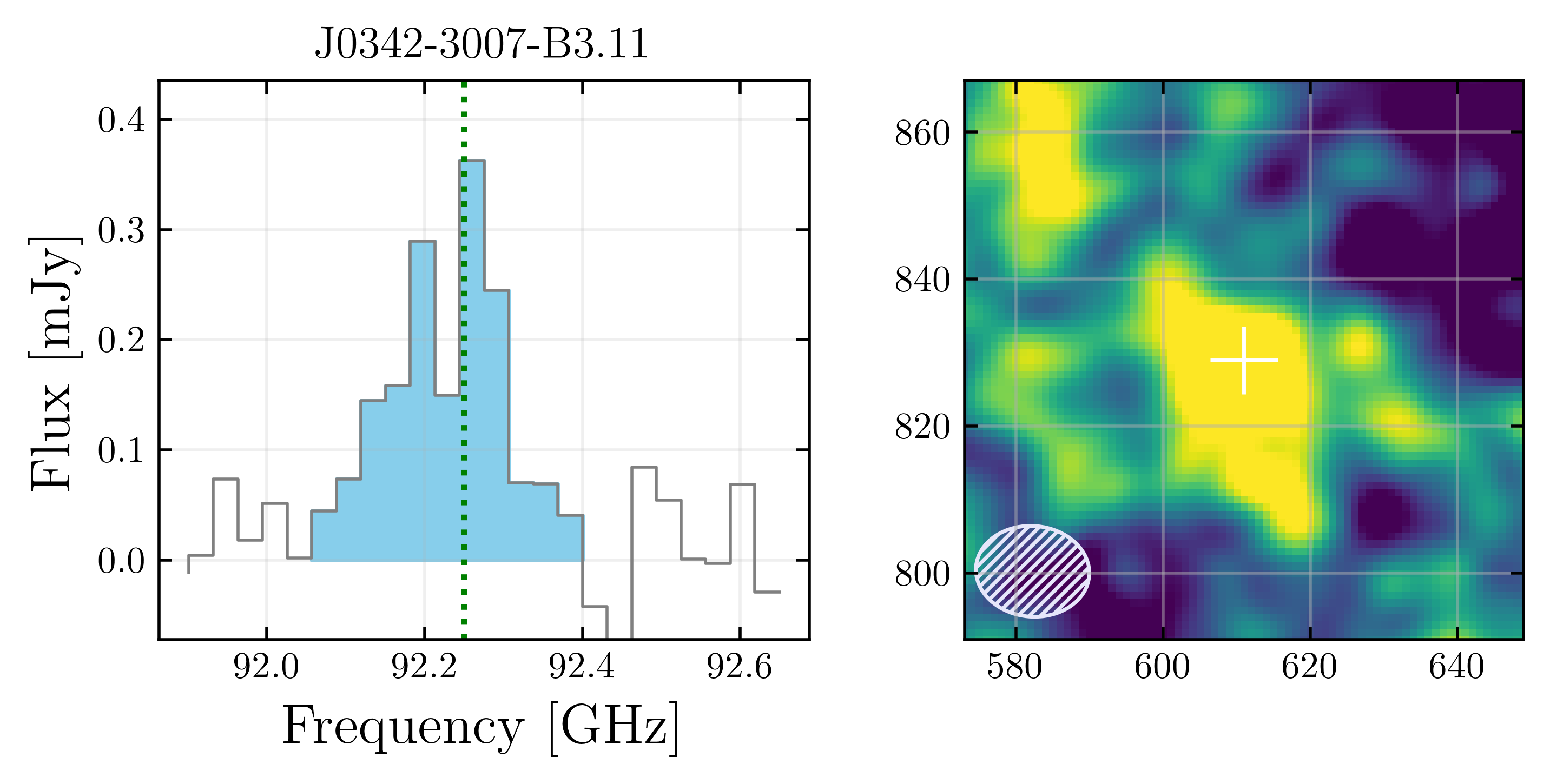}

    \includegraphics[width=0.32\textwidth]{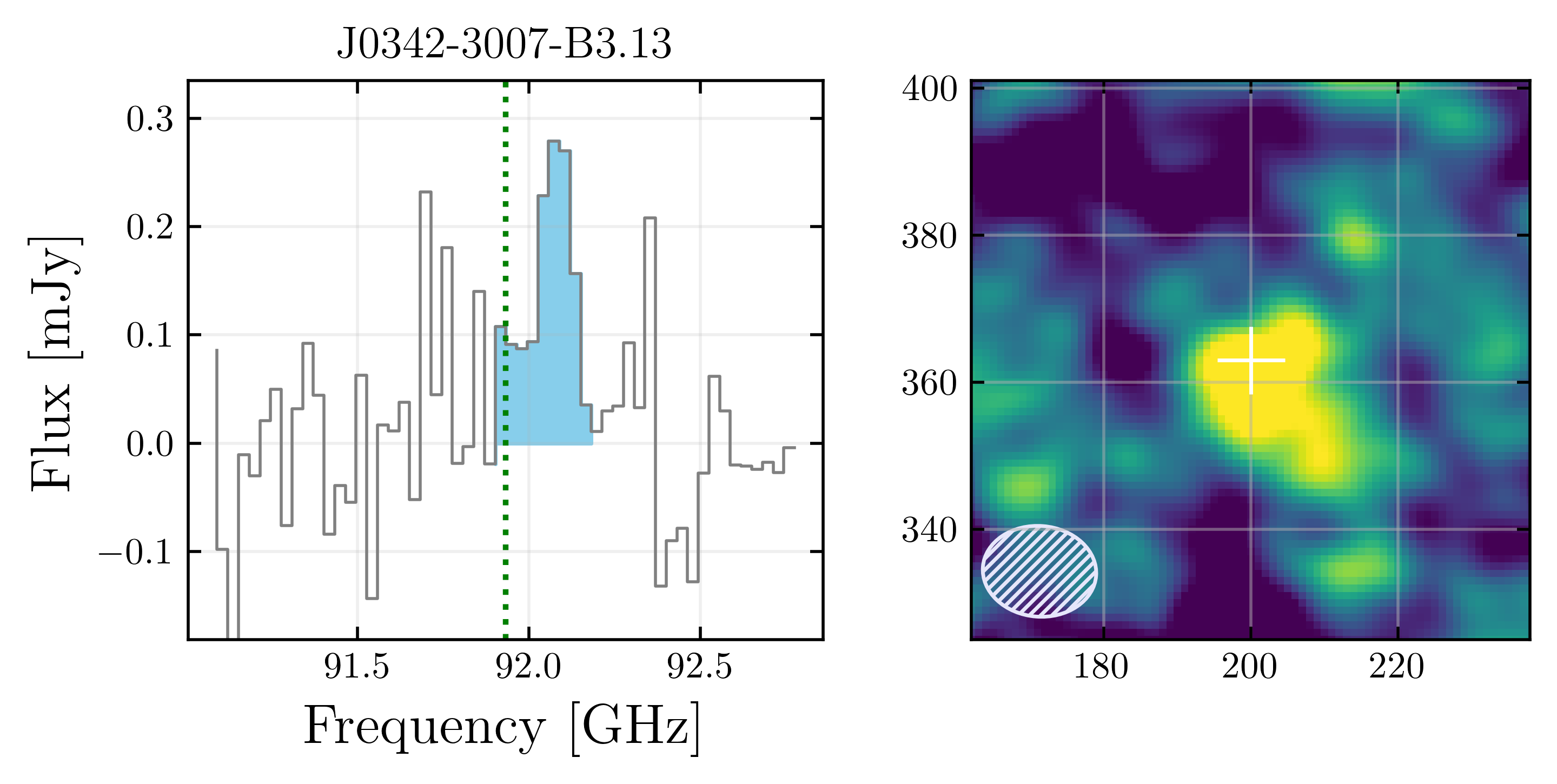}
    \includegraphics[width=0.32\textwidth]{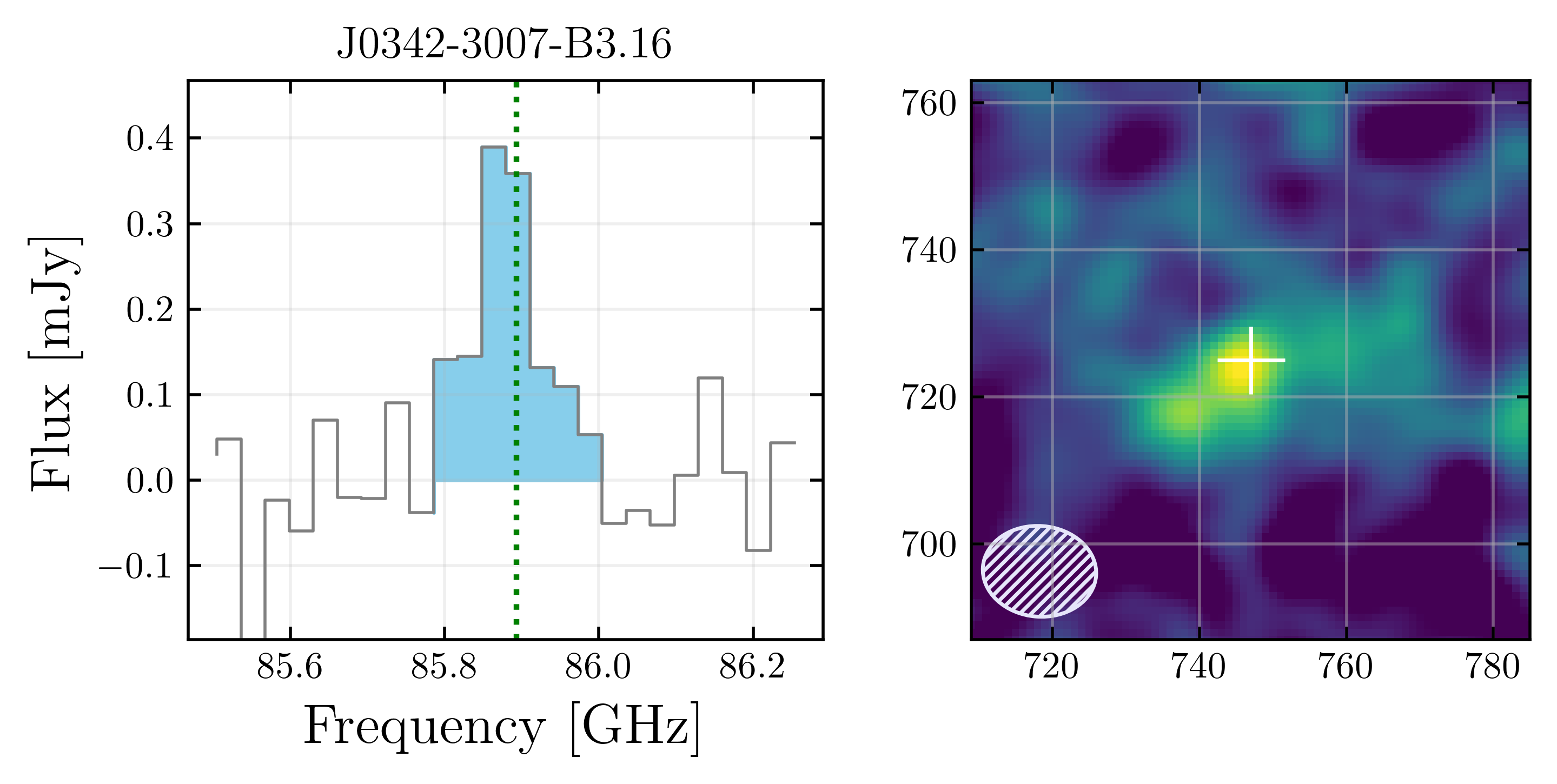}
    \includegraphics[width=0.32\textwidth]{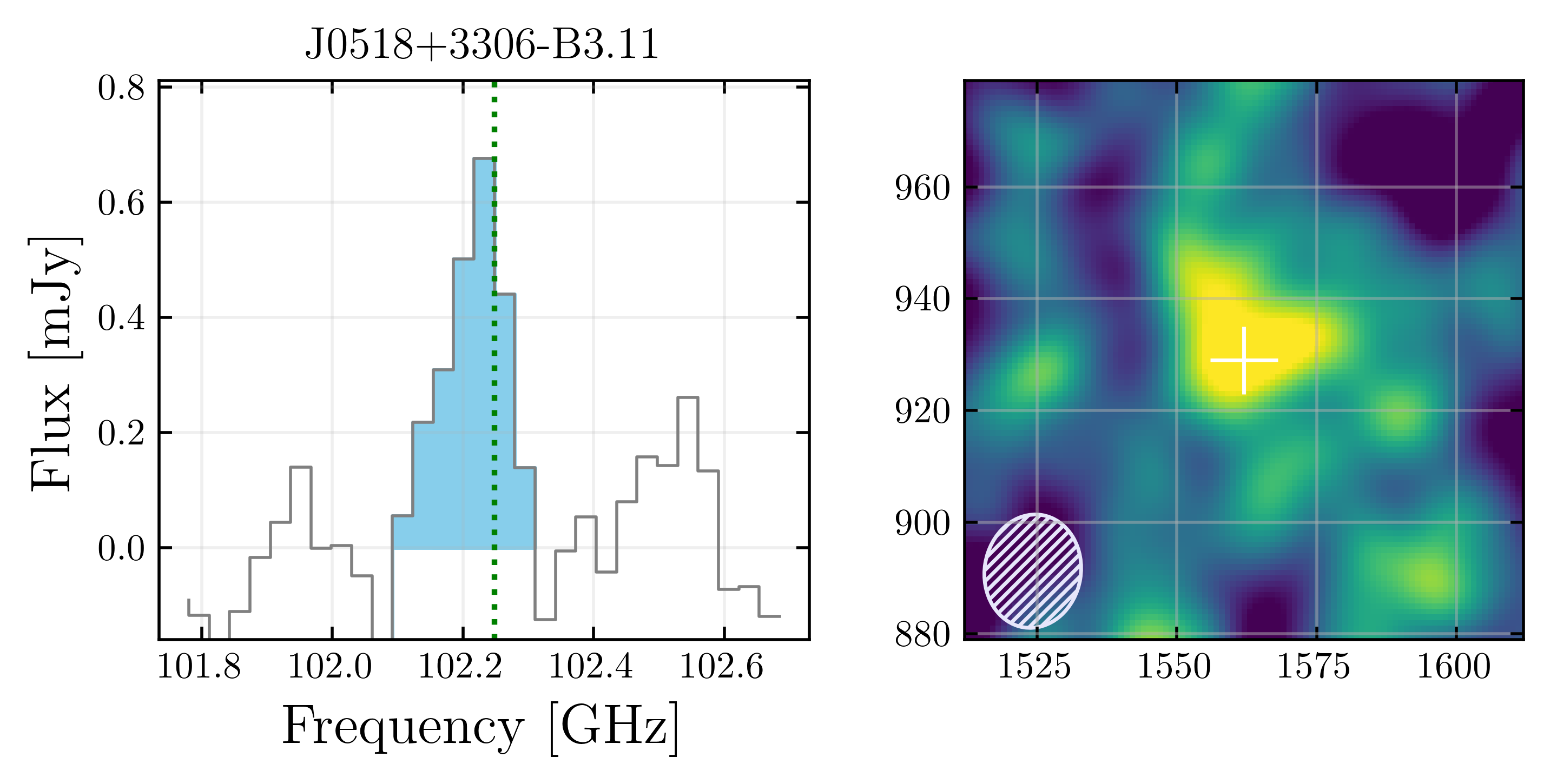}

    \includegraphics[width=0.32\textwidth]{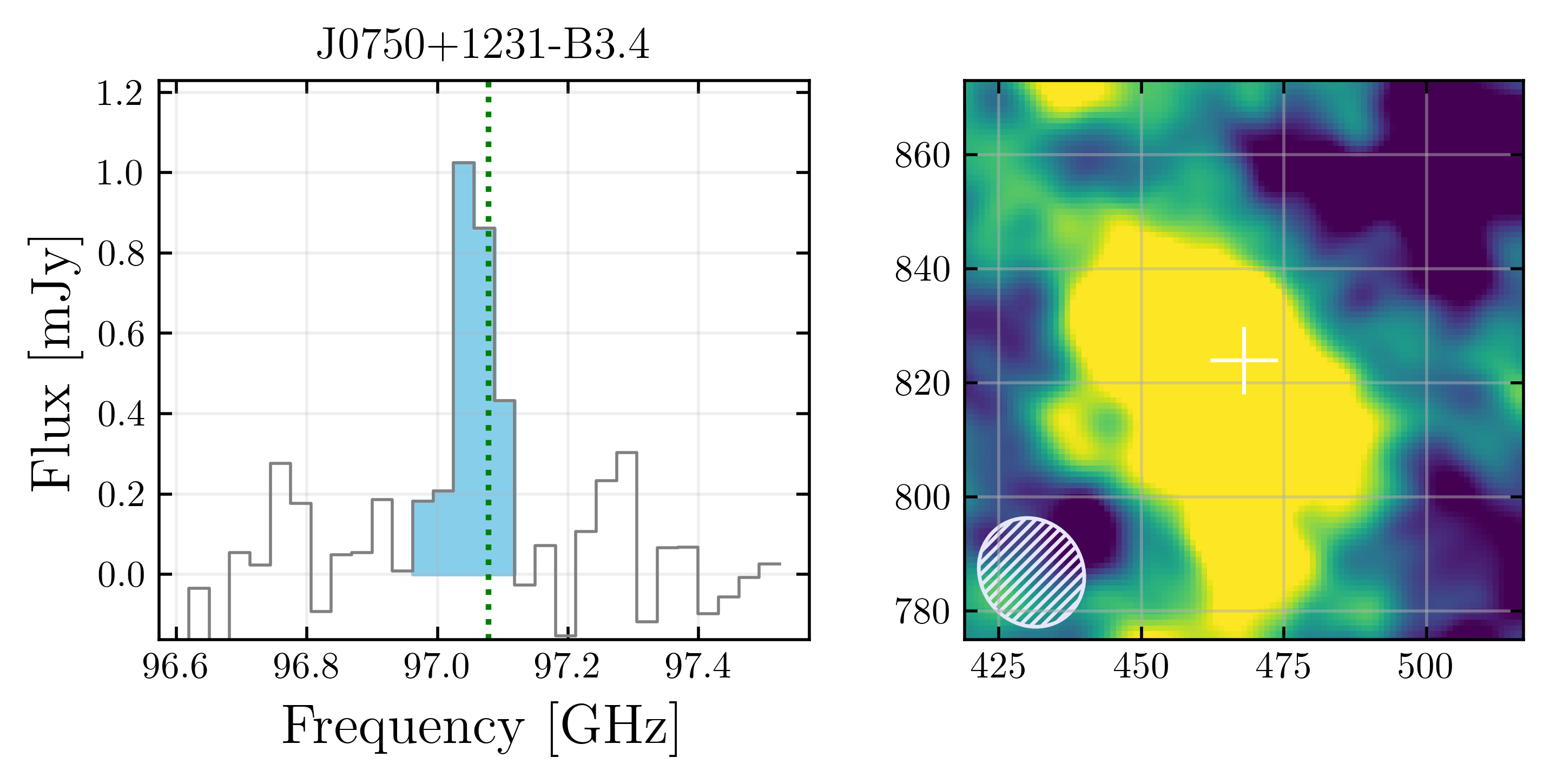}
    \includegraphics[width=0.32\textwidth]{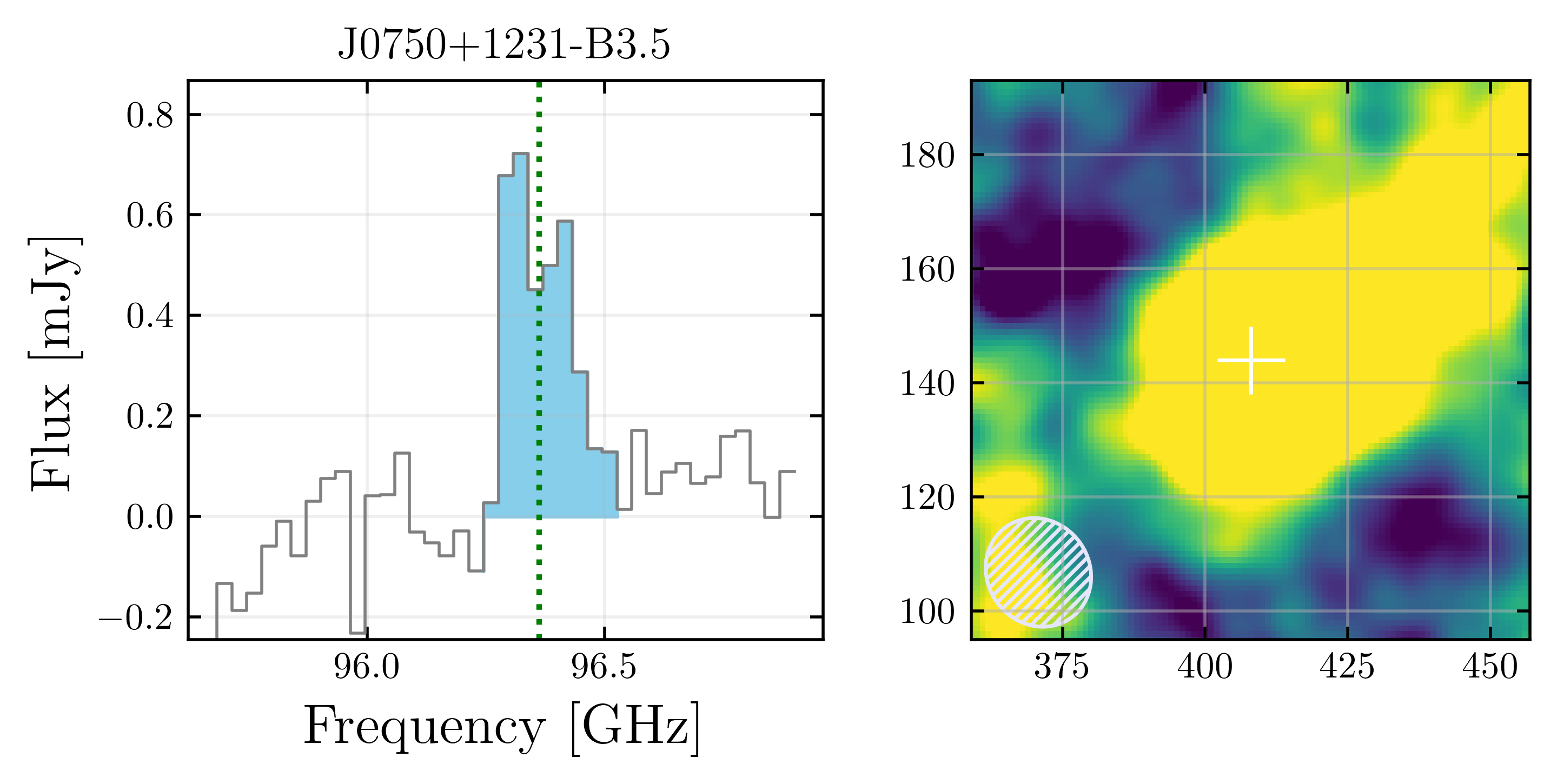}
    \includegraphics[width=0.32\textwidth]{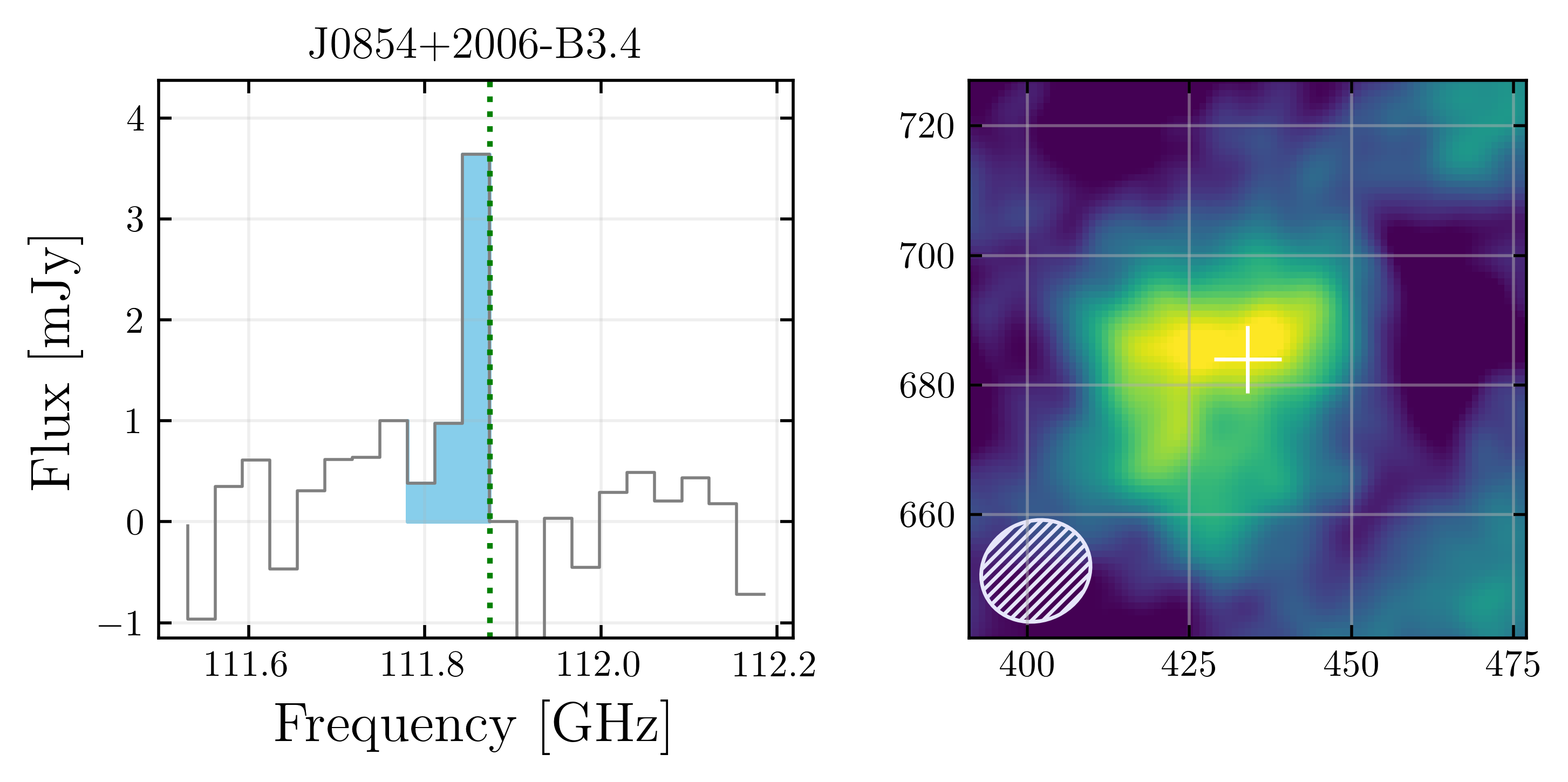}

    \includegraphics[width=0.32\textwidth]{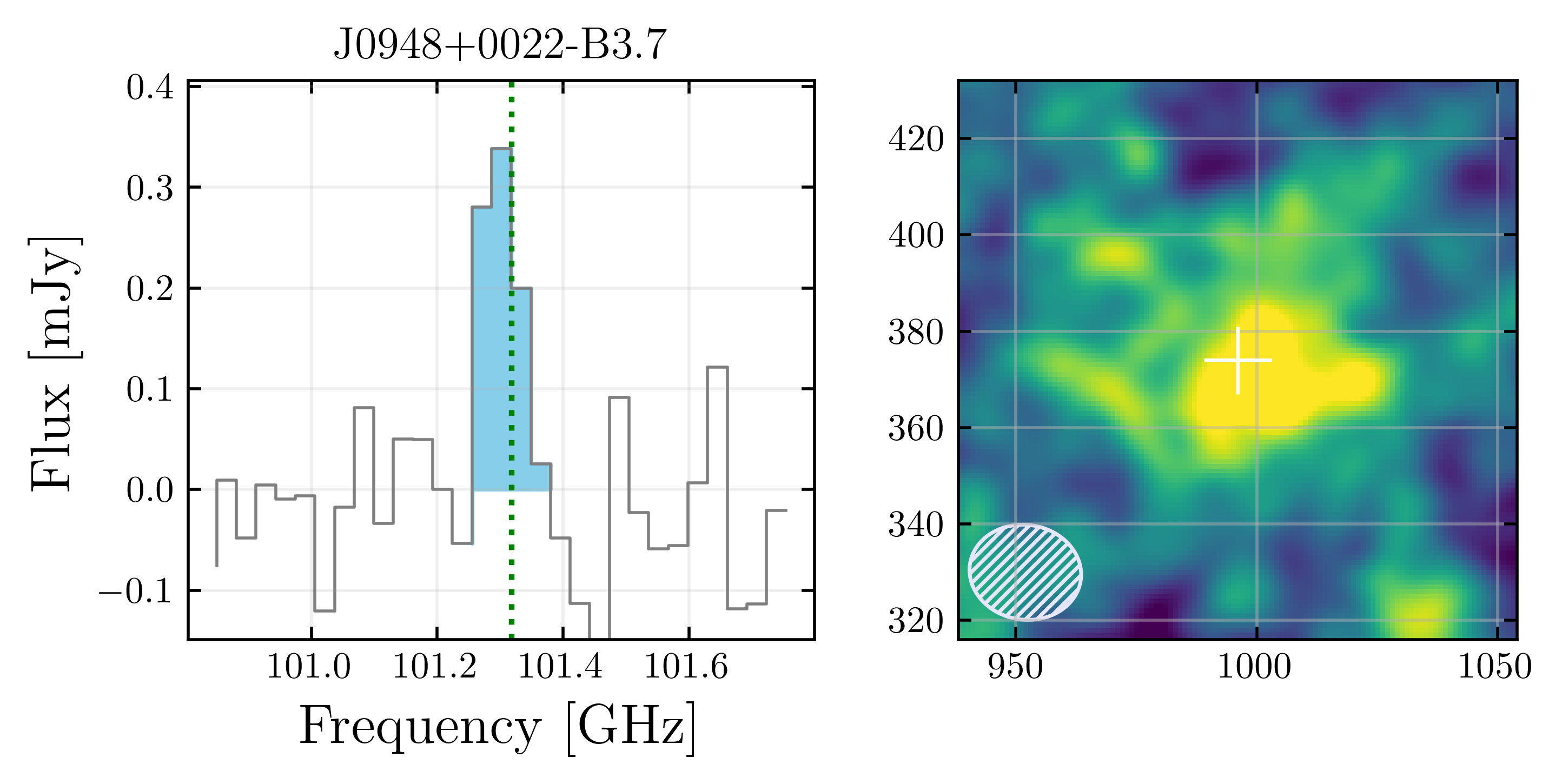}
    \includegraphics[width=0.32\textwidth]{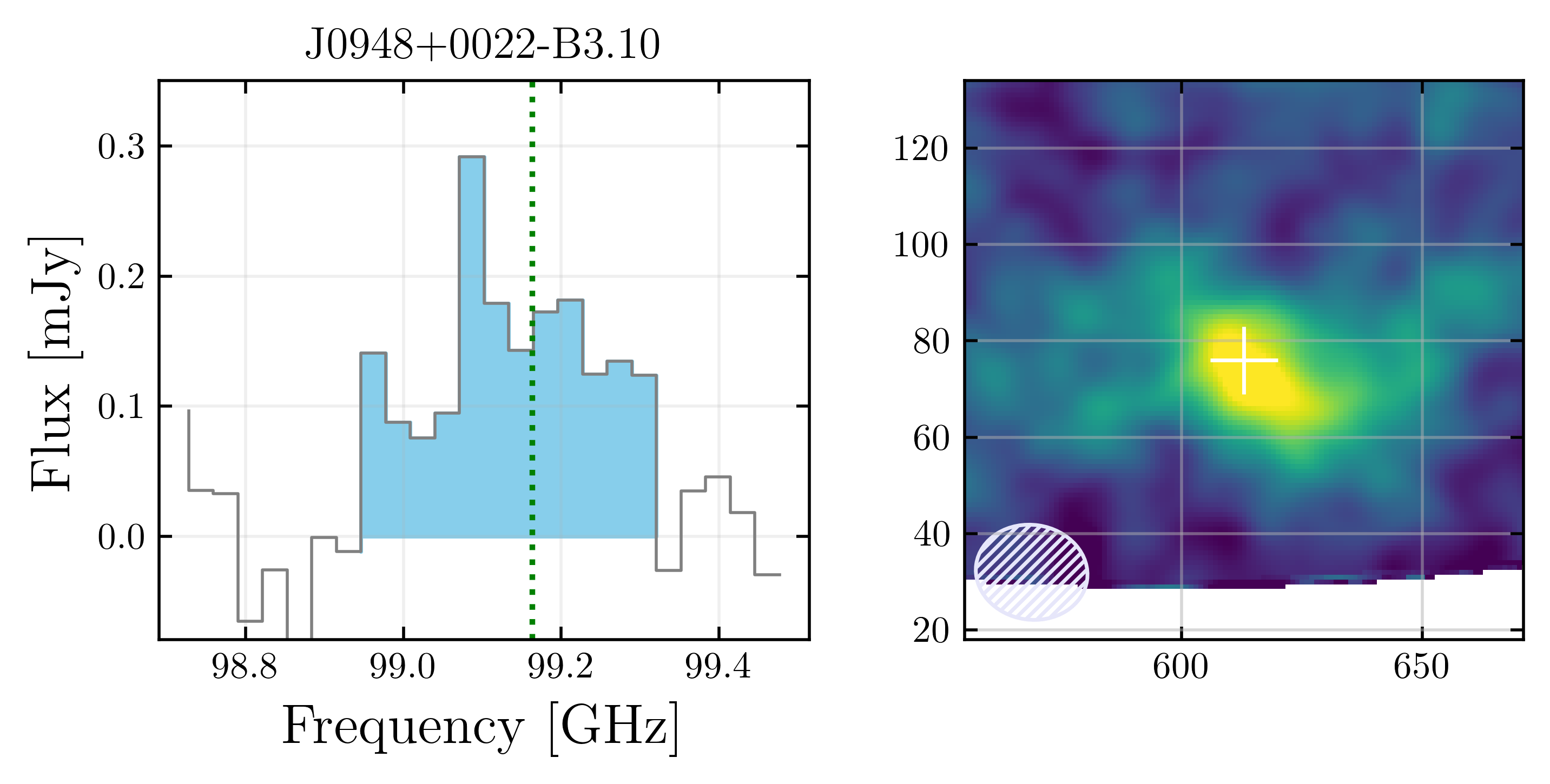}
    \includegraphics[width=0.32\textwidth]{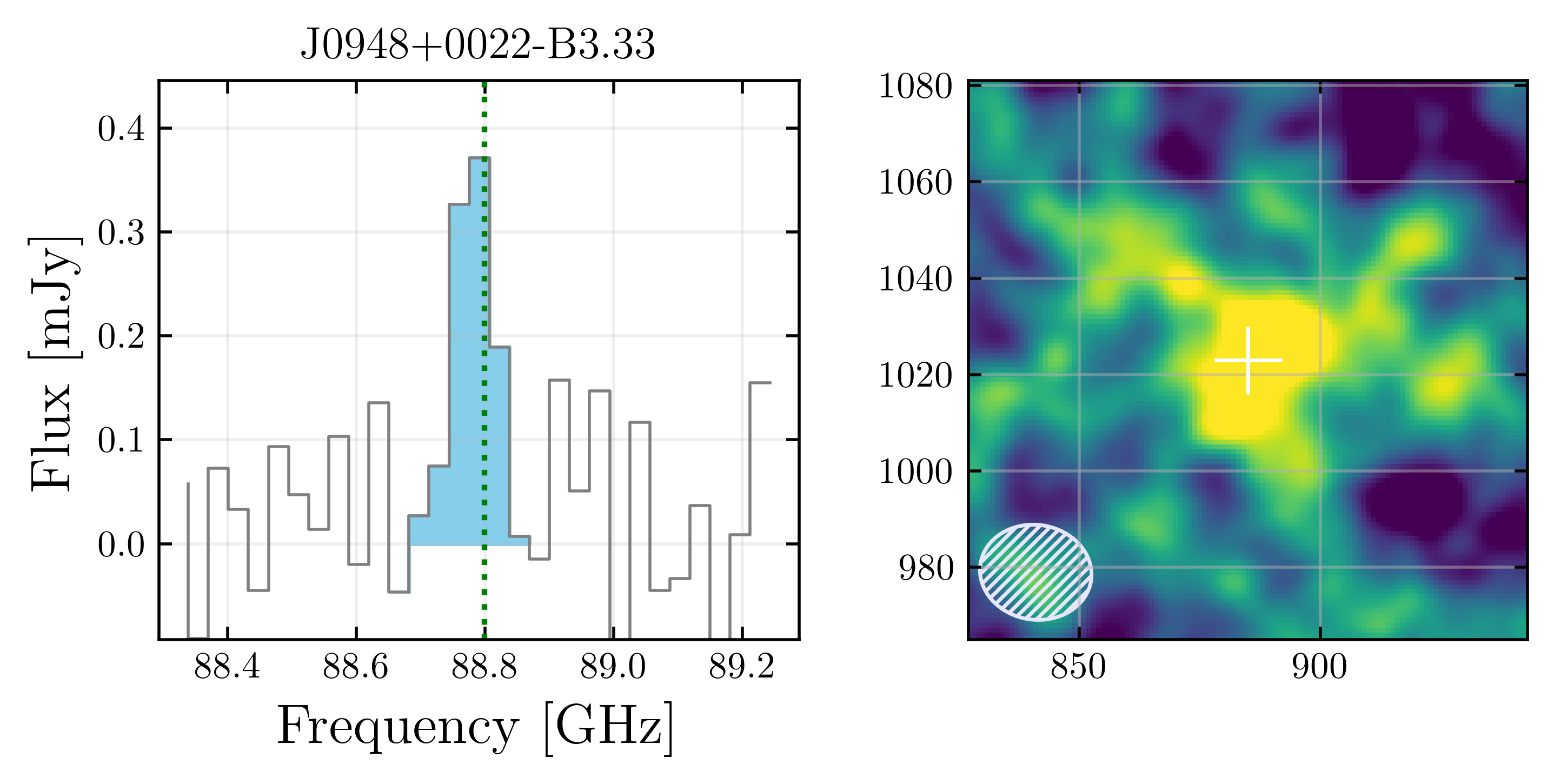}

    \includegraphics[width=0.32\textwidth]{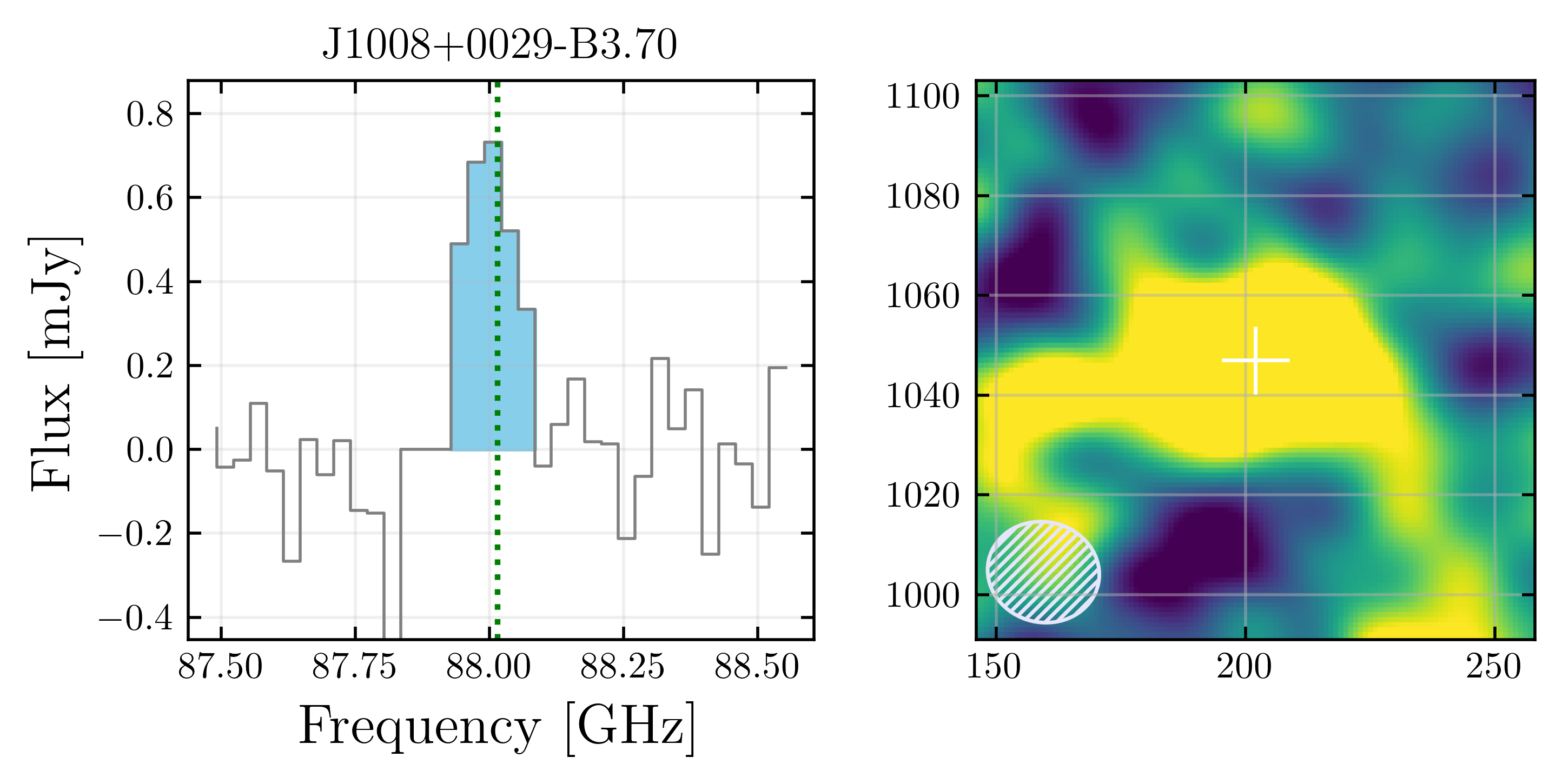}
    \includegraphics[width=0.32\textwidth]{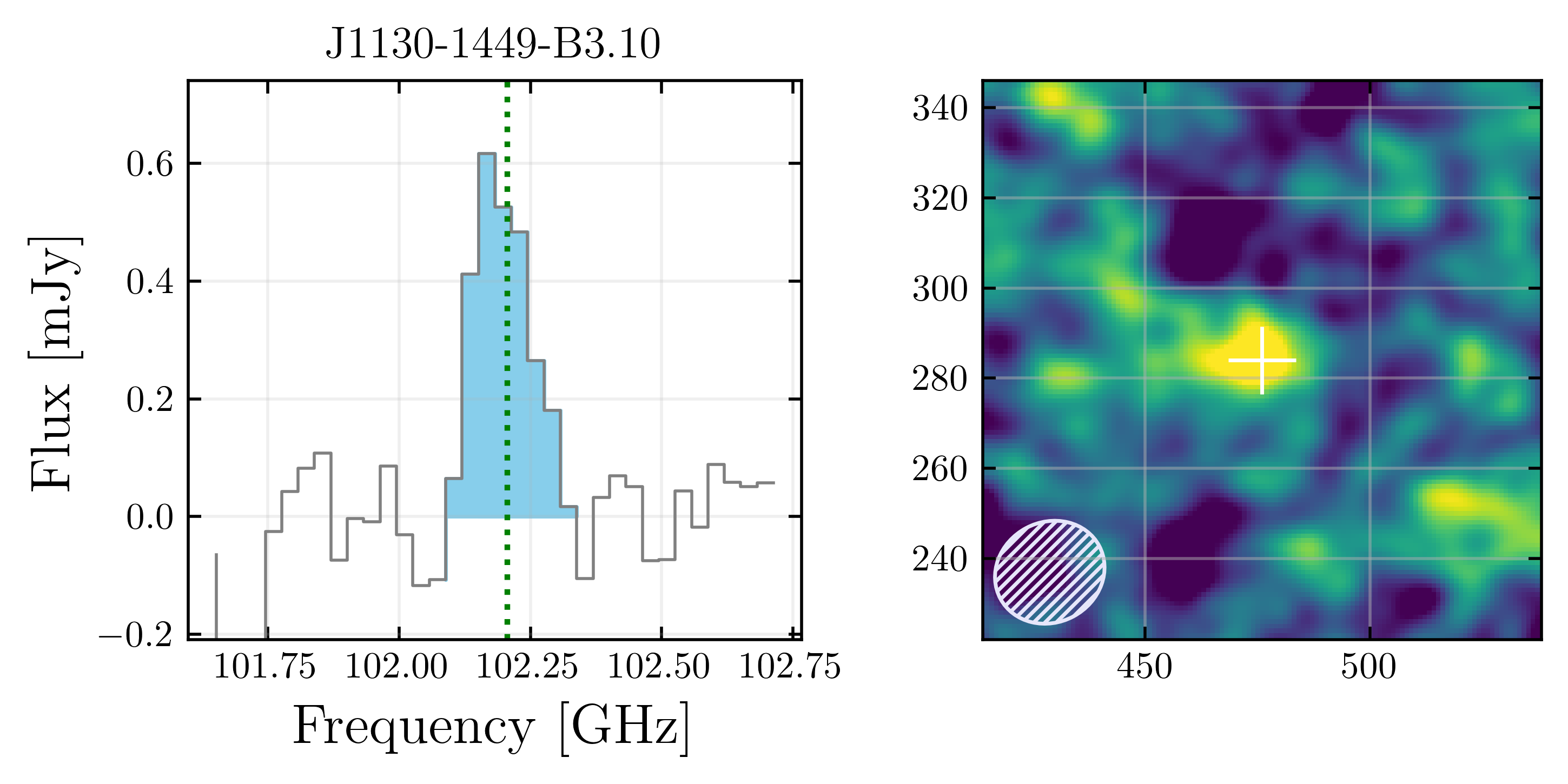}
    \includegraphics[width=0.32\textwidth]{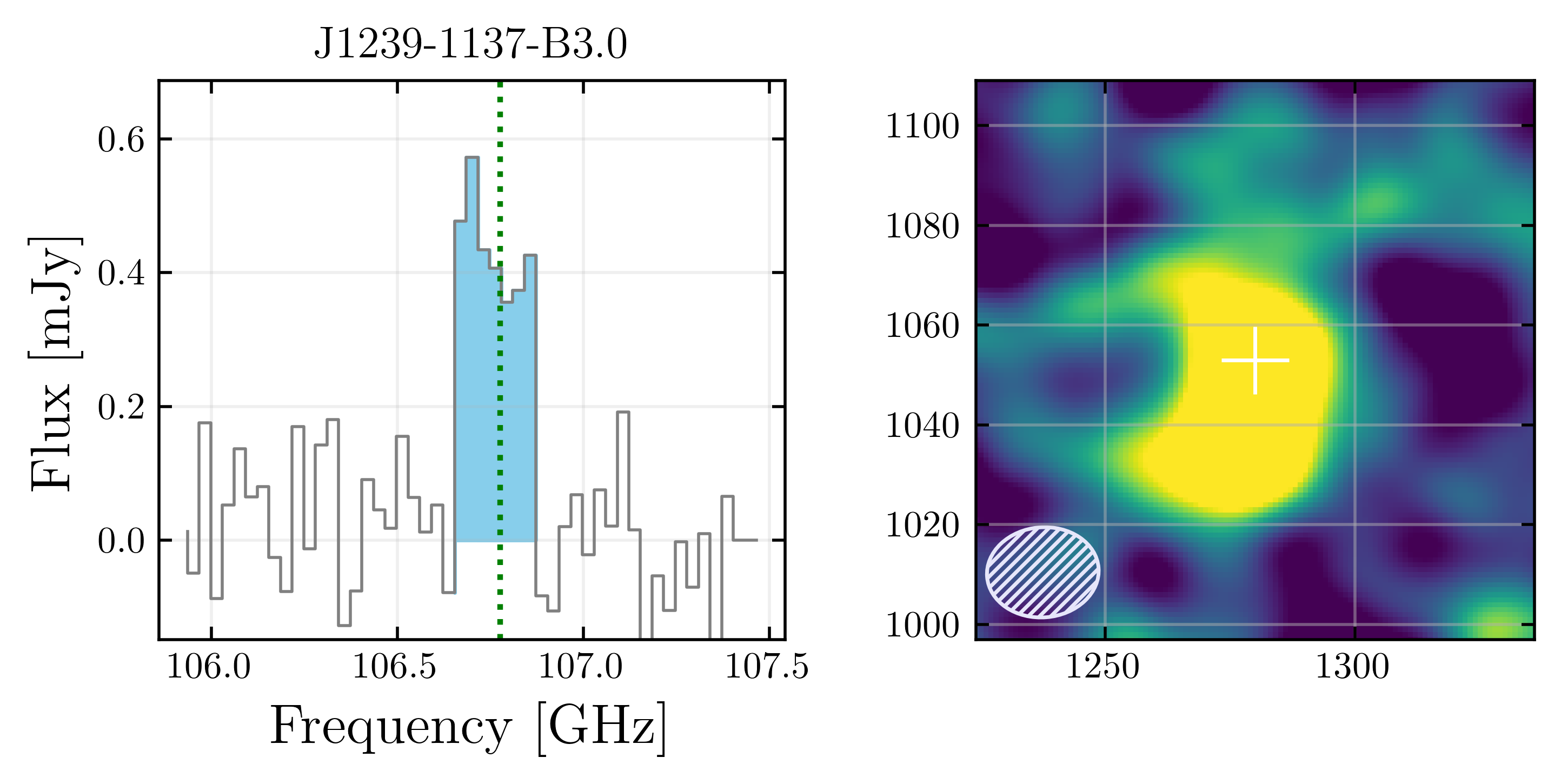}

    \includegraphics[width=0.32\textwidth]{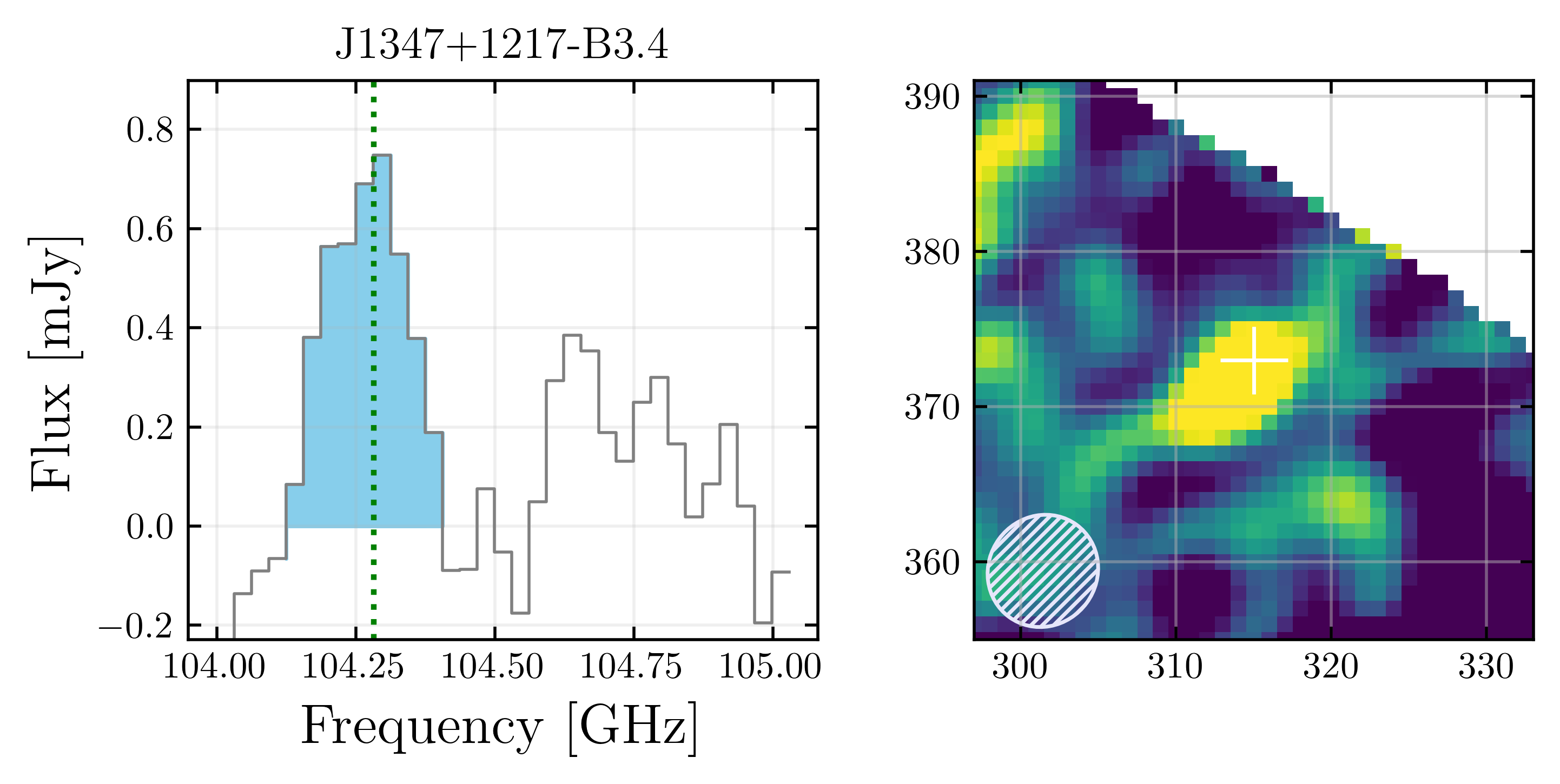}
    \includegraphics[width=0.32\textwidth]{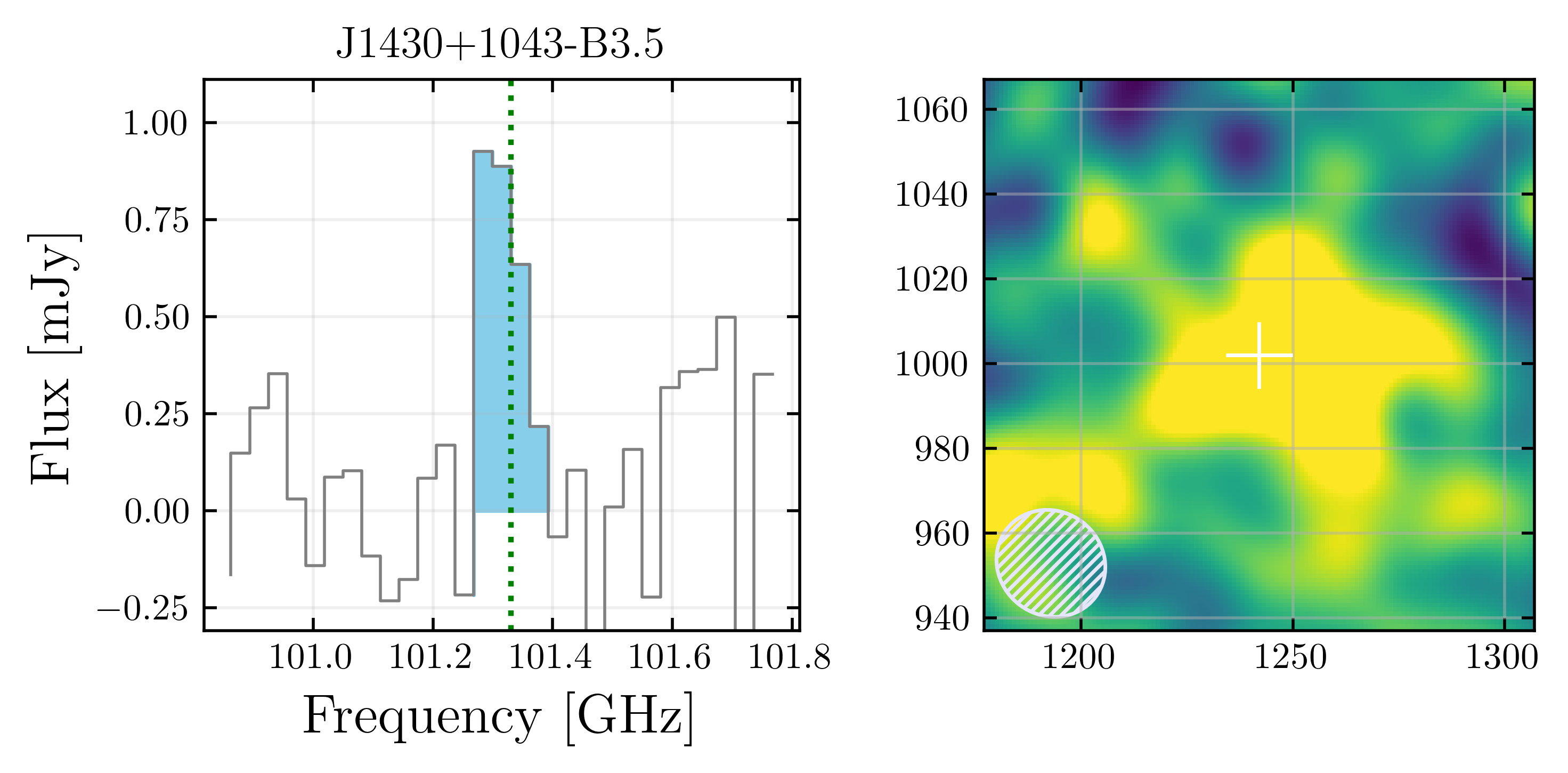}
    \includegraphics[width=0.32\textwidth]{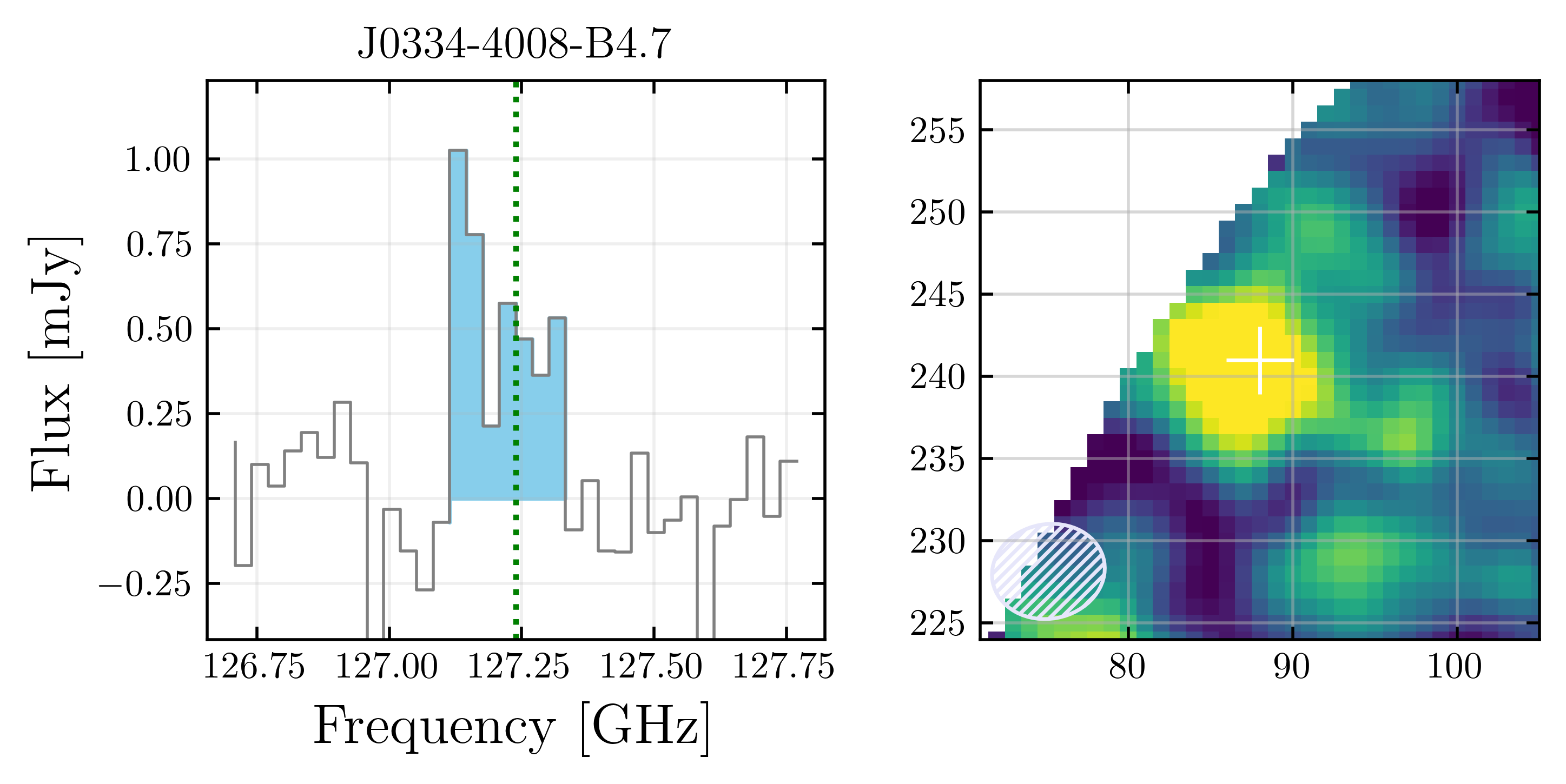}

    \includegraphics[width=0.32\textwidth]{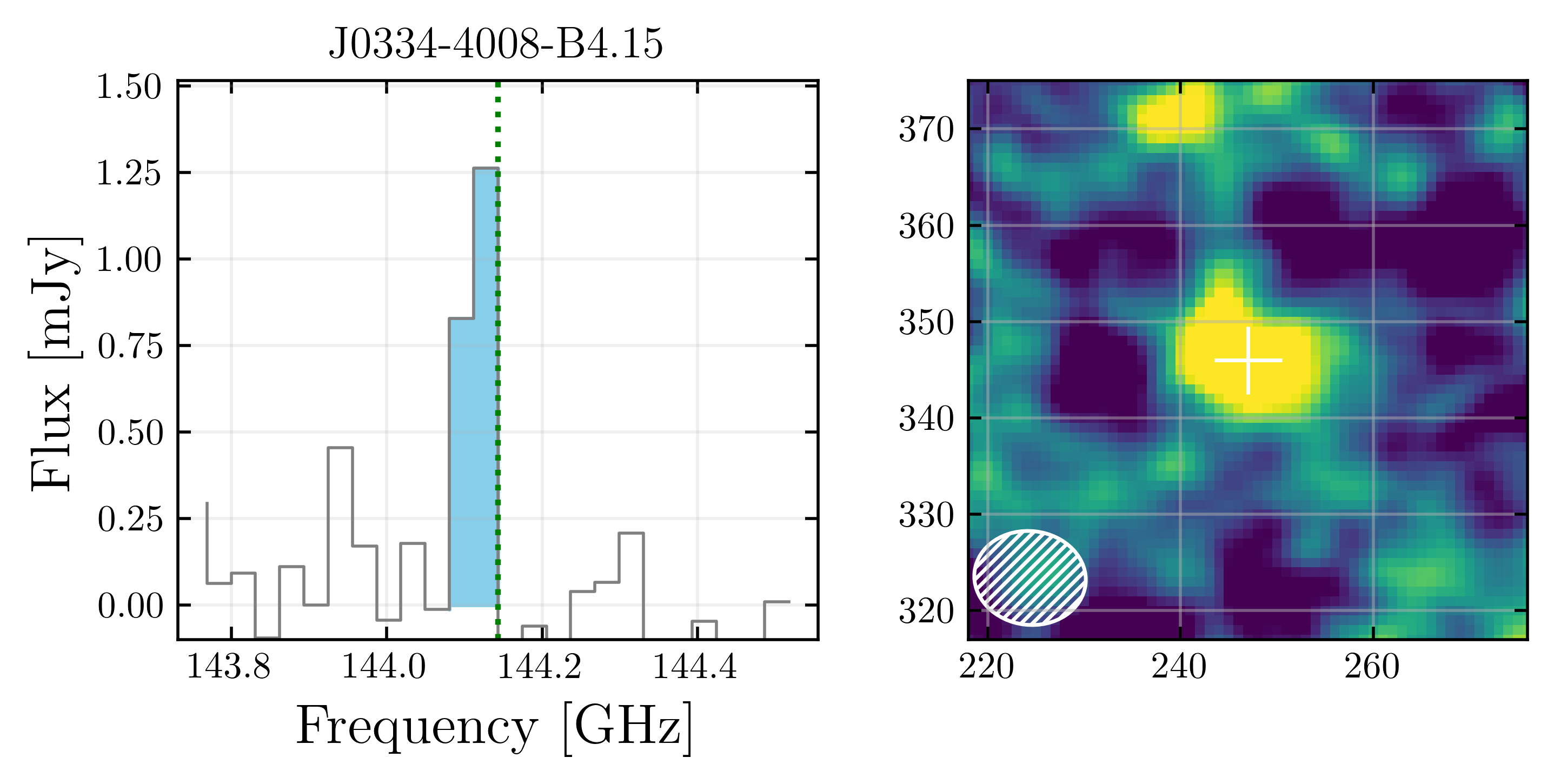}
    \includegraphics[width=0.32\textwidth]{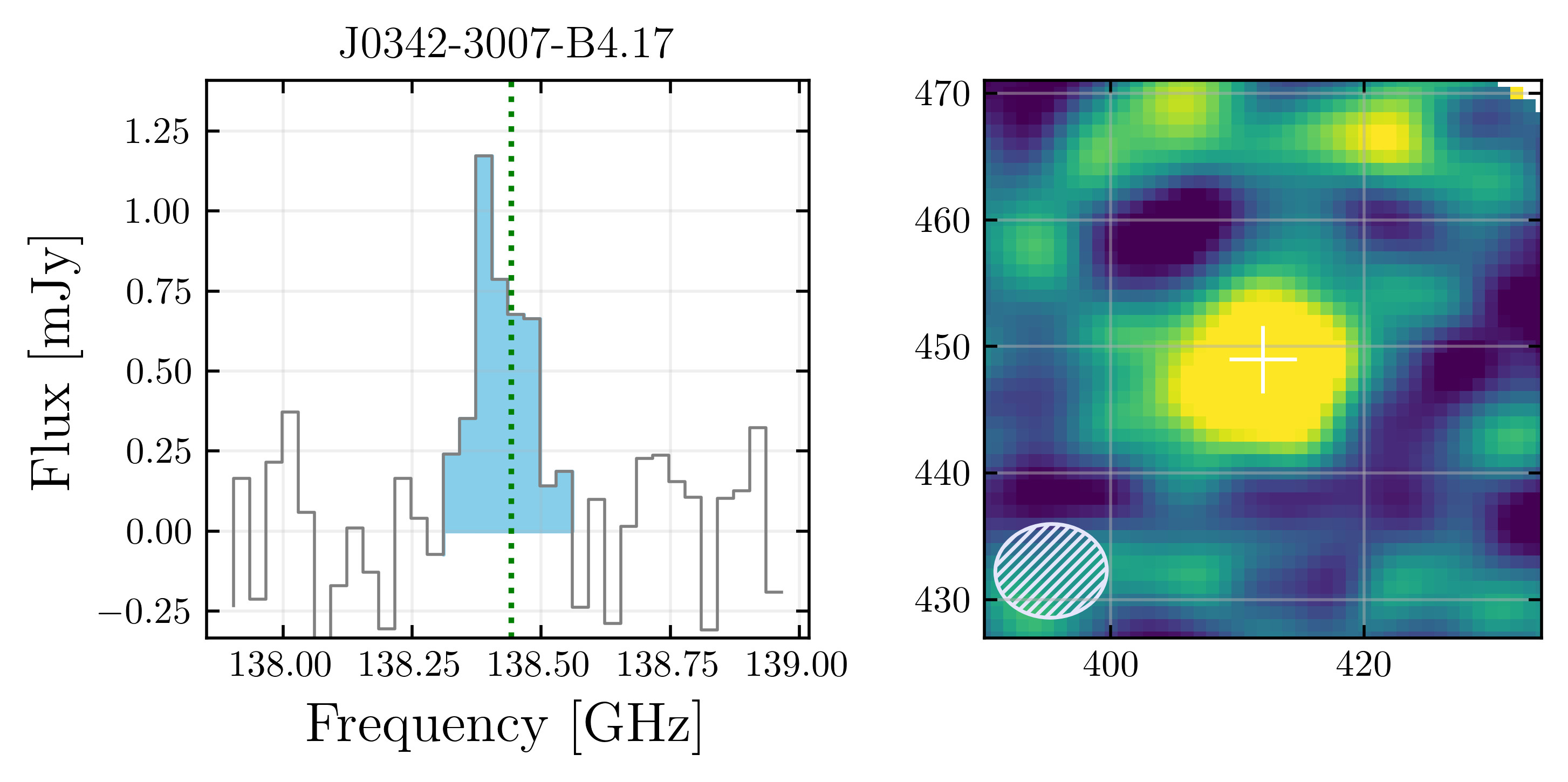}
    \includegraphics[width=0.32\textwidth]{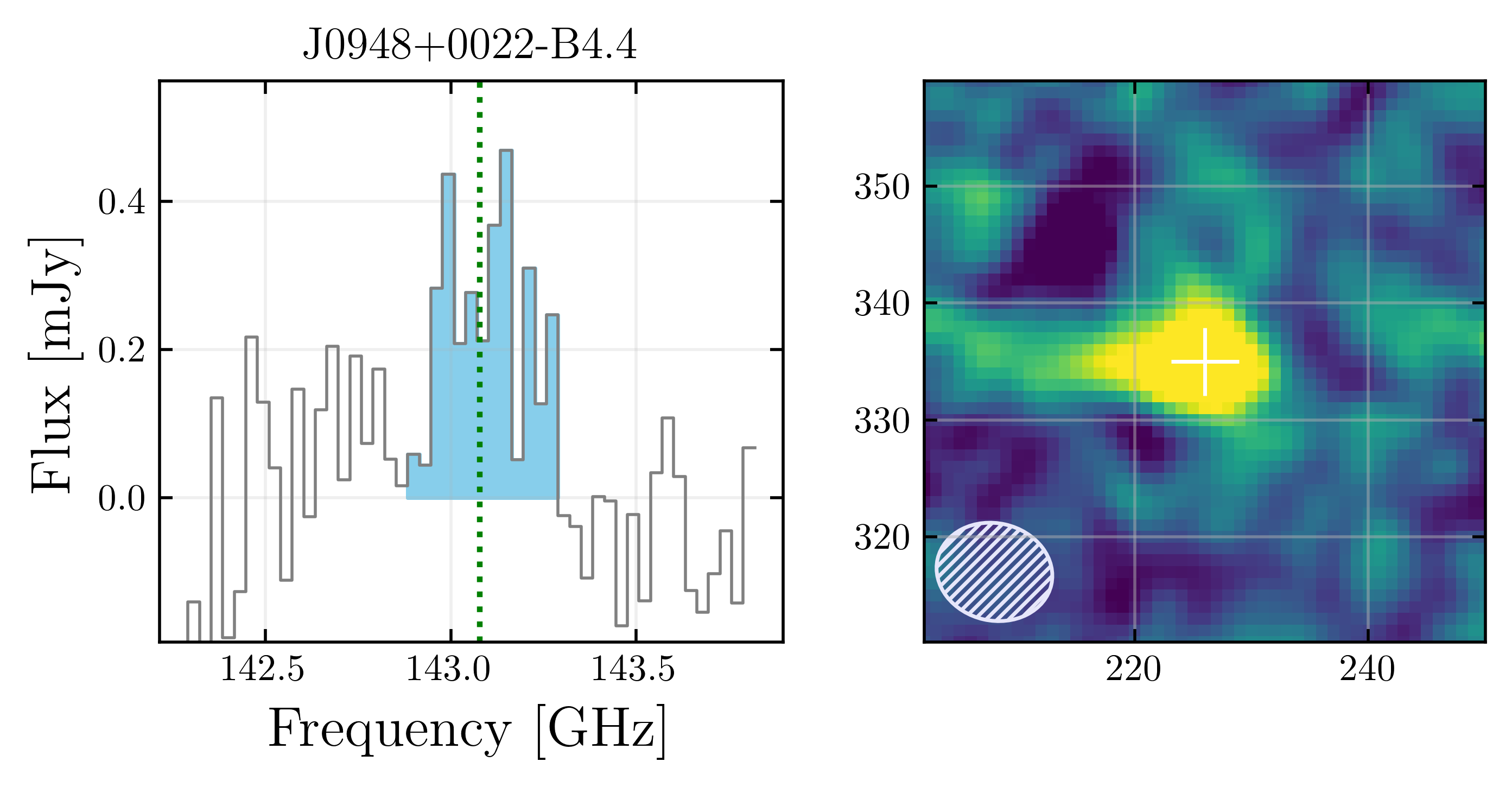}

     \includegraphics[width=0.32\textwidth]{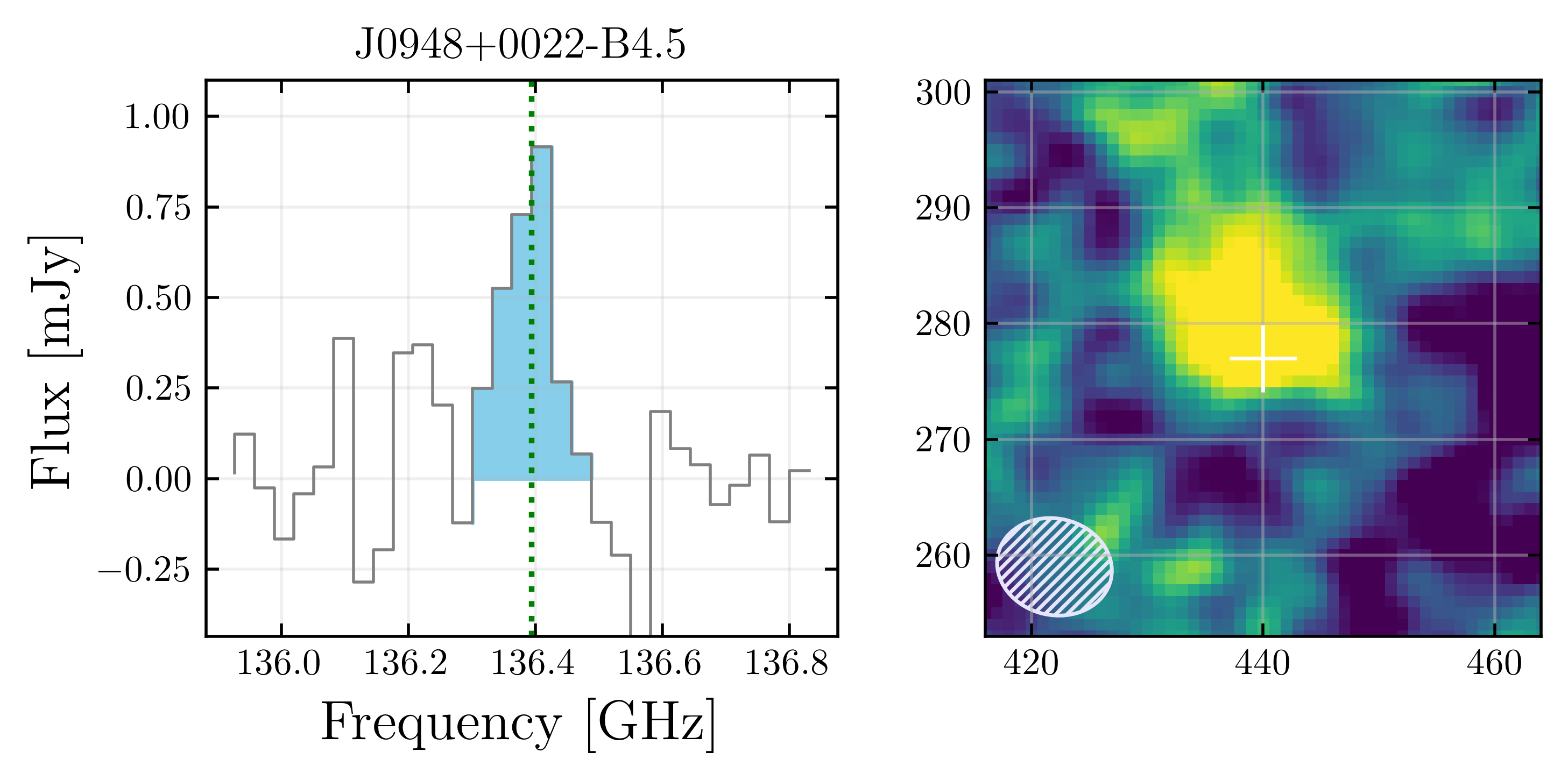}
    \includegraphics[width=0.32\textwidth]{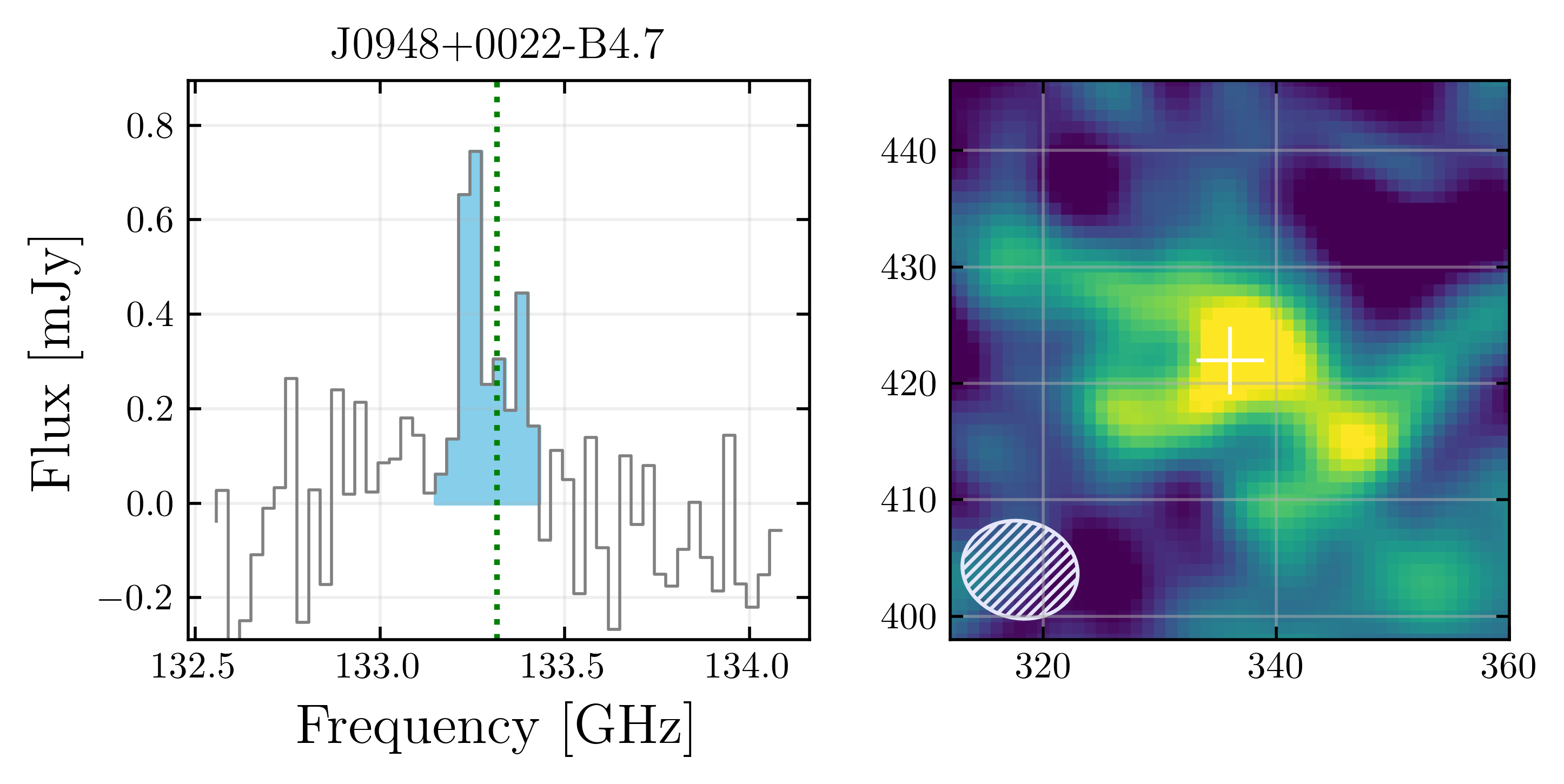}
    \includegraphics[width=0.32\textwidth]{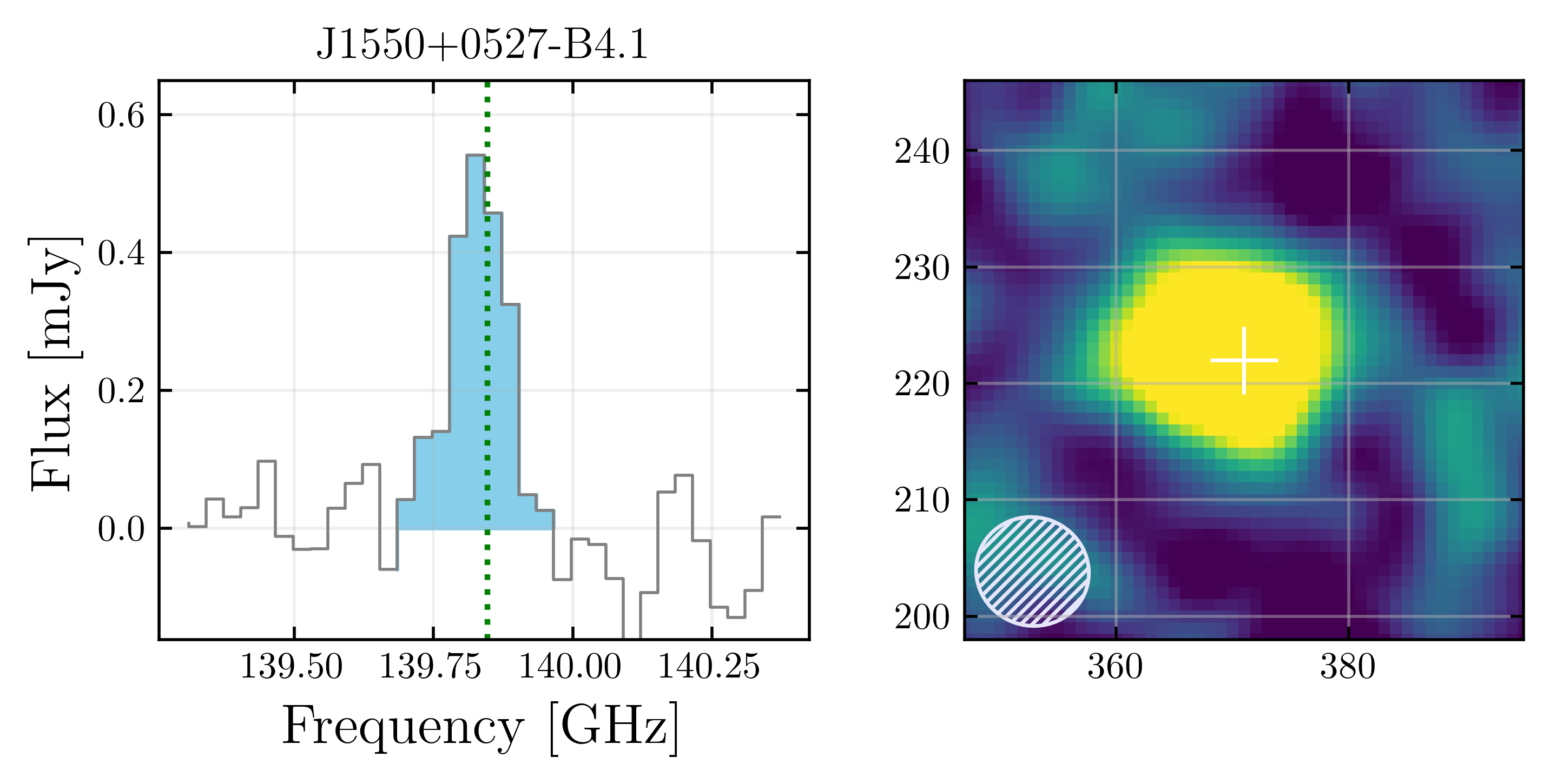}

    \caption{Continuation.}
    \label{fig:detections2}
    
\end{figure}

\begin{figure}
\centering
     \includegraphics[width=0.32\textwidth]{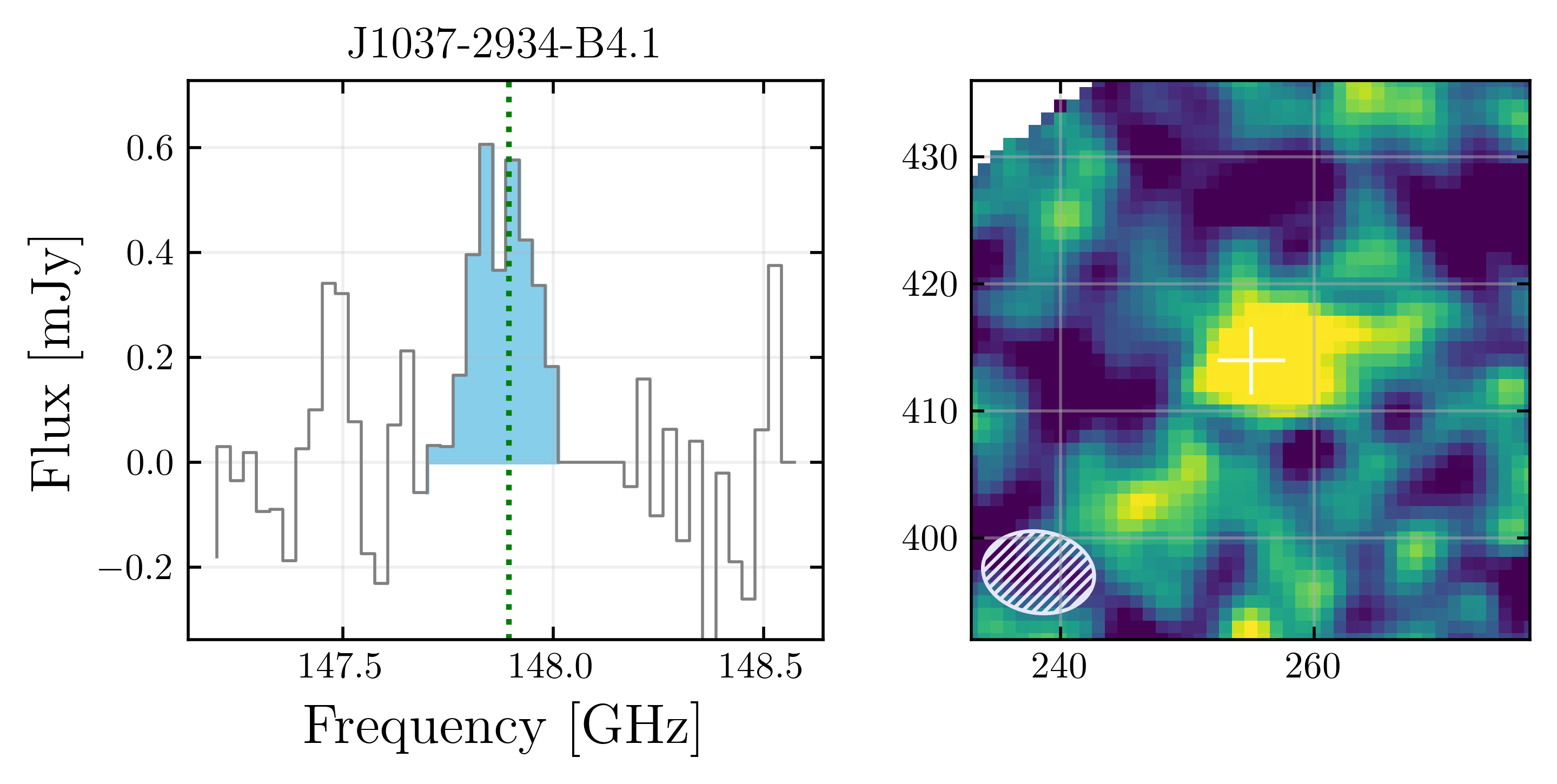}
    \includegraphics[width=0.32\textwidth]{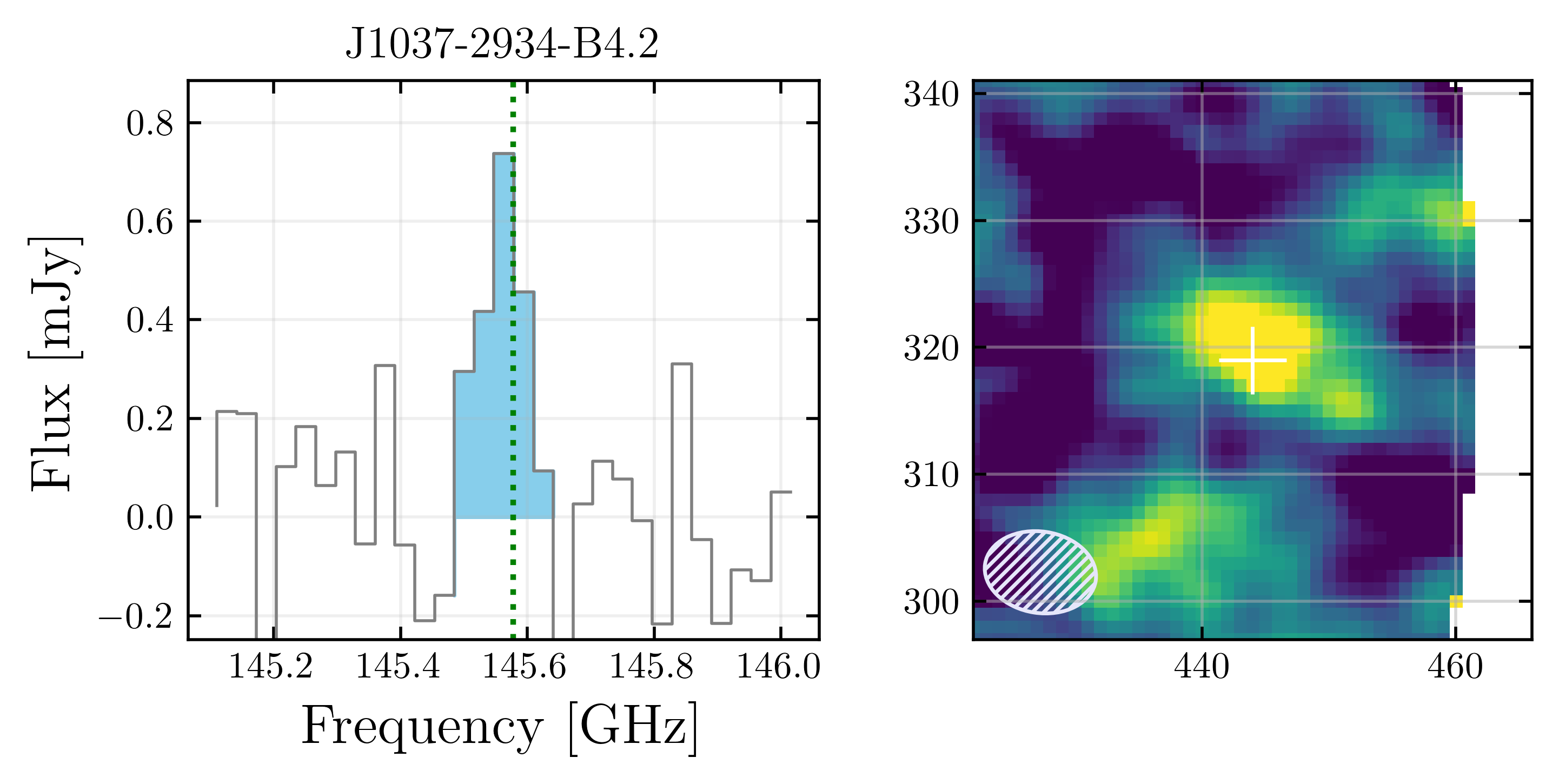}
    \includegraphics[width=0.32\textwidth]{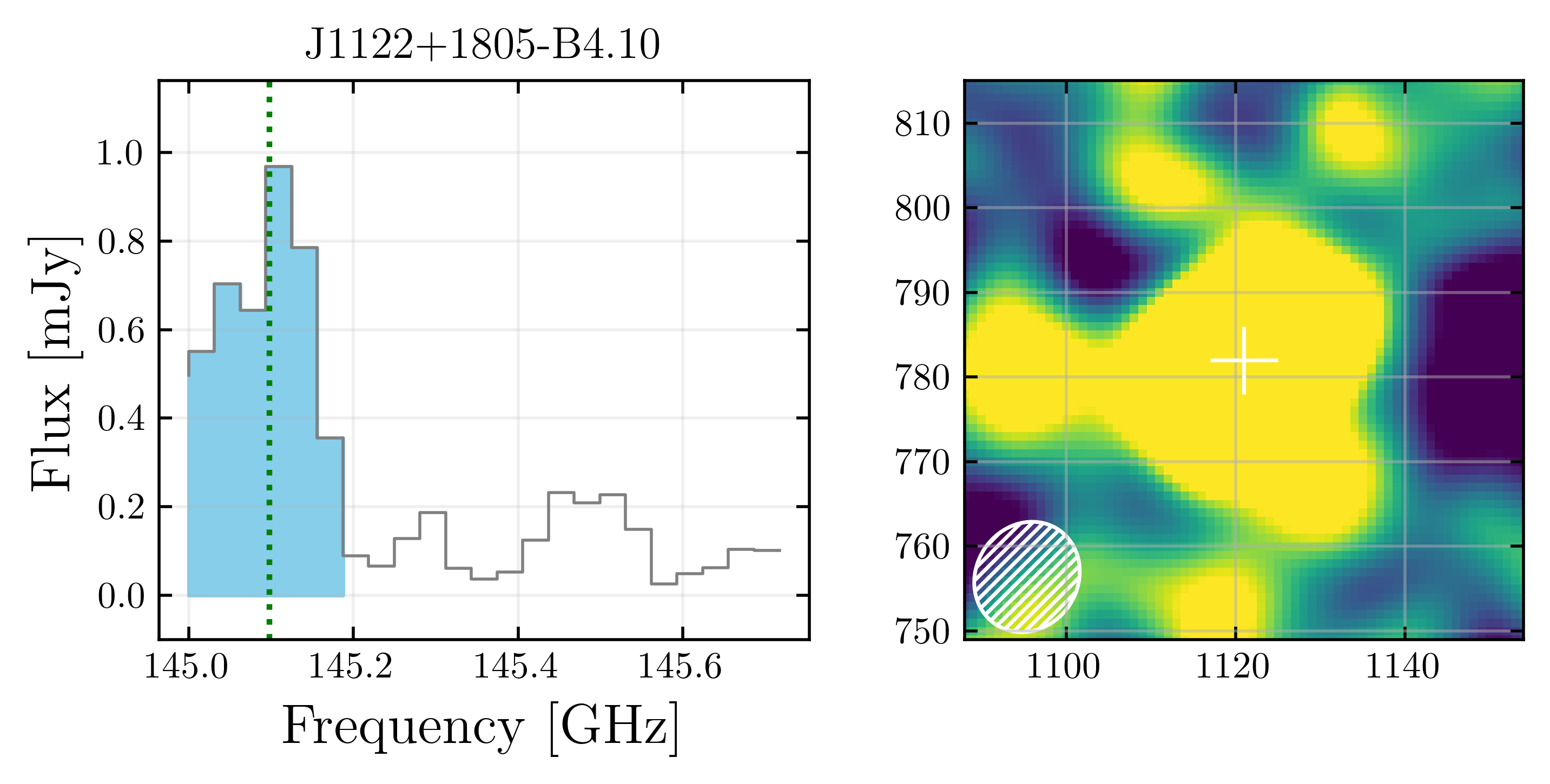}
    
    \includegraphics[width=0.32\textwidth]{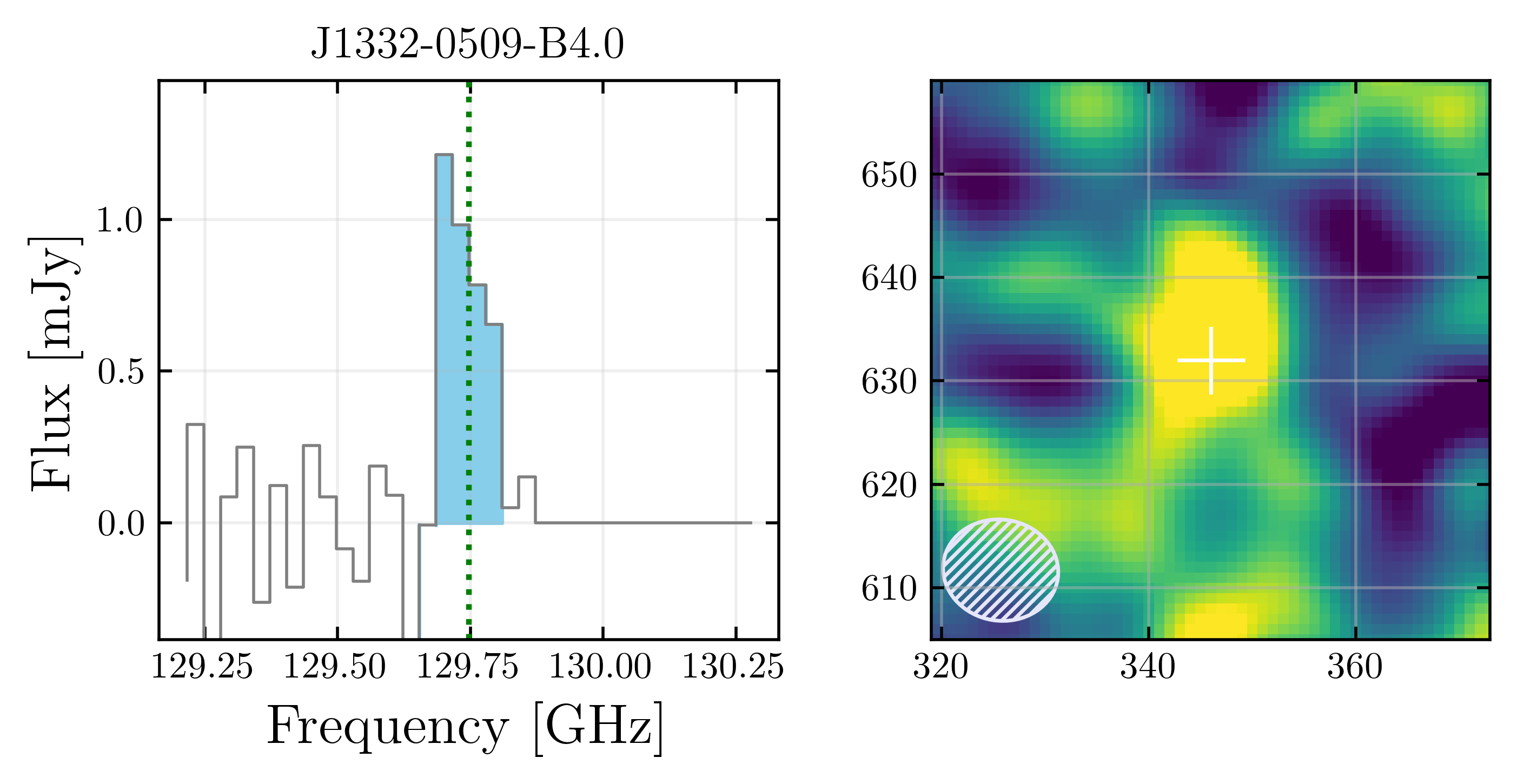}
    \includegraphics[width=0.32\textwidth]{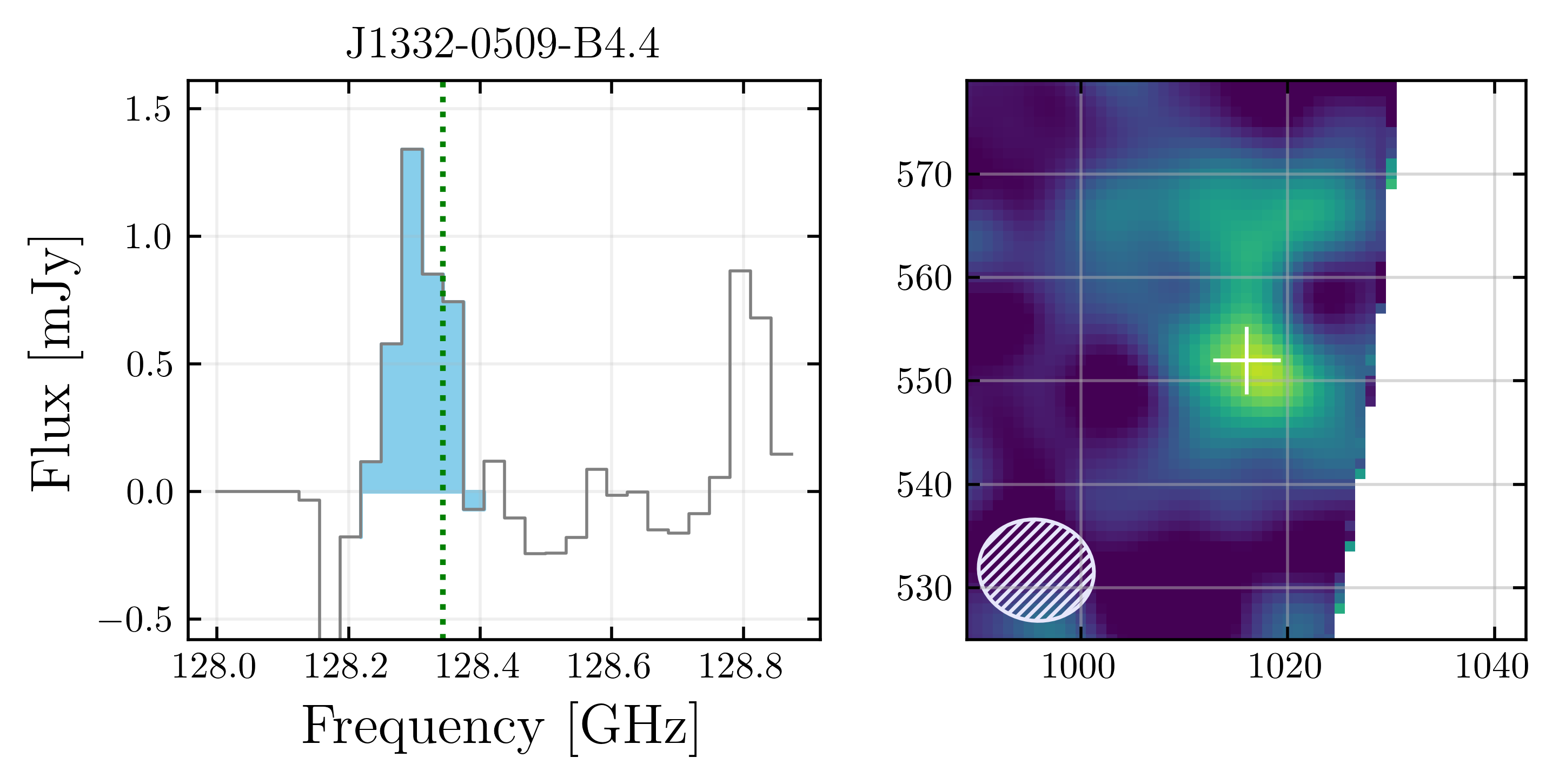}
    \includegraphics[width=0.32\textwidth]{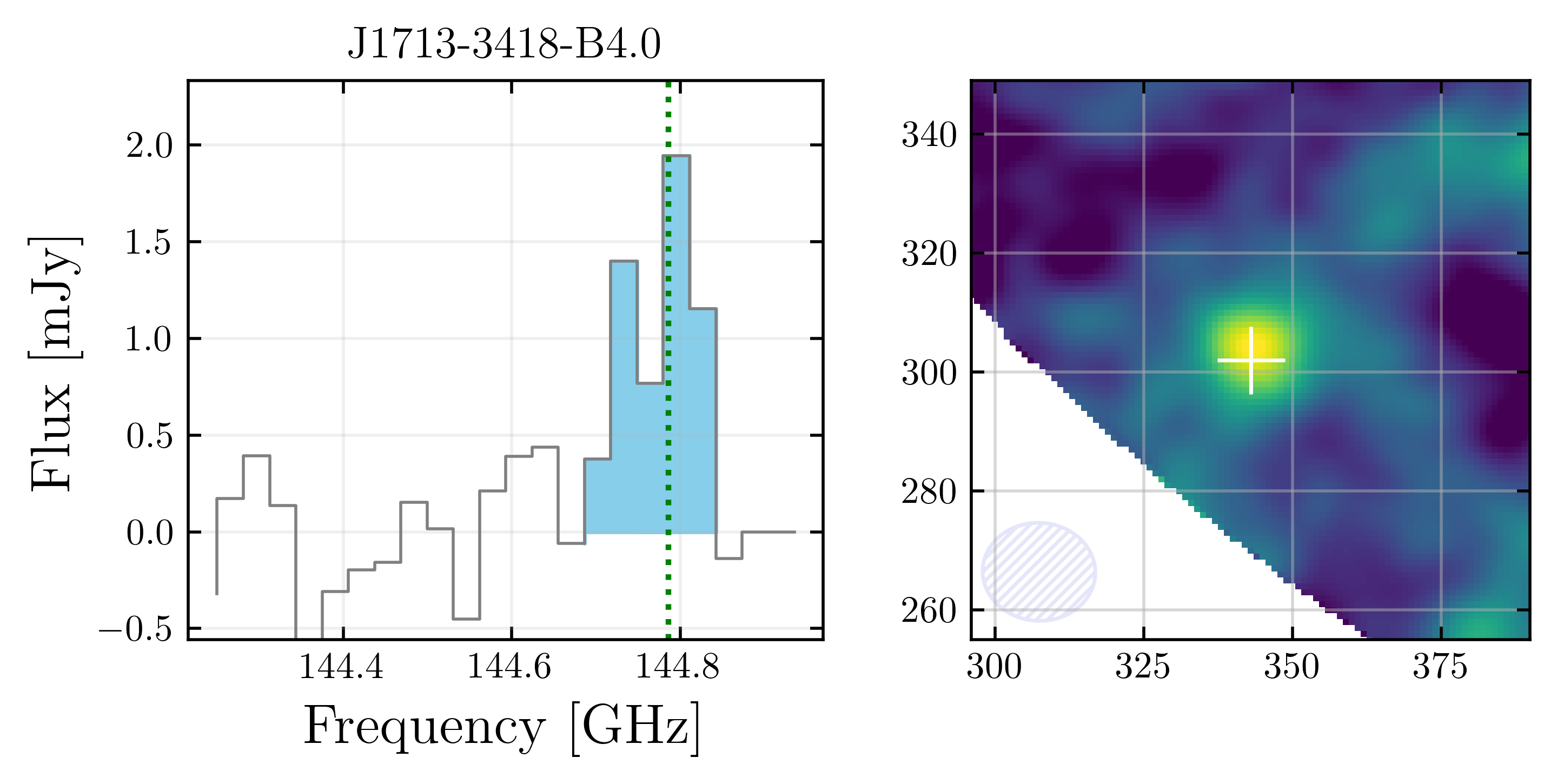}

    \includegraphics[width=0.32\textwidth]{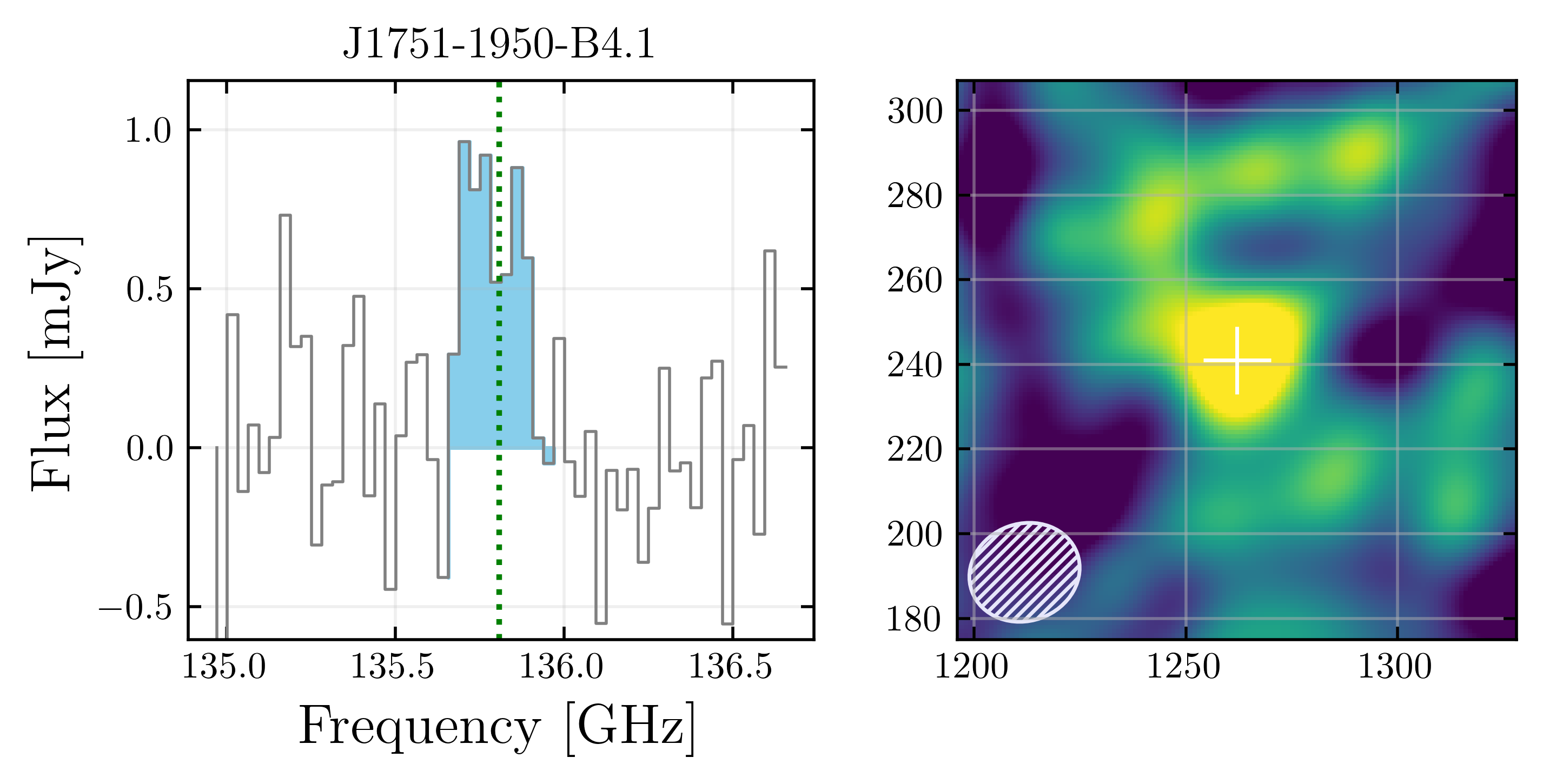}
    \includegraphics[width=0.32\textwidth]{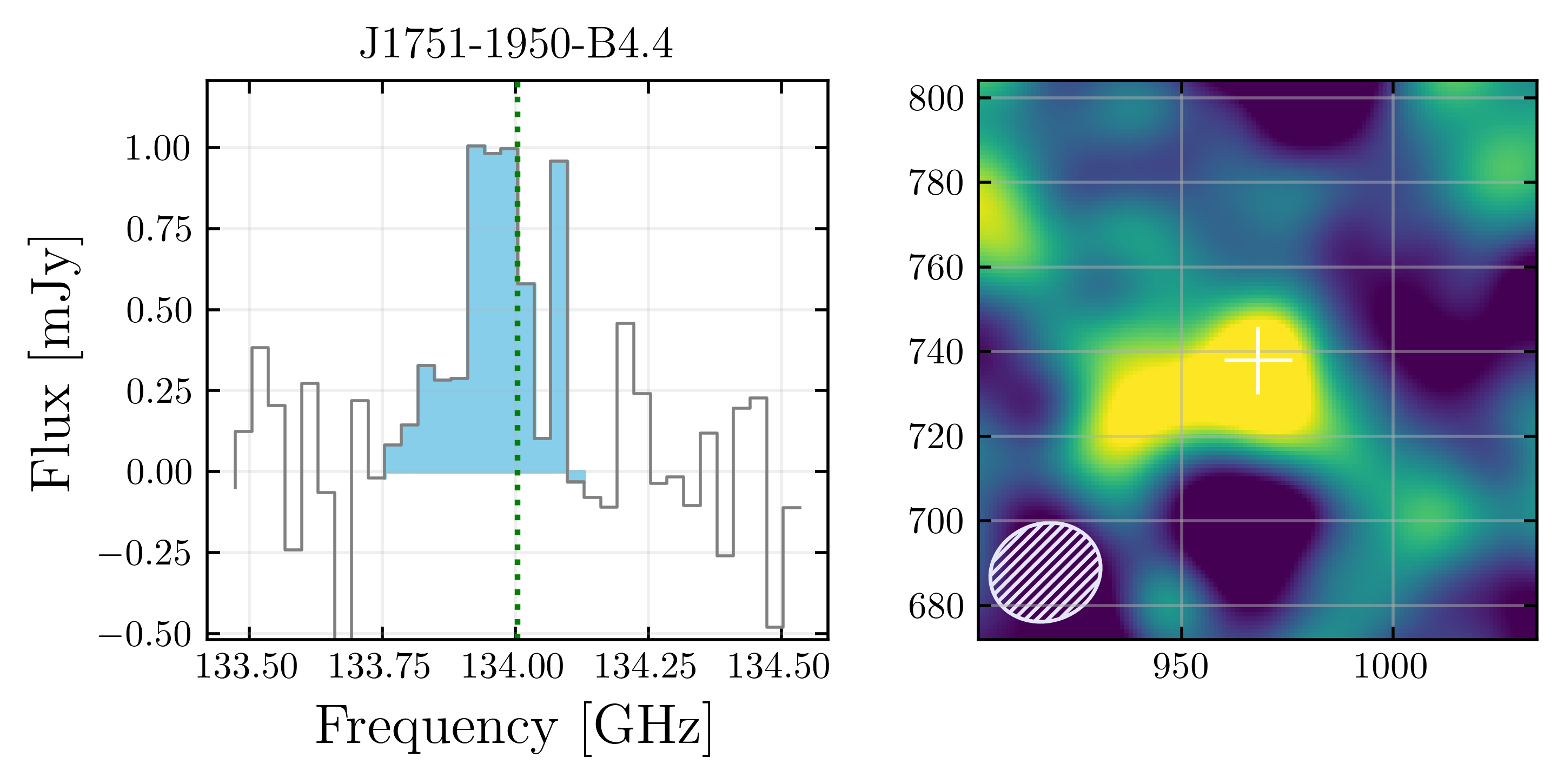}
    \includegraphics[width=0.32\textwidth]{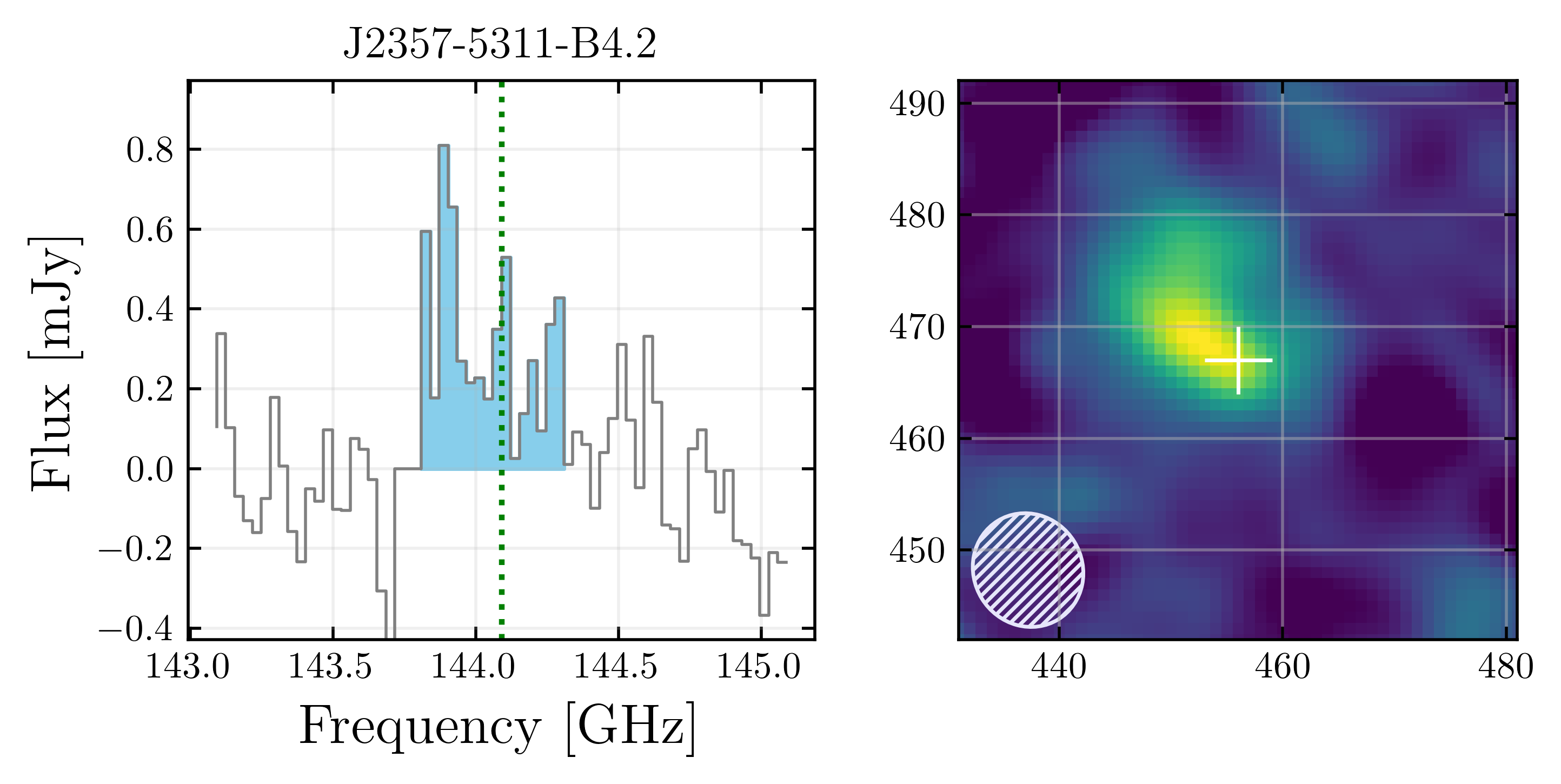}

    \includegraphics[width=0.32\textwidth]{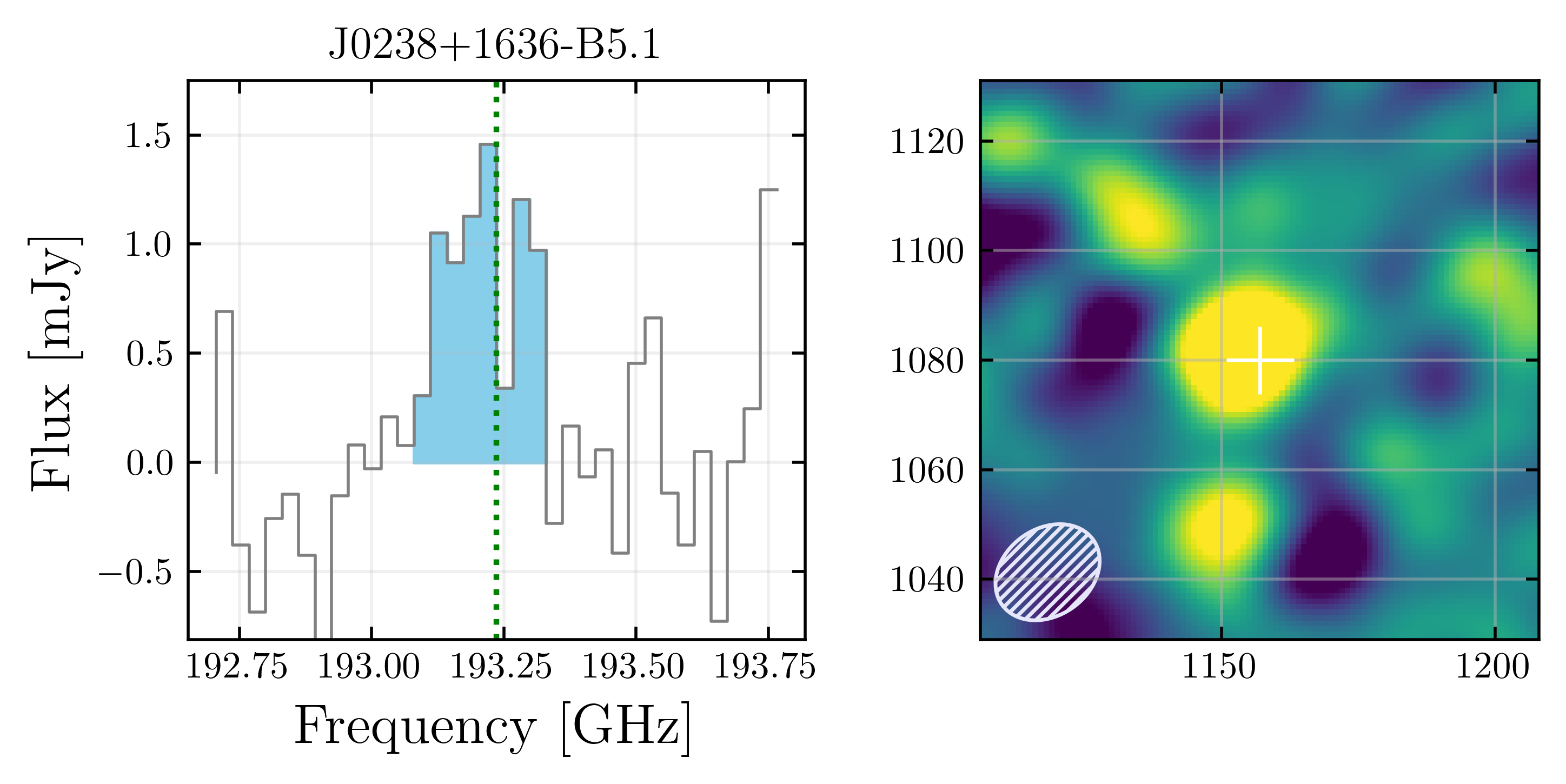}
    \includegraphics[width=0.32\textwidth]{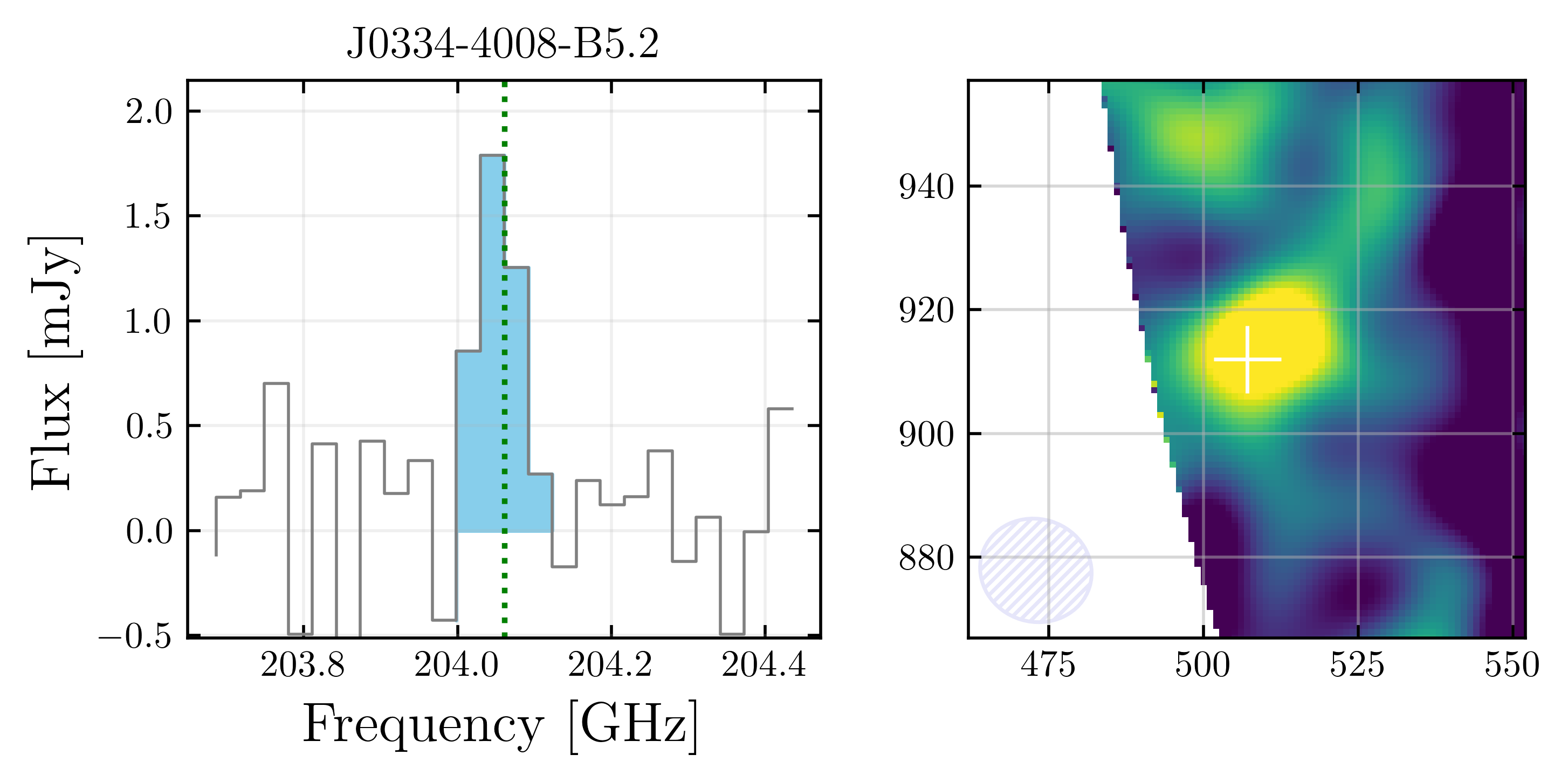}
    \includegraphics[width=0.32\textwidth]{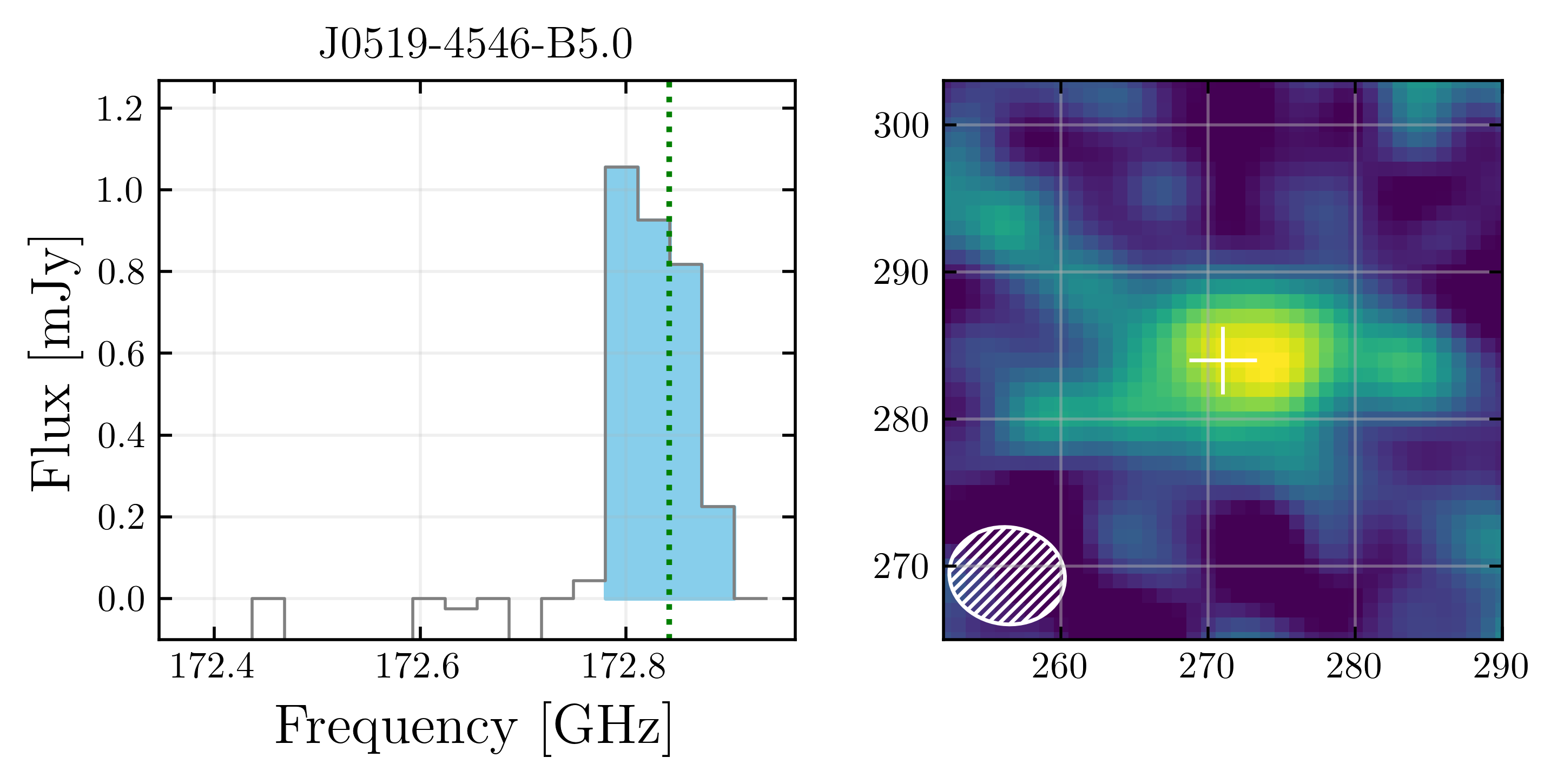}

    \includegraphics[width=0.32\textwidth]{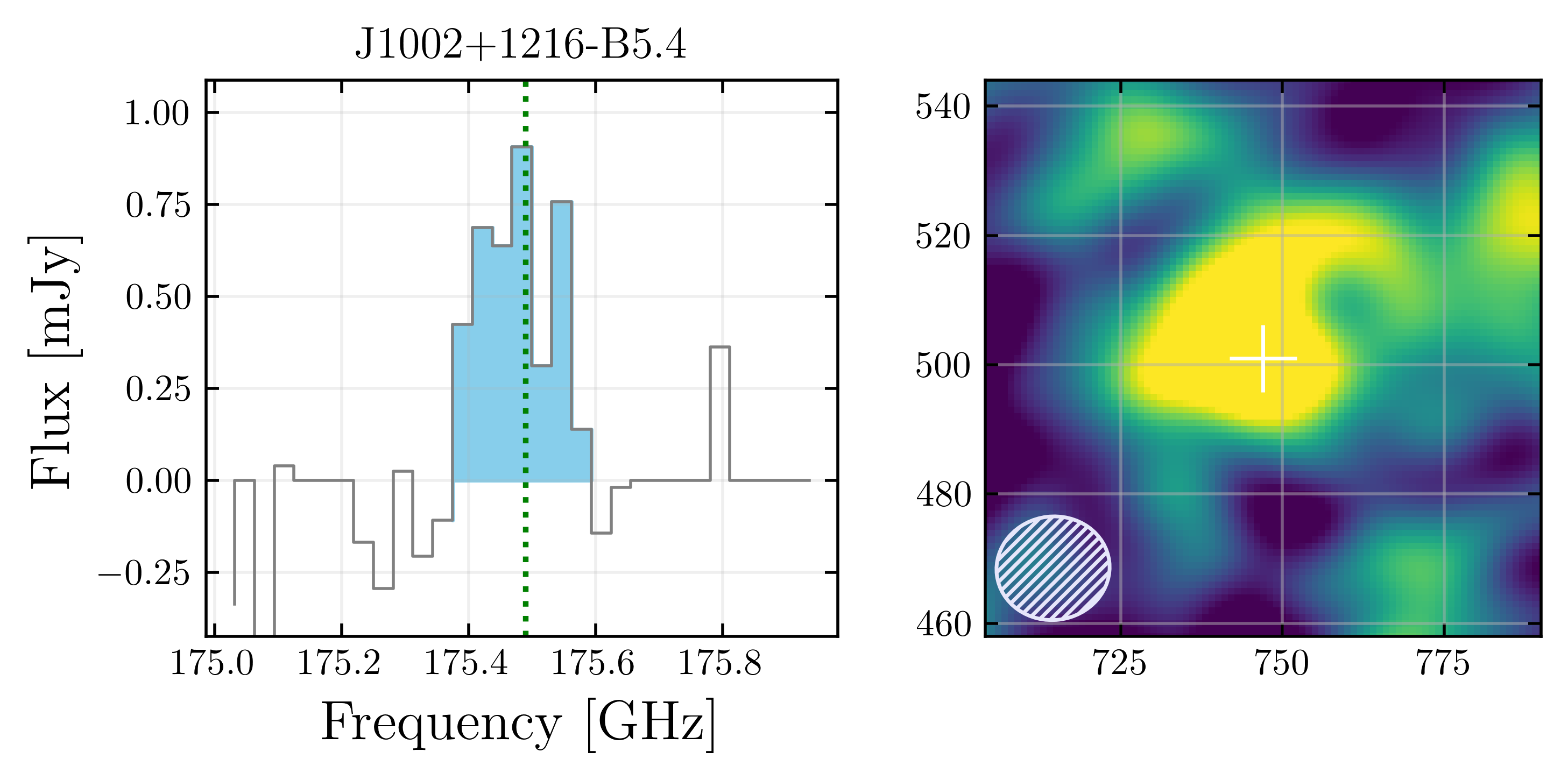}
    \includegraphics[width=0.32\textwidth]{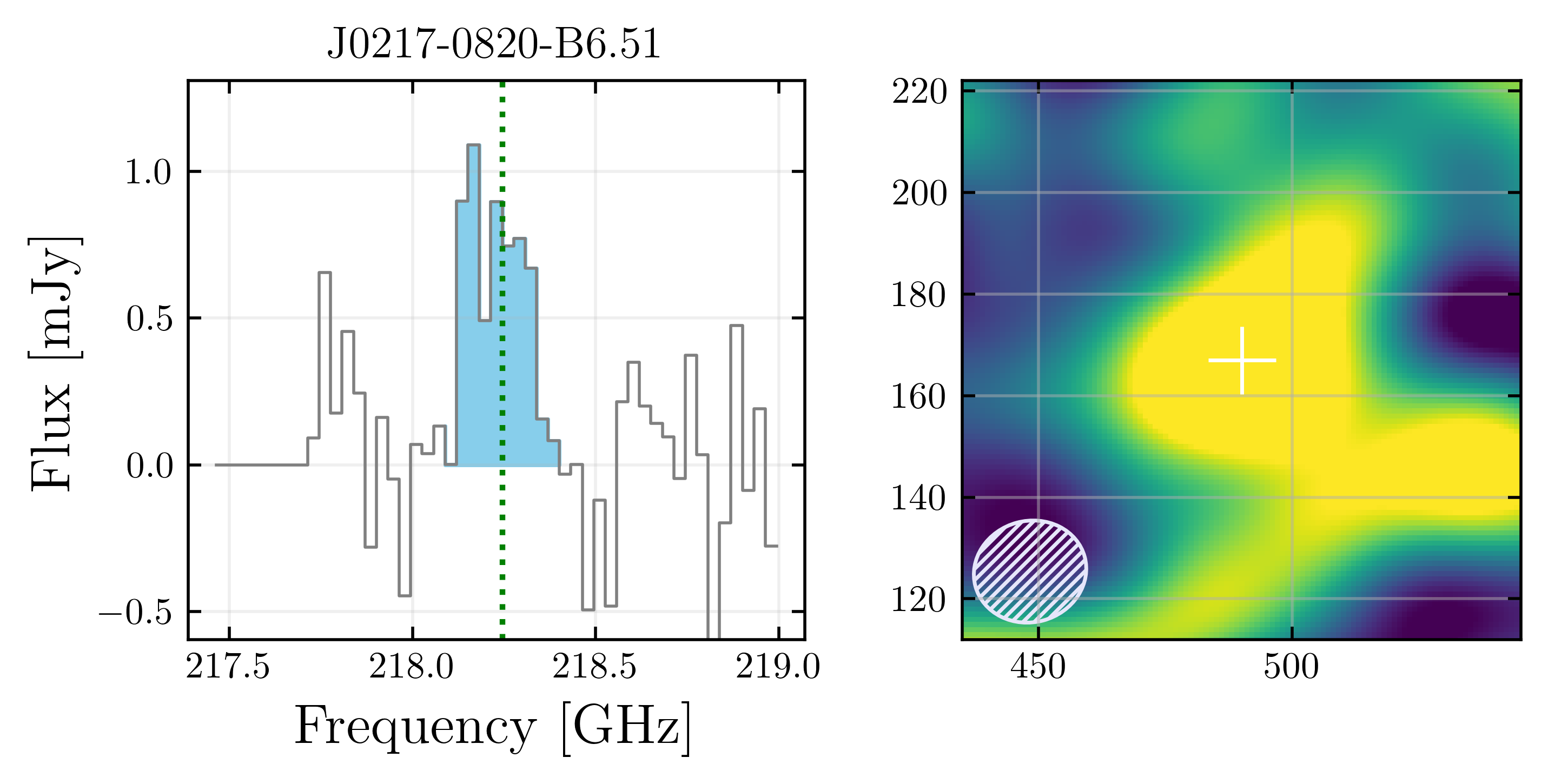}
    \includegraphics[width=0.32\textwidth]{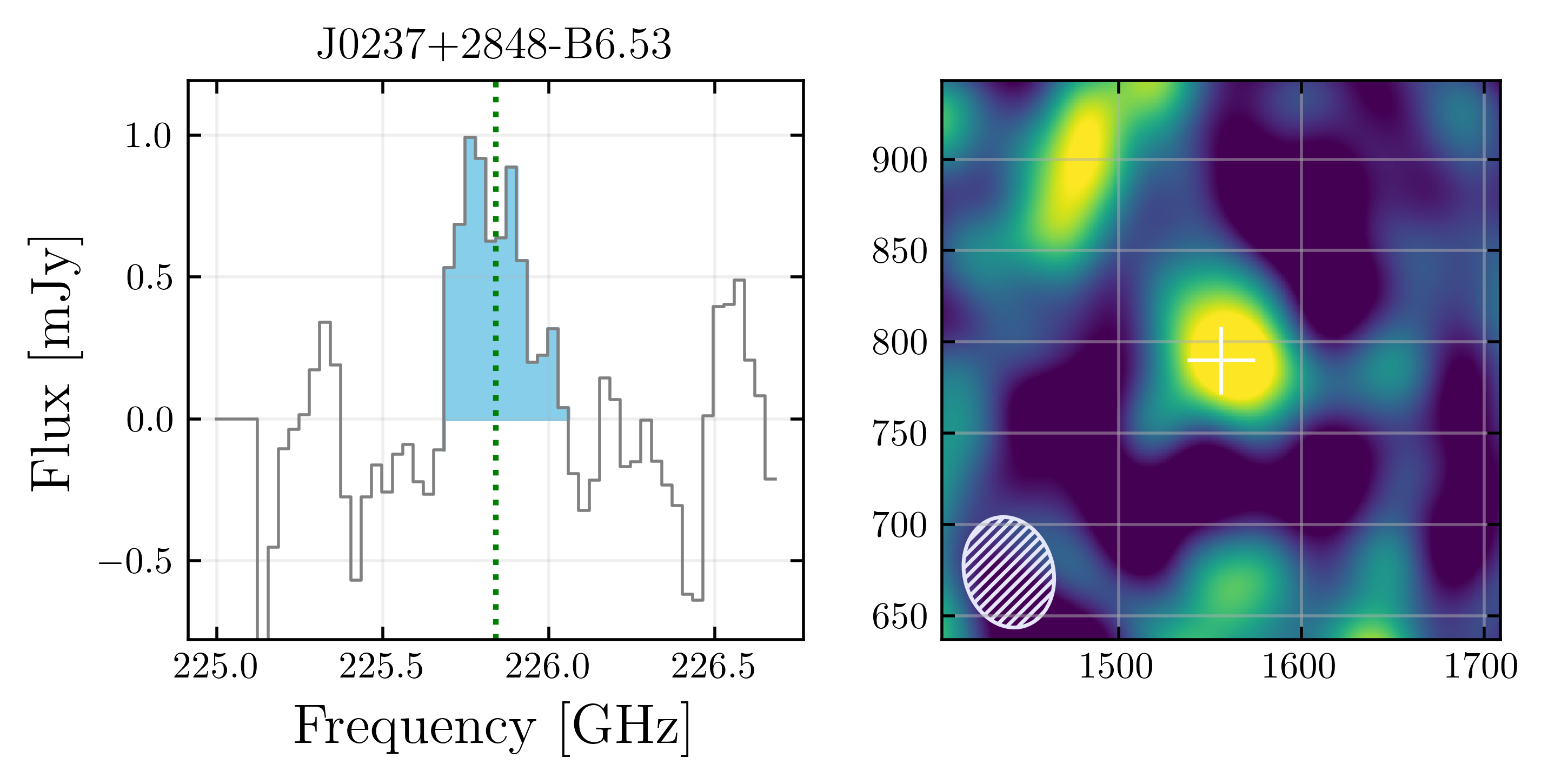}

    \includegraphics[width=0.32\textwidth]{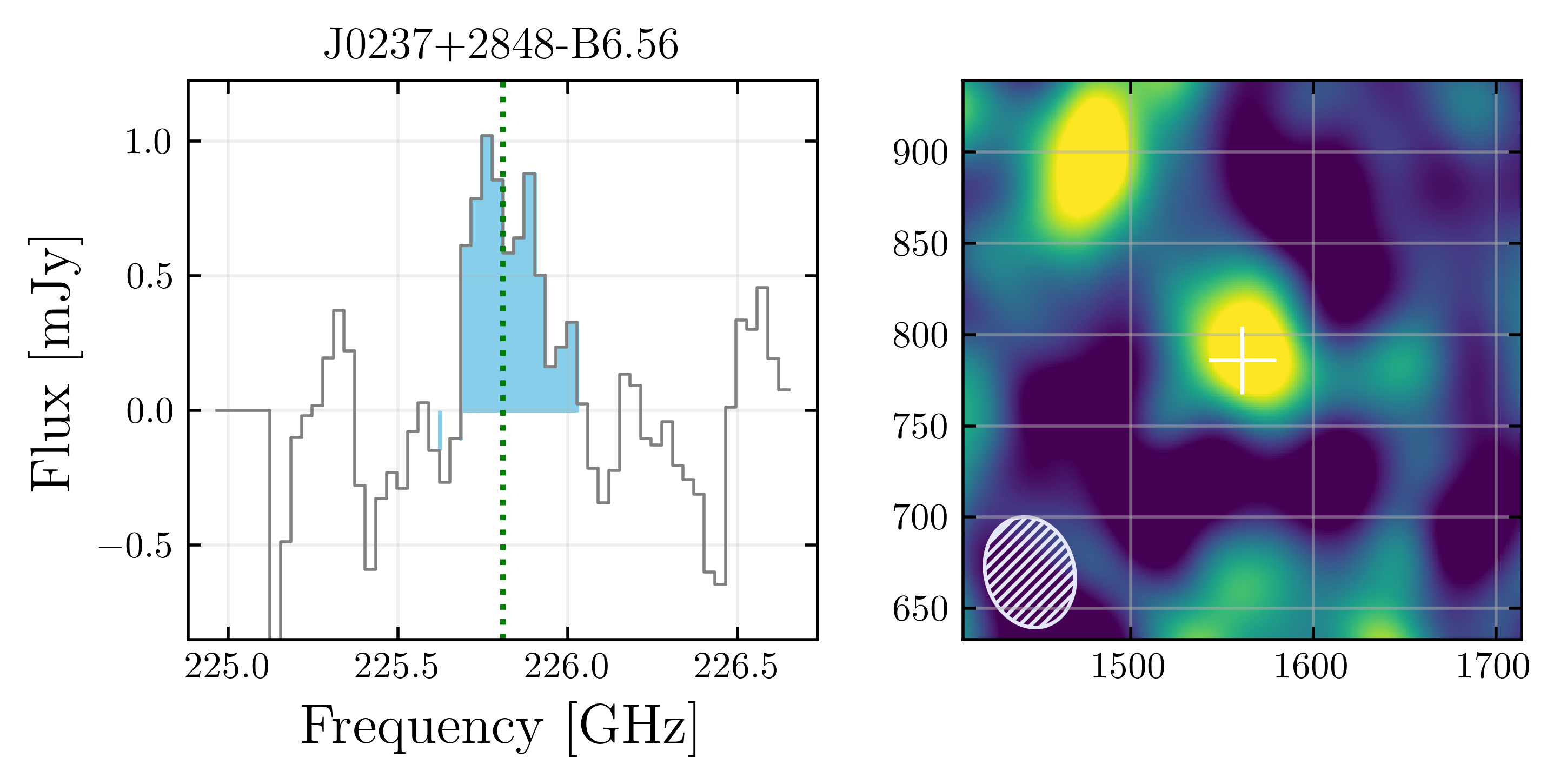}
    \includegraphics[width=0.32\textwidth]{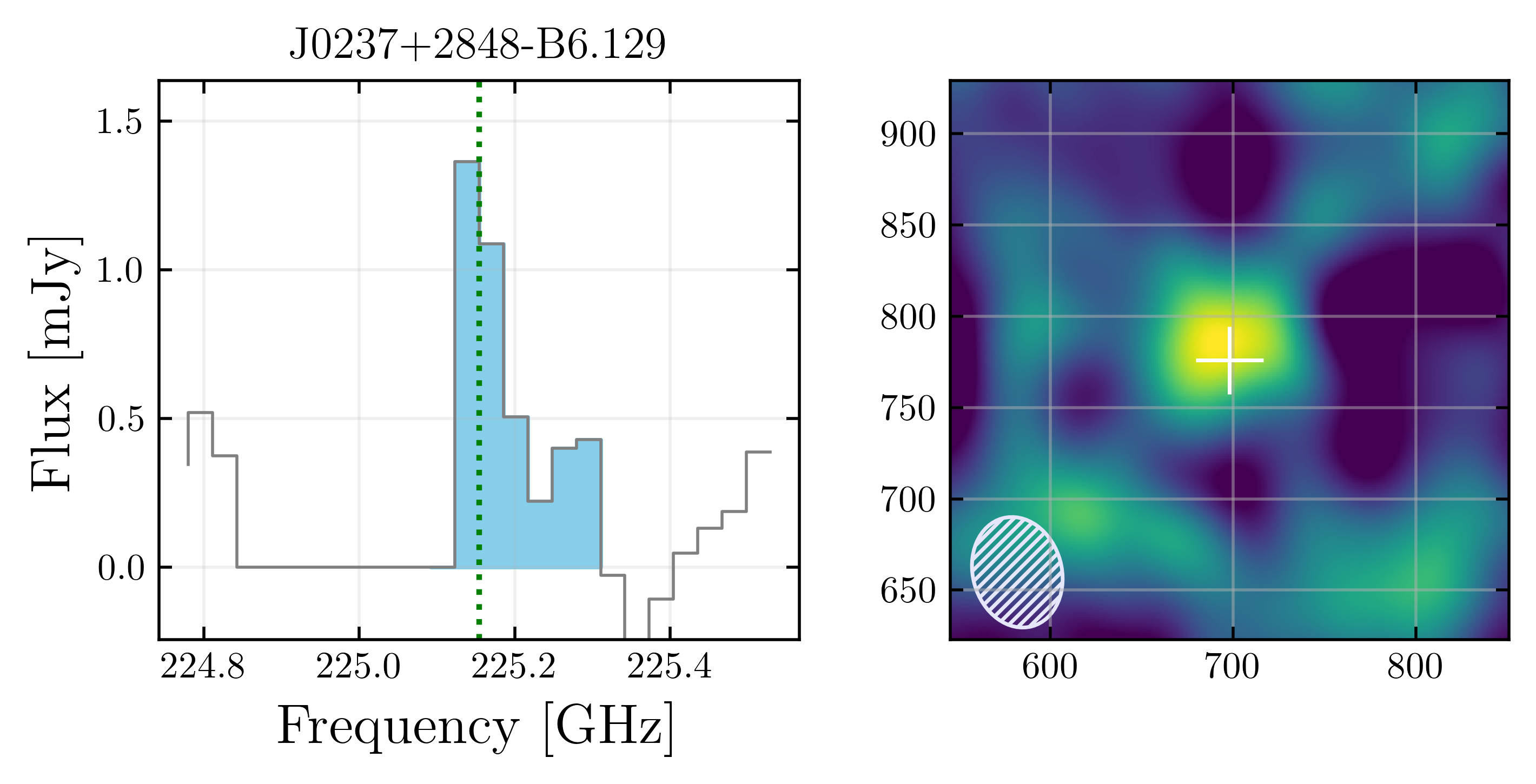}
    \includegraphics[width=0.32\textwidth]{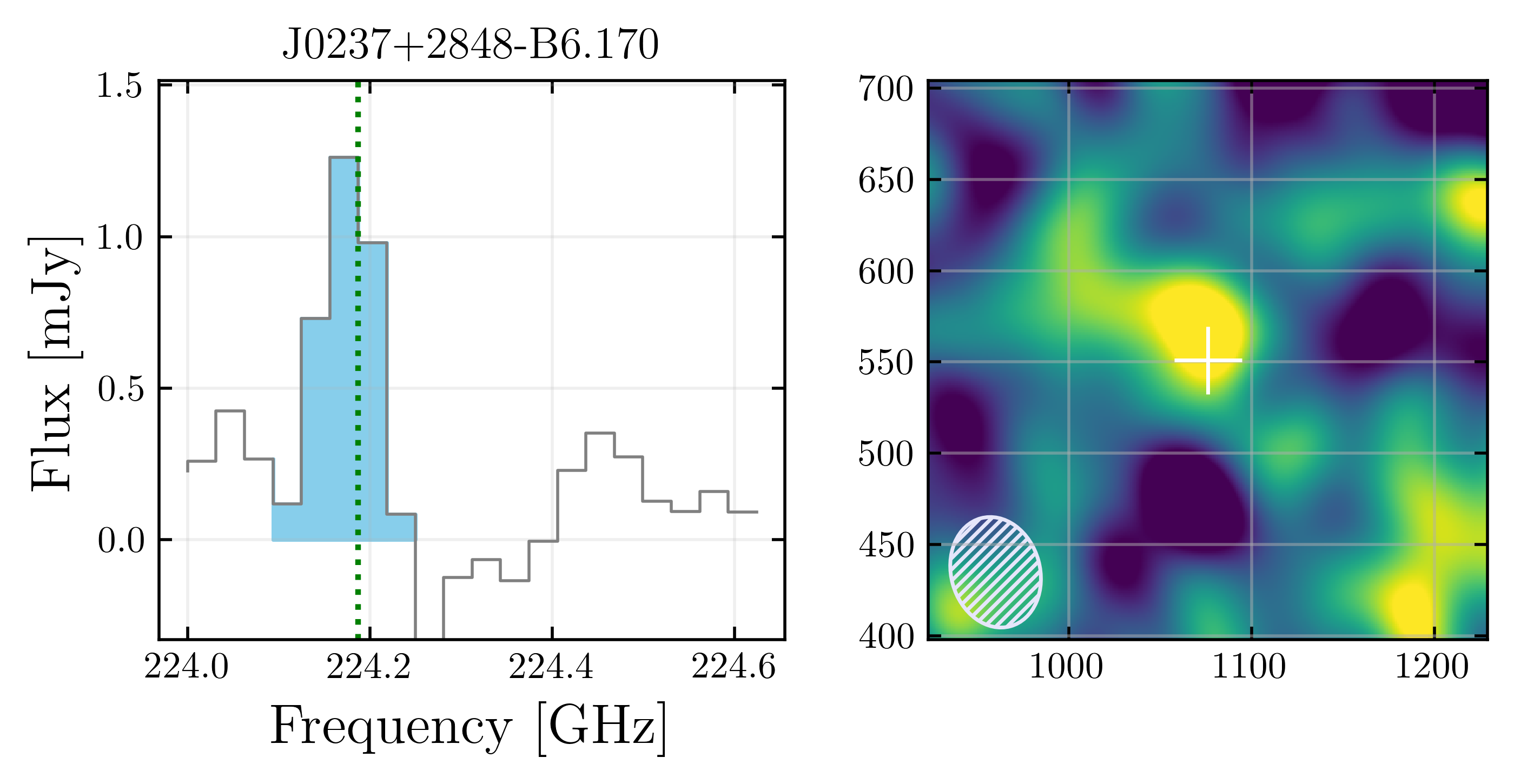}

    \includegraphics[width=0.32\textwidth]{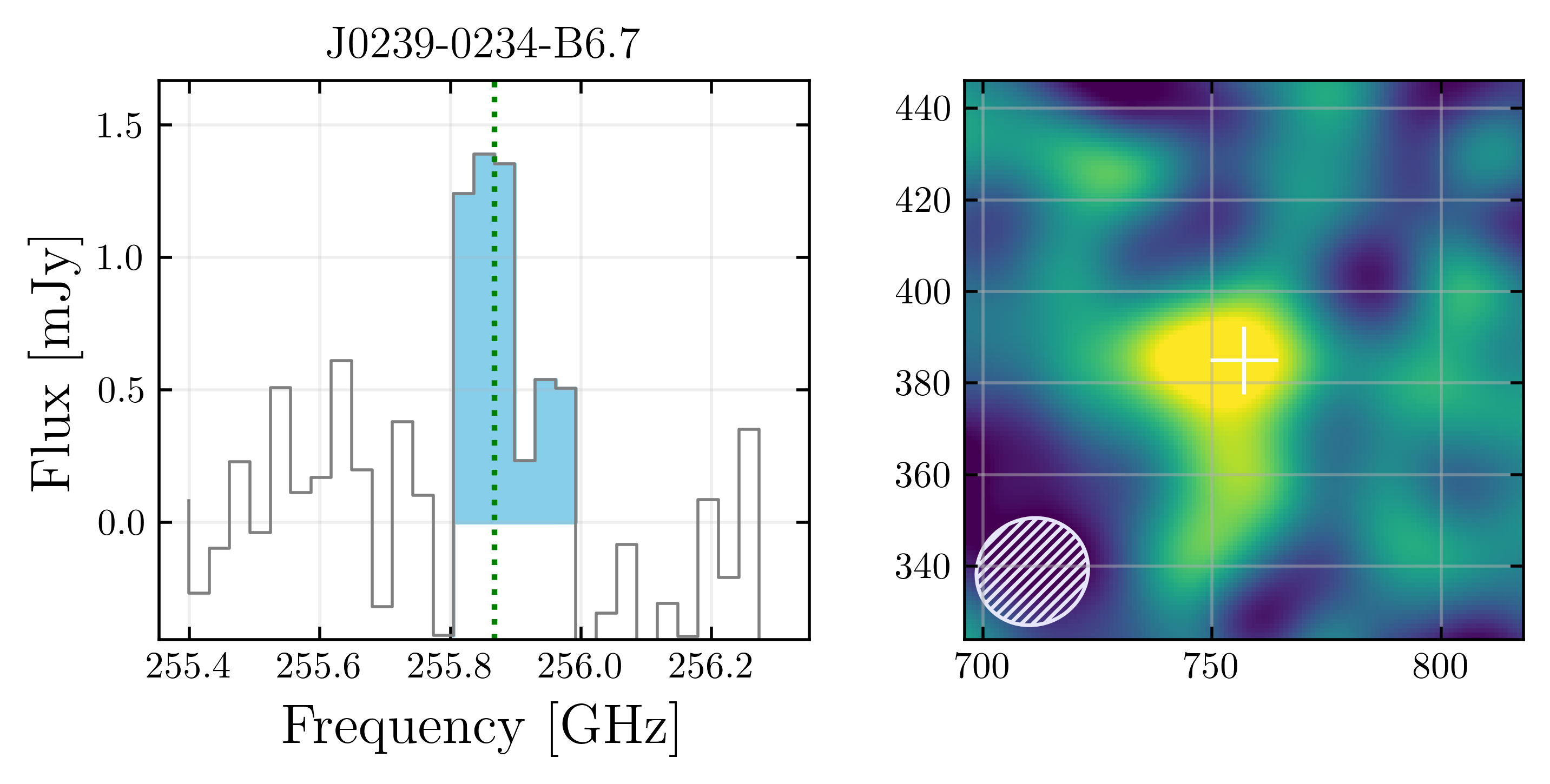}
    \includegraphics[width=0.32\textwidth]{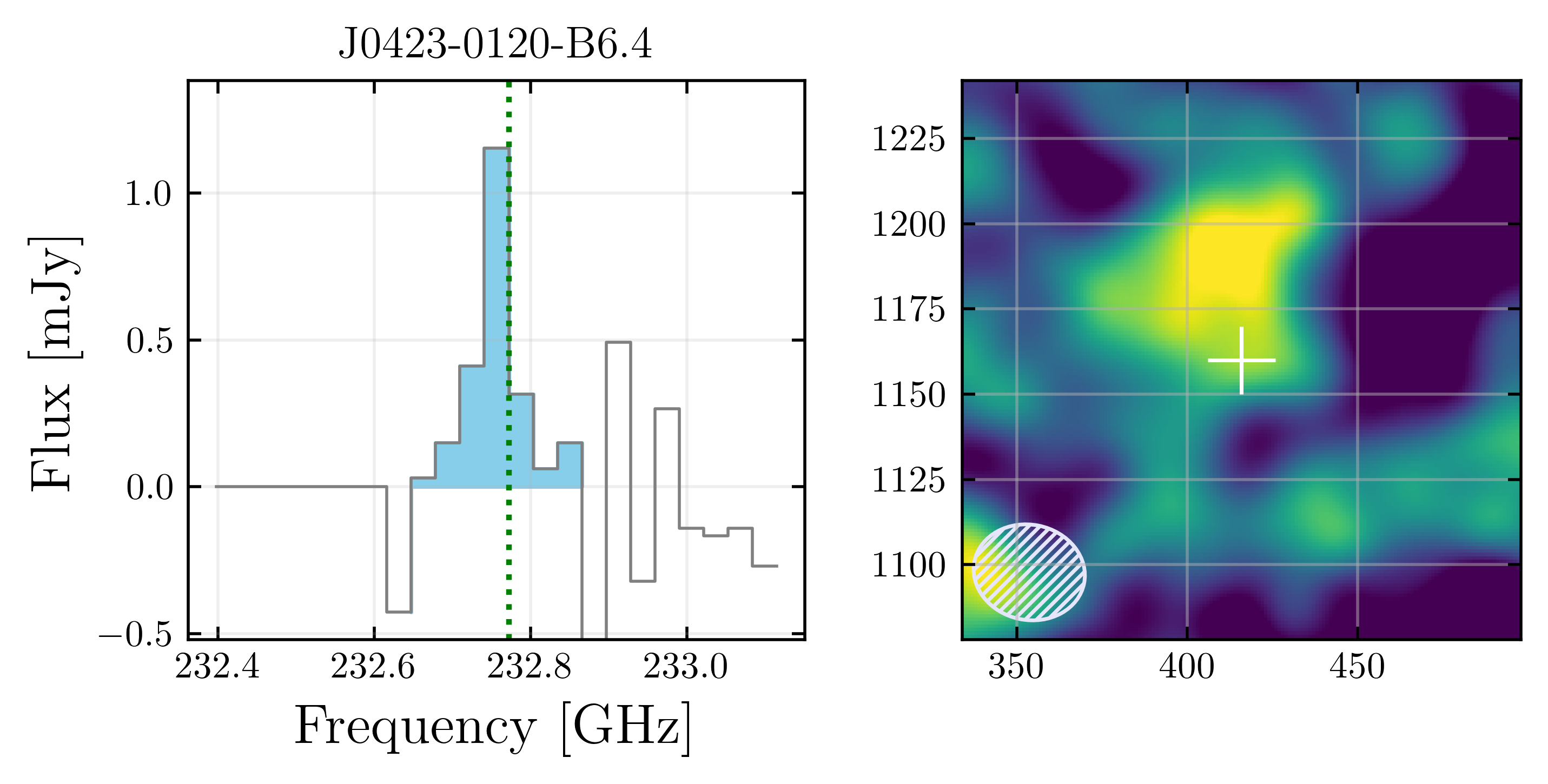}
    \includegraphics[width=0.32\textwidth]{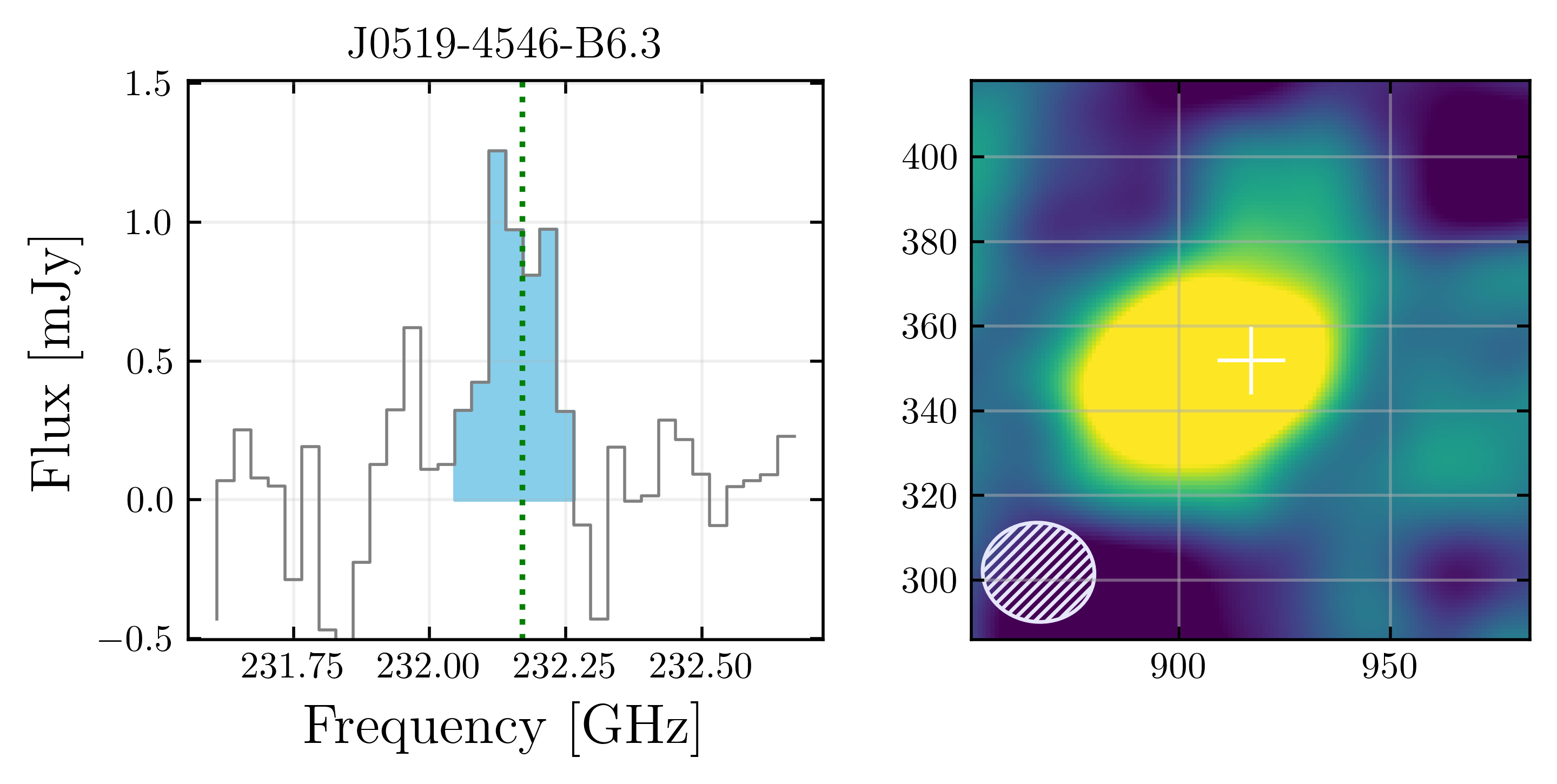}

    \includegraphics[width=0.32\textwidth]{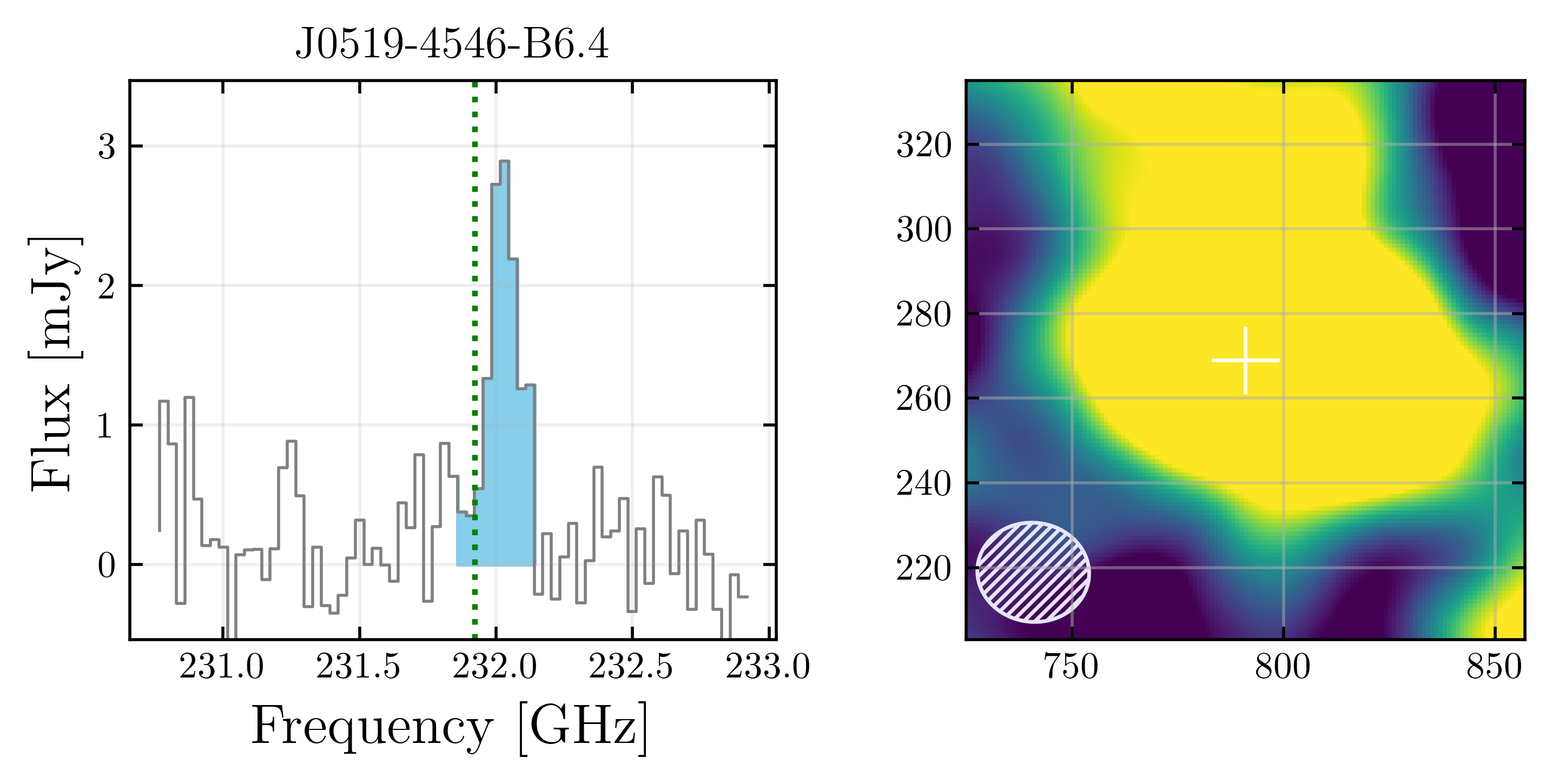}
    \includegraphics[width=0.32\textwidth]{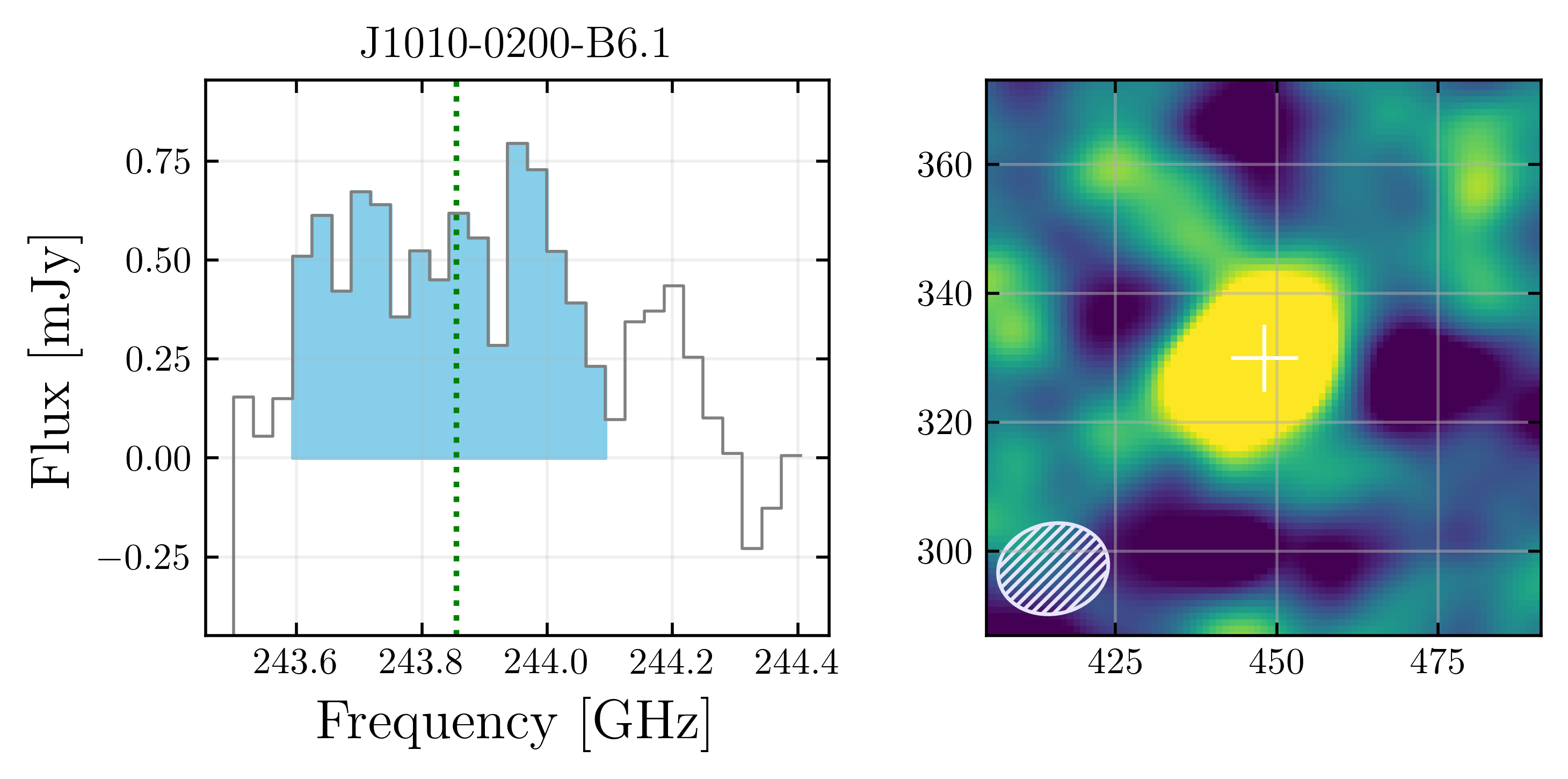}
    \includegraphics[width=0.32\textwidth]{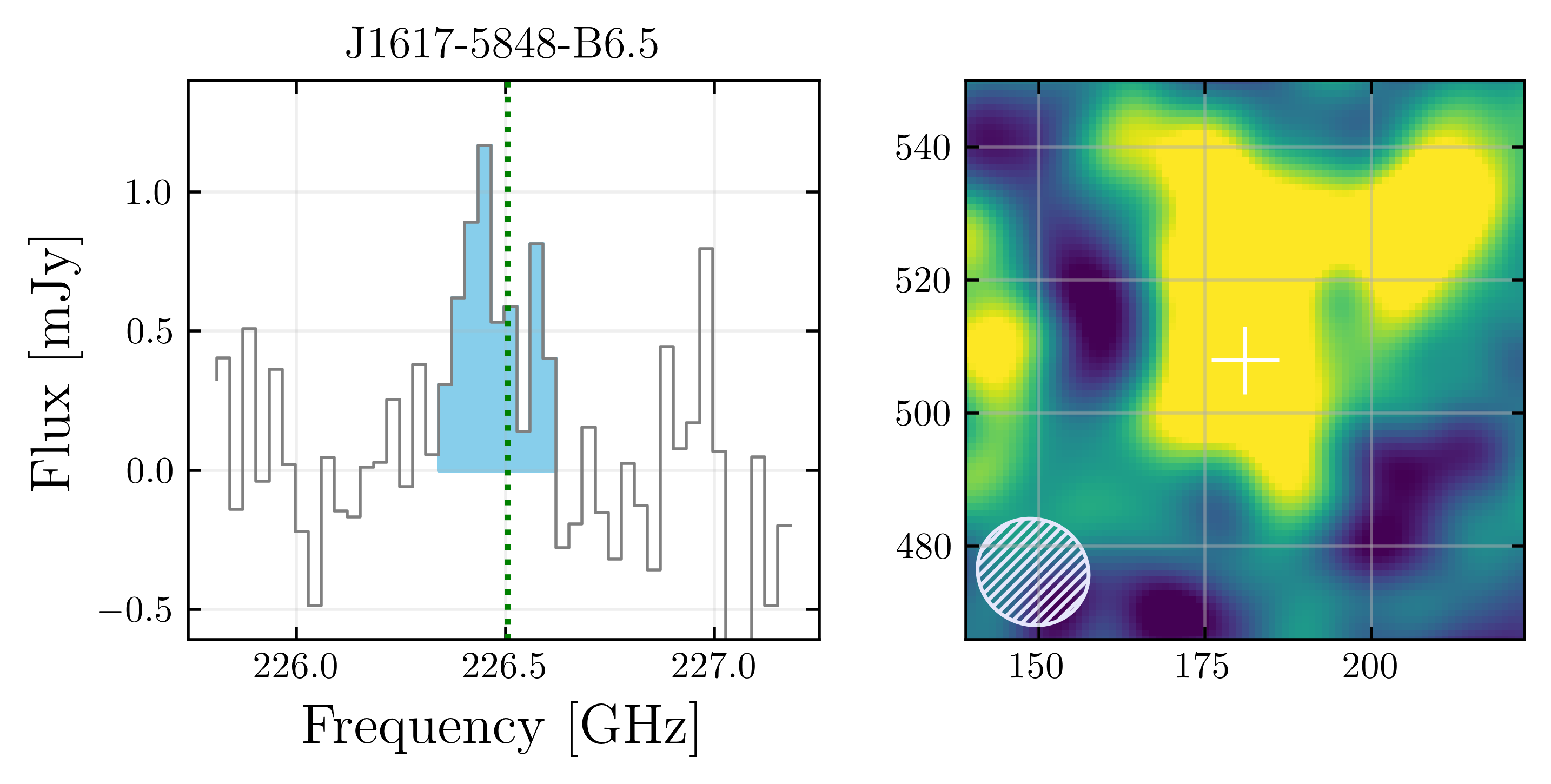}

    \caption{Continuation.}
    \label{fig:detections3}
\end{figure}

\begin{figure}
\centering
    \includegraphics[width=0.32\textwidth]{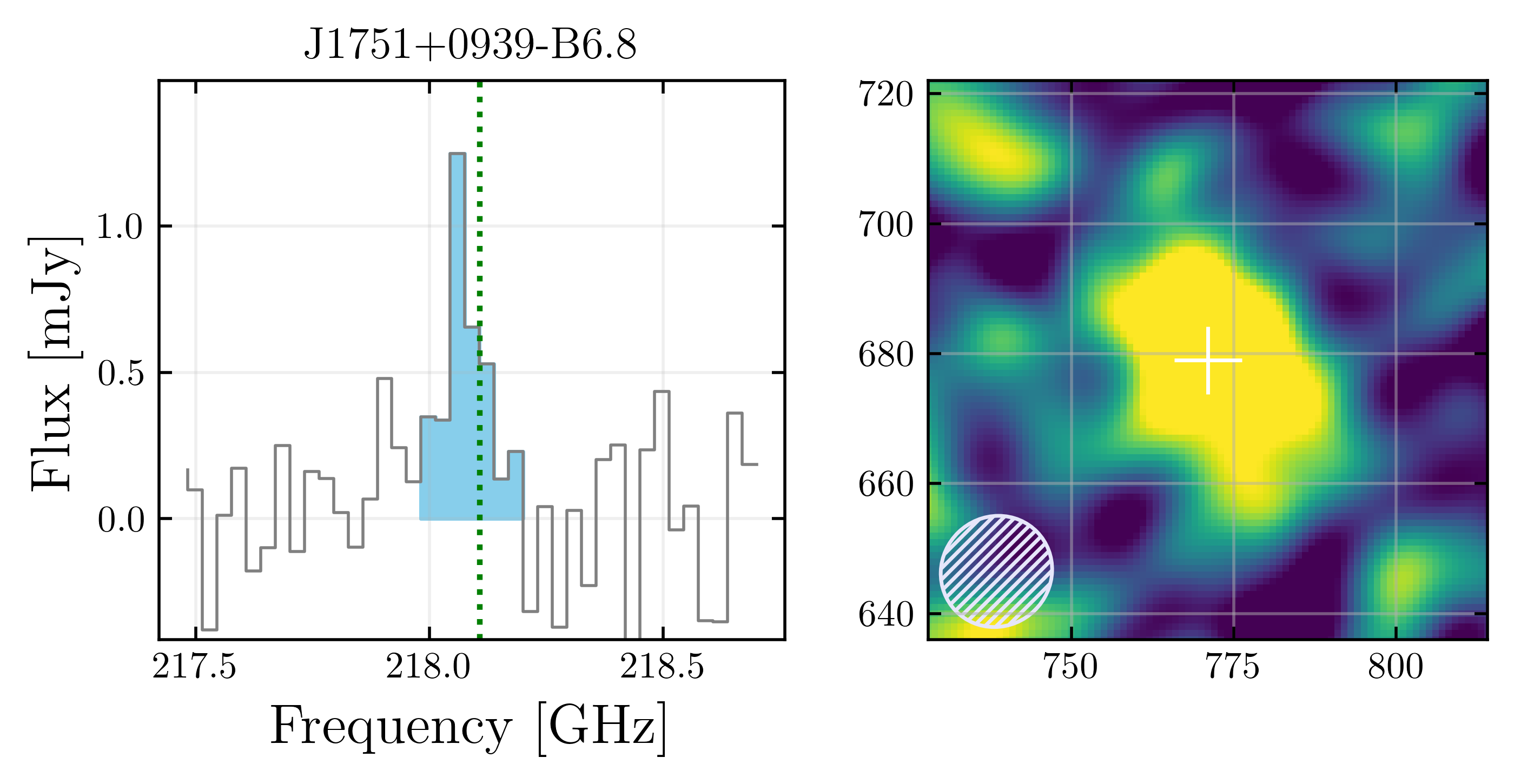}
    \includegraphics[width=0.32\textwidth]{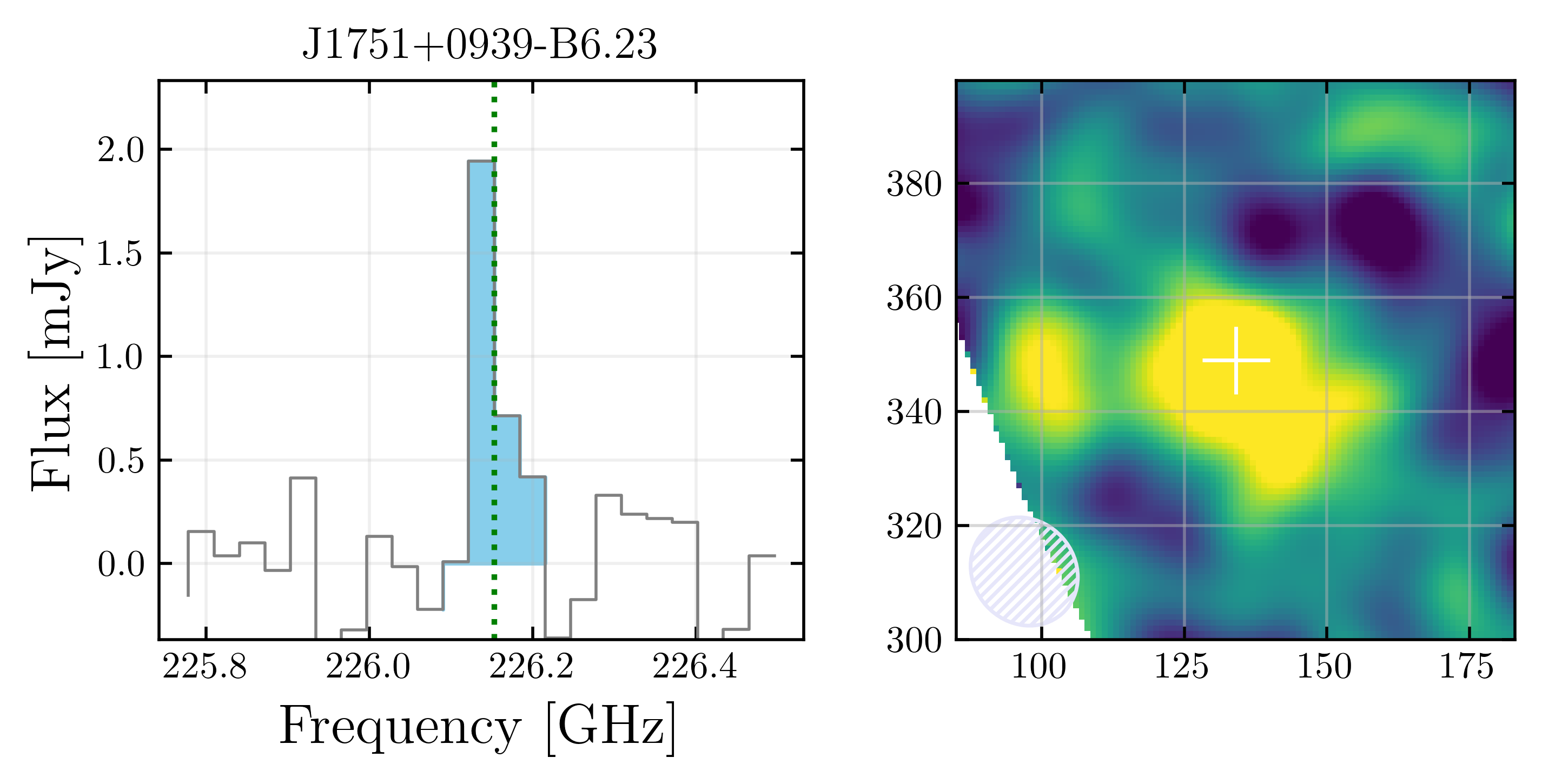}
    \includegraphics[width=0.32\textwidth]{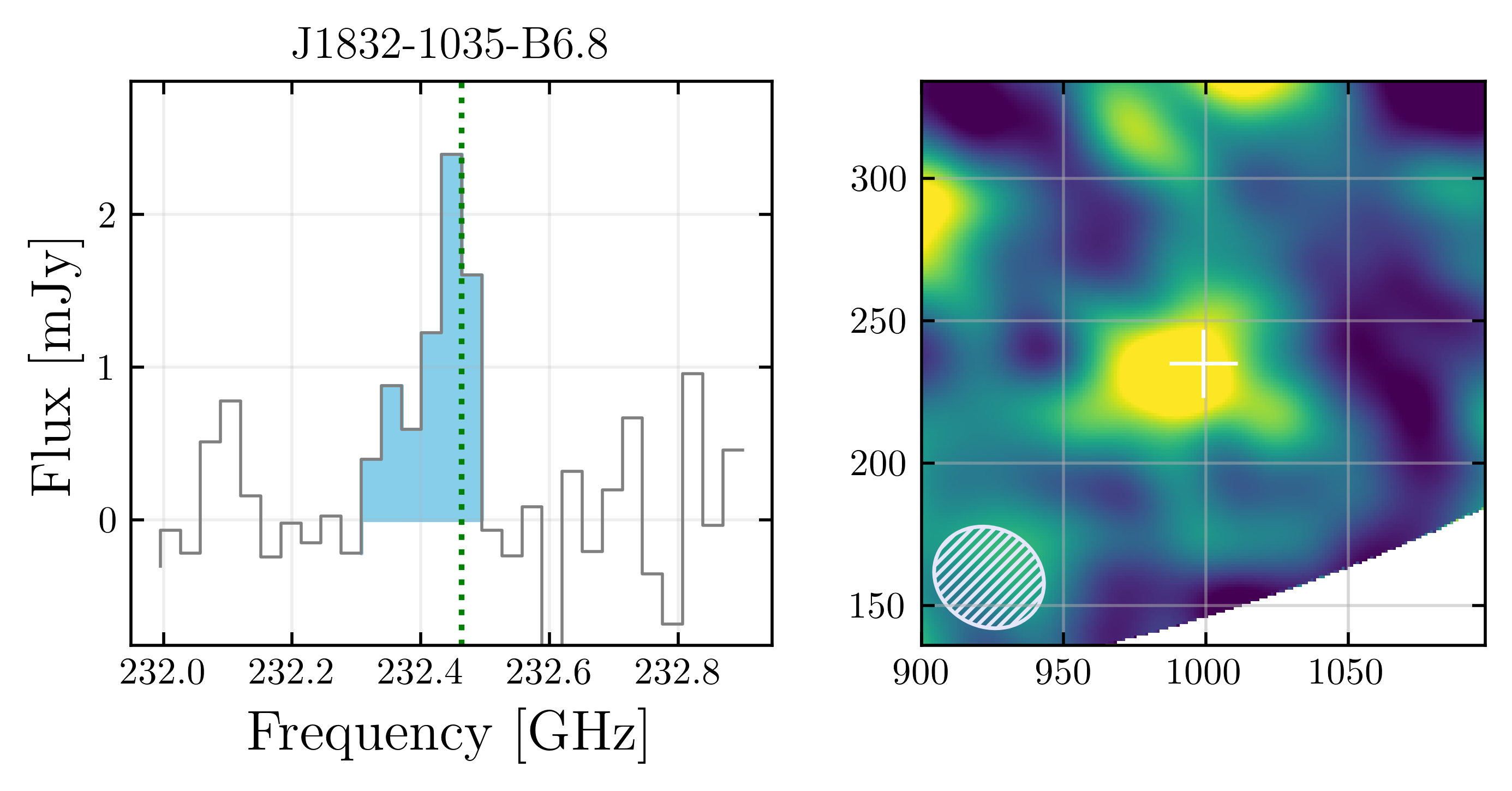}
    
     \includegraphics[width=0.32\textwidth]{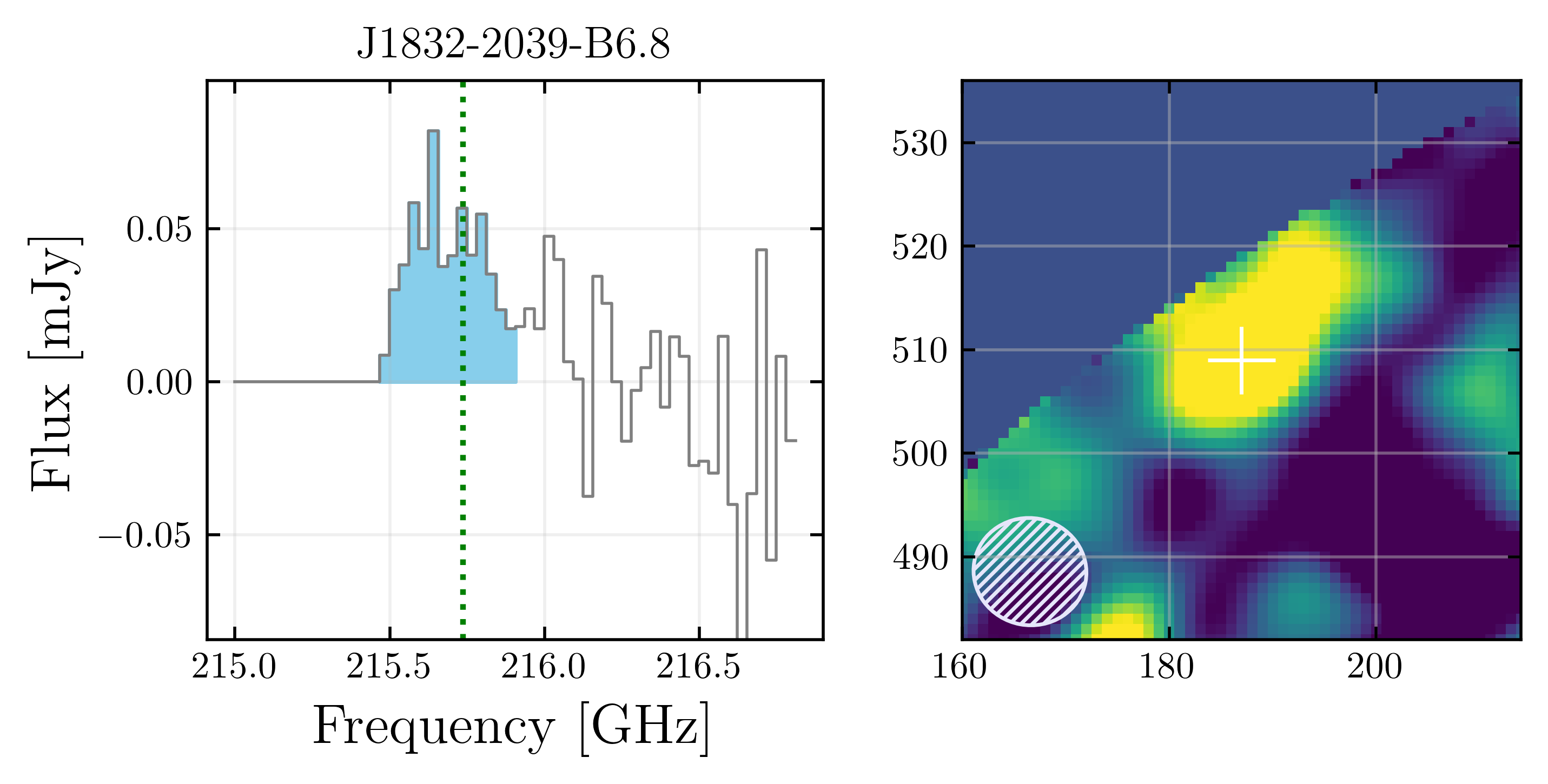}
    \includegraphics[width=0.32\textwidth]{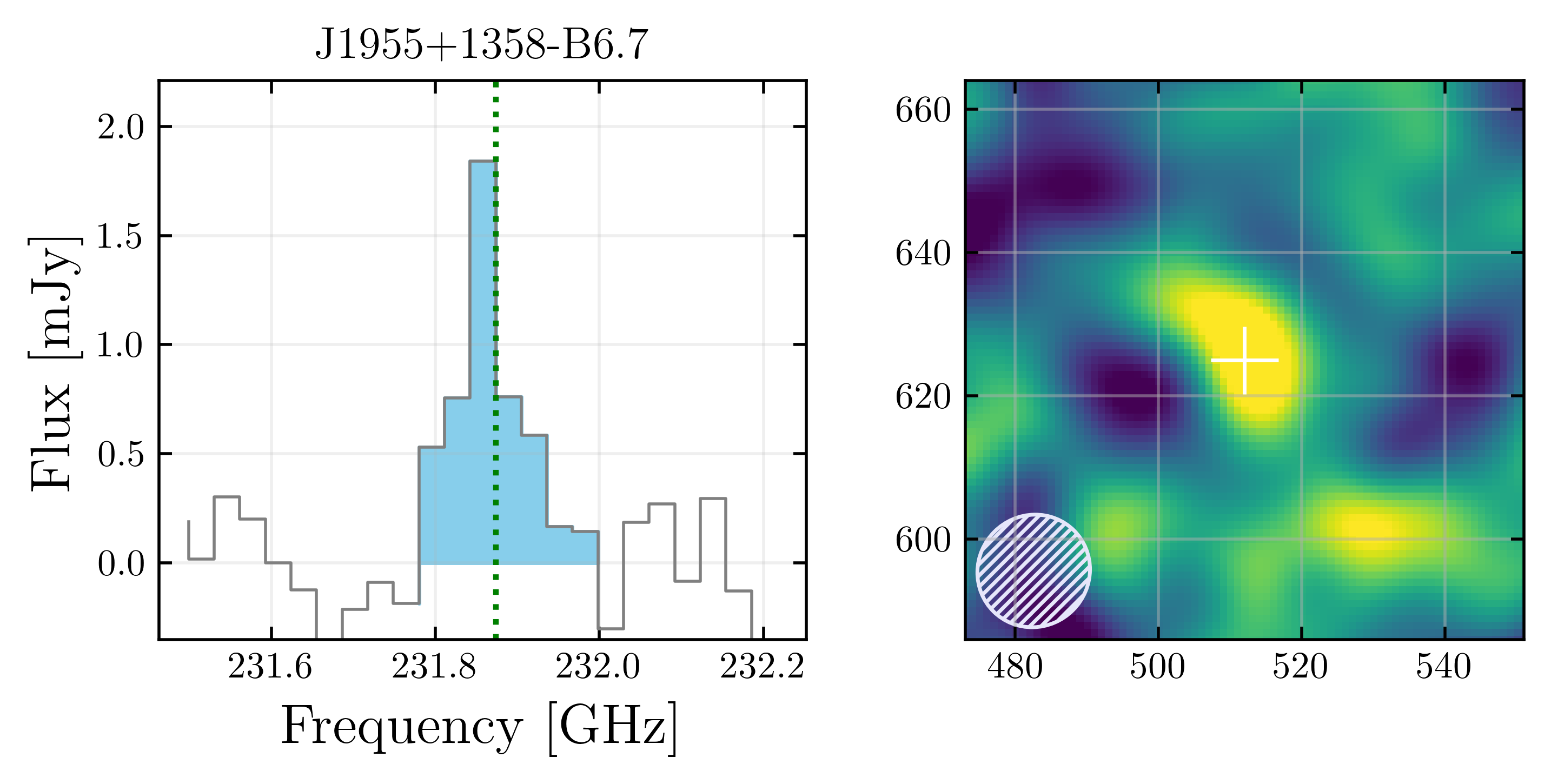}
    \includegraphics[width=0.32\textwidth]{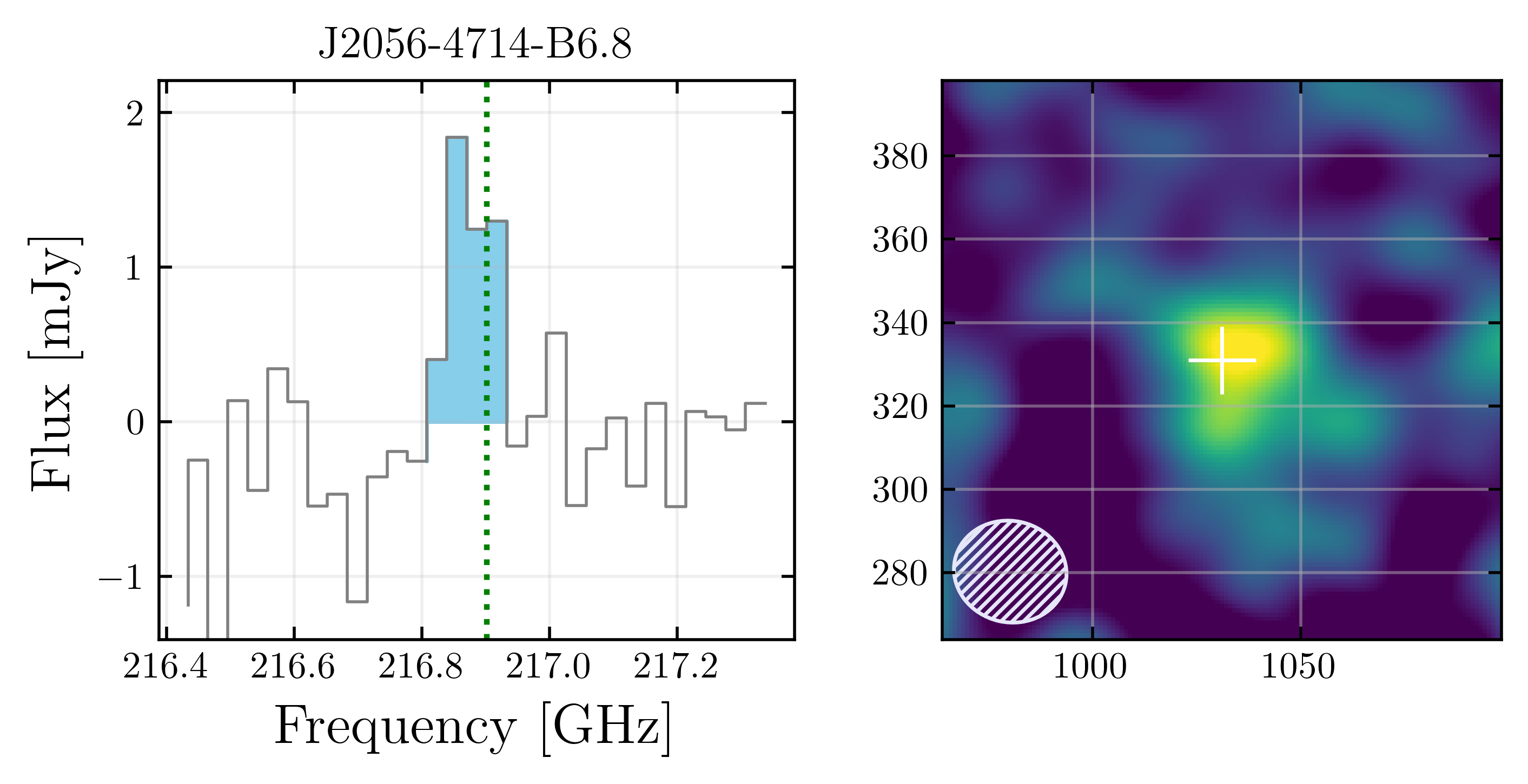}

    \includegraphics[width=0.32\textwidth]{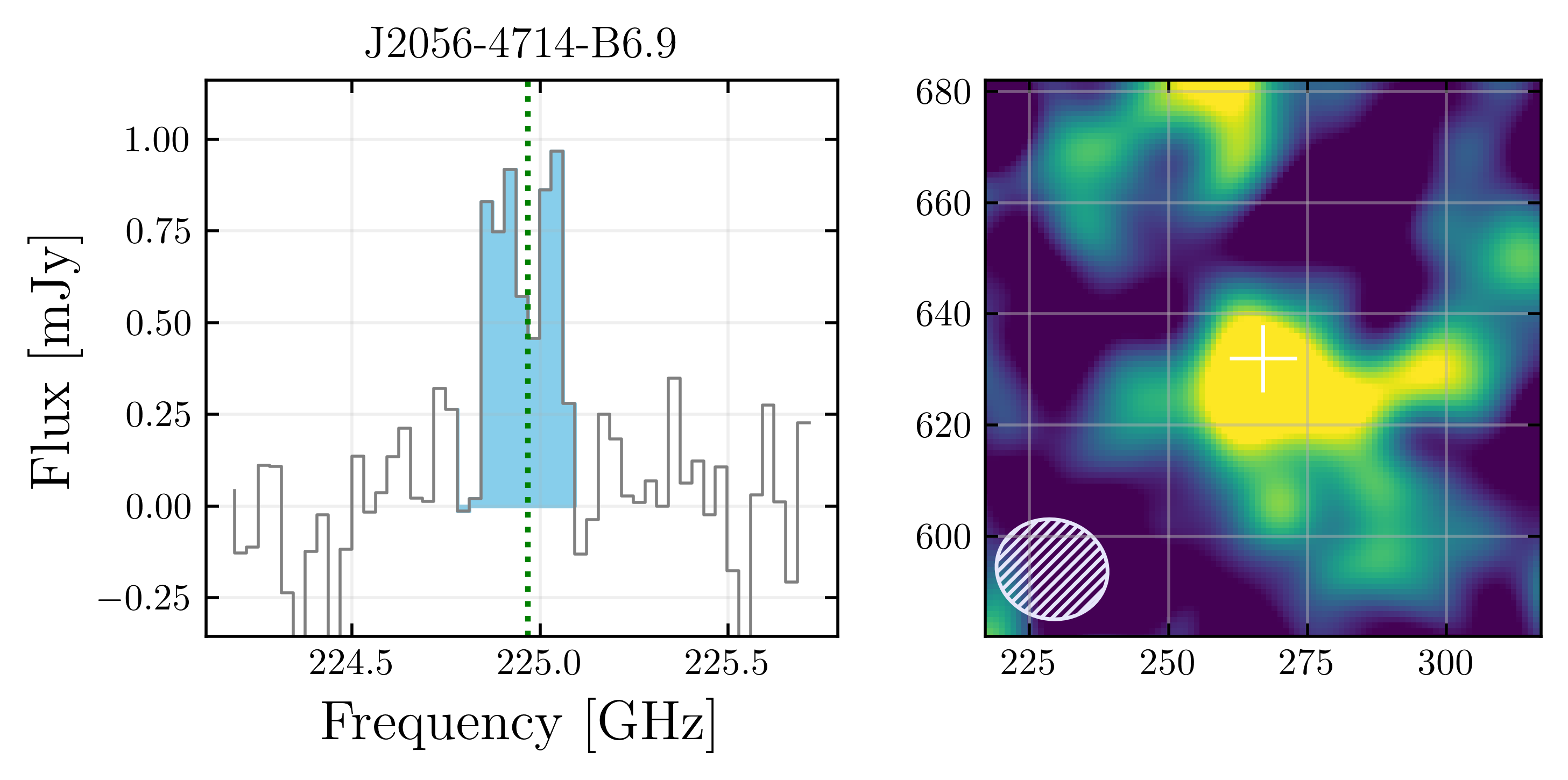}
    \includegraphics[width=0.32\textwidth]{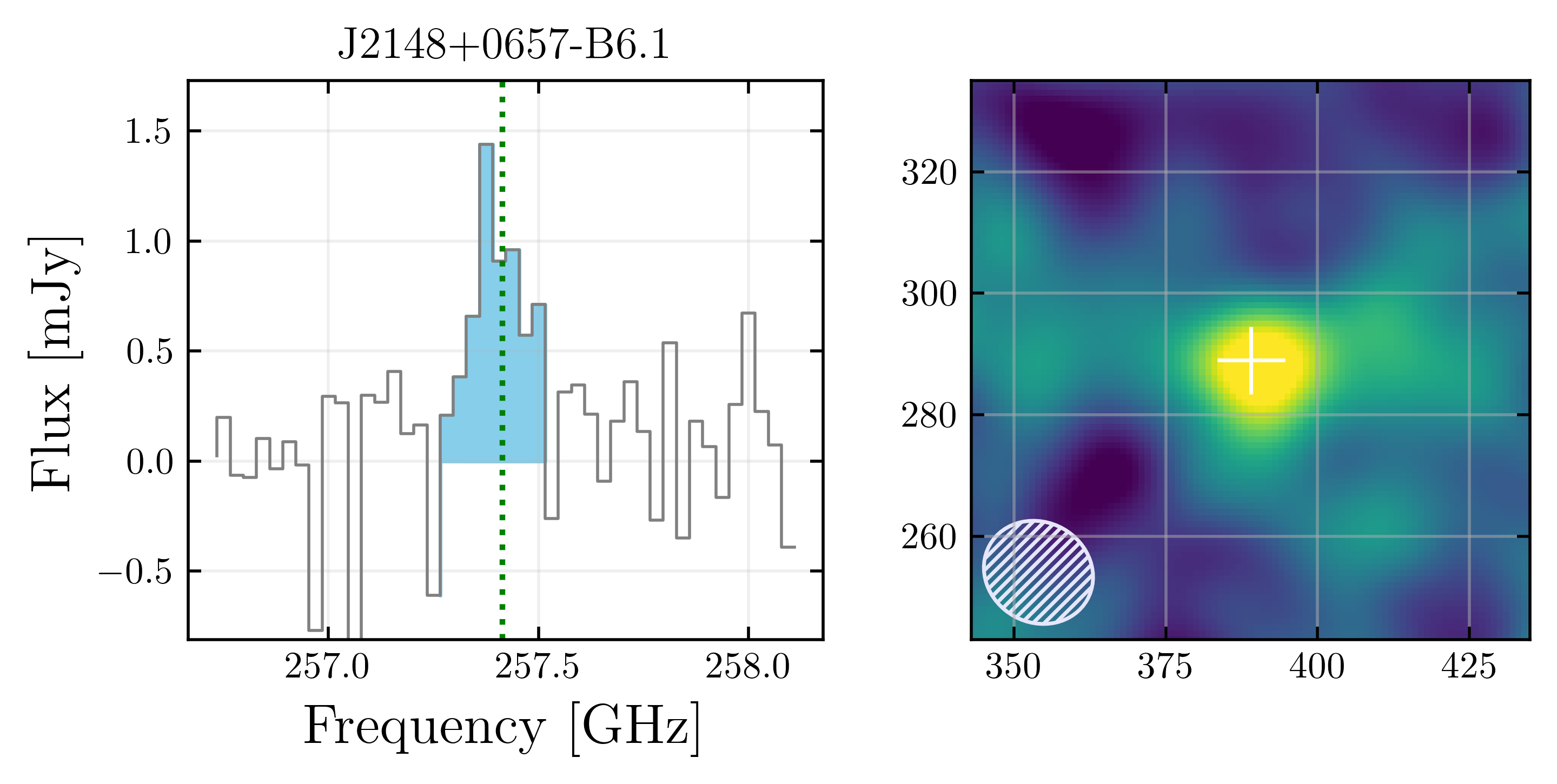}
    \includegraphics[width=0.32\textwidth]{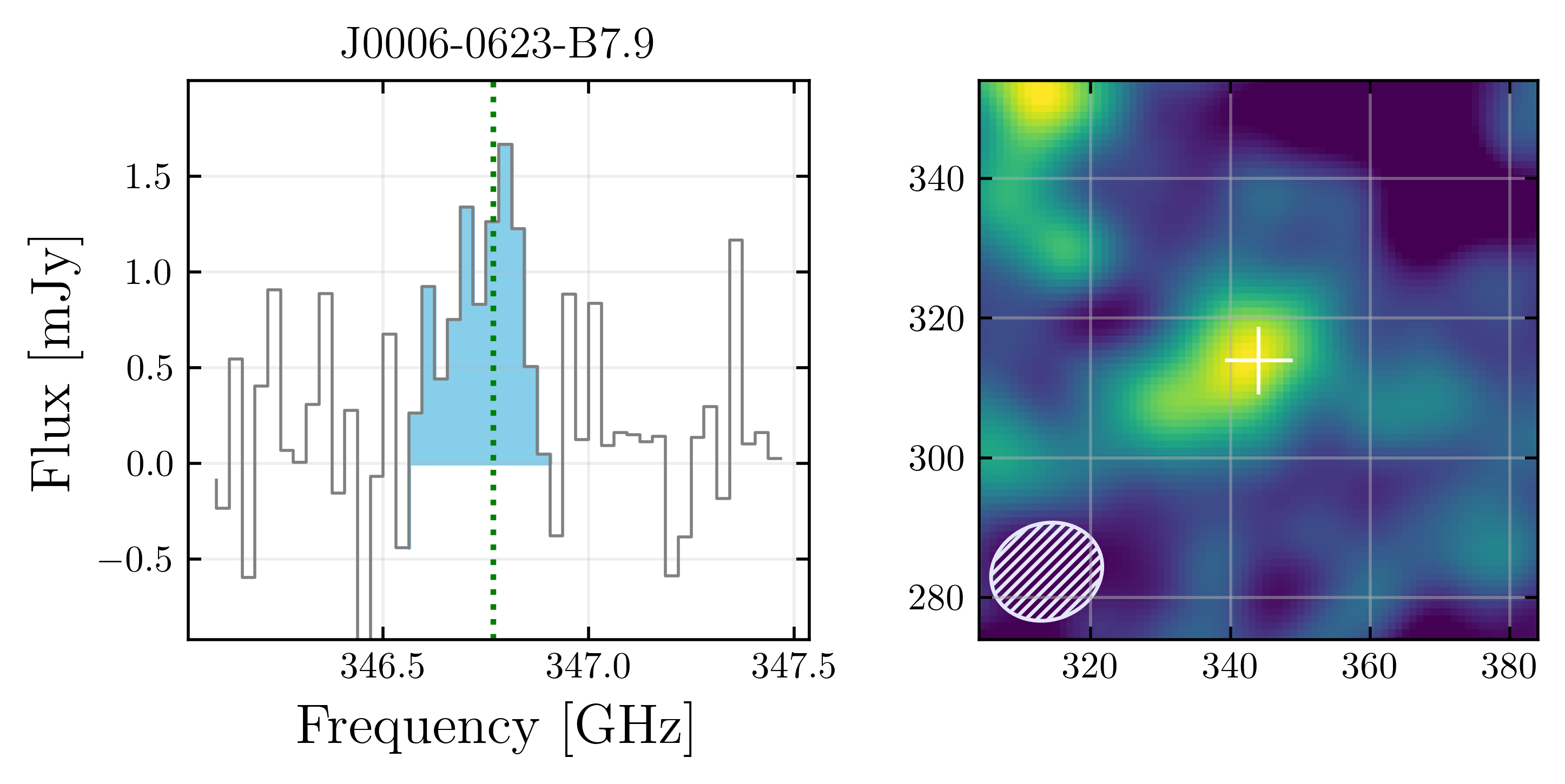}

    \includegraphics[width=0.32\textwidth]{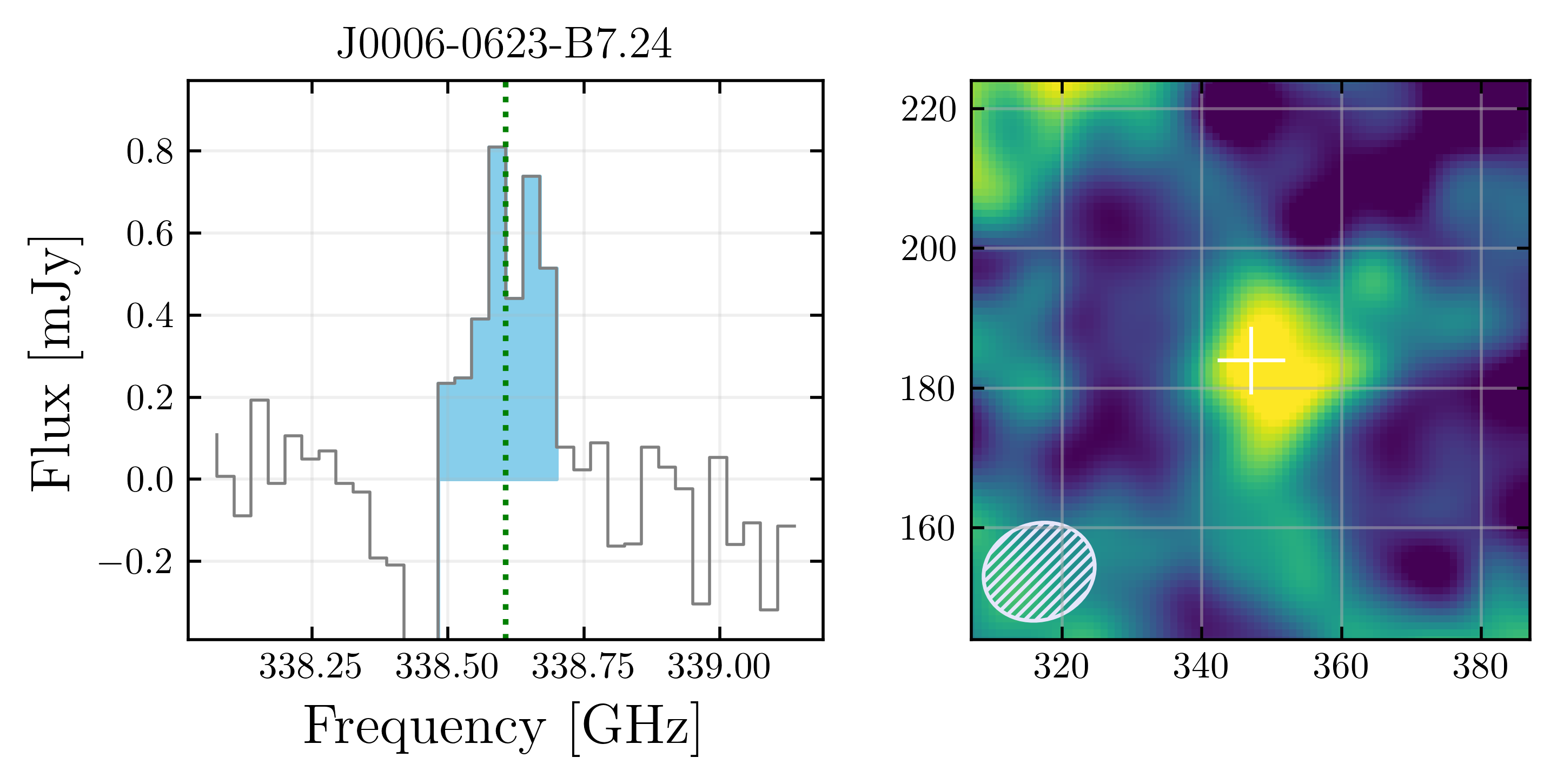}
    \includegraphics[width=0.32\textwidth]{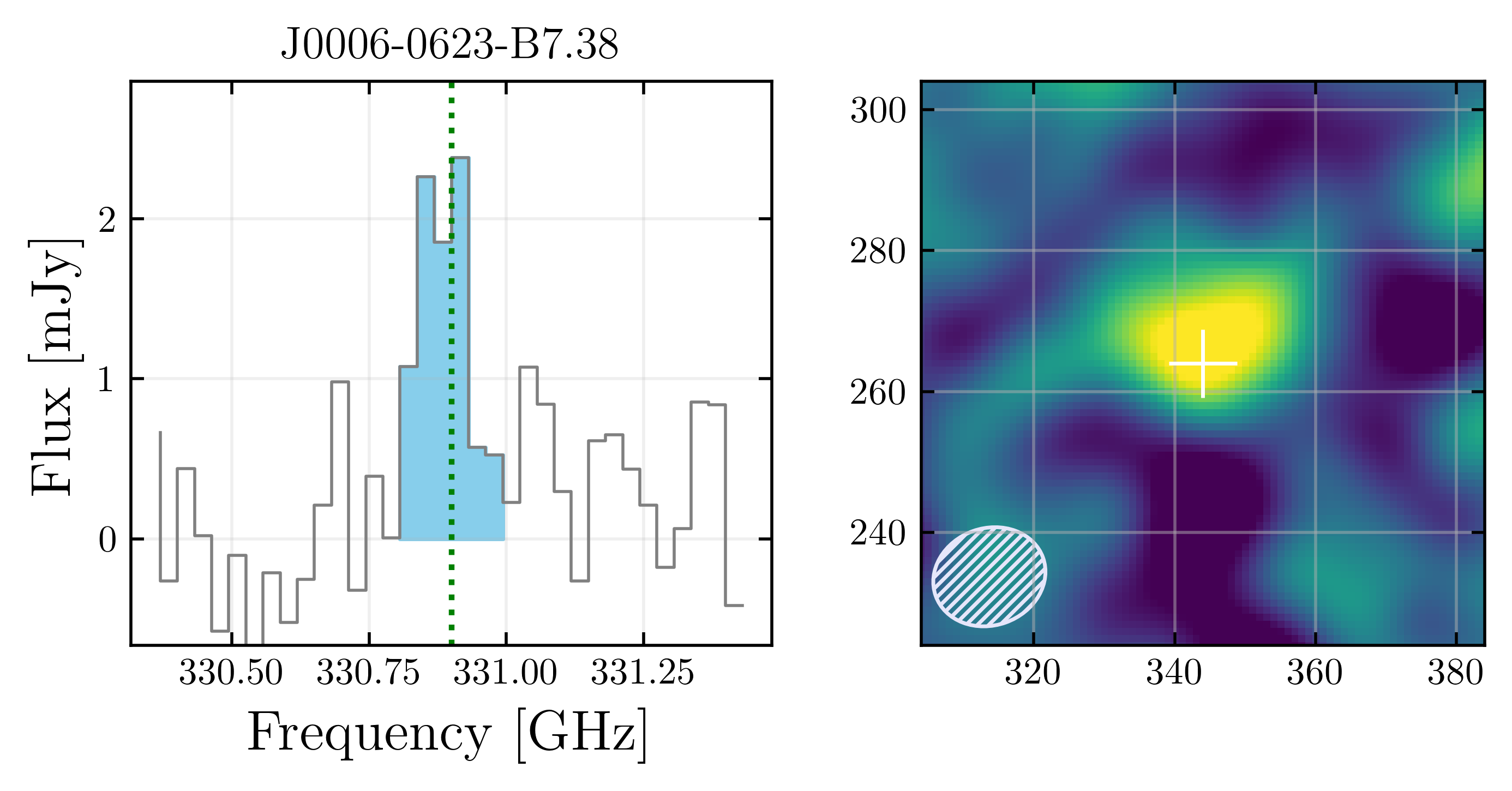}
    \includegraphics[width=0.32\textwidth]{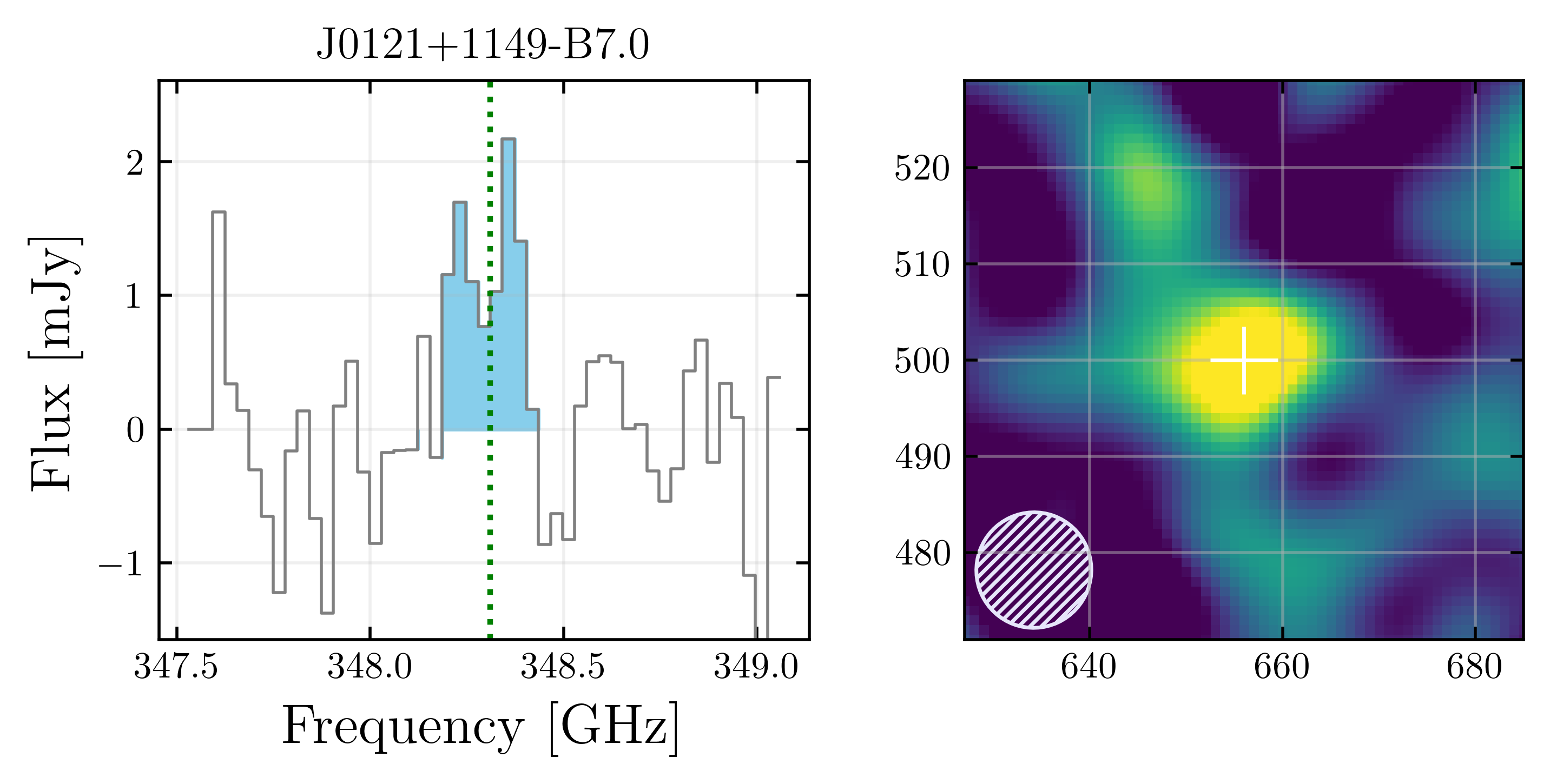}

    \includegraphics[width=0.32\textwidth]{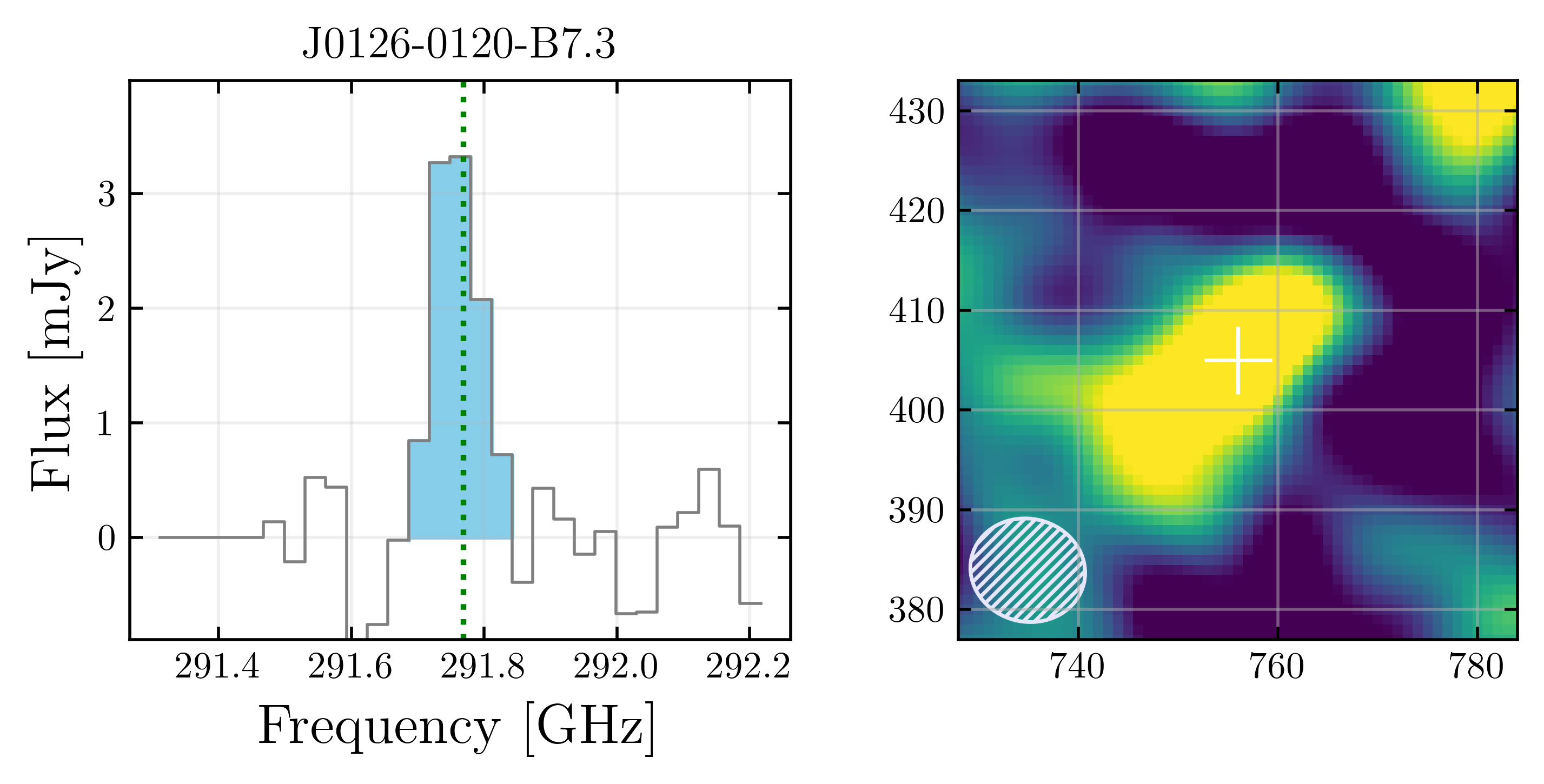}
    \includegraphics[width=0.32\textwidth]{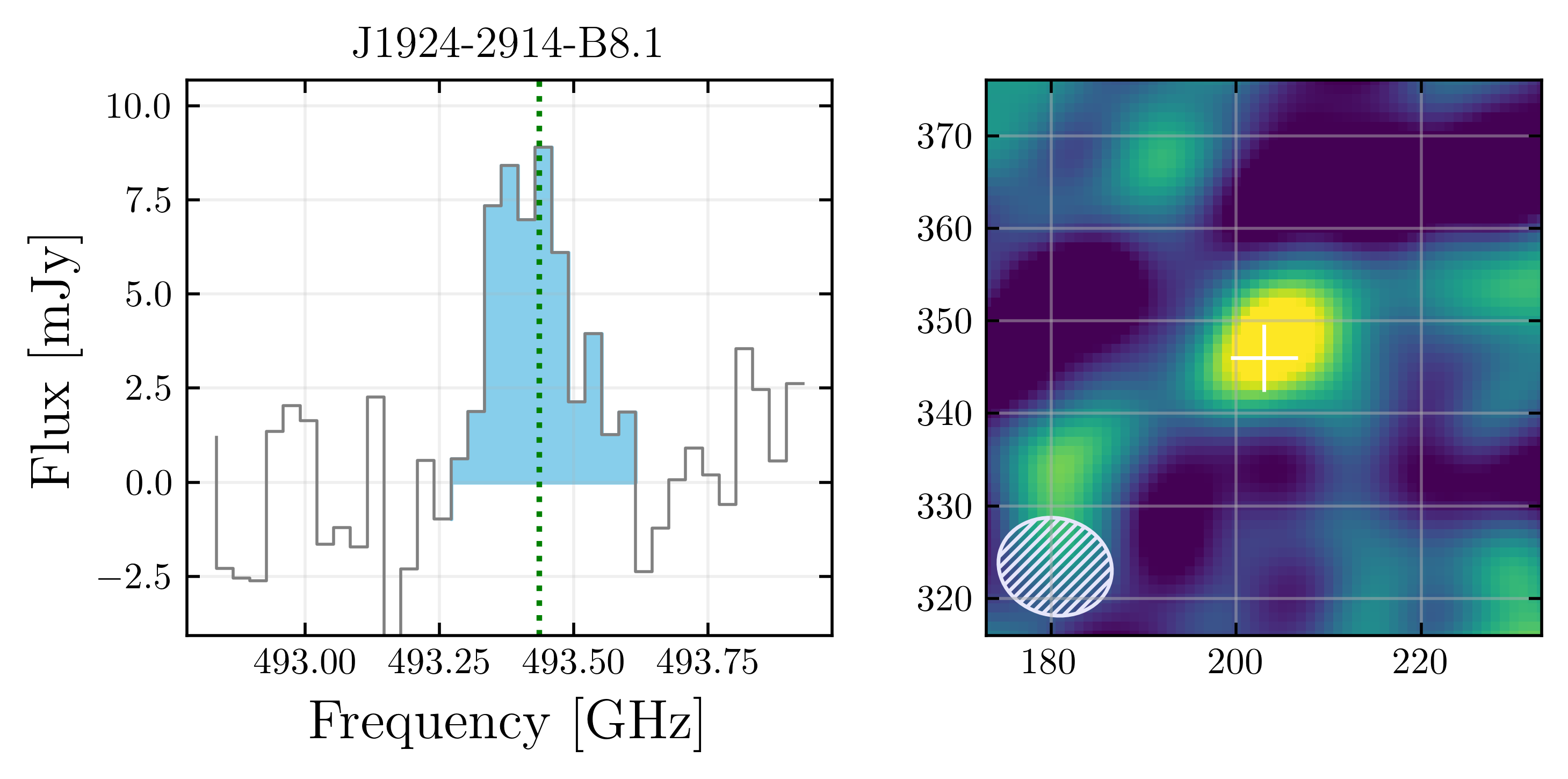}
    \includegraphics[width=0.32\textwidth]{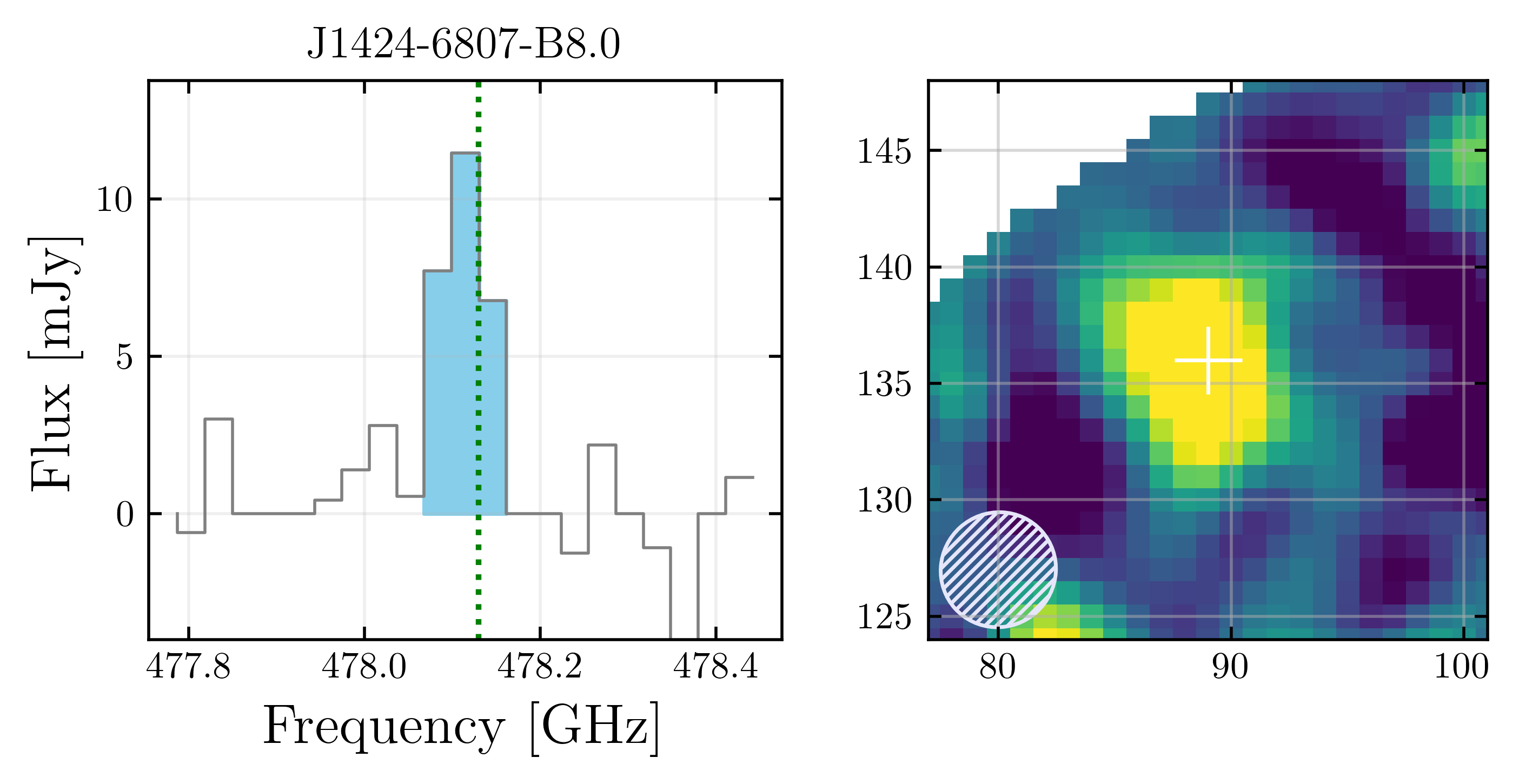}

    \includegraphics[width=0.32\textwidth]{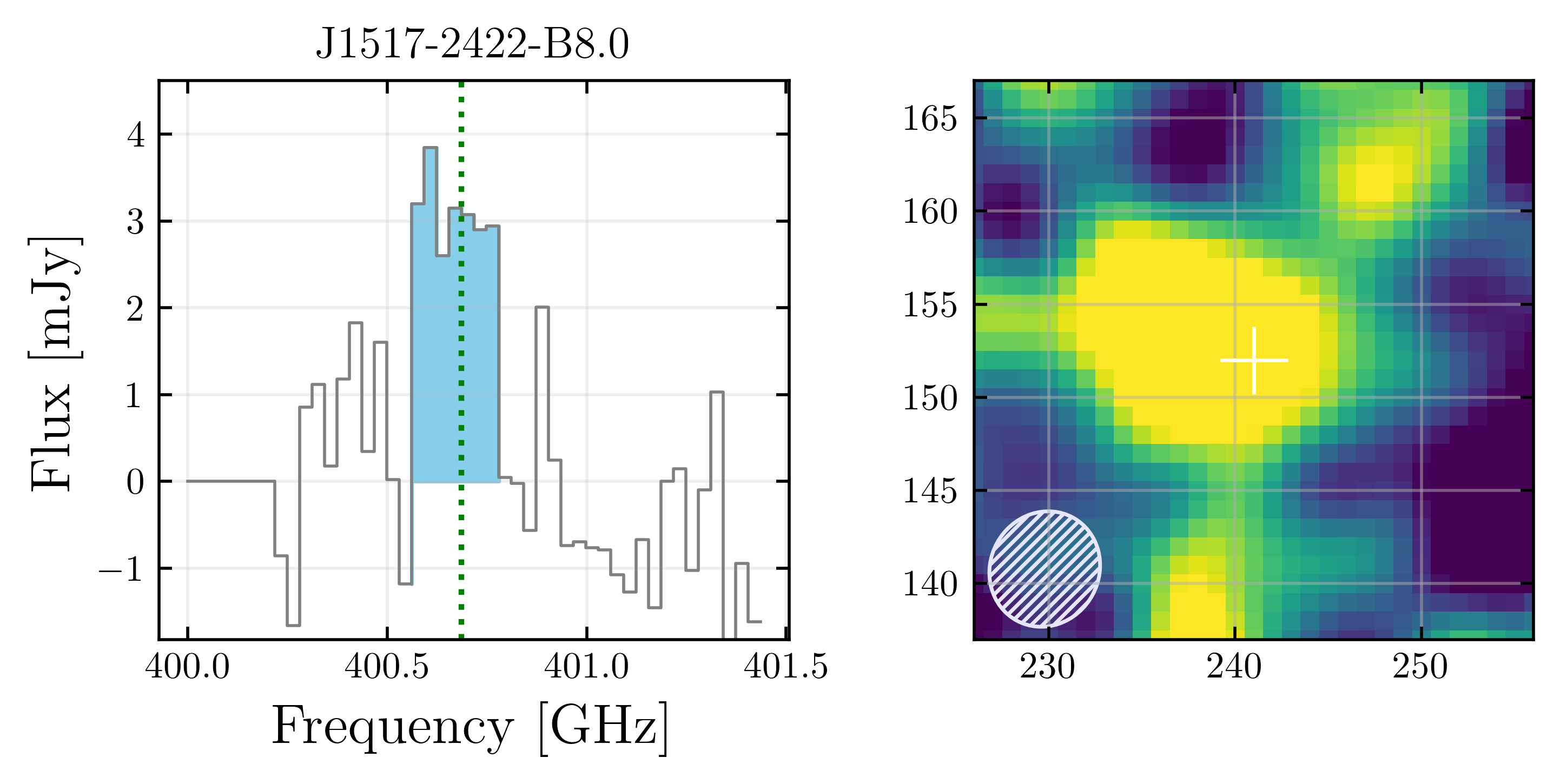}
    \includegraphics[width=0.32\textwidth]{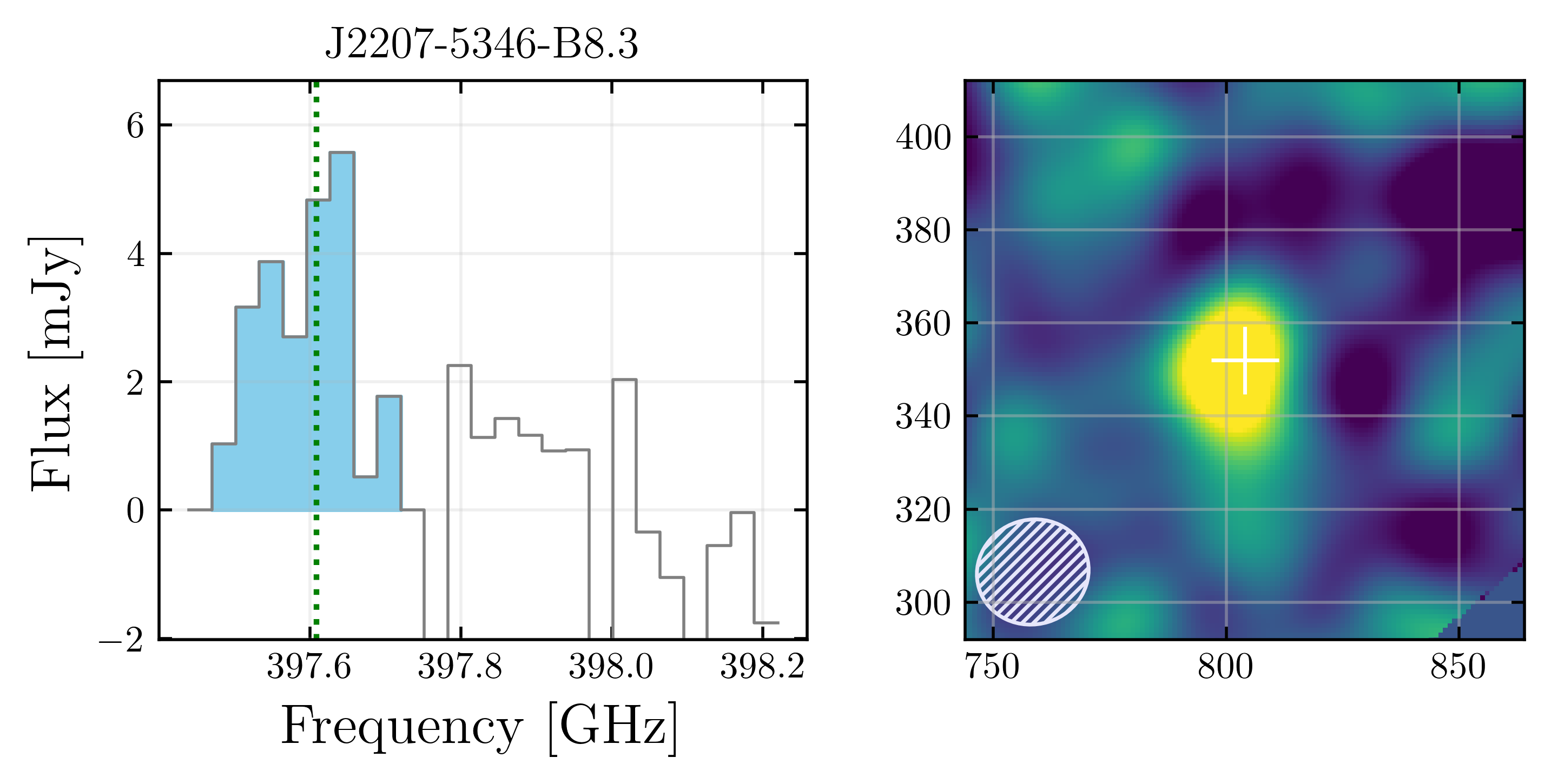}
    \includegraphics[width=0.32\textwidth]{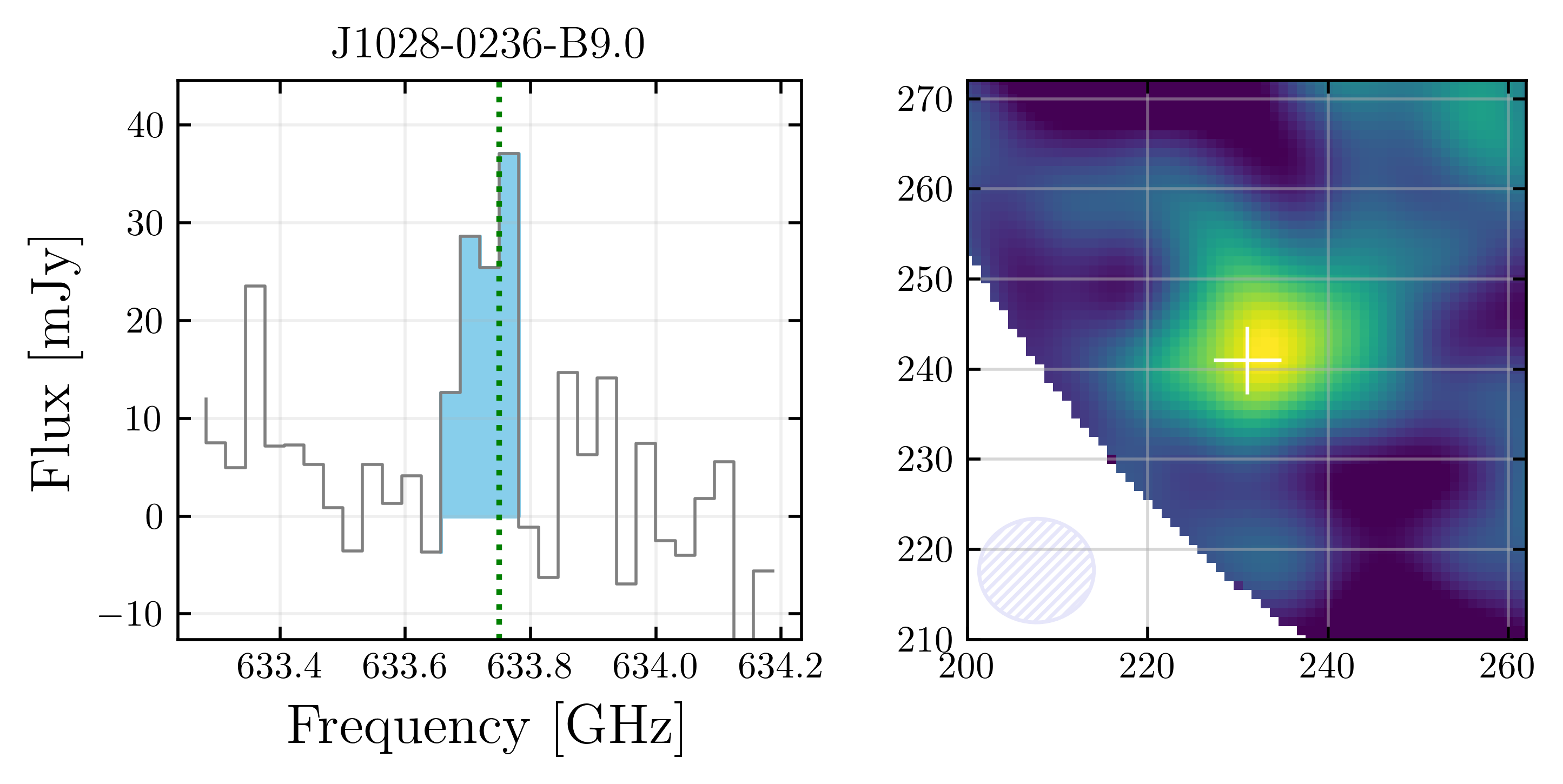}

    \caption{Continuation.}
    \label{fig:detections4}
\end{figure}

\begin{table*}[] \label{tab:detections}
\centering
\caption{Physical properties of the ALMACAL$-22$ detections}
\begin{tabular}{cccccccc}
\hline 
\rule{0pt}{2.5ex} Name.ID & Frequency & Flux & FWHM & S/N & Completeness & Reliability \\ [0.1cm]
 & [GHz] & [mJy km s$^{-1}$] & [km/s] &  & \% &  \% \\ [0.1cm]
\hline 
\rule{0pt}{2.5ex}J0030-4224.2 & 91.4 & $643^{+260}_{-560}$ & $318^{+34}_{-28}$ & 7.24 & 0.78 & 0.27 \\ [0.2cm]
J0030-4224.7 & 91.1 & $919^{+27}_{-37}$ & $202^{+18}_{-18}$ & 7.96 & 0.81 & 0.19 \\ [0.2cm]
J0030-4224.2 & 98.3 & $942^{+19}_{-26}$ & $287^{+42}_{-36}$ & 6.43 & 0.80 & 0.27 \\ [0.2cm]
J0038-2459.1 & 111.8 & $159^{+789}_{-159}$ & $182^{+26}_{-28}$ & 8.60 & 0.83 & 0.39 \\ [0.2cm]
J0108+0135.42 & 99.6 & $946^{+22}_{-26}$ & $338^{+33}_{-33}$ & 8.55 & 0.82 & 0.52 \\ [0.2cm]
J0215-0222.6 & 104.2 & $1000^{+0}_{-13}$ & $53^{+51}_{-3}$ & 10.48 & 0.81 & 0.25 \\ [0.2cm]
J0215-0222.15 & 102.6 & $796^{+94}_{-252}$ & $275^{+261}_{-69}$ & 10.10 & 0.81 & 0.42 \\ [0.2cm]
J0217-0820.0 & 106.7 & $854^{+62}_{-96}$ & $384^{+69}_{-66}$ & 12.43 & 0.79 & 0.71 \\ [0.2cm]
J0257-1212.4 & 107.2 & $394^{+521}_{-394}$ & $303^{+164}_{-125}$ & 7.42 & 0.78 & 0.53 \\ [0.2cm]
J0257-1212.7 & 105.6 & $929^{+56}_{-260}$ & $300^{+253}_{-72}$ & 5.05 & 0.81 & 0.37 \\ [0.2cm]
J0257-1212.10 & 105.2 & $900^{+27}_{-38}$ & $473^{+60}_{-50}$ & 6.73 & 0.81 & 0.57 \\ [0.2cm]
J0334-4008.1 & 104.4 & $929^{+23}_{-36}$ & $442^{+39}_{-32}$ & 13.67 & 0.82 & 0.92 \\ [0.2cm]
J0334-4008.2 & 101.7 & $991^{+9}_{-30}$ & $53^{+62}_{-14}$ & 10.62 & 0.81 & 1.00 \\ [0.2cm]
J0334-4008.3 & 97.3 & $935^{+36}_{-77}$ & $285^{+63}_{-40}$ & 9.10 & 0.81 & 1.00 \\ [0.2cm]
J0334-4008.4 & 89.8 & $860^{+23}_{-36}$ & $1756^{+92}_{-96}$ & 6.37 & 0.75 & 1.00 \\ [0.2cm]
J0334-4008.9 & 86.5 & $839^{+90}_{-316}$ & $461^{+49}_{-41}$ & 6.46 & 0.83 & 0.40 \\ [0.2cm]
J0334-4008.10 & 86.4 & $907^{+54}_{-116}$ & $378^{+59}_{-57}$ & 9.93 & 0.79 & 0.66 \\ [0.2cm]
J0342-3007.0 & 104.3 & $732^{+212}_{-538}$ & $255^{+16}_{-19}$ & 9.39 & 0.80 & 0.49 \\ [0.2cm]
J0342-3007.2 & 103.8 & $627^{+208}_{-118}$ & $419^{+100}_{-171}$ & 12.68 & 0.80 & 0.40 \\ [0.2cm]
J0342-3007.5 & 99.9 & $993^{+3}_{-4}$ & $54^{+2}_{-2}$ & 12.24 & 0.87 & 0.17 \\ [0.2cm]
J0342-3007.6 & 99.0 & $349^{+651}_{-345}$ & $1009^{+1214}_{-863}$ & 5.43 & 0.76 & 0.34 \\ [0.2cm]
J0342-3007.7 & 97.9 & $911^{+62}_{-196}$ & $203^{+38}_{-32}$ & 13.00 & 0.86 & 0.42 \\ [0.2cm]
J0342-3007.9 & 95.7 & $1000^{+0}_{-148}$ & $61^{+476}_{-4}$ & 4.31 & 0.78 & 0.33 \\ [0.2cm]
J0342-3007.11 & 92.3 & $427^{+167}_{-256}$ & $549^{+67}_{-63}$ & 9.76 & 0.75 & 0.39 \\ [0.2cm]
J0342-3007.13 & 91.9 & $983^{+12}_{-67}$ & $443^{+220}_{-88}$ & 4.52 & 0.77 & 0.53 \\ [0.2cm]
J0342-3007.16 & 85.9 & $275^{+488}_{-273}$ & $328^{+78}_{-80}$ & 9.48 & 0.80 & 0.36 \\ [0.2cm]
J0518+3306.11 & 102.2 & $879^{+80}_{-87}$ & $295^{+58}_{-65}$ & 8.52 & 0.82 & 0.38 \\ [0.2cm]
J0750+1231.4 & 97.1 & $907^{+85}_{-321}$ & $221^{+74}_{-67}$ & 8.55 & 0.83 & 0.39 \\ [0.2cm]
J0750+1231.5 & 96.3 & $888^{+23}_{-21}$ & $485^{+36}_{-35}$ & 8.94 & 0.76 & 1.00 \\ [0.2cm]
J0854+2006.4 & 111.9 & $973^{+20}_{-74}$ & $81^{+35}_{-33}$ & 8.46 & 0.78 & 0.05 \\ [0.2cm]
J0948+0022.7 & 101.3 & $983^{+7}_{-9}$ & $192^{+19}_{-16}$ & 9.48 & 0.81 & 0.50 \\ [0.2cm]
J0948+0022.10 & 99.1 & $348^{+48}_{-50}$ & $917^{+103}_{-100}$ & 8.23 & 0.74 & 0.22 \\ [0.2cm]
J0948+0022.33 & 88.8 & $931^{+53}_{-300}$ & $245^{+29}_{-29}$ & 8.45 & 0.83 & 0.37 \\ [0.2cm]
J1008+0029.70 & 88.0 & $935^{+27}_{-60}$ & $367^{+25}_{-26}$ & 9.90 & 0.80 & 0.66 \\ [0.2cm]
J1130-1449.10 & 102.2 & $814^{+115}_{-369}$ & $348^{+44}_{-49}$ & 9.77 & 0.81 & 0.65 \\ [0.2cm]
\hline
\end{tabular}
\end{table*}

\begin{table*}[]
\centering
\caption{Continuation.}
\begin{tabular}{cccccccc}
\hline
\rule{0pt}{2.5ex}Name.NID & Frequency & Flux & FWHM & S/N & Completeness & Reliability \\ [0.2cm]
 & [GHz] & [mJy km s$^{-1}$] & [km/s] &  & \% & \% \\ [0.2cm]
\hline
\rule{0pt}{2.5ex}J1239-1137.0 & 106.7 & $921^{+50}_{-74}$ & $471^{+95}_{-146}$ & 9.91 & 0.76 & 0.86 \\ [0.2cm]
J1347+1217.4 & 104.3 & $763^{+173}_{-503}$ & $464^{+27}_{-26}$ & 6.92 & 0.81 & 0.72 \\ [0.2cm]
J1430+1043.5 & 101.3 & $975^{+22}_{-56}$ & $213^{+73}_{-90}$ & 7.62 & 0.79 & 0.30 \\ [0.2cm]
J0334-4008.7 & 127.2 & $8^{+555}_{-8}$ & $111^{+300}_{-64}$ & 4.38 & 0.80 & 0.48 \\ [0.2cm]
J0334-4008.15 & 144.1 & $1000^{+0}_{-4}$ & $39^{+54}_{-2}$ & 9.29 & 0.75 & 0.34 \\ [0.2cm]
J0342-3007.17 & 138.4 & $530^{+347}_{-492}$ & $272^{+35}_{-28}$ & 5.18 & 0.81 & 0.35 \\ [0.2cm]
J0948+0022.4 & 143.1 & $479^{+214}_{-388}$ & $634^{+114}_{-98}$ & 5.04 & 0.81 & 0.38 \\ [0.2cm]
J0948+0022.5 & 136.4 & $959^{+32}_{-52}$ & $192^{+37}_{-46}$ & 5.40 & 0.81 & 0.34 \\ [0.2cm]
J0948+0022.7 & 133.3 & $24^{+776}_{-24}$ & $318^{+177}_{-191}$ & 3.98 & 0.82 & 0.15 \\ [0.2cm]
J1550+0527.1 & 139.8 & $874^{+80}_{-211}$ & $238^{+37}_{-34}$ & 9.91 & 0.79 & 1.00 \\ [0.2cm]
J1037-2934.1 & 147.9 & $374^{+480}_{-364}$ & $346^{+33}_{-28}$ & 4.09 & 0.82 & 0.12 \\ [0.2cm]
J1037-2934.2 & 145.6 & $965^{+15}_{-18}$ & $159^{+13}_{-12}$ & 7.96 & 0.81 & 0.32 \\ [0.2cm]
J1122+1805.10 & 145.1 & $830^{+58}_{-83}$ & $375^{+72}_{-49}$ & 12.52 & 0.79 & 0.77 \\ [0.2cm]
J1332-0509.0 & 129.7 & $952^{+21}_{-24}$ & $214^{+34}_{-44}$ & 6.75 & 0.79 & 0.31 \\ [0.2cm]
J1332-0509.4 & 128.3 & $978^{+5}_{-5}$ & $201^{+11}_{-9}$ & 6.25 & 0.80 & 0.25 \\ [0.2cm]
J1713-3418.0 & 144.8 & $136^{+725}_{-136}$ & $222^{+17}_{-14}$ & 3.86 & 0.80 & 0.54 \\ [0.2cm]
J1751-1950.1 & 135.8 & $819^{+108}_{-342}$ & $428^{+46}_{-34}$ & 4.18 & 0.80 & 0.48 \\ [0.2cm]
J1751-1950.4 & 134.0 & $776^{+161}_{-137}$ & $372^{+122}_{-163}$ & 6.59 & 0.81 & 0.29 \\ [0.2cm]
J2357-5311.2 & 144.1 & $34^{+803}_{-34}$ & $302^{+495}_{-69}$ & 4.56 & 0.78 & 0.94 \\ [0.2cm]
J0238+1636.1 & 193.2 & $743^{+109}_{-113}$ & $277^{+80}_{-66}$ & 6.67 & 0.79 & 0.30 \\ [0.2cm]
J0334-4008.2 & 204.0 & $920^{+15}_{-15}$ & $92^{+5}_{-6}$ & 7.78 & 0.78 & 0.20 \\ [0.2cm]
J0519-4546.0 & 172.8 & $959^{+36}_{-82}$ & $138^{+51}_{-55}$ & 5.44 & 0.81 & 0.33 \\ [0.2cm]
J1002+1216.4 & 175.5 & $744^{+130}_{-162}$ & $228^{+54}_{-56}$ & 5.97 & 0.80 & 0.20 \\ [0.2cm]
J0217-0820.51 & 218.2 & $899^{+26}_{-28}$ & $266^{+23}_{-23}$ & 6.68 & 0.79 & 0.69 \\ [0.2cm]
J0237+2848.53 & 225.8 & $932^{+28}_{-35}$ & $264^{+33}_{-35}$ & 4.45 & 0.81 & 0.52 \\ [0.2cm]
J0237+2848.56 & 225.8 & $925^{+31}_{-37}$ & $268^{+35}_{-35}$ & 6.08 & 0.80 & 0.76 \\ [0.2cm]
J0237+2848.129 & 225.1 & $950^{+48}_{-350}$ & $90^{+121}_{-62}$ & 7.38 & 0.79 & 0.22 \\ [0.2cm]
J0237+2848.170 & 224.2 & $979^{+19}_{-371}$ & $97^{+108}_{-25}$ & 6.39 & 0.79 & 0.22 \\ [0.2cm]
J0239-0234.7 & 255.9 & $973^{+15}_{-42}$ & $93^{+25}_{-12}$ & 10.14 & 0.81 & 0.57 \\ [0.2cm]
J0423-0120.4 & 232.8 & $924^{+36}_{-69}$ & $62^{+16}_{-10}$ & 9.46 & 0.78 & 0.50 \\ [0.2cm]
J0519-4546.3 & 232.1 & $830^{+69}_{-100}$ & $174^{+22}_{-17}$ & 10.86 & 0.82 & 0.22 \\ [0.2cm]
J0519-4546.4 & 231.9 & $998^{+1}_{-1}$ & $170^{+5}_{-6}$ & 12.11 & 0.82 & 0.71 \\ [0.2cm]
J1010-0200.1 & 244.0 & $252^{+147}_{-167}$ & $574^{+942}_{-219}$ & 4.44 & 0.78 & 0.99 \\ [0.2cm]
J1617-5848.5 & 226.5 & $760^{+106}_{-311}$ & $257^{+34}_{-40}$ & 6.06 & 0.80 & 0.47 \\ [0.2cm]
J1751+0939.8 & 218.1 & $987^{+13}_{-51}$ & $141^{+46}_{-44}$ & 7.08 & 0.79 & 0.49 \\ [0.2cm]
\hline
\end{tabular}

\end{table*}

\begin{table*}[]
\centering
\caption{Continuation.}
\begin{tabular}{cccccccc}
\hline
 \rule{0pt}{2.5ex}Name.NID & Frequency & Flux & FWHM & S/N & Completeness & Reliability \\ [0.2cm]
 & [GHz] & [mJy km s$^{-1}$] & [km/s] &  & \% & \% \\ [0.2cm]
\hline
\rule{0pt}{2.5ex}J1751+0939.23 & 226.1 & $997^{+3}_{-17}$ & $43^{+12}_{-19}$ & 11.57 & 0.84 & 0.18 \\ [0.2cm]
J1832-1035.8 & 232.4 & $981^{+13}_{-60}$ & $92^{+29}_{-13}$ & 10.60 & 0.82 & 0.57 \\ [0.2cm]
J1832-2039.8 & 215.7 & $454^{+291}_{-418}$ & $513^{+39}_{-30}$ & 8.71 & 0.77 & 1.00 \\ [0.2cm]
J1955+1358.7 & 231.9 & $783^{+116}_{-101}$ & $107^{+26}_{-36}$ & 11.02 & 0.82 & 0.21 \\ [0.2cm]
J2056-4714.8 & 216.9 & $935^{+19}_{-23}$ & $111^{+11}_{-10}$ & 7.26 & 0.78 & 0.27 \\ [0.2cm]
J2056-4714.9 & 225.0 & $849^{+52}_{-77}$ & $297^{+58}_{-39}$ & 5.09 & 0.81 & 0.39 \\ [0.2cm]
J2148+0657.1 & 257.4 & $883^{+85}_{-242}$ & $161^{+42}_{-36}$ & 8.99 & 0.83 & 1.00 \\ [0.2cm]
J0006-0623.9 & 346.8 & $267^{+449}_{-263}$ & $181^{+117}_{-56}$ & 6.38 & 0.78 & 1.00 \\ [0.2cm]
J0006-0623.24 & 338.6 & $834^{+53}_{-45}$ & $121^{+14}_{-14}$ & 7.30 & 0.79 & 0.37 \\ [0.2cm]
J0006-0623.38 & 330.9 & $907^{+81}_{-408}$ & $115^{+154}_{-40}$ & 5.49 & 0.81 & 0.30 \\ [0.2cm]
J0121+1149.0 & 348.3 & $925^{+42}_{-48}$ & $153^{+23}_{-24}$ & 3.12 & 0.75 & 0.15 \\ [0.2cm]
J0126-0120.3 & 291.8 & $837^{+126}_{-463}$ & $83^{+7}_{-7}$ & 7.65 & 0.78 & 0.30 \\ [0.2cm]
J1924-2914.1 & 493.4 & $635^{+231}_{-497}$ & $95^{+11}_{-10}$ & 6.75 & 0.77 & 0.44 \\ [0.2cm]
J1424-6807.0 & 478.1 & $715^{+97}_{-160}$ & $42^{+16}_{-9}$ & 4.65 & 0.73 & 0.70 \\ [0.2cm]
J1517-2422.0 & 400.7 & $910^{+47}_{-133}$ & $139^{+55}_{-23}$ & 4.60 & 0.78 & 0.29 \\ [0.2cm]
J2207-5346.3 & 397.6 & $809^{+93}_{-285}$ & $112^{+10}_{-8}$ & 6.89 & 0.78 & 0.46 \\ [0.2cm]
J1028-0236.0 & 633.7 & $939^{+20}_{-29}$ & $42^{+4}_{-3}$ & 5.15 & 0.74 & 0.41 \\ [0.2cm]
\hline
\end{tabular}

\end{table*}

\end{appendix}

\end{document}